%% file: WARP_QST_revised.tex
\pdfoutput=1
\documentclass[twocolumn,10pt]{article}

\usepackage[margin=1.8cm]{geometry} \usepackage{amsmath,amssymb} \usepackage{braket}
\usepackage{graphicx} \usepackage{tikz} \usepackage{quantikz} \usepackage{booktabs}
\usepackage{natbib} \usepackage{xcolor} \usepackage{authblk}
\usepackage[colorlinks=true,linkcolor=blue,citecolor=blue,urlcolor=blue]{hyperref}
\usepackage{pdflscape}      
\usepackage{longtable}      
\usepackage{array}          
\usepackage{caption}  \usepackage{enumitem}

\usetikzlibrary{arrows.meta,positioning,shapes.geometric}

\makeatletter
\newcounter{warppart}
\renewcommand{\thewarppart}{\Roman{warppart}}
\newcommand{\warppart}[2]{%
  \par\addvspace{1.1\baselineskip}%
  \refstepcounter{warppart}%
  \noindent\begingroup
    \centering
    {\footnotesize\bfseries\scshape Part \thewarppart}\\[1pt]%
    {\large\bfseries #1}\\[3pt]%
    \parbox{\columnwidth}{\small\itshape\centering #2}\par
  \endgroup
  \par\nobreak\addvspace{0.35\baselineskip}%
  \hrule height 0.4pt
  \addvspace{0.9\baselineskip}%
  \addtocontents{toc}{\protect\warptocpart{\the\value{warppart}}{#1}}%
  \@afterheading}

\newcommand{\warptocpart}[2]{%
  \par\addvspace{2pt}%
  \noindent{\bfseries\scshape Part \@Roman{#1}\ \ \textup{\bfseries #2}}\par
  \addvspace{0.5pt}}
\makeatother

\makeatletter
\renewcommand{\l@section}[2]{%
  \ifnum \c@tocdepth >\z@
    \addpenalty\@secpenalty
    \addvspace{0.3pt \@plus\p@}%
    \setlength\@tempdima{1.6em}%
    \begingroup
      \parindent\z@ \rightskip\@pnumwidth
      \parfillskip-\@pnumwidth
      \leavevmode\bfseries
      \advance\leftskip\@tempdima \hskip-\leftskip
      #1\nobreak\hfil\nobreak\hb@xt@\@pnumwidth{\hss #2}\par
    \endgroup
  \fi}
\renewcommand{\l@subsection}{\@dottedtocline{2}{1.6em}{2.4em}}
\makeatother

\renewcommand{\d}{\mathrm{d}} \newcommand{\ee}{\mathrm{e}}
\newcommand{\Ha}{\,\mathrm{Ha}} \newcommand{\Eo}{E_0}

\title{\bfseries Warm-Start Quantum Computation:\\Spectral-Weight Resources for Energy Readout\\and State Preparation}

\author{Tien D. Kieu} \affil{Centre for Quantum Technology Theory and Optical Sciences Centre \\ Swinburne University of Technology, Australia}
\date{\today}

\begin{document}
\twocolumn[
\begin{@twocolumnfalse}
\maketitle
\begin{abstract}
Estimating a ground-state energy and preparing a ground state both split into a
\emph{front-end} supplying an approximate state $|\chi\rangle$ and a \emph{back-end}
reading or refining it. We introduce a warm-start architecture, WARP, whose
transferable resource is the spectral weight distribution
$p_n=|\langle E_n|\chi\rangle|^2$ over the eigenstates of the problem Hamiltonian $H_P$.
The ground-state weight $p_0$ controls access to the target; the remaining weight
sets the spectral complexity a readout must resolve. The front-end can therefore be
exchanged without altering the back-end, preparation costs stated separately.

Four back-ends spend that resource: a spectral readout of
$g(t)=\langle\chi|e^{-iH_Pt}|\chi\rangle$ by ESPRIT, adiabatic preparation anchored
on a projector driver, amplitude amplification, and a double-bracket imaginary-time
recursion. The readout reaches chemical accuracy on H$_2$ from $N=40$ samples
against $N=2048$ for a Fourier baseline, cutting record length and Trotter depth
$50\times$. Seven substantially different front-ends are substituted across one
interface.

We recover overlap weights four orders below the shot-noise floor, and with them
symmetry-forbidden transition amplitudes and their phases. We obtain a
closed-form admission test for filtering, a measured exchange rate between front-end
effort and adiabatic time, a halting index no bound supplies, and a convergence
estimate for a recursion whose published guarantee errs by six orders of magnitude.
On stretched N$_2$ a sample-based diagonalisation is $79\,$mHa high (50$x$ above the chemical accuracy); reading the state it has
already prepared returns $E_0$ to $1.4\times10^{-6}\,$Ha (three orders of magnitude below the chemical accuracy). A differential contrast between the
two Hadamard-test quadratures injects mirror lines no fixed model order can hold, and
interleaving the settings removes it. WARP does
not manufacture overlap, and the demonstrations carry Trotter and shot noise but no
full hardware error model.

\end{abstract}
\end{@twocolumnfalse}
]

{\footnotesize\linespread{0.96}\selectfont
\setlength{\parskip}{0pt}
\noindent{\normalsize\bfseries Contents}\par\vspace{1pt}
\makeatletter\@starttoc{toc}\makeatother
}
\vspace{0.6em}

\section{Introduction}
\label{sec:intro}

Estimating eigenenergies and preparing useful eigenstates are central tasks in quantum
simulation and quantum chemistry~\citep{aspuru2005simulated,cao2019chemistry,mcardle2020quantum}.
The dimension of the underlying Hilbert space grows as $2^n$ in the number of spin-orbitals
$n$, so exact diagonalisation and exact real-time propagation become classically intractable
at a few tens of qubits, while a quantum computer can encode the state in $n$ qubits and
implement the propagator $\ee^{-iHt}$ with a circuit whose size is polynomial in $n$
for standard Hamiltonian simulation~\citep{lloyd1996universal}. A practical quantum workflow,
however, rarely begins from an arbitrary state. It begins from a state that some front-end has
already prepared and then asks a back-end to extract information from, or further refine, that
state.

The central physical observation of this work is that such a warm start carries more structured
information than a single variational energy or ground-state overlap. Write $H_P$ for the
problem Hamiltonian, whatever its origin, with eigenpairs
\begin{equation}
H_P\ket{E_n}=E_n\ket{E_n},\qquad E_0\le E_1\le E_2\le\cdots,
\label{eq:eigen}
\end{equation}
so that $\ket{E_0}$ is the ground state and $E_0$ the ground-state energy, and the
eigenvectors are taken orthonormal. The molecular Hamiltonian of \S\ref{sec:ham} is one
instance of $H_P$, the spin and combinatorial Hamiltonians of later sections are others, and
nothing below depends on which. If
$\ket{\chi}=\sum_n c_n\ket{E_n}$, then its spectral weights
$ p_n=|c_n|^2 $ determine both which eigenenergies are visible to coherent spectral readout
and how much target-state weight is available to preparation procedures. We therefore treat the
spectral-weight distribution as a measurable resource shared by otherwise different quantum
back-ends. That is the organising principle of the architecture set out here, rather than an
assumption that any one algorithm is universally preferable.

Two families of algorithms dominate the literature, and they occupy opposite ends of the
hardware-cost spectrum. The variational quantum eigensolver
(VQE)~\citep{peruzzo2014vqe,mcclean2016theory} prepares a parametrised trial state
$\ket{\psi(\boldsymbol\theta)}$, measures the energy expectation value $\braket{\psi|H|\psi}$
as a sum of Pauli terms, and minimises it on a classical optimiser. VQE uses shallow
circuits and tolerates certain errors~\citep{omalley2016scalable}, making it the
workhorse of the noisy intermediate-scale quantum (NISQ) era~\citep{preskill2018nisq}.
Its defining limitation is structural. VQE returns
an upper bound, the Rayleigh quotient which is a \emph{weighted
average} of eigenvalues,  whose tightness is capped by the ansatz. Its optimisation is beset by
local minima and barren plateaus~\citep{mcclean2018barren,cerezo2021vqa,tilly2022vqe}.
Quantum phase estimation
(QPE)~\citep{kitaev1995qpe,abrams1999quantum,nielsen2010quantum} sits at the other
extreme. It returns the \emph{true} eigenvalue at the Heisenberg limit, but needs
coherent evolution to long times and many ancillae and a quantum
Fourier transform, well beyond near-term hardware.

Between those regimes sits a family of classical \emph{time-series} or \emph{statistical}
phase estimation methods~\citep{obrien2019qpe,somma2019time,lin2022heisenberg}, which move the
hard estimation off the device and so need one ancilla and can trade circuit depth for
measurement repetitions. That is the resource profile of \emph{early fault-tolerant}
hardware~\citep{lin2022heisenberg,ding2023even,wan2022randomized,zhang2022computing,dong2022ground},
and it is one of the routes this paper uses.

What is missing from all of them is an account of what a warm start actually is. Every method
above takes a prepared state as given and prices its own performance against a single number,
usually the overlap with the target. That number is not what the state supplies and it is not
what the methods spend. Identifying the quantity that is, measuring it where it is too small to
see by direct means, and then spending it on results that are not otherwise available, is the
work of this paper. The architecture it builds, and the way its pieces feed one another, is
collected in Fig.~\ref{fig:arch}.

We name the arrangement \textbf{WARP}, for \emph{WArm start, Readout and
state Preparation}. The readout back-end returns spectral information,
while the preparation back-ends return an improved state. Where the
adiabatic construction needs a separate name we use WA-ASP, for
warm-start-anchored adiabatic state preparation. The notation $U_\chi$
denotes a circuit that prepares the supplied warm-start state and does not
encode which front-end produced it. Front-end interchangeability is thus an
interface statement, not an assertion of equal preparation or compilation
cost.

\subsection{Contributions}
\label{sec:contributions}

The organising idea of this paper is that a warm start is a resource whose form is more
informative than a single scalar measure of state quality, and that once its form is identified
it can be measured and spent on questions that are otherwise out of reach. With
\begin{equation}
    \ket{\chi}=\sum_n c_n\ket{E_n},\qquad
    p_n=|c_n|^2=|\braket{E_n|\chi}|^2,
\end{equation}
the warm-start state enters the spectral readout through the weights $p_n$.
The estimator does not need to know how $\ket\chi$ was produced. The
ground-state weight $p_0=a^2$ controls the leading overlap dependence of the
state-preparation procedures, whereas the distribution of the remaining
weight determines the spectral complexity encountered by energy readout.
Thus two warm starts with the same $p_0$ can have different downstream costs.
This motivates a front-end/back-end separation in which the front-end supplies
the spectral distribution and the back-end spends it. The useful resource
description is therefore the distribution $\{p_n\}$ itself, or, for the cost
laws studied here, a pair such as $(p_0,\mathcal C_{\rm spec})$;
$\mathcal C_{\rm spec}$ is a task-dependent measure rather than a universal
scalar quality of the state.

The specific contributions are as follows.

\begin{enumerate}[leftmargin=2.0em,itemsep=0.45em]

\item \textbf{Identification of the transferable resource.}
We formulate WARP as a modular architecture in which the front-end supplies
a warm state and the back-end consumes its spectral weight. The resource is
therefore associated with the resulting distribution $\{p_n\}$ rather than
with the algorithm that produced $\ket\chi$. For ground-state preparation,
$p_0$ is the directly relevant overlap resource; for spectral readout, the
structure of the full distribution controls the number and relative strength
of spectral components that must be resolved. The same interface can
therefore connect different state-preparation mechanisms to different
quantum procedures. That separation is not only a convenience of exposition. Seitz
\emph{et al.}~\citep{seitz2026workflows} run a front-end and a back-end on two different
machines of two different qubit modalities, five hundred kilometres apart, passing between them
only the classical description of $\ket\chi$, which is the strongest form the claim can
take.

\item \textbf{Warm-start spectral energy readout.}
The autocorrelation of the supplied state,
\begin{equation}
    g(t)=\langle\chi|e^{-iH_Pt}|\chi\rangle
    =\sum_n p_n e^{-iE_nt},
    \label{eq:gt}
\end{equation}
is a sum of pure complex exponentials whose frequencies are exactly the eigenvalues of
Eq.~\eqref{eq:eigen} and whose amplitudes are exactly the resource. We sample it with a single-ancilla Hadamard test and extract its frequencies by classical
parametric estimation. In particular, we use ESPRIT as a super-resolution
back-end and show that, for the tested instances, the required energy
accuracy can be obtained from substantially shorter coherent records than
with a conventional discrete Fourier transform. ESPRIT itself is not
introduced as a new estimator here. The contribution is its integration with
the warm-start resource interface, together with explicit treatment of model
order, spectral complexity, quadrature imbalance, and noise, and its reuse
by the preparation back-ends.

\item \textbf{A common interface to several preparation back-ends.}
The same warm-start resource is used for WA-ASP, discrete amplitude
amplification, and recursive double-bracket imaginary-time refinement. These procedures consume the
supplied spectral weight in different ways. In particular, $p_0$ controls
the leading overlap dependence of procedures whose purpose is to increase
ground-state weight, while the detailed spectral distribution enters the
readout cost. WARP therefore separates what the front-end supplies from how
a particular back-end spends it.

\item \textbf{Eigenvalue information beyond a variational expectation value.}
We demonstrate that spectral readout can recover an eigenvalue from a state
whose variational energy remains an upper bound above the ground-state
energy. In the stretched-N$_2$ sample-based diagonalisation calculation,
the variational estimate is approximately $79$~mHa above the ground-state
energy, which is 50$x$ above the chemical accuracy. Whereas spectral readout of the already prepared state recovers
$E_0$ to approximately $1.4\times10^{-6}$~Ha, three orders of magnitude below the chemical accuracy. These quantities answer
different questions: the variational energy is the expectation value of the
supplied state, whereas spectral readout identifies eigenvalues carrying
spectral weight in that state. The readout therefore complements the state-preparation method.

\item \textbf{Front-end interchangeability.}
We test the interface with warm starts obtained from variational preparation,
QAOA, classical heuristics, lattice-reduction methods, generative models,
sample-based diagonalisation, and machine-learned states. These examples
are not intended to rank the front-ends. Rather, they test whether the same
back-end constructions can operate on states produced by substantially
different mechanisms. The results support spectral weight as an algorithmic
interface while making explicit that front-end preparation, compilation, and
overlap-generation costs remain problem- and platform-dependent.

\item \textbf{Metering the resource, and the operating conditions.}
Because the cost laws are denominated in the weights, the weights must be measurable when they
are scarce, and we show that they are. The same interference acquisition that reads an
eigenvalue returns the weight carried by lines far below the noise floor of a direct
measurement, which is what makes a halting index and an admission probe available at all. We
also make the principal conditions governing the architecture explicit. For
spectral readout these include the supplied overlap, spectral complexity,
model-order selection, coherent acquisition time, and shot budget. For
WA-ASP, the implementation cost of the projector onto $\ket\chi$, and hence
of a circuit $U_\chi$ that prepares the warm start, must be included in the
resource accounting; an improved adiabatic path does not make preparation
of an arbitrary warm start free. Similarly, amplification and recursive
refinement remain overlap-dependent. The numerical conditions are stated in \S\ref{sec:scope}.

\end{enumerate}

Taken together, these contributions identify the spectral-weight distribution of a warm start as
a common, physically measurable resource for quantum readout and state preparation. WARP does not
replace the individual algorithms used within it, nor does it manufacture ground-state overlap.
Its contribution is to expose what the prepared state already supplies, to measure the parts of
that resource relevant to downstream tasks, and to show how the same state can be passed between
different quantum back-ends with preparation, coherent-evolution, measurement and
classical-processing costs stated separately. What follows from the identification is the
substance. A resource that can be measured below its own noise floor yields admission tests,
halting indices, exchange rates and impossibility results, none of which is available to a
description that prices a warm start by one number.

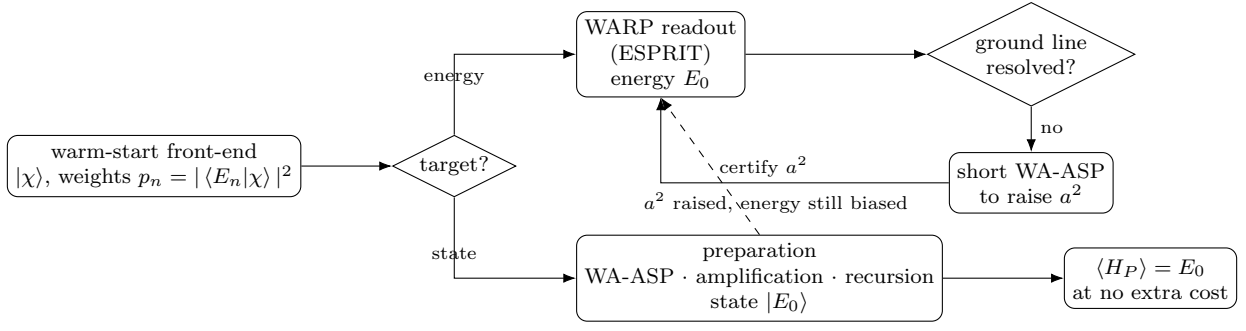
\begin{figure*}[t]\centering
\begin{tikzpicture}[node distance=7mm and 12mm,>=Latex,
 box/.style={draw,rounded corners,align=center,font=\footnotesize,minimum height=8mm,inner sep=3pt},
 dec/.style={draw,diamond,aspect=2,align=center,font=\footnotesize,inner sep=1pt}]
\node[box](v){warm-start front-end\\$\ket{\chi}$, weights $p_n=|\braket{E_n|\chi}|^2$};
\node[dec,right=of v](d){target?};
\node[box,above right=of d](a){WARP readout\\(ESPRIT)\\energy $E_0$};
\node[box,below right=of d](b){preparation\\WA-ASP $\cdot$ amplification $\cdot$ recursion\\state $\ket{E_0}$};
\node[box,right=16mm of b](hb){$\langle H_P\rangle=E_0$\\at no extra cost};
\node[dec,right=24mm of a](d2){ground line\\resolved?};
\node[box,below=6mm of d2](amp){short WA-ASP\\to raise $a^2$};
\draw[->](v)--(d);
\draw[->](d)|-(a) node[pos=0.25,above,font=\scriptsize]{energy};
\draw[->](d)|-(b) node[pos=0.25,below,font=\scriptsize]{state};
\draw[->](b)--(hb);
\draw[->,dashed](b.north)--(a.south) node[midway,right,font=\scriptsize]{certify $a^2$};
\draw[->](a)--(d2);
\draw[->](d2)--(amp) node[midway,right,font=\scriptsize]{no};
\draw[->](amp)-|(a) node[pos=0.3,below,font=\scriptsize]{$a^2$ raised, energy still biased};
\end{tikzpicture}
\caption{\textbf{Which back-end, and when.} Both branches are paid for out of the same
account, the front-end's weight distribution $p_n=|\braket{E_n|\chi}|^2$. The ground-state
weight $p_0=a^2$ sets the leading term in every cost law on the diagram. The spread of the weight left over, measured by $r_{\rm eff}=(\sum_n
p_n^2)^{-1}$, prices the acquisition alongside it, so the rates below are quoted against
the pair $(p_0,r_{\rm eff})$ and not against $p_0$ alone (\S\ref{sec:resource}). The readout returns an energy and leaves no state behind. The preparation back-end returns a state, from
which the energy follows. Three routes run down that branch, the adiabatic schedule of
\S\ref{sec:aqc}, the discrete amplification of \S\ref{sec:grover} and the recursion of
\S\ref{sec:dbqite}, and \S\ref{sec:dbqite} also gives the rule that chooses among them. When the ground line is too weak to resolve, a
partial schedule spends time to raise $a^2$ and hands back to the readout---the exchange
rate quantified in \S\ref{sec:qaoa} and \S\ref{sec:lattice}. That return hands back an unconverged state, whose energy is still biased, which is why the readout. The dashed return is the certificate of
\S\ref{sec:stopping}, where the readout run on the prepared state gives the ground-line weight
$a^2$, so each back-end diagnoses the other.}
\label{fig:selection}
\end{figure*}

\subsection{Arrangement of the paper.} The front-end/back-end split organises the paper
as well as the argument. Part~\ref{part:readout} treats the readout back-end alone,
Part~\ref{part:prep} the preparation back-end alone, each with its own illustrations and
its own failure modes, and neither needing the other to be read first.
Part~\ref{part:together} is where they meet. Part~\ref{part:frontends} then swaps the front-end seven times over, which is the experiment the swappability claim deserves.
Part~\ref{part:summary} closes. Table~\ref{tab:cases} lists every case studied, the
component it exercises, the conventional method it is measured against and the result,
including the final block, which records the architecture's limitations.

\section{Setup, and the resource the front-end supplies}
\label{sec:setup}

\subsection{The electronic Hamiltonian and its encoding}
\label{sec:ham}

Within the Born--Oppenheimer approximation the electronic Hamiltonian in second
quantisation is
\begin{equation}
H=\sum_{pq} h_{pq}\,a_p^\dagger a_q
 +\tfrac12\sum_{pqrs} h_{pqrs}\,a_p^\dagger a_q^\dagger a_r a_s,
\label{eq:Hsecond}
\end{equation}
with one- and two-electron integrals $h_{pq},h_{pqrs}$ obtained from a classical
quantum-chemistry package. A fermion-to-qubit mapping (Jordan--Wigner or
Bravyi--Kitaev)~\citep{jordan1928wigner,bravyi2002fermionic,seeley2012bravyi}
turns Eq.~\eqref{eq:Hsecond} into a qubit operator. For H$_2$ in a minimal
STO-3G basis at the equilibrium bond length $R=0.735\,$\AA, symmetry reduction
leaves a two-qubit Hamiltonian~\citep{omalley2016scalable}
\begin{align}
H =\;& g_0 I + g_1 Z_0 + g_2 Z_1 + g_3 Z_0 Z_1 \nonumber\\
 & {} + g_4 X_0 X_1 + g_5 Y_0 Y_1,
\label{eq:Hpauli}
\end{align}
with the representative coefficients (in Hartree) $g_0=-1.052373$,
$g_1=0.397937$, $g_2=-0.397937$, $g_3=-0.011280$, $g_4=g_5=0.180931$. Exact
diagonalisation of Eq.~\eqref{eq:Hpauli} gives eigenvalues
$\{-1.915370,\,-1.063653,\,-1.063653,\,-0.166816\}\Ha$. The ground-state energy
that all methods below target is $\Eo=-1.915370\Ha$. The doubly degenerate level
at $-1.063653\Ha$ corresponds to the uncoupled basis states $\ket{00},\ket{11}$
(uncoupled because $g_4=g_5$). This degeneracy plays a role in
\S\ref{sec:modelorder}.

Throughout we quote errors relative to \emph{chemical accuracy},
$1\,\mathrm{kcal/mol}=1.594\times10^{-3}\Ha$, the threshold below which computed
energetics become predictive for reaction rates.

\subsection{The simulated experiment}
\label{sec:experiment}

All spectral methods consume the same signal $g(t)$ of Eq.~(\ref{eq:gt}), which on hardware is obtained as follows. We simulate every step at the statevector level, including shot noise, rather
than evaluating $g(t)$ by matrix exponentiation.

\paragraph{Time evolution by Trotterisation.} The propagator is approximated by a
product formula~\citep{lloyd1996universal,childs2021trotter}. Grouping
Eq.~\eqref{eq:Hpauli} into its commuting families $A=\{Z_0,Z_1,Z_0Z_1\}$ and
$B=\{X_0X_1,Y_0Y_1\}$, one Trotter step of size $\delta$ is the ordered product of
single-Pauli rotations $\ee^{-i c_j P_j \delta}=\cos(c_j\delta)I-i\sin(c_j\delta)P_j$,
with the identity term contributing a global phase $\ee^{-i g_0\delta}$ that becomes
observable once placed under control. We sample $g$ at $t_k=k\delta$, one Trotter step
per sample,  so the circuit depth at sample $k$ is exactly $k$ steps, making ``signal
length'' and ``maximum circuit depth'' synonymous, the central trade-off of this work.

\begin{figure}[t]
\centering
\begin{quantikz}[column sep=0.32cm,row sep=0.36cm]
\lstick{$\ket{0}_a$} & \gate{H} & \gate{S^\dagger} & \ctrl{1} & \gate{H} & \meter{} \\
\lstick[2]{$\ket{i}$} & \qw & \qw & \gate[2]{\ee^{-iHt}} & \qw & \qw \\
& \qw & \qw & & \qw & \qw
\end{quantikz}
\caption{Hadamard test for one sample of $g(t)=\braket{\chi|\ee^{-iHt}|\chi}$. Omitting
the $S^\dagger$ gate measures $\mathrm{Re}\,g(t)$; including it measures
$\mathrm{Im}\,g(t)$. The controlled propagator is compiled by Trotterisation. Its depth grows linearly with $t$.}
\label{fig:hadamard}
\end{figure}
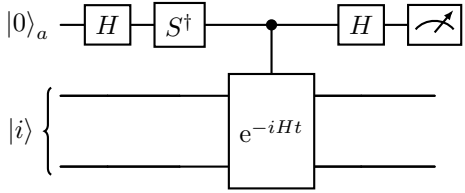

\paragraph{Hadamard-test readout.} The real and imaginary parts of $g(t)$ are measured
with one ancilla (Fig.~\ref{fig:hadamard}). Preparing the ancilla in $\ket{+}$, applying
a controlled $\ee^{-iHt}$, and a final Hadamard gives
$\mathrm{Re}\,g(t)$. Inserting an $S^\dagger$ gate on the ancilla before
the controlled evolution gives $\mathrm{Im}\,g(t)$; here
$S=\mathrm{diag}(1,i)$ is the phase gate, so $S^\dagger=\mathrm{diag}(1,-i)$ rotates the
ancilla from $\ket{+}$ to $(\ket 0-i\ket 1)/\sqrt2$ and thereby exchanges the two
quadratures of $g$. Each expectation is estimated from a finite number of shots, giving
the binomial noise present in any real experiment. Its sampling statistics, the
signal-to-noise ratio, and the resulting Cram\'er--Rao bound are given in
Appendix~\ref{app:shot}.

\subsection[Overlap and spectral complexity]{Overlap and spectral complexity: the resource actually spent}
\label{sec:resource}

The ground-state overlap is the leading resource, but it is not the whole resource, as
the model-order experiment of \S\ref{sec:modelorder} will show. Write the warm start in the
eigenbasis of $H$ as
\begin{equation}
|\chi\rangle=\sum_n c_n|E_n\rangle,\qquad
p_n=|c_n|^2,\qquad \sum_n p_n=1 .
\label{eq:spectralweights}
\end{equation}
The signal is then
\begin{equation}
g(t)=\sum_n p_n\,\ee^{-iE_nt}.
\end{equation}
The readout therefore sees the complete spectral measure $\{E_n,p_n\}$, not only
$p_0$. Two warm starts with the same ground-state overlap can consequently have
different readout costs if one spreads its remaining weight over many resolvable
lines. This is the distinction between the transferable resource and the cost of
spending it.

A convenient summary of the latter is the effective spectral rank
\begin{equation}
r_{\rm eff}=\frac{1}{\sum_n p_n^2},
\label{eq:reff}
\end{equation}
or, equivalently, the spectral entropy
\begin{equation}
S_{\rm spec}=-\sum_n p_n\log p_n .
\label{eq:specentropy}
\end{equation}
Neither quantity replaces the estimator's model order---the number of exponentials it is
told to fit, introduced in \S\ref{sec:esprit} and written $p$ throughout---or the level
spacings, and neither is intended as a universal complexity measure. They are useful diagnostics. A large
$r_{\rm eff}$ or $S_{\rm spec}$ signals that a fixed amount of ground-state overlap is
shared among a more complicated spectrum. The dense TFIM spectra in
\S\ref{sec:scaling} illustrate precisely this transition, where the many-body frequency
forest progressively removes the short-signal advantage of ESPRIT.

This suggests a two-component resource description,
\begin{equation}
\boxed{\qquad
\text{warm-start resource}
=
\bigl(p_0,\ {\cal C}_{\rm spec}\bigr),
\qquad}
\label{eq:resourcepair}
\end{equation}
where ${\cal C}_{\rm spec}$ denotes spectral complexity through model order, line
separation, and weight distribution. For spectral readout, increasing $p_0$ makes the
target line easier to detect, while increasing ${\cal C}_{\rm spec}$ makes the
subspace identification on which the estimator of \S\ref{sec:esprit} rests, and the
selection of its order, harder. For WA-ASP, the leading dependence
is instead the ground-state overlap through the minimum gap. The same warm start can
therefore be cheap for one back-end and expensive for the other. The first coordinate is
recognised independently on hardware. Reporting a warm-started phase estimation,
Seitz \emph{et al.}~\citep{seitz2026workflows} state its accuracy to be set by the sampling
parameters and by the population of the ground state in the input, which is $p_0$ under another
name.

\begin{figure}[t]
\centering
\includegraphics[width=\linewidth]{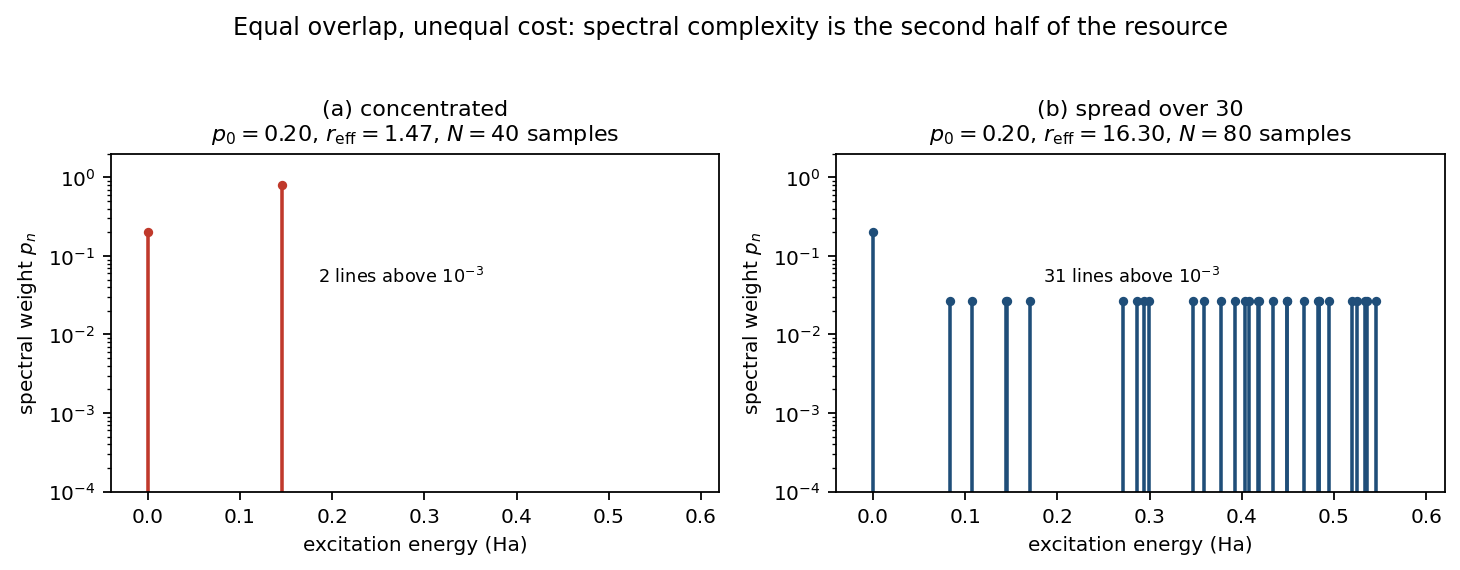}
\caption{\textbf{Overlap is not the whole resource.} Two warm starts with the same
$p_0=0.20$ on SO$_2$ CAS(6e,6o). The concentrated start needs half the acquisition length
of the spread one for the same target accuracy, and $r_{\rm eff}$, not $p_0$, is what
distinguishes them.}
\label{fig:reffcost}
\end{figure}

Take two warm starts on the CAS(6e,6o) space of
\S\ref{sec:so2} with \emph{identical} ground-state overlap $p_0=0.20$, one placing its
remaining weight on a single excited state and one spreading it evenly over thirty
(Fig.~\ref{fig:reffcost}). Their effective spectral ranks are $r_{\rm eff}=1.47$ and
$r_{\rm eff}=16.30$, and the acquisitions they require to reach $10^{-4}\Ha$ differ by a
factor of two in length, $N=40$ samples against $N=80$. The transferable resource is the same in
both cases; the cost of spending it is not. This is the distinction
Eq.~\eqref{eq:resourcepair} draws, made quantitative on a molecular Hamiltonian rather
than argued from the dense TFIM spectra alone.

That construction holds $p_0$ fixed deliberately, because fixing it is what isolates the second
coordinate. Families of warm starts supplied by convention do not span the plane so freely. Along
a ladder of Hartree--Fock and CASSCF references at a sequence of bond lengths, a better reference
is both heavier on the ground line and more concentrated on the lines it does populate, and the
two improvements arrive together, so $r_{\rm eff}$ then carries little that $p_0$ does not. The
second coordinate earns its keep where the start is engineered rather than inherited --- a
filtered state (\S\ref{sec:filter}), a pool of circuits of identical variational energy
(\S\ref{sec:gen}), a mixture (\S\ref{sec:mixedstart}) --- which is where the pair above comes
from.

This viewpoint also changes how a front end should be optimised. Minimising
$\langle H\rangle$ is not in general equivalent to maximising $p_0$, and maximising
$p_0$ alone need not minimise spectral-readout cost. A useful front-end objective is
therefore not a single variational number but a task-dependent functional of
$\{p_n\}$.

\subsection{Notation}
\label{sec:notation}

We write $\tau=N\delta$ throughout for the length of an acquired signal, equivalently
the maximum coherent evolution time the readout must sustain, and reserve $T$ for the length of
an adiabatic schedule in \S\ref{sec:aqc} onward; the two are different resources spent by
different back-ends and are never interchanged. We write $p$ for the model order of the spectral
estimator and $p_n$ with a subscript for a spectral weight, the order of a product formula
being written $q$, and $N_s$ for shots per sample. We write $g$ for the correlation function
alone, the adiabatic interpolation profiles being $f_I,f_P$ and a catalyst $f_C$, with $f$
left for a fidelity. We write $k$ for a sample index, the constant of the closed-form gap law
being $\eta$, and $L$ and $M$ for the two dimensions of a Hankel matrix. We write $D$ for the
depth of a variational circuit, $\Omega$ for a
spectral bandwidth, and the subscripts $W$ and $S$ for a weak and a strong
reference state in \S\ref{sec:homodyne}. The upright $S=\mathrm{diag}(1,i)$ of the
Hadamard-test circuits is the phase gate. We write $K_{\rm sp}$ for the sparsity of a state expressed over computational-basis
strings. And matrices carrying a problem's
data, a lattice basis $\mathbf B$ and its Gram matrix $\mathbf G$ in \S\ref{sec:lattice}, are
set in bold.

The mean excitation energy of the population a warm start leaks onto excited states is
$\bar\mu$, carrying a bar throughout, so that a bare $\mu$ is always a dipole or other
transition operator. The two iterative preparation routes carry different indices. The
discrete amplification of \S\ref{sec:grover} is indexed by $\ell$, with $\ell^\star$ for its
halting index; the recursion of \S\ref{sec:dbqite} is indexed by $m$, with $m^\dagger$ and
$\hat m$ for its guaranteed and predicted counts, and its step size is $\xi_m$ with
$\hat\xi_m$ for the value the measured moments anchor it on --- $s$ being reserved to the
adiabatic interpolation parameter and $s^\star$ to the bottleneck location along it. The scaling applied to every Gaussian
exponent of a basis in \S\ref{sec:basis} is $\zeta$, and the acquisition priced there is
$r_{\rm eff}/p_0^{2}$, which we call the \emph{priced cost}. It is a stated pricing, combining
the inverse fourth power in the overlap amplitude with the model order $r_{\rm eff}$ sets, and
not a measured runtime. We write $\sigma^2$ for the energy variance of a state throughout, $\sigma^2_\phi$ for that of a
classical reference in \S\ref{sec:notpriced}, and $\sigma_g^2$ for the per-sample shot-noise
variance of an acquired $g$ in App.~\ref{app:shot}. The relative
error of an estimated overlap is $\epsilon_a$. And the
transition moment recovered in \S\ref{sec:homodyne} is $\mathcal A$, since $A_n$ is the
residue of the damped spectral model. Landau order is written $\mathcal O$, leaving $O$ for
the operator whose transition amplitude is measured.

\warppart{The readout back-end}{An acquisition on the warm start, and an
estimator that returns a true eigenvalue from it.}
\label{part:readout}

\section{Spectral readout of a warm start}
\label{sec:readout}

The readout back-end does one thing. It samples $g(t)$ on the device and hands the samples
to a classical estimator, which returns eigenvalues of $H_P$ together with the weights the
warm start places on them. Nothing about the state is assumed beyond Eq.~\eqref{eq:gt},
and nothing about how it was made is used. This section builds that back-end on the
two-qubit H$_2$ Hamiltonian and the simulated experiment of \S\ref{sec:experiment}, where
every quantity can be checked against exact diagonalisation, and establishes its cost, its
one hyperparameter and its failure modes.

\subsection{The Fourier baseline}
\label{sec:fft}

\begin{figure}[t]
\centering
\includegraphics[width=0.49\linewidth]{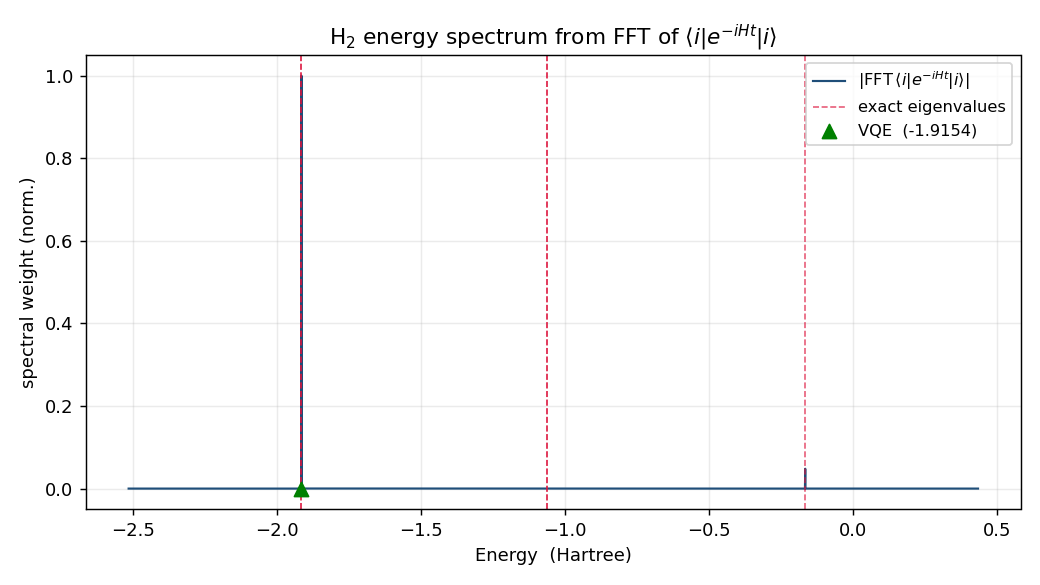}\hfill
\includegraphics[width=0.49\linewidth]{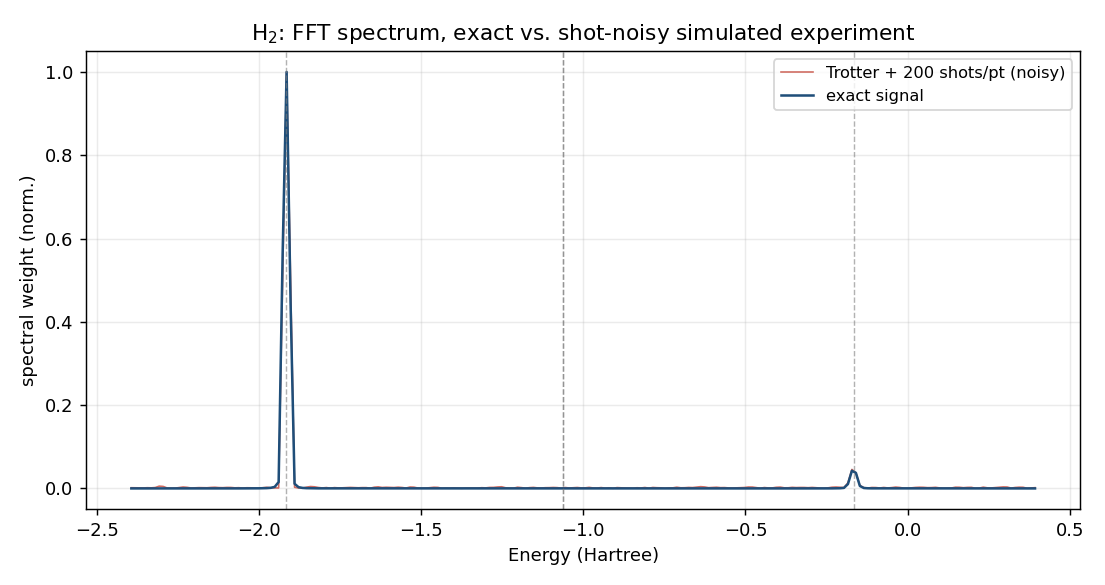}
\caption{\textbf{Fourier baseline.} Left: spectrum from the exact signal with
the HF reference. The FFT peak sits on $\Eo$, the VQE estimate (triangle)
coincides, and the degenerate sector carries zero weight. Right: spectrum from
the full simulated experiment (Trotter + Hadamard test + $200$ shots/point,
red) overlaid on the exact-signal spectrum (blue). Shot noise raises the floor
but barely shifts the peak.}
\label{fig:fft}
\end{figure}

Given uniformly spaced samples $g(t_k)$, the discrete Fourier transform
$\hat g(\omega)=\sum_k g(t_k)\ee^{-i\omega t_k}$ exposes the spectrum of
Eq.~\eqref{eq:gt}. With that kernel the line at energy $E_n$ appears at
$\omega=-E_n$, so energies are read as $E=-\omega$ throughout, with heights
$\propto|c_n|^2$. The same convention gives the estimator of \S\ref{sec:esprit}
its modes $z_n=\ee^{-iE_n\delta}$ and its energies $E_n=-\arg(z_n)/\delta$. The
ground-state energy is the lowest peak with nonzero weight. Figure~\ref{fig:fft} (left)
shows the spectrum obtained from the Hartree--Fock reference $\ket{\chi}=\ket{10}$, with a
dominant peak at $\Eo$, a small peak at the excited partner $-0.167\Ha$, and \emph{zero}
weight at the degenerate $-1.064\Ha$ level. This lies in a symmetry sector orthogonal to
the reference.

The DFT is simple but resource-hungry. Its frequency resolution is
$\Delta\omega=2\pi/\tau$ with total evolution time $\tau=N\delta$, so sharp peaks
demand long signals and accordingly deep circuits.

Decomposing the error of the
recovered $\Eo$ from the full simulated experiment
(Fig.~\ref{fig:fft}, right) we find, with $N=2048$, $\delta=0.25$:
\begin{center}
\small
\begin{tabular}{lr}
\toprule
error source (cumulative) & $\Eo$ error (Ha) \\
\midrule
DFT resolution only (exact signal) & $+2.4\times10^{-4}$ \\
\;+ Trotter error ($\infty$ shots) & $+9.8\times10^{-4}$ \\
\;+ shot noise ($8000$ shots/point) & $+9.8\times10^{-4}$ \\
\bottomrule
\end{tabular}
\end{center}
The peak \emph{location} is robust to shot noise (it is set collectively by all
$N$ points). Shot noise instead raises the spectral floor, which is what would
bury a weak peak. Reaching this accuracy cost a maximum depth of $2047$ Trotter
steps and up to $3.3\times10^{7}$ shots, an extravagant budget for a two-qubit
molecule, and thus the motivation for the parametric estimator of \S\ref{sec:esprit}.

\subsection{ESPRIT: super-resolution from short signals}
\label{sec:esprit}

Equation~\eqref{eq:gt} is exactly the model that parametric estimators are designed
for: a sum of complex exponentials. Rather than reading peaks off a grid, ESPRIT,
Estimation of Signal Parameters via Rotational Invariance Techniques~\citep{roy1989esprit}, closely related to the matrix-pencil and Prony
methods~\citep{hua1990matrix}, recovers the complex nodes $z_n=\ee^{-iE_n\delta}$
algebraically. One forms an $L\times M$ Hankel matrix from the samples and takes its
singular value decomposition; the matrix $\mathbf U_s\in\mathbb C^{L\times p}$ collects the $p$
left singular vectors of largest singular value, whose columns span the signal subspace.
Writing $\mathbf U_s^{\uparrow}$ and $\mathbf U_s^{\downarrow}$ for $\mathbf U_s$ with its last and its first row
deleted respectively, the Vandermonde structure of the model forces the shift-invariance
$\mathbf U_s^{\uparrow}\boldsymbol\Psi=\mathbf U_s^{\downarrow}$ for some $p\times p$ matrix $\boldsymbol\Psi$. The eigenvalues
of $\boldsymbol\Psi=(\mathbf U_s^{\uparrow})^{+}\mathbf U_s^{\downarrow}$, with $(\cdot)^{+}$ the Moore--Penrose
pseudoinverse, are the $z_n$, whence
$E_n=-\arg(z_n)/\delta$. Because this solves an algebraic problem rather than binning a
grid, ESPRIT resolves frequencies far below the $2\pi/\tau$ Fourier limit, i.e.\ from a far
shorter signal. A self-contained derivation of the estimator and a pseudocode listing
are given in Appendix~\ref{app:esprit}.

\paragraph{What is new in its use here.} ESPRIT itself is a classical spectral-estimation method and is
not introduced as a new estimator. The contribution is its placement inside a warm-start quantum
interface: the state supplied by the front-end determines the spectral weights, the quantum device
acquires the short coherent signal, and the classical estimator converts that signal into eigenvalue
information. The accompanying treatment of quadrature imbalance, adaptive model order, noise-aware
admissibility, and reuse of the same warm start across preparation and readout is part of this
architecture. The resource comparison with Fourier readout is therefore a comparison of two ways of
spending the same quantumly acquired signal, not a claim that ESPRIT is universally superior to Fourier
analysis.

\begin{figure}[t]
\centering
\includegraphics[width=\linewidth]{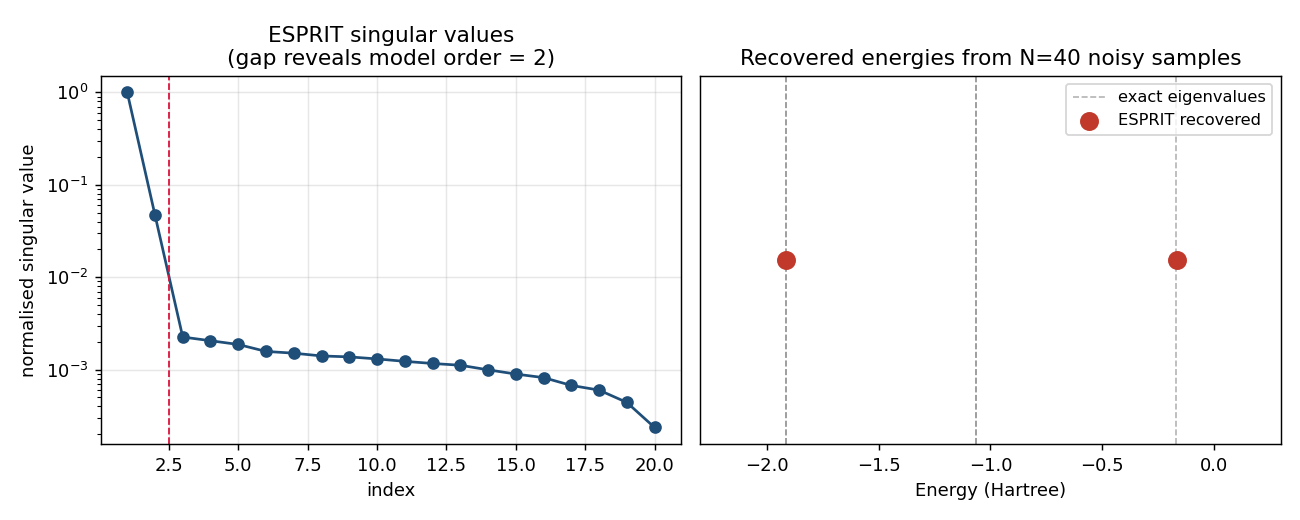}
\caption{\textbf{ESPRIT from $N=40$ noisy samples.} Left: the Hankel singular
values drop by two orders of magnitude after index $2$, revealing the model
order $p=2$. Right: the two recovered energies sit on the two eigenvalues with
nonzero overlap. The degenerate sector (middle dashed line) has no weight.}
\label{fig:esprit}
\end{figure}

This matters because, in our sampling convention, signal length equals circuit depth.
With the HF reference, two distinct frequencies present, ESPRIT reaches chemical
accuracy from $N=40$ samples, maximum depth $39$ Trotter steps, versus $2047$ for the
DFT, a $\sim\!50\times$ reduction (Table~\ref{tab:fftvsesprit}; Fig.~\ref{fig:esprit}).
Over $200$ shot-noise realisations at $\delta=0.15$, $32000$ shots/point, $100\%$ of
runs land within chemical accuracy, with a systematic bias of $+3\times10^{-4}\Ha$ (the
$\mathcal O(\delta^2)$ Trotter floor, see below) and a shot-noise standard deviation of
$5.4\times10^{-4}\Ha$.

\begin{table}[t]
\centering\small
\caption{ESPRIT versus the Fourier baseline at chemical accuracy on the same
H$_2$ Hamiltonian and HF reference.}
\label{tab:fftvsesprit}
\begin{tabular}{lcc}
\toprule
 & DFT & ESPRIT \\
\midrule
samples $N$ & $2048$ & $40$ \\
max.\ circuit depth (steps)& $2047$ & $39$ \\
max.\ evolution time $\tau$ & $511.8$& $5.85$ \\
total shots & $3.3\times10^{7}$ & $2.6\times10^{6}$ \\
$\Eo$ error (Ha) & $+9.8\times10^{-4}$ & $-8.9\times10^{-4}$ \\
\bottomrule
\end{tabular}
\end{table}

\paragraph{Two error channels, cleanly separated.} The recovered energy is the
eigenphase of the single-step Trotter unitary, an \emph{effective} Hamiltonian differing
from $H$ by a step-size-dependent term. For Eq.~\eqref{eq:Hpauli}, the
two-commuting-group structure makes this bias $\mathcal O(\delta^2)$ even for a first-order
product formula. Numerically it falls from $1.4\times10^{-3}\Ha$ at $\delta=0.30$ to
$3.6\times10^{-4}\Ha$ at $\delta=0.15$. This systematic floor is fixed by $\delta$ and
is \emph{independent of signal length}, so shortening the circuit does not worsen it.
The statistical channel is the shot-noise spread, scaling as $1/\sqrt{\mathrm{shots}}$.
We verified the trade-off directly, increasing shots from $8\mathrm{k}$ to
$32\mathrm{k}$ per point cut the standard deviation from $1.1\times10^{-3}$ to
$5.6\times10^{-4}\Ha$ and raised the chemical-accuracy success rate from $85\%$ to
$100\%$.

\subsection{The model order}
\label{sec:modelorder}

\begin{figure}[t]
\centering
\includegraphics[width=\linewidth]{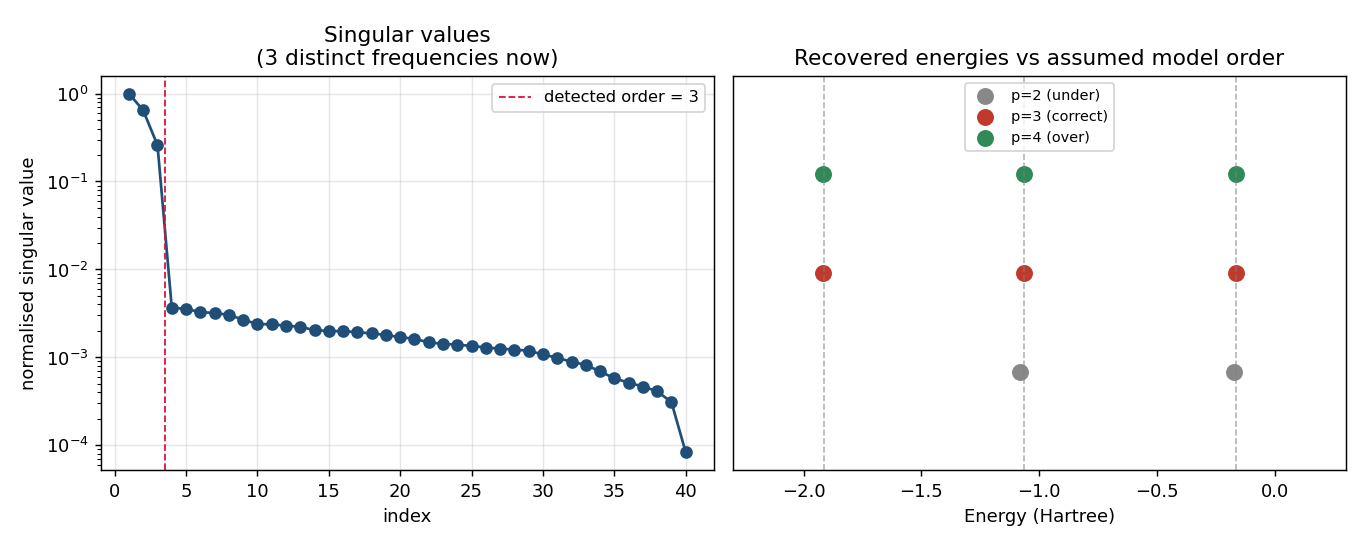}
\caption{\textbf{Model-order stress test with $\ket{+}\ket{+}$.} Left: singular
values now drop after index $3$ (three distinct frequencies with the degenerate pair
merges), and the detector selects $p=3$. Right: recovered energies for assumed
order $p=2$ (under, gray, \emph{no marker at the ground energy}), $p=3$
(correct, red), $p=4$ (over, green). The spurious fourth root lies off-scale and
is pruned by the unit-circle test.}
\label{fig:modelorder}
\end{figure}

ESPRIT needs the number of exponentials $p$. To probe its robustness we switch to
$\ket{\chi}=\ket{+}\ket{+}$, which overlaps all eigenstates. Two effects appear
(Fig.~\ref{fig:modelorder}).

\paragraph{Degeneracy is invisible.} Although four eigenstates are populated, the
exactly degenerate pair contributes a \emph{single} exponential at $-1.063653\Ha$
carrying their combined weight, so only \emph{three} distinct frequencies exist. ESPRIT estimates frequencies rather than states, so a degenerate level presents as one line. The singular
spectrum shows a gap after three, and an automatic detector, largest consecutive
singular-value ratio, selects $p=3$ in all $200$ independent noise realisations.

\paragraph{Mis-specification fails asymmetrically.} With $\ket{+}\ket{+}$ the ground
state has the \emph{smallest} weight ($0.147$, versus $0.500$ for the degenerate level
and $0.353$ for the excited state). Under-specifying ($p=2$) keeps the two strongest
frequencies and silently discards the ground state. The lowest reported energy becomes
$-1.08\Ha$, an error of $0.85\Ha$, failing in $100\%$ of runs. Over-specifying ($p=4,5$)
is benign but must be pruned. The genuine frequencies remain pinned to the unit circle
($|z_n|=1.000$) while spurious roots drift off it ($|z_n|\approx0.96$) and carry
vanishing amplitude. A combined unit-circle/amplitude filter removes them reliably. The
practical lesson is to over-specify and prune rather than risk dropping a weak ground
state.

\paragraph{Overlap is the true currency.} With the correct order, recovering the
low-weight ground state to chemical accuracy needed either more shots or more samples. Tellingly, \emph{samples beat shots}. Doubling $N$ from $80$ to $160$ cut the error
standard deviation by $\sim\!3\times$, while quadrupling the shots cut it by only
$\sim\!2\times$, consistent with the $N^{-3/2}$ versus $\mathrm{shots}^{-1/2}$ scaling
of frequency estimation~\citep{stoica1989music}, derived in Appendix~\ref{app:shot}.
Circuit depth buys precision faster than measurement does in this acquisition model. More precisely,
the frequency-estimation variance improves as $N^{-3}$ in the single-tone asymptotic regime, whereas
shot averaging contributes the usual $N_s^{-1}$ variance scaling. This is a statement about coherent
acquisition-time scaling versus measurement averaging, not a claim of a metrological quantum advantage
over the standard quantum limit.

\subsection{Warm start $\to$ readout: below the variational bound}
\label{sec:hybrid}

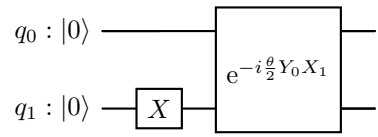
\begin{figure}[t]
\centering
\begin{quantikz}[column sep=0.45cm,row sep=0.40cm]
\lstick{$q_0:\ket{0}$} & \qw & \gate[2]{\ee^{-i\frac{\theta}{2}Y_0 X_1}} & \qw \\
\lstick{$q_1:\ket{0}$} & \gate{X} & & \qw
\end{quantikz}
\caption{One-parameter unitary-coupled-cluster ansatz for H$_2$. An $X$ prepares
the Hartree--Fock reference $\ket{10}$, and a single Givens-type rotation mixes
the $\{\ket{01},\ket{10}\}$ sector. At $\theta^\star$ it spans the exact ground
state.}
\label{fig:vqe}
\end{figure}

The two limitations highlighted above, VQE's variational bias and ESPRIT's appetite for
overlap, are complementary, which suggests composing them. Use a shallow VQE circuit
(Fig.~\ref{fig:vqe}) purely as a \emph{state-preparation} routine to obtain a
high-overlap $\ket{\psi(\boldsymbol\theta^\star)}$, then read the true eigenvalue off
its time signal with ESPRIT. VQE need only get the
\emph{state} about right, the energy it reports is not used. It is what makes the front-end interchangeable. If only the \emph{state} is kept, then any procedure
yielding a comparable state serves equally well, an annealer sample, a QAOA circuit, a
mean-field or tensor-network state, an imaginary-time preparation. We use VQE here
because it makes the central comparison vivid, the readout returns an energy below the
bound the front-end reported,  not because the readout needs it. Section~\S\ref{sec:qaoa} runs \S\ref{sec:apps} from non-variational
starts and the identical protocol from a QAOA front-end.

\begin{figure}[t]
\centering
\includegraphics[width=\linewidth]{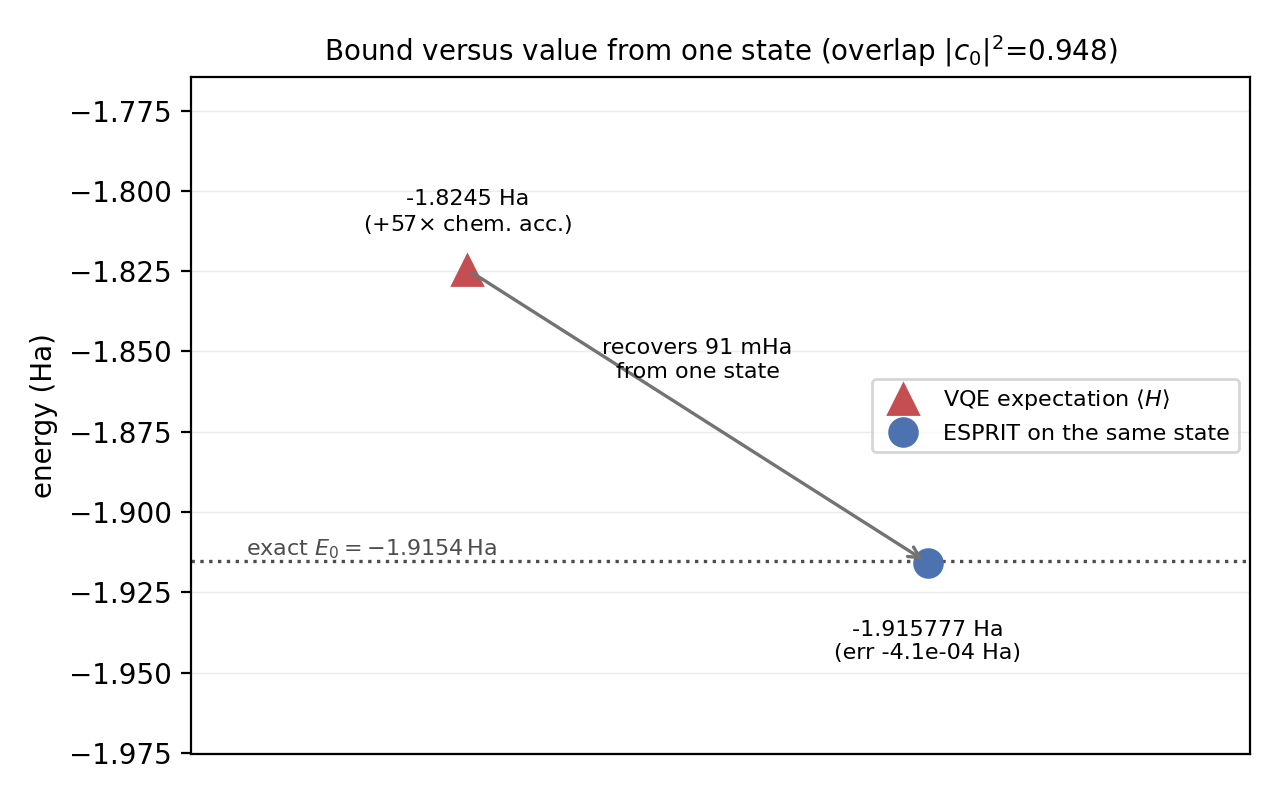}
\caption{\textbf{Bound versus value from one state.} For an imperfect ansatz
state, the VQE expectation value (triangle, $-1.824\Ha$) is a loose upper bound,
while ESPRIT on the identical state (circle) recovers the true eigenvalue
$-1.915\Ha$ to $10^{-6}\Ha$.}
\label{fig:hybrid}
\end{figure}

The mechanism is the contrast between Eq.~\eqref{eq:gt} and the Rayleigh quotient. The
expectation value $\braket{\psi|H|\psi}=\sum_n|c_n|^2 E_n$ blurs the eigenvalues into a
weighted average and is bounded below by $\Eo$. The time signal exposes each $E_n$ as a
discrete frequency, so the ground frequency can be extracted.
We make this concrete with a deliberately under-converged ansatz. It is obtained by
detuning the single ansatz parameter by a fixed $0.46\,$rad from its variational optimum,
which leaves the overlap high, $|c_0|^2=0.948$, while making the bound loose. Its energy
is $-1.824491\Ha$, an upper bound lying
$+0.091\Ha$ (about $57\times$ chemical accuracy) above the truth. ESPRIT applied to the
\emph{same} state returns $-1.915371\Ha$, an error of $-1.0\times10^{-6}\Ha$
(Fig.~\ref{fig:hybrid}). Here the recovered line is selected by amplitude, the ground line
being dominant for this warm start. Where the target is not dominant, as on the combinatorial and lattice instances  of \S\ref{sec:qaoa} and \S\ref{sec:lattice}, selection
is instead by lowest recovered line. The spectral readout recovers $91\,\mathrm{mHa}$ of binding
energy that the expectation value could not see. The variational bias of the ansatz
becomes inessential.

\paragraph{Why not simply measure the state?} If the front-end already places weight
$a^2$ on the ground state, it is natural to ask why one does not measure and read $\Eo$
off directly, repeating until the ground state is drawn. Whether that works is decided
by a single question: does $H_P$ commute with the measurement basis?

For the combinatorial and lattice problems of \S\ref{sec:qaoa} and \S\ref{sec:lattice} where the cost operator is
diagonal, each shot collapses to a basis string, evaluating the cost on that string
returns an \emph{exact} eigenvalue, and $\mathcal O(1/a^2)$ shots with the minimum retained
suffice. This is why the spectral back-end has no content there (\S\ref{sec:aqc}) and
why the certificate of \S\ref{sec:stopping} is free in that setting.

Where it does not, as for any molecular $H_P$, no such measurement exists. A
computational-basis shot returns a string that is not an eigenstate of $H_P$. What is measurable is each Pauli
term, $\pm1$ per shot, and the $E_n$ are not among the outcomes of any of them. One
never samples the eigenvalue distribution, only its mean. 
That mean is the Rayleigh quotient, and it carries a deficit
\begin{equation}
\braket{H}-\Eo=\Delta E=\sum_{n>0}|c_n|^2(E_n-E_0)=(1-a^2)\,\bar\mu,
\label{eq:deficit}
\end{equation}
with $\bar\mu$ the mean excitation energy of the leaked population. Section  \S\ref{sec:qml} makes this the basis of a quantitative comparison.
The consequence is quantitative and severe. With
$\bar\mu\sim2\Ha$, driving that deficit below chemical accuracy demands
$a^2>0.9992$, whereas the readout returns exact eigenvalues from $a^2=0.222$ with the
target only the third brightest of twelve lines (Table~\ref{tab:svpread}).

The operation that
genuinely projects onto $\ket{E_n}$ with probability $|c_n|^2$ is phase estimation, that
is, the coherent circuit whose cost motivated this work. ESPRIT recovers the same line
positions from $g(t)$ statistically, paying repetitions in place of ancillae and a
coherent transform.

High overlap is advantageous. It collapses the signal toward a
single frequency, shrinking the effective model order and the sampling budget.
Table~\ref{tab:overlap} quantifies the $1/|c_0|^2$ precision cost
(Appendix~\ref{app:shot}) across three initial states at fixed $N=40$. A near-perfect
VQE state succeeds with effective order one. The low-overlap $\ket{+}\ket{+}$ fails at
this short length and needs the longer signals of \S\ref{sec:modelorder}.

\begin{table}[t]
\centering\small
\caption{Effect of ground-state overlap on spectral readout
($N=40$, $\delta=0.15$, $32\mathrm{k}$ shots/point; generous order then
unit-circle/amplitude pruning). Success $=$ fraction of $150$ runs within
chemical accuracy.}
\label{tab:overlap}
\begin{tabular}{lccc}
\toprule
initial state & $|c_0|^2$ & real freqs.\ & success \\
\midrule
$\ket{+}\ket{+}$ & $0.147$ & $3$ & $4\%$ \\
HF $\ket{10}$ & $0.955$ & $2$ & $99\%$ \\
VQE $\ket{\psi^\star}$ & $1.000$ & $1$ & $97\%$ \\
\bottomrule
\end{tabular}
\end{table}

\subsection{Adaptive acquisition and a noise-aware model}
\label{sec:adaptive}

The present calculations use uniformly spaced times and a fixed model order. These
choices make the comparison with the Fourier baseline transparent, but neither is
required by the spectral model. A practical implementation can instead make the
sampling and the model order adaptive.

For a first pass, take a short uniform sequence and estimate a provisional spectrum.
The singular values of the Hankel matrix then provide an estimate of the number of
resolvable modes. Rather than committing to one value of $p$, one can fit a sequence
$p=1,2,\ldots$ and accept the smallest order for which the reconstruction residual is
consistent with the measured shot-noise floor. The retained roots should also be stable
under changes of the Hankel dimensions and under removal of individual time points.
Bootstrap resampling of the measured quadratures supplies confidence intervals for both
the roots and their weights. The unit-circle and amplitude filters used above then
become stability tests rather than the sole protection against spurious low-energy
roots. This is deliberately conservative: underestimating $p$ can silently remove a
weak ground line, whereas a modest overestimate can be pruned.

The time grid can likewise be made nonuniform. After the short-time pass has located
the broad spectral support, additional samples can be placed where the estimated
Fisher information for the target energy is largest. In resource notation the
experiment can be viewed as choosing $\{t_k,N_{s,k}\}$ to minimise the uncertainty in
$E_0$ subject to a total shot budget,
\begin{equation}
\min_{\{t_k,N_{s,k}\}}\operatorname{Var}(\widehat E_0),
\qquad
\sum_k N_{s,k}\leq\mathcal M,\quad
0\leq t_k\leq \tau_{\max}.
\label{eq:adaptive_design}
\end{equation}
A short-time block followed by a small number of long-time samples is a particularly
simple implementation. The short block identifies the modes; the long samples resolve
close frequencies. This separates the need for spectral identification from the need
for high phase resolution and avoids paying the maximum coherent depth at every point.

Real devices also introduce decay, for which the undamped model
$|z_n|=1$ is no longer exact. The natural extension is the damped model
\begin{equation}
g(t)=\sum_n A_n\ee^{-(\gamma_n+iE_n)t},
\qquad
z_n=\ee^{-(\gamma_n+iE_n)\delta},
\label{eq:dampedmodel}
\end{equation}
so that ESPRIT estimates both the energy $E_n$ and an effective decay rate
$\gamma_n$. In this setting a root slightly inside the unit circle is not automatically
spurious, and a noise-aware implementation should replace the hard unit-circle cut with a
confidence region determined from calibration or bootstrap statistics
(Fig.~\ref{fig:admiss}).

\input{figcircle}

The per-mode form of $\gamma_n$ has to be taken seriously, because the decay is not in general one number. Write
$\gamma_n=\gamma_{\rm com}+\gamma_n^{\rm res}$. Ancilla dephasing, and any channel acting as a
depolarisation of fixed strength per Trotter step, multiply the whole interferometric signal
by $\ee^{-\gamma_{\rm com}t}$. Every pole is displaced radially by the same amount, and no
eigenvalue and no ratio of weights is disturbed. That part is a property of the device and
the circuit, scales with gate count, and is what a calibration can remove. Amplitude damping toward a fixed
computational state, leakage out of a symmetry sector, and coherent gate errors are not of that
kind. They act differently on different eigenstates, so $\gamma_n^{\rm res}$ depends on $n$ and,
through the eigenvectors, on $H_P$ itself. It follows that a calibration performed on a
\emph{different} Hamiltonian cannot remove the residual, and that subtracting a single calibrated
rate from every mode would corrupt the recovered weights. What the
calibration should do is estimate $\gamma_{\rm com}$, test whether the calibration rates agree
across modes within their bootstrap uncertainty, and report
$\gamma_n^{\rm res}$ as an output, since it is a physical statement about how the device acts on
the eigenstates of the problem at hand. The construction that avoids the transfer question
altogether is to calibrate on $H_P$ itself, repeating the acquisition at several deliberately
amplified noise levels and extrapolating the recovered rates to zero noise, whereupon what scales
with the amplification is $\gamma_{\rm com}$ and what does not is the residual.

The resulting readout is an adaptive sequence,
\begin{equation}
\begin{gathered}
\text{short signal}\ \rightarrow\ \text{order and spectrum}\\
\rightarrow\ \text{targeted samples}\ \rightarrow\ \text{stable estimate},
\end{gathered}
\end{equation}
rather than a fixed FFT-like acquisition. No new quantum primitive is required by
this construction; the additional work is classical model selection and allocation of
shots.

The SQD study of \S\ref{sec:sqd} below requires both and fails without either, in the two
ways reported there. The adaptive and noise-aware machinery is therefore load-bearing at low
budgets, which is the regime hardware experiments will occupy first.

\subsection{Both quadratures in one sweep}
\label{sec:quadrature}

The Hadamard test of Fig.~\ref{fig:hadamard} returns one quadrature of $g(t)$ per measurement
setting. Omitting the $S^\dagger$ gate returns $\mathrm{Re}\,g$; including it returns
$\mathrm{Im}\,g$. An acquisition is therefore two passes over the same sample times, differing
only in that gate, and nothing in the construction requires the interferometric contrast of the
two passes to be equal. It is worth asking what happens when it is not, because the same question, asked of a radio receiver, has a
sixty-year-old answer that turns out to transfer exactly.

Write the acquired parts as $c(1+e)\,\mathrm{Re}\,g$ and $c(1-e)\,\mathrm{Im}\,g$, splitting the
contrast into a common factor $c$ and a differential $e$. The reconstructed signal is then
\begin{equation}
\tilde g(t) \;=\; c\,\bigl[\,g(t) \;+\; e\,g^{*}(t)\,\bigr],
\label{eq:mirror}
\end{equation}
which is not a distorted version of $g$ but $g$ plus a \emph{conjugate} component. This is
precisely the I/Q imbalance of a quadrature radio or optical receiver, where the same algebra
admits an \emph{image} at the mirror frequency and where estimating and compensating the mismatch
is a mature art~\citep{valkama2001iq}. The transfer is exact and we claim no novelty in the
mechanism. What differs is the consequence, because a receiver's downstream processing is a
transform and ours is a parametric estimator with a committed line count. Since
$g^{*}(t)=\sum_n p_n \ee^{+iE_nt}$, the imbalance injects a set of \emph{mirror lines} at
$-E_n$ carrying relative weight $e$. The common factor $c$ is harmless: it rescales every
residue equally, leaves the roots $z_n$ on their circle and is absorbed by the damping
calibration of \S\ref{sec:adaptive}. The differential part is not harmless, because it presents
the estimator with lines that are not in the spectrum.

\begin{figure*}[t]\centering
\includegraphics[width=\textwidth]{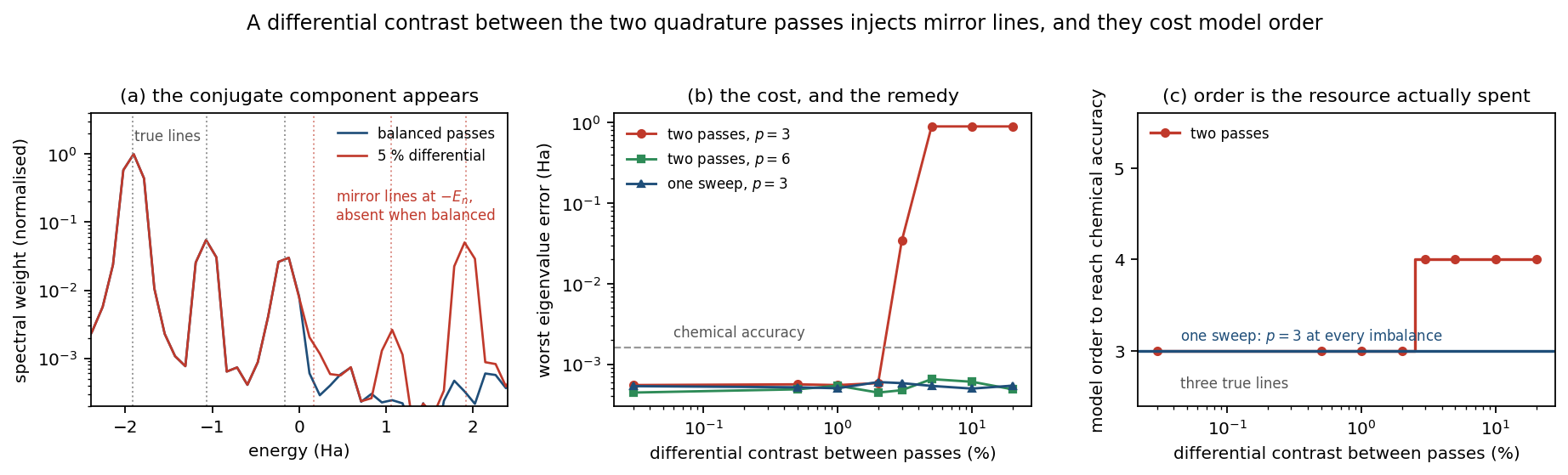}
\caption{\textbf{The two quadrature passes must share a contrast, or the estimator pays for it in
model order.} \textbf{(a)} A five per cent differential contrast between the passes adds the
conjugate component of Eq.~\eqref{eq:mirror} to the $H_2$ acquisition, putting mirror lines at
$-E_n$ (red dotted) where the balanced acquisition has none. \textbf{(b)} Worst error over the
three true lines, at $2\times10^{7}$ shots and $N=96$. At the order chosen for the true lines the
two-pass acquisition fails abruptly beyond two per cent; doubling the order restores it; an
acquisition whose settings are interleaved within one sweep never incurs the defect.
\textbf{(c)} The order needed to hold chemical accuracy. Interleaving buys no variance --- every
scheme has the same statistical efficiency at a fixed budget --- and costs no ancilla, no gate and
no extra shot; what it holds is $p=3$ at every imbalance, and order is the resource
\S\ref{sec:modelorder} and \S\ref{sec:resource} show to be scarce.}
\label{fig:quadrature}
\end{figure*}

Figure~\ref{fig:quadrature}(a) shows them arriving. At five per cent differential contrast the
$H_2$ acquisition carries three spurious peaks at $+1.92$, $+1.06$ and $+0.17\Ha$, mirroring the
three true ones, at a weight that puts the largest of them above the second true line.

The consequence is a failure of the estimator, not a loss of precision, and it is abrupt.
Figure~\ref{fig:quadrature}(b) gives the worst error over the three true lines against the
imbalance, at $2\times10^{7}$ shots and $N=96$ samples. At a model order $p=3$ chosen for the
three true lines the error holds at $5.5\times10^{-4}\Ha$ up to two per cent imbalance, rises to
$3.4\times10^{-2}\Ha$ at three per cent, and saturates at $0.89\Ha$ beyond five, the estimator
having locked onto a mirror line in place of a true one. There is no graceful degradation: the
subspace of dimension three cannot hold six lines, and which three it returns is decided by their
weights.

The remedy within a two-pass acquisition is the one that works. Raising the order to
$p=6$, so that the estimator fits the mirror lines, restores
the accuracy at every imbalance tested, returning $6.5\times10^{-4}\Ha$ at five per cent and
$4.9\times10^{-4}\Ha$ at twenty. That comparison is why the finding is
about model order. Figure~\ref{fig:quadrature}(c) puts it plainly:
holding chemical accuracy on this spectrum needs $p=3$ below about two per cent imbalance and $p=4$ above it, while the worst case is a doubling of the order, because the conjugate component doubles the line count.

Model order is not a free parameter in this architecture. It is the estimator's single
hyperparameter, whose misspecification \S\ref{sec:modelorder} identifies as the dominant failure
mode; it is what \S\ref{sec:adaptive} spends a scouting acquisition to determine; and it is half
of the two-component resource $(p_0, r_{\rm eff})$ that \S\ref{sec:resource} prices. Spending it
on artefacts of the measurement rather than on the spectrum is a real cost.

\paragraph{One sweep, and no new hardware.} It is worth being precise about what fixes this. What admits the conjugate component is the \emph{separation
in time} of the two settings, not the choice of settings. Interleave them --- select
$V\in\{\mathbf 1, S^\dagger\}$ independently at each repetition and record the outcome against the
setting that produced it --- and a contrast drift is sampled identically by both, so $e$ is no
longer a systematic quantity but a residual of order (drift)$/\sqrt{N_s}$. The circuit of Fig.~\ref{fig:hadamard} is unaltered, no ancilla is added, no gate is added, and the
shot count is unchanged.

An acquisition of this shape is already being run. The phase estimation of Seitz
\emph{et al.}~\citep{seitz2026workflows} draws its quadrature phase
$\beta\in\{0,\pi/2\}$ afresh at every repetition, so its two settings are interleaved by
construction and the differential never forms. We note it because the protection is incidental
there rather than argued for, and a two-pass version of the same experiment would carry the
defect with no outward sign. It also settles the question of implementability. The remedy is
not a proposal, it is a shot order that hardware runs today.

Two variants are worth recording. The selection can be made coherently, by a second ancilla in
$\ket{+}$ that controls the quadrature gate and is itself measured to label the shot; circuits of
that family are known~\citep{imamura2026parallel}, motivated there by the reduction in distinct
circuit types. Or the settings can be taken as
$V\in\{T^\dagger,T\}$, whereupon each measures $(\mathrm{Re}\,g\pm\mathrm{Im}\,g)/\sqrt2$ and
every shot carries information about \emph{both} quadratures, each unsharp by $1/\sqrt2$ as a
joint measurement of conjugate observables must be~\citep{arthurs1965conjugate}.

That last variant carries a trap. Run as two passes it
does not remove the conjugate component but rotates it. The reconstructed signal becomes
$c[g+ieg^{*}]$ rather than $c[g+eg^{*}]$, of the same magnitude, with mirror lines of the same
weight. A diagonal-pair two-pass acquisition fails at twenty per cent drift exactly as the
ordinary two-pass does, at $0.89\Ha$, while the interleaved acquisition holds
$8\times10^{-3}\Ha$. Sharing a shot between the quadratures is neither necessary nor sufficient;
interleaving the settings is both.

None of this is more efficient than two passes. The two quadratures are expectations of
anticommuting ancilla observables, so no arrangement of settings or ancillas alters the total
information per repetition. At a fixed budget the summed
variances agree across the schemes to within two per cent. What interleaving buys is the absence
of a differential to begin with, and Fig.~\ref{fig:quadrature} shows it flat at $p=3$ and
$5\times10^{-4}\Ha$ across four decades of imbalance.

\paragraph{And the phase.} A second benefit is specific to the interference acquisition of
\S\ref{sec:homodyne} and does not arise here. The residues of an autocorrelation are the weights
$p_n$, real and non-negative, with no phase to protect. The residues of an interference
acquisition are the products $c_S^*c_W$, whose phase that section returns as a measured quantity
and uses twice, for the transition amplitudes and for the frame of the amplification reflection.
Because interleaving lets both quadratures sample the same contrast, a factor common to them
cancels in their ratio. Under a five per cent root-mean-square contrast drift between
passes the error on the returned phase falls from $0.042$ to $0.028$ radians, under ten per cent
from $0.072$ to $0.028$, and under twenty per cent from $0.138$ to $0.029$. The single-sweep value
is flat in the drift while the two-pass value grows in proportion to it, and at zero drift the
two agree to within one per cent. The modulus is not protected, the common factor surviving there, and it continues to be handled by the damping calibration.

\paragraph{What this does not settle.} The imbalance is imposed here as a stated differential
gain, not derived from a device model, so the figure answers what an imbalance costs and not how
large an imbalance a given processor has. That is a calibration measurement, by acquiring a known single-line signal in both settings and comparing the recovered amplitudes
returns $e$ directly. The two-qubit gate of the second arrangement is counted but its own
infidelity is not modelled. And on a spectrum already dense enough that the order is set by line
count, the saving is proportionately smaller, since the mirror lines are
then a fixed addition to a large number rather than a doubling of a small one.

\section{The readout at scale, and on real molecules}
\label{sec:readoutscale}

Everything above was established on two qubits, where the exact answer is available by
inspection. We now hold the pipeline of \S\ref{sec:readout} fixed and change only the
Hamiltonian: to a strongly correlated spin chain at $16$ qubits, to two molecules in
active spaces with an explicit Trotter circuit, to excited states rather than the ground
state, and finally to a vibronic band, where the object read out of the same acquisition
is a spectrum rather than a line. The point of the sequence is that the estimator, the
circuit and the resource accounting do not change across it. Only the spectrum does, and
the places where the method degrades are places where the spectrum has
become harder.

\subsection{Sixteen qubits: a strongly correlated benchmark}
\label{sec:scaling}

H$_2$ isolates every mechanism but cannot show the challenges that emerge at scale. We therefore repeat the entire hybrid analysis on a system that is both genuinely many-body
and representative of the early fault-tolerant benchmark class. That system is the
one-dimensional transverse-field Ising model (TFIM),
$H=-J\sum_i Z_iZ_{i+1}-h\sum_i X_i$, $Z_i \in \{-1, +1\}$ on an open chain at $h/J=1.1$,
just into the paramagnetic phase, where the ground state is strongly entangled and the gap
is small. The same two-commuting-group structure as
Eq.~\eqref{eq:Hpauli} keeps the evolution matrix-free. We use a Hamiltonian-variational
ansatz~\citep{wecker2015progress} for VQE, a Lanczos reference for the exact low-lying
spectrum, and the identical Trotter$+$Hadamard$+$ESPRIT/FFT pipeline of
\S\ref{sec:experiment}--\ref{sec:hybrid}.

\begin{figure}[t]
\centering
\includegraphics[width=\linewidth]{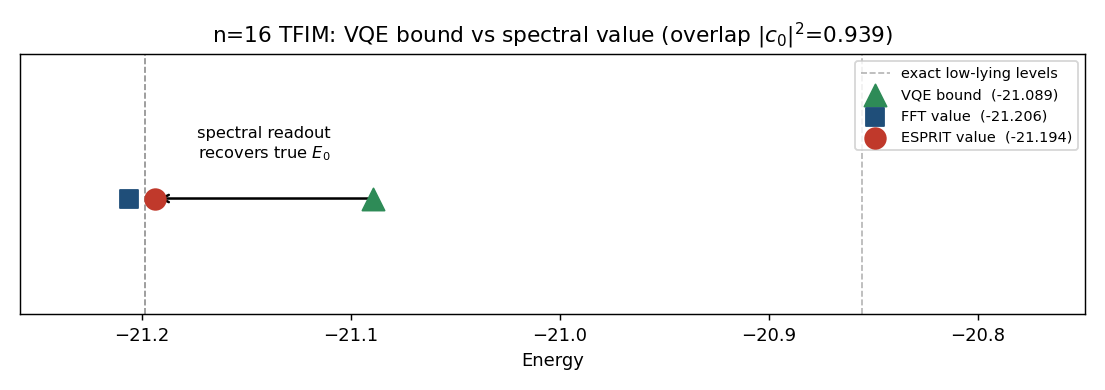}
\caption{\textbf{Bound versus value at $n=16$.} The five-layer VQE bound
(triangle) lies $0.11$ above the exact energy. ESPRIT (circle) and FFT (square)
on the same VQE state recover $E_0$ to $\sim\!10^{-2}$.}
\label{fig:tfim_bv}
\end{figure}

\paragraph{The hybrid survives at scale.} At $n=16$ the exact ground energy bound is
$E_0=-21.198$ with a gap of $0.34$. A five-layer VQE returns a \emph{loose} bound of
$-21.089$ (error $+0.109$, about $70\times$ chemical accuracy) while securing an overlap
$|c_0|^2=0.939$. Feeding that VQE state into the spectral readout recovers the true
energy far below the bound (Fig.~\ref{fig:tfim_bv}): ESPRIT returns $-21.194$ (error
$+0.005$) and the FFT $-21.206$ (error $-0.008$). As on H$_2$, VQE need only secure the
state. The spectrum supplies the value.

\begin{figure}[t]
\centering
\includegraphics[width=\linewidth]{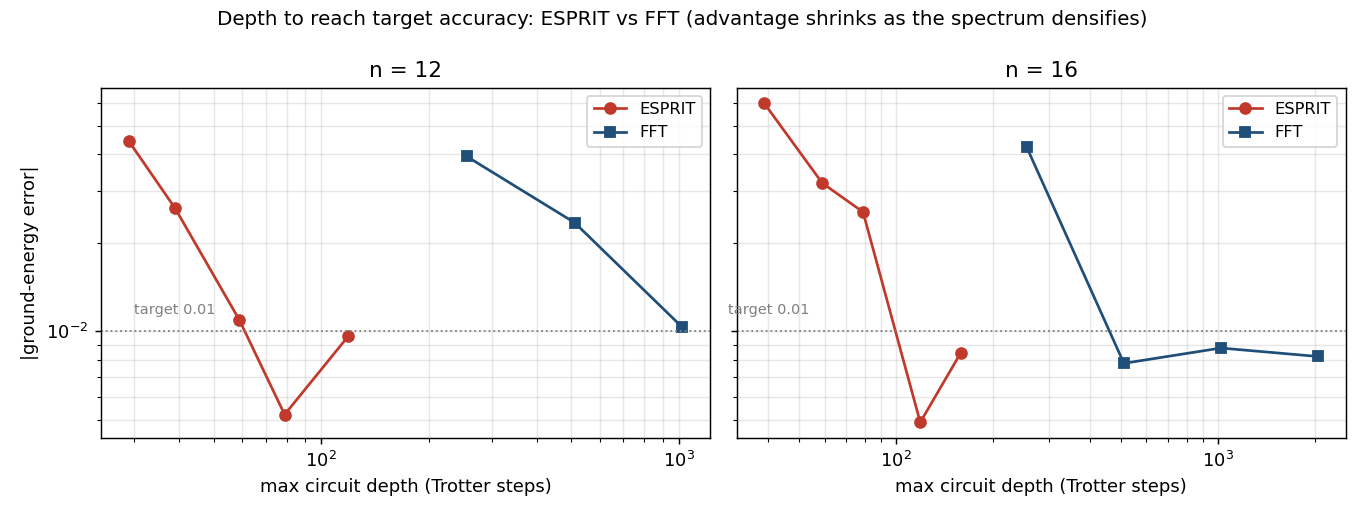}
\caption{\textbf{Depth to reach target accuracy.} ESPRIT (red) crosses the
$0.01$ target at roughly an order of magnitude shallower depth than the FFT
(blue) at both sizes, but the advantage contracts from $n=12$ to $n=16$ as the
spectrum densifies.}
\label{fig:tfim_depth}
\end{figure}

\paragraph{ESPRIT stays shallower, but its margin shrinks.} ESPRIT again reaches the
target accuracy at far smaller depth than the FFT (Fig.~\ref{fig:tfim_depth}). At $n=12$
it crosses $|{\rm error}|=0.01$ near depth $40$ versus $\sim\!1000$ for the FFT (a
$\sim\!25\times$ saving), and at $n=16$ near depth $120$ versus $\sim\!500$ (a
$\sim\!4$--$6\times$ saving). The shrinking margin is the central scaling lesson. A
many-body initial state populates a \emph{dense forest} of small excited-state
amplitudes, which act as correlated noise for a low-order exponential model. ESPRIT's
super-resolution edge is largest when the spectrum is sparse and erodes as it densifies,
while the FFT degrades more gracefully. Thus, a higher-overlap warm start concentrates weight on the ground line and
restores ESPRIT's advantage.

\paragraph{A robust ground selector follows from the warm start.} Taking the
\emph{lowest} recovered frequency, natural for H$_2$, is unsafe at scale. Shot noise
occasionally produces a spurious line below $E_0$ that corrupts the estimate (we
observed catastrophic $\sim\!0.7$ outliers). Because the VQE warm start makes the ground
the \emph{dominant} overlap, the correct rule here is to select the largest-amplitude
on-circle frequency. This single change removes the outliers and is, again, a direct
dividend of the hybrid structure. (For a \emph{minimisation} objective the ground line
can instead be subdominant, and the selector switches to the lowest resolvable on-circle
line; see \S\ref{sec:qaoa}.) We also note the Trotter bias grows with system size as
$\sim\!n\delta^2$, requiring $\delta\!\approx\!0.02$ at $n=16$ to keep it near
$10^{-3}$.



\subsection{SO\texorpdfstring{$_2$}{2} in an active space}
\label{sec:so2}

The TFIM is many-body but \emph{local}. A molecular electronic Hamiltonian is dense and
non-local, with $\mathcal O(n^4)$ terms in the number $n$ of spin orbitals, and is the setting the
methods are ultimately meant for. Throughout, $n$ denotes spin orbitals, equivalently
qubits under Jordan--Wigner.
We therefore close the empirical study with a genuine molecular instance, sulphur dioxide. In a
minimal STO-3G basis SO$_2$ spans $19$ spatial orbitals, i.e.\ $38$ qubits under
Jordan--Wigner, beyond exact classical statevector simulation.
The standard remedy is an active space. From a restricted
Hartree--Fock solution (total energy $-540.602\,$Ha) we freeze the lowest $13$ orbitals
and retain a CAS(6e,6o) window straddling the HOMO--LUMO gap, yielding a $12$-qubit
Hamiltonian. We build it with PySCF and
OpenFermion~\citep{sun2018pyscf,mcclean2020openfermion}. The Jordan--Wigner operator has
$1.1\times10^{5}$ nonzero matrix elements, two orders of magnitude denser than the TFIM. Its exact CAS ground energy is $E_0=-540.7215\,$Ha with a gap
of $0.089\,$Ha and a near-threefold-degenerate excited cluster at $-540.633\,$Ha.

The full basis of SO$_2$ is not
simulable, forcing the active-space reduction. But as a \emph{methodological} benchmark
it is far more representative than a spin chain. The readout must contend with a dense
molecular spectrum and an initial state whose overlap is set by real correlation rather
than a tunable knob. On hardware the corresponding cost is severe, the dense
Hamiltonian's $\sim\!10^3$ Pauli terms make each Trotter step far more expensive than
the TFIM's $\mathcal O(n)$ terms, so we evolve exactly here and note that Trotterisation would
add the $\mathcal O(\delta^2)$ systematic floor of \S\ref{sec:esprit}, now with a much larger
per-step gate count.

\begin{figure}[t]
\centering
\includegraphics[width=\linewidth]{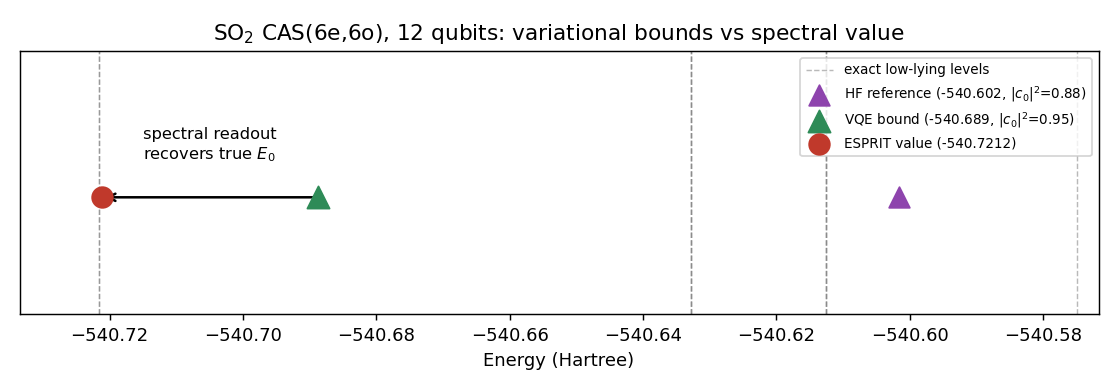}
\caption{\textbf{SO$_2$ CAS(6e,6o), 12 qubits.} The Hartree--Fock and pair-UCCD
VQE variational bounds (triangles) lie $0.12$ and $0.033\,$Ha above $E_0$.  ESPRIT
on the VQE state (circle) recovers the true ground energy.}
\label{fig:so2_bv}
\end{figure}

\paragraph{Bound versus value.} The two natural warm starts bracket the hybrid's premise
(Fig.~\ref{fig:so2_bv}). The Hartree--Fock reference, the zeroth-order variational
state, gives a bound of $-540.602\,$Ha, fully $0.120\,$Ha above
$E_0$, with overlap $|c_0|^2=0.88$. A compact nine-parameter pair-UCCD VQE tightens this
to $-540.689\,$Ha (still $+0.033\,$Ha, $\sim\!20\times$ chemical accuracy) and raises
the overlap to $0.954$. Feeding the VQE state into the spectral readout recovers
$-540.7212\,$Ha, within chemical accuracy of the truth and $0.033\,$Ha below the VQE
bound. The variational error is again bypassed as the spectrum
supplies the value.

\begin{figure}[t]
\centering
\includegraphics[width=0.49\linewidth]{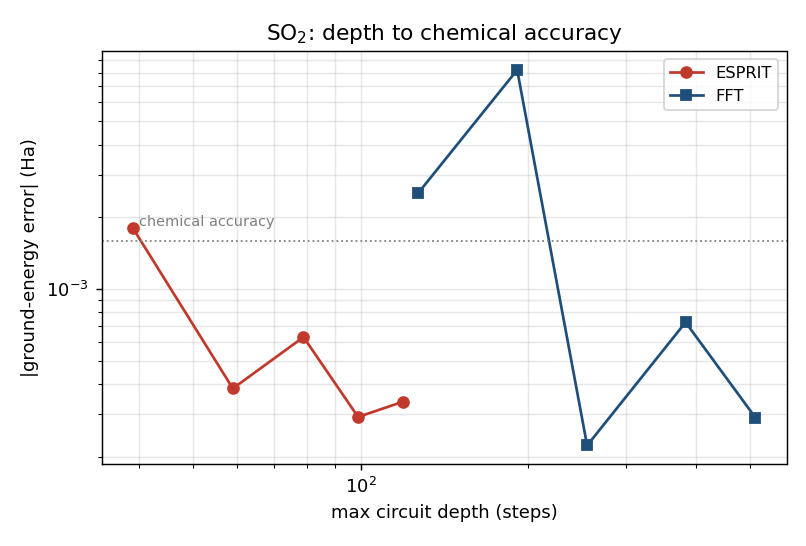}\hfill
\includegraphics[width=0.49\linewidth]{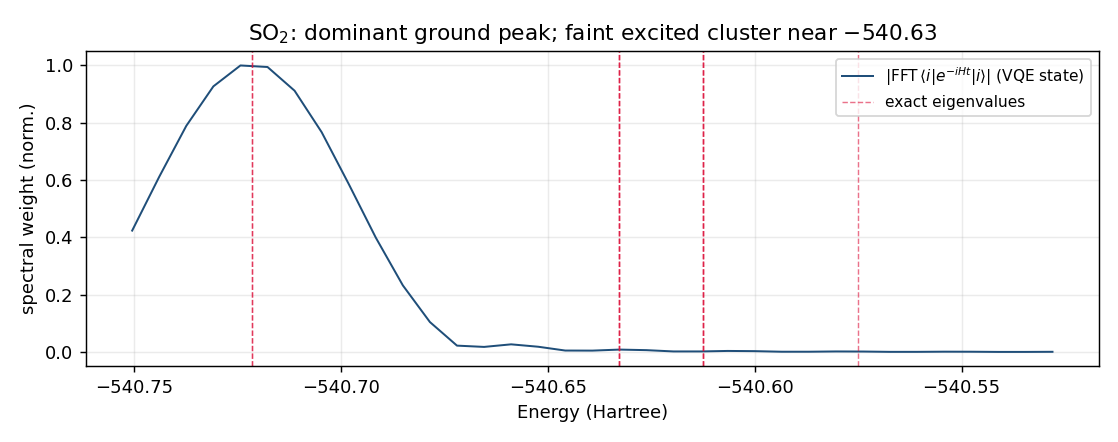}
\caption{\textbf{Left:} depth to chemical accuracy on SO$_2$. ESPRIT (red)
crosses near depth $59$, the FFT (blue) near $255$. \textbf{Right:} the spectrum
of $\langle i|\ee^{-iHt}|i\rangle$ from the VQE state, a dominant ground peak on
$E_0$ with a faint excited cluster near $-540.63\,$Ha.}
\label{fig:so2_depth}
\label{fig:so2_spec}
\end{figure}

\paragraph{ESPRIT, FFT, and the spectrum.} Both estimators recover $E_0$ to within
chemical accuracy, ESPRIT at maximum depth $\sim\!59$ steps and the FFT at $\sim\!255$
(Fig.~\ref{fig:so2_depth}), a $\sim\!4\times$ depth saving, smaller than the sparse
two-frequency H$_2$ case but consistent with the dense-spectrum trend of
\S\ref{sec:scaling}. The high VQE overlap makes the ground line dominant for both
methods (Fig.~\ref{fig:so2_spec}). This is why even the FFT performs well once the warm
start is good. The FFT's larger scatter near its resolution threshold is the price of
its nonparametric peak-picking. The lower-overlap HF warm start ($|c_0|^2=0.88$) still
gives the ground energy to $5\times10^{-4}\,$Ha at depth $119$, confirming that the
spectral readout tolerates an imperfect, but not arbitrary, initial state.

\subsubsection{An explicit Trotter circuit}
\label{sec:trotter}
\begin{figure}[t]
\centering
\includegraphics[width=0.66\linewidth]{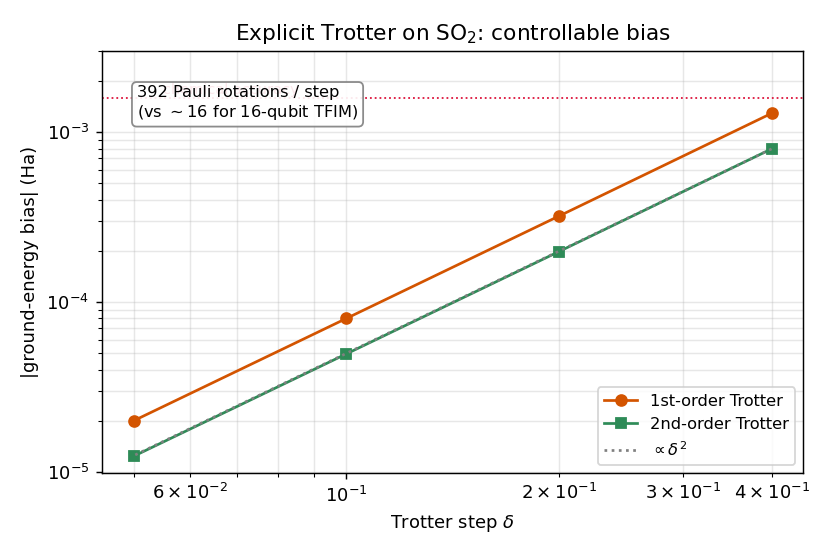}
\caption{\textbf{Explicit Trotter on SO$_2$.} The recovered ground-energy bias
scales as $\delta^2$ and stays below chemical accuracy. Each step compiles to
$392$ Pauli rotations, some twenty-five times more than a local spin-model step.}
\label{fig:trotter}
\end{figure}

On hardware one Trotterises, we therefore compile
an explicit second-order Trotter circuit for the $12$-qubit SO$_2$ Hamiltonian,
splitting it into a diagonal layer (all $Z$-only Pauli terms, applied as a single
phase per basis state) and the off-diagonal Pauli rotations
$\ee^{-ic_jP_j\delta}$. One second-order step compiles to {392 off-diagonal
Pauli rotations} plus the diagonal layer, against the $\sim\!16$ gates of a
$16$-qubit TFIM step (\S\ref{sec:scaling}). This twenty-five-fold gap
\emph{is} the central practical message for chemistry. The dense molecular
Hamiltonian, not the classical estimator, dominates the near-term cost. The
recovered ground-energy bias scales as $\delta^2$ for both first- and
second-order formulas (Fig.~\ref{fig:trotter}), the second-order $\sim\!1.6\times$
smaller, and stays below chemical accuracy for $\delta\lesssim0.1$. A full
explicit-Trotter run with shot noise ($\delta=0.1$, depth $119$,
$1.6\times10^4$ shots) recovers $E_0$ to $1.5\times10^{-3}\,$Ha, confirming that
ESPRIT operates unchanged on a genuinely Trotterised signal~\citep{childs2021trotter}.

\subsubsection{Beyond the single reference: stretched SO\texorpdfstring{$_2$}{2}}
\label{sec:so2mr}
\begin{figure}[t]
\centering
\includegraphics[width=\linewidth]{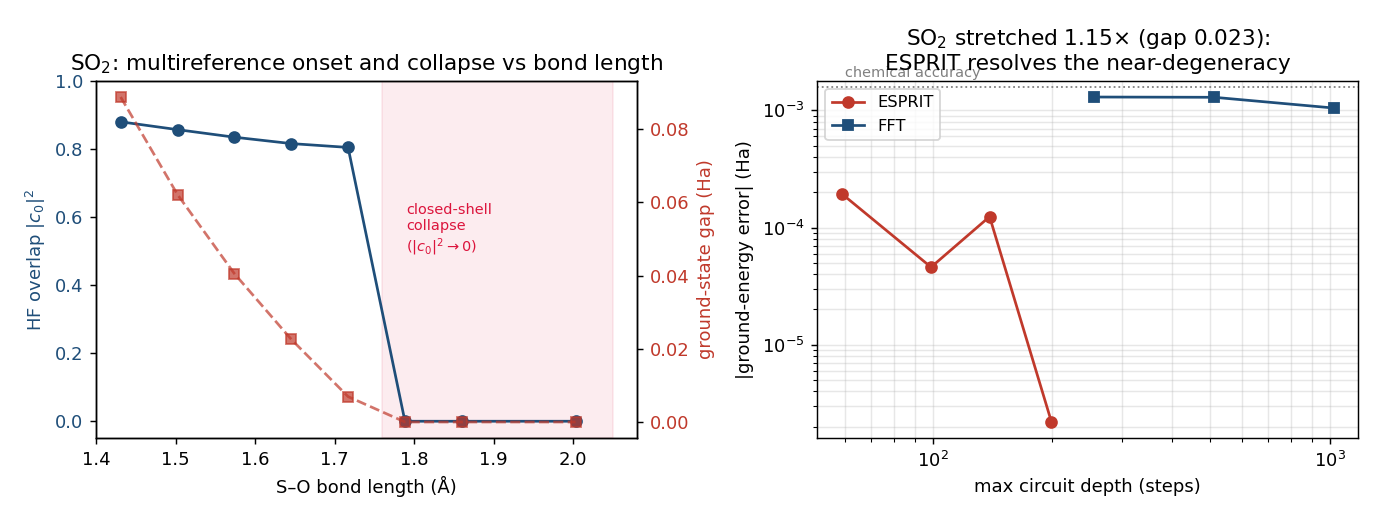}
\caption{\textbf{Stretched SO$_2$.} \textbf{Left:} the gap closes and the
closed-shell overlap declines with bond length, collapsing to zero past a state
crossing ($\sim\!1.22\times$). \textbf{Right:} at the multireference onset
($1.15\times$) ESPRIT resolves the near-degenerate gap to $10^{-5}$--$10^{-6}$
where the FFT cannot ($\sim\!10^{-3}$).}
\label{fig:so2mr}
\end{figure}

The equilibrium ground state is single-reference, so the high HF overlap.
Stretching both S--O bonds drives the system into the static-correlation
regime~\citep{lyakh2012multireference}; the gap closes and the closed-shell
overlap declines (Fig.~\ref{fig:so2mr}, left). Two qualitatively different things
follow. At a mild stretch ($1.15\times$, gap $0.023\,$Ha) the spectrum becomes
\emph{near-degenerate}, and this is exactly where ESPRIT's super-resolution is
decisive. It reaches $10^{-5}$--$10^{-6}\,$Ha while the FFT, unable to resolve the
tight gap without $\tau\gtrsim270$, stalls near $10^{-3}$ (Fig.~\ref{fig:so2mr},
right). The multireference \emph{onset} is the friendliest case for the parametric
estimator. Past a state crossing near $1.22\times$, however, the ground state
becomes an open-shell diradical in a different spatial-symmetry sector. The
closed-shell HF and pair-UCCD overlaps collapse to \emph{zero} and the hybrid
\emph{fails}, ESPRIT misses $E_0$ by $0.021\,$Ha because a ground line carrying no
amplitude is invisible. This is the operating boundary of the approach. The spectral
readout inherits the warm start's overlap problem and cannot manufacture a ground
component the preparation circuit never supplied. A genuinely multireference
reference would be needed, and its overlap, not the estimator, would again be
the currency.

The SO$_2$ benchmark so reproduces the full hybrid narrative on a genuine, dense
molecular Hamiltonian. A loose VQE bound, an unbiased spectral value below it, and an
ESPRIT depth advantage that survives but narrows as the spectrum densifies, while
underscoring that the dominant near-term cost for chemistry is the Trotterisation of the
molecular Hamiltonian itself, not the classical estimator that follows.

\subsection{Water, and the multireference frontier}
\label{sec:h2o}
\begin{figure}[t]
\centering
\includegraphics[width=\linewidth]{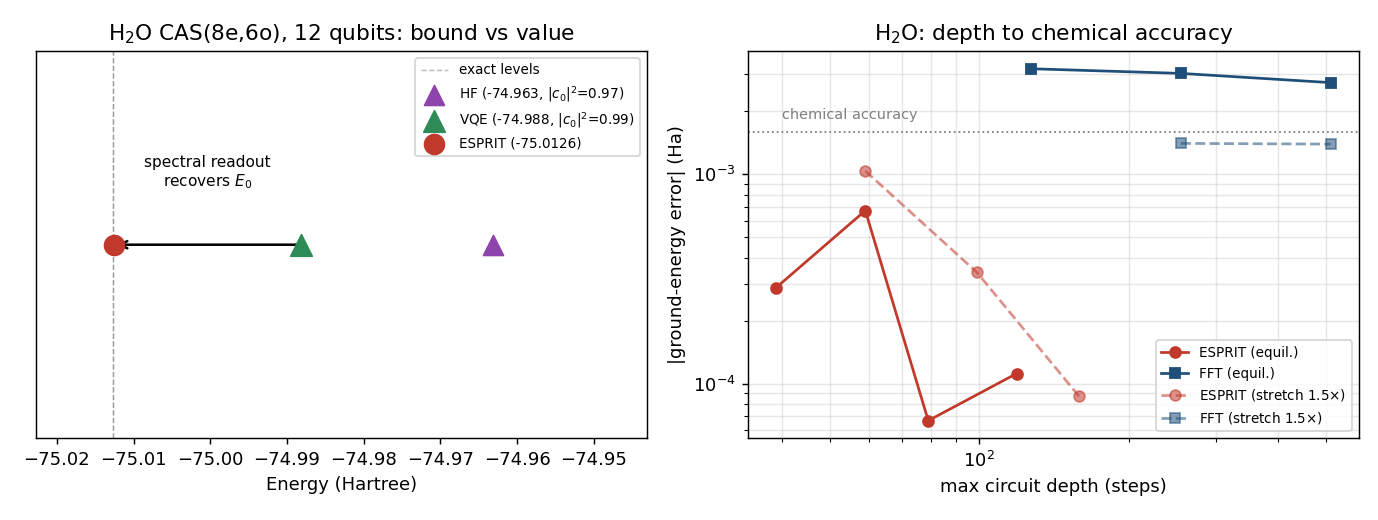}
\caption{\textbf{H$_2$O CAS(8e,6o), 12 qubits.} \textbf{Left:} loose HF and
pair-UCCD VQE bounds. ESPRIT on the VQE state recovers $E_0$. \textbf{Right:}
depth to chemical accuracy at equilibrium (solid) and at a $1.5\times$ symmetric stretch (dashed). The hybrid degrades gracefully into the multireference regime.}
\label{fig:h2o}
\end{figure}

To check that nothing is special to SO$_2$, we repeat the entire pipeline on a
second molecule, H$_2$O, in a CAS(8e,6o) active space (freezing the O$\,1s$ core),
again $12$ qubits, from an RHF reference at $-74.963\,$Ha. At equilibrium the
ground energy is $-75.0127\,$Ha with a comfortable $0.318\,$Ha gap, and the story
matches equilibrium SO$_2$ (Fig.~\ref{fig:h2o}, left). The HF reference is a loose
bound ($+0.050\,$Ha, overlap $0.973$), the nine-parameter pair-UCCD
VQE~\citep{limacher2013pccd} tightens it to $+0.025\,$Ha (overlap $0.987$), and
ESPRIT on the VQE state recovers the true energy below the bound. ESPRIT reaches
$\sim\!10^{-4}\,$Ha by depth $40$ while the FFT plateaus near $3\times10^{-3}$
(Fig.~\ref{fig:h2o}, right), a clean win on a well-separated spectrum.

Symmetrically stretching both O--H bonds to $1.5\times$ gives water its textbook
multireference character. The HF overlap falls to $0.84$ and the gap shrinks, but,
unlike SO$_2$, no nearby symmetry crossing intervenes, so the degradation is
\emph{gradual}. The pair-UCCD VQE (overlap $0.92$) remains a usable warm start and the
hybrid still recovers the ground energy to chemical accuracy at depth $\sim\!100$
(dashed curves, Fig.~\ref{fig:h2o}, right). SO$_2$ and H$_2$O thus display the two faces
of multireference character: a smooth loss of single-reference quality that the hybrid
tolerates, versus an abrupt state crossing that defeats any single-determinant warm
start. In both, the determining quantity is the overlap that state preparation can
secure.

\subsection{Excited states, and the per-line currency}
\label{sec:excited}
Every result so far has used one number out of $g(t)$ and discarded the rest. But
Eq.~\eqref{eq:gt} carries the \emph{whole} spectrum. ESPRIT estimates 
$E_n$ and  $|c_n|^2$ jointly.
Excited states thus cost no extra circuits, only a decision to read the lines one
has already paid for. This matters because excitation energies, not ground energies,
are what spectroscopy measures~\citep{higgott2019excited,ollitrault2020eom}.

\begin{figure*}[t]\centering
\includegraphics[width=\textwidth]{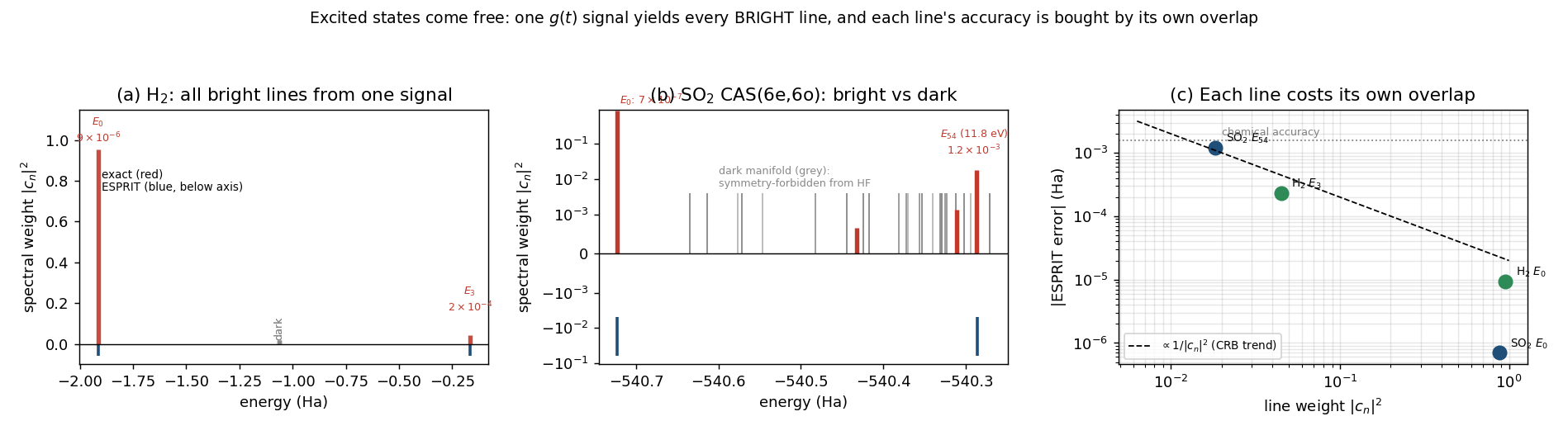}
\caption{\textbf{Excited states from the same signal.} (a) H$_2$: both bright lines
recovered from one $g(t)$. The two triplets are dark from a closed-shell reference.
(b) SO$_2$ CAS(6e,6o). Only four of the sixty lowest states are bright from HF; $E_0$
lands to $7\times10^{-7}\Ha$ and the brightest excited state to chemical accuracy,
while the dark manifold (grey) is invisible at any depth from this reference. (c) Per-line currency. A
line's accuracy is bought by its own weight, following the Cram\'er--Rao trend.}
\label{fig:excited}
\end{figure*}

\paragraph{H$_2$: the full bright spectrum from one signal.} From the Hartree--Fock
start, a single $g(t)$ record ($N=200$, $2^{17}$ shots) returns both bright lines:
$E_0=-1.915370$ to $9\times10^{-6}\Ha$ and $E_3=-0.166816$ to $2\times10^{-4}\Ha$
(Fig.~\ref{fig:excited}a). Their difference, the physical observable, is the excitation
energy $1.748553\Ha$, recovered as $1.748796\Ha$, an error of $2.4\times10^{-4}\Ha$, comfortably inside chemical accuracy.

\paragraph{Selection rules are overlap.} The two degenerate states at
$-1.063653\Ha$ are \emph{invisible}. The closed-shell HF reference has rigorously zero
overlap with them, so they contribute no amplitude and no estimator can recover them. A state is
readable exactly when the warm start overlaps it, so the warm start plays the role that
a transition dipole plays in optical spectroscopy, partitioning the spectrum into
\emph{bright} and \emph{dark}. What to do about a dark state is the subject of
\S\ref{sec:multistart}.

\paragraph{SO$_2$: bright lines are rare, and dim lines are expensive.} On the 12-qubit CAS(6e,6o) the picture sharpens. Of the $60$ lowest eigenstates only
\emph{four} are bright from HF. The six lowest excited states are dark to $10^{-28}$
(Fig.~\ref{fig:excited}b). ESPRIT returns $E_0$ to $7\times10^{-7}\Ha$ and the brightest
excited state ($|c_n|^2=0.018$, an $11.8\,$eV excitation) to $1.2\times10^{-3}\Ha$,
within chemical accuracy, The two dim bright lines ($|c_n|^2\sim10^{-3}$) sit at the
shot-noise floor and are not resolved. The remaining recovered lines are effective,
absorbing the residual weight spread over the higher manifold. Plotting error against
weight (Fig.~\ref{fig:excited}c) shows the trade plainly. Accuracy per line degrades as
its weight falls, consistent with the $\mathrm{Std}(\hat E)\propto|c_n|^{-2}$ of
Appendix~\ref{app:shot}. The currency is thus not only per-state but \emph{per-line}. The
ground energy is cheap because the warm start is designed to overlap it, and each
excited state must purchase its own accuracy with its own overlap. Reading more lines
also demands a model order at least as large as their number, so the hyperparameter of
\S\ref{sec:modelorder} becomes a deliberate setting.

\subsection{Several warm starts, and dark states}
\label{sec:multistart}

A single warm start cannot reveal an eigenstate for which $c_n=0$, and no amount of signal
processing recovers a line that carries no amplitude. The remedy is not a cleverer estimator
but a second acquisition from a reference that lies in the sector concerned, read exactly as
in \S\ref{sec:esprit}. Symmetry, spin or particle number decide which reference sees which
manifold, so a small set of complementary references, each read on its own, covers a spectrum
that none of them covers alone.

The accounting generalises accordingly. With non-negative acquisition weights $\nu_j$ the
relevant resource is
\begin{equation}
p_n^{\rm multi}=\sum_j \nu_j\,|\langle E_n|\chi_j\rangle|^2,
\end{equation}
and a few complementary references trade front-end preparation cost against the probability of
missing a target line, and it is what illuminates the dark manifold of \S\ref{sec:excited}
above. What it costs is one acquisition per reference; what it buys is coverage, and coverage
is the only thing it buys.

\subsection[Vibronic spectra: a band rather than a line]{Vibronic spectra: reading a band rather than a line}
\label{sec:vibronic}

Everything above concerns a Hamiltonian whose eigenvalues we want. There is a second class of
problem in which the same signal appears for a quite different reason, and where the estimator
matters more than it does here, because the sampling window is not cheap.

Simulations of non-adiabatic molecular dynamics propagate a wavepacket under a vibronic
Hamiltonian of K\"oppel--Domcke--Cederbaum form~\citep{koppel1984multimode}, and the
spectroscopic observable is the absorption spectrum, obtained as the Fourier transform of the
dipole autocorrelation $\int\!\d t\,\ee^{-i\omega t}\langle\psi|\ee^{iH_{\rm vib}t}\mu
\ee^{-iH_{\rm vib}t}\mu|\psi\rangle$. Writing $|\chi\rangle=\mu|\psi\rangle/\lVert\mu|\psi\rangle
\rVert$, which is the state such a simulation prepares anyway, and taking $|\psi\rangle$ to be the
vibronic ground state of energy $E_g$, that correlation is
$\lVert\mu\psi\rVert^2\ee^{iE_gt}\langle\chi|\ee^{-iH_{\rm vib}t}|\chi\rangle$. It is $g(t)$
exactly, with the dipole-excited state as the warm start. The overlap question that governs
everything else in this paper does not arise, since $\chi$ is a Franck--Condon state and cheap;
what is expensive is the propagation. The propagation time of such a simulation \emph{is} the
acquisition length $\tau$ of the readout, and we use the two names for the one quantity, the first
because it is what the simulation literature calls it and the second because it is what
\S\ref{sec:fft} calls it.

That changes the arithmetic. Read with a transform, the resolution is $2\pi\hbar/\tau$, which at
$\tau=100\,$fs is $41\,$meV, or $334\,\mathrm{cm}^{-1}$ --- coarser than a
vibrational quantum, and far coarser than the tunnelling and conical-intersection splittings such
simulations are run to see. A recent resource estimate for this
problem~\citep{motlagh2024vibronic} costs the propagation as $t^{1+1/q}$ for a $q$th-order product
formula, and its own tables bear the exponents out, a fivefold increase in $t$ costing
$12.8\times$ against $5^{3/2}=11.2$ at second order, and a tenfold tightening of the Trotter
error tolerance $\epsilon_{\rm T}$ costing $3.16\times$ against the
$\epsilon_{\rm T}^{-1/2}$ that formula predicts. The propagation time is therefore
amplified to the power $3/2$ in the gate count, and every factor saved on $\tau$ is worth more than
a factor.

\begin{figure*}[t]
\centering
\includegraphics[width=0.92\textwidth]{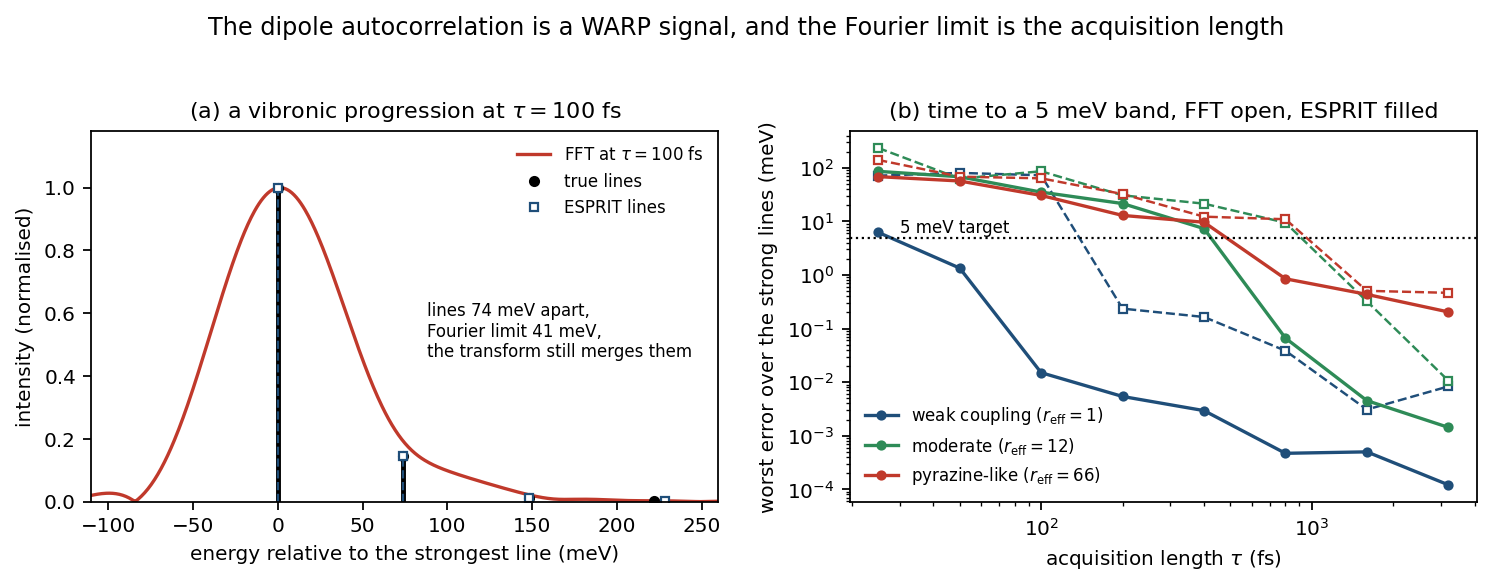}
\caption{\textbf{Vibronic dynamics.} \textbf{Left:} at $\tau=100\,$fs the windowed transform merges
a $74\,$meV splitting into one band with a shoulder, while ESPRIT places both lines on the same
samples, the stronger to $0.002\,$meV and the weaker to $0.04\,$meV. \textbf{Right:} worst error
over the strong lines against acquisition length, open squares for the transform and filled circles
for ESPRIT.}
\label{fig:vibronic}
\end{figure*}

Figure~\ref{fig:vibronic} measures what a parametric estimator saves on exactly this signal. We
build two-state linear vibronic coupling models, take the Franck--Condon state as $\chi$, sample
its autocorrelation under the binomial statistics of the interferometer at $10^6$ shots, and ask
what acquisition length each estimator needs to place \emph{every} line carrying at least two per
cent of the band to within $5\,$meV.

\begin{table}[t]
\centering
\footnotesize
\setlength{\tabcolsep}{4pt}
\begin{tabular}{lrrrr}
\hline
model & $r_{\rm eff}$ & lines & FT & ESPRIT \\
\hline
two modes, weak     &  1.3 &  2 &  $200\,$fs & $50\,$fs \\
two modes, moderate & 11.9 & 14 & $1600\,$fs & $800\,$fs \\
pyrazine-like       & 66.0 &  7 & $1600\,$fs & $800\,$fs \\
\hline
\end{tabular}
\caption{Propagation time needed to place every line carrying at least two per cent of the band
to within $5\,$meV, for the Fourier transform (FT) and for ESPRIT. The time ratios of $4$ and $2$
are gate-count ratios of $8$ and $2.8$ under the $t^{3/2}$ scaling of a second-order product
formula.}
\label{tab:vibronic}
\end{table}

The gain is real and it is bounded. On the sparse progression the transform needs four times the
acquisition length, which is eight times the gates. On congested bands the advantage falls to a
factor of two in time and under three in gates (Table~\ref{tab:vibronic}). The reason for the ceiling is  the same resource argument as \S\ref{sec:resource} in another guise. The
sampling interval must resolve the whole band, so the number of samples available is
$N\approx \tau\Omega/\pi\hbar$ for a bandwidth $\Omega$, and a subspace estimator can fit at most $N/2$
modes. A band of $1.5\,$eV sampled for $100\,$fs affords about a hundred samples and therefore at
most fifty lines, whatever the estimator. Super-resolution buys the resolution of a line, never
the number of them. What decides whether a given molecule falls on the useful side is the
spectral complexity of its Franck--Condon state, which is $r_{\rm eff}$, computable classically
from the vibronic parameters before any quantum resource is spent. On these three models it
ordered the outcomes correctly, which suggests it as a triage criterion for the candidate
screening such pipelines are built around. Three systems are too few to establish the rule,
so we offer it as a hypothesis worth testing rather than as a result.

Three further pieces of the machinery in this paper transfer to that setting, two of them
developed above and one below. The transition-amplitude measurement of \S\ref{sec:homodyne},
given in Part~\ref{part:together}, takes the spin--orbit operator of a
spin-vibronic Hamiltonian as its $O$, so an intersystem-crossing amplitude, whose square is
below the floor precisely because the transition is forbidden, is read at linear sensitivity by
the same acquisition that resolves the band. The population traces
of a non-adiabatic simulation are sums of damped exponentials at difference frequencies, and the
rates fitted from them are the physical observable those pipelines exist to produce, so the
damped model of \S\ref{sec:adaptive} is the estimator for them. And on hardware the fitted decay is the physical rate plus the device's own decoherence, so
a rate read off a noisy device without a separation step is not a rate. That separation is only
partly a matter of calibration. The common-mode part of the decoherence can be estimated and
removed, best by amplifying the noise on the vibronic Hamiltonian itself and extrapolating
to zero. But any part that acts differently on different diabatic populations does not
transfer from a calibration on another Hamiltonian. That part has to be reported alongside
the rate.

None of this is a new idea about signals. Extracting eigenvalues from short segments of a
molecular autocorrelation by filter diagonalisation is a mature classical
technique~\citep{wall1995filter,mandelshtam1997harmonic}, developed for this exact observable and
for this exact reason. What is new is the setting, in which the signal is acquired on a quantum
processor at a cost that grows as $t^{3/2}$, so that the estimator's efficiency in $\tau$ is
converted into gate count at an exponent greater than one.

\warppart{The preparation back-end}{The same warm start, spent instead on a
state. The driver is anchored on whatever the front-end produced, and the schedule
length is fixed by the same overlap.}
\label{part:prep}

\section[Warm-start-anchored adiabatic preparation]{Warm-start-anchored adiabatic state preparation}
\label{sec:aqc}

Part~\ref{part:readout} spent the warm start on an \emph{energy} and left no state
behind. The natural question is whether the same overlap can instead be spent to
\emph{prepare} the true ground state, from which every observable follows. It can, through
an adiabatic construction that is the exact preparation-side mirror of the readout hybrid
and that exposes the overlap dependence even more transparently.

\paragraph{Construction.} Take any warm start $\ket{\chi}$. In the illustrations that
follow it is a converged variational state $|\chi\rangle=U_{\chi}|\mathrm{ref}\rangle$, but
the construction needs only a state, not a variational one, and it is anchored on
Hartree--Fock, annealer-style, Boltzmann and QAOA states elsewhere in this paper. Define
the driver Hamiltonian as the projector
\begin{equation}
H_I \;=\; \mathbf{1} - |\chi\rangle\langle\chi|,
\end{equation}
whose unique ground state (eigenvalue $0$) is $|\chi\rangle$ itself, with
the entire orthogonal complement degenerate at eigenvalue $1$. As the projector in general does not commute with $H_P$ (except pathological cases), we start in
$|\chi\rangle$, interpolate $H(s)=(1-s)\,H_I + s\,H_P$ from $s\;{=}\;0$ to $1$. The adiabatic theorem (Appendix~\ref{app:adiabatic}) then
delivers the ground state of the (shifted, rescaled) problem Hamiltonian $H_P$.
The driver is anchored to the supplied warm state, so
\emph{warm-start-anchored adiabatic state preparation} (WA-ASP). Crucially the
projector is not an abstract operator. Since $|\chi\rangle=U_{\chi}
|\mathrm{ref}\rangle$,
\begin{equation}
\ee^{\,i\theta|\chi\rangle\langle\chi|}
= U_{\chi}\,\ee^{\,i\theta|\mathrm{ref}\rangle\langle\mathrm{ref}|}\,U_{\chi}^\dagger,
\end{equation}
a phase on a single computational-basis state sandwiched between the $U_{\chi}$ circuit
and its inverse~\citep{rolandcerf2002}. The cost of $U_{\chi}$ and $U_{\chi}^\dagger$, together
with the implementation of the controlled phase, is part of the end-to-end resource cost; the
shorter adiabatic path does not make preparation of an arbitrary warm start free. Here, that idea
feeds an adiabatic state preparation~\citep{aspuru2005simulated,farhi2000adiabatic,albash2018adiabatic}.

\paragraph{Relation to warm-started QAOA.} Anchoring the driver on the warm start has a
direct precursor in combinatorial optimisation. The authors of~\citep{egger2021warm} warm-start QAOA from the solution $c^\star$ of a continuous
relaxation, and crucially they also replace the standard mixer $-\sum_i\hat X_i$ by a
modified one constructed so that the warm start is its ground state. They invoke the
adiabatic theorem for convergence, and their Trotterised annealing reaches the minimum in
less total time and fewer steps than the equal-superposition mixer. That is similar to the structural idea as our projector above.

Three things separate the present construction, and they are what the chemistry demands.
First, their mixer is a \emph{sum of single-qubit terms}, so the state it anchors on is a
product state $\bigotimes_i\hat R_Y(\theta_i)|0\rangle$. That is forced, since a relaxation
returns $c^\star\in[0,1]^n$ coordinate by coordinate. Ours is a rank-one projector onto an
\emph{arbitrary preparable} $|\chi\rangle$, entangled or not. For a correlated molecular
ground state their product reference has vanishing overlap. Our generalisation also removes their reachability pathology, though on a diagonal cost
Hamiltonian it does not remove every obstruction of that kind. When $c_i^\star\in\{0,1\}$
the single-qubit mixer term becomes $\pm\hat Z_i$, which commutes with a diagonal cost
Hamiltonian, so that qubit is frozen for all time; this is why they introduce the
regularisation $\varepsilon$. Our projector $-|\chi\rangle\langle\chi|$ is not a product of
local terms and has no such single-qubit fixed point. It does, however, inherit a weaker
obstruction in the same setting. A diagonal $\hat H_P$ preserves computational-basis
support, and $|\chi\rangle\langle\chi|$ maps into $\mathrm{span}\{|\chi\rangle\}$, so
evolution from $|\chi\rangle$ is confined to
$\mathrm{span}\{\hat H_P^{\,m}|\chi\rangle\}\subseteq\mathrm{supp}(\chi)$ for every
schedule and every total time. Any basis state on which $\chi$ has strictly zero amplitude
is unreachable, and a warm start regularised in the spirit of $\varepsilon$ is needed
there too.\footnote{Both statements are numerically exact in the diagonal case. Writing
$H(s)=s\hat H_P-(1-s)|\chi\rangle\langle\chi|$ with $\hat H_P$ diagonal, the dynamics from
$|\chi\rangle$ closes on $\mathrm{span}\{\hat H_P^{\,m}|\chi\rangle\}$, whose dimension is
the number of \emph{distinct} cost values rather than $2^n$. On Max-Cut instances at
$n=12$ this is a collapse from $4096$ to $23$, and split-step evolution in the reduced
space agrees with the full $2^n$ evolution to $2\times10^{-16}$. With a marginal saturated
at a value that no optimum shares, both the single-qubit mixer and the projector return
success probability identically zero out to $T=2000$.} The support argument is specific to
diagonal cost Hamiltonians and does not apply at its first step for the molecular $H_P$ of
\S\ref{sec:aqc}, which does not preserve computational-basis support. There the projector
construction is unobstructed. Second, their cost Hamiltonian is diagonal, so the
eigenstates are computational-basis strings and the target is a bitstring. There is no
spectrum to read, and no analogue of the readout back-end exists in that setting. The
spectral half of this paper has content only when $H_P$ does not commute with the
measurement basis. Third, warm-started QAOA \emph{inherits a classical
approximation guarantee}. Their $\varepsilon$, interpolating
continuously between warm and cold, is the closest thing in that work to our  overlap dial
of \S\ref{sec:qaoa}, and their annealing comparison is an empirical precursor of the gap
law derived next.

\begin{figure}[t]
\centering
\includegraphics[width=\linewidth]{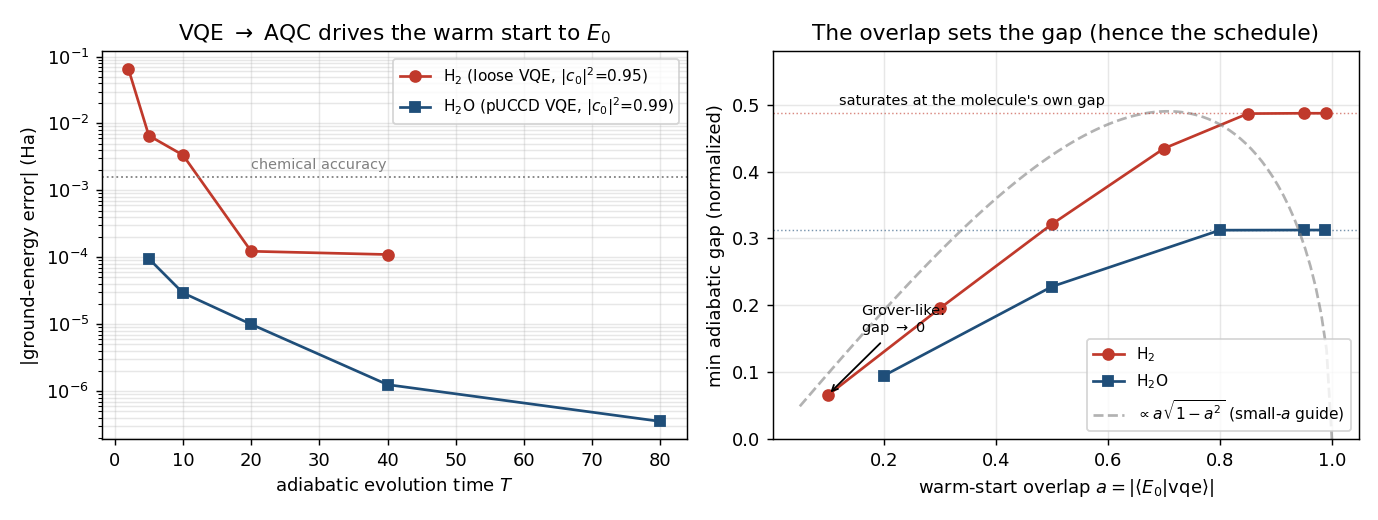}
\caption{\textbf{Warm-start-anchored adiabatic preparation.} \textbf{Left:}
the VQE warm start is driven to $E_0$ (overlap $\to1$) below chemical accuracy.
\textbf{Right:} the minimum adiabatic gap rises with warm-start overlap and
saturates at the molecule's intrinsic gap; at low overlap it collapses
(Grover-like).}
\label{fig:aqc}
\end{figure}

\begin{figure}[t]\centering
\includegraphics[width=\linewidth]{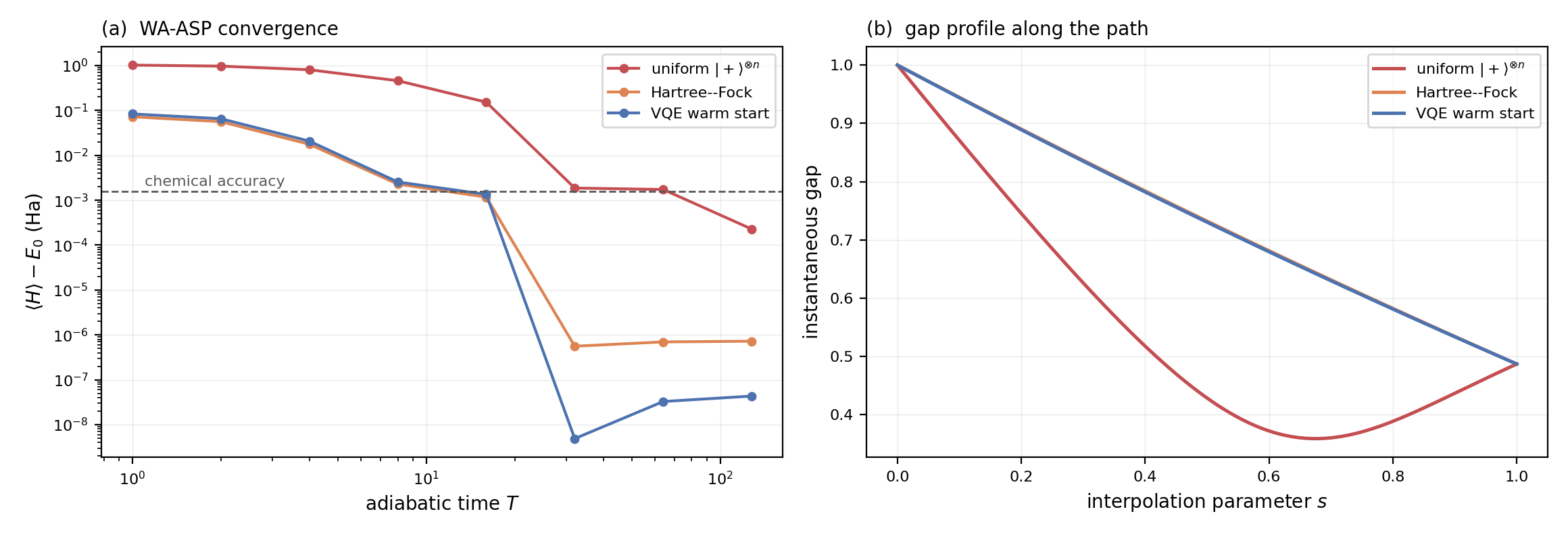}
\caption{\textbf{Pure vs warm-started AQC (H$_2$).} \textbf{Left:} the transverse-field/uniform start needs ${\sim}8\times$ the adiabatic time of either warm start. Schedule times here are the first member of $\{1,2,4,\dots,128\}$ to reach chemical accuracy, so the ratio is resolved only to the nearest power of two ($128$ against $16$), and this comparison uses the linear schedule rather than the local-adiabatic one of \S\ref{sec:schedule}. \textbf{Right:} its gap profile dips into a bottleneck ($0.36$) while the projector warm starts ride a flat, high gap ($0.49$) to the molecular value.}
\label{fig:pureaqc}
\end{figure}

\paragraph{The overlap sets the gap.} \label{sec:gaplaw} Restricting to the
two-dimensional span of the true ground $|E_0\rangle$ and the orthogonal part of the
warm start, with overlap amplitude $a=|\langle E_0|\chi\rangle|$, the
avoided-crossing gap near the crossing is
$\Delta_{\min}\!\approx\!2(1-s^\star)\,a\sqrt{1-a^2}$, where $s^\star$ is the position of the minimum gap. Numerically (Fig.~\ref{fig:aqc},
right) the minimum gap along the path \emph{rises} from near zero at small $a$, the
adiabatic-Grover regime where the schedule blows up, and \emph{saturates at the
molecule's own gap} once $a$ is good ($0.49$ for H$_2$, $0.31$ for H$_2$O in normalised
units). A converged VQE thus does not only help, it places one in the flat region where
the schedule length is set by the molecule, not by the warm start. With the loose H$_2$
state ($|c_0|^2{=}0.95$, $57\times$ chemical accuracy) and the H$_2$O pair-UCCD state
($|c_0|^2{=}0.99$), a plain linear schedule drives the energy below chemical accuracy by
$T\!\approx\!20$ and $T\!\approx\!5$ respectively, with the overlap climbing to unity
(Fig.~\ref{fig:aqc}, left). The state is genuinely repaired. A
cold, Hartree--Fock--anchored, adiabatic run is the natural control. On stretched H$_2$O
it needs about twice the adiabatic time for chemical accuracy ($T\!\approx\!40$) as the
VQE-anchored version ($T\!\approx\!20$), the advantage growing with correlation as the
reference overlap drops. A textbook \emph{pure} adiabatic run, transverse-field driver
$H_I=\sum_i(\mathbf{1}-X_i)/2$ with the uniform $\ket{+}^{\otimes n}$ start, carrying no
problem knowledge, fares far worse. On H$_2$ that start overlaps the ground by only
$0.15$, opening a gap bottleneck (Fig.~\ref{fig:pureaqc}) that pushes the time to
chemical accuracy to $T\!\approx\!128$ against $T\!\approx\!16$ for either warm start.
The hierarchy thus is: pure $\ll$ HF $\approx$ VQE for single-reference systems, the HF/VQE
distinction emerging only under multireference correlation.

\paragraph{The gap law off the molecules.} The derivation above assumes only a
two-dimensional avoided crossing, so it should hold wherever the driver is anchored,
including on problems this paper does not otherwise treat. Max-Cut is the sharpest
available test: the cost Hamiltonian is diagonal, the readout back-end has no content
there (\S\ref{sec:aqc}), and the overlap can be dialled continuously by tempering the
Goemans--Williamson marginals $c^\star$ toward $1/2$ before building
$|\chi\rangle=\bigotimes_i\hat R_Y(\theta_i)|0\rangle$, which sweeps $a^2$ over three
decades at fixed instance. Two features of the diagonal case are helpful. A rank-one
perturbation of a diagonal matrix has its entire spectrum given by the roots of the
secular equation $1=(1-s)\sum_k|\chi_k|^2/(s\,d_k-E)$, where $d_k=\bra{z_k}H_P\ket{z_k}$
is the cut value of computational basis state $\ket{z_k}$ and $\chi_k=\braket{z_k|\chi}$ is
the amplitude the warm start places on it, so no eigensolver is required; and
the dynamics closes on $\mathrm{span}\{\hat H_P^{\,m}|\chi\rangle\}$, of dimension equal
to the number of distinct cut values rather than $2^n$.

\begin{figure}[t]\centering
\includegraphics[width=\linewidth]{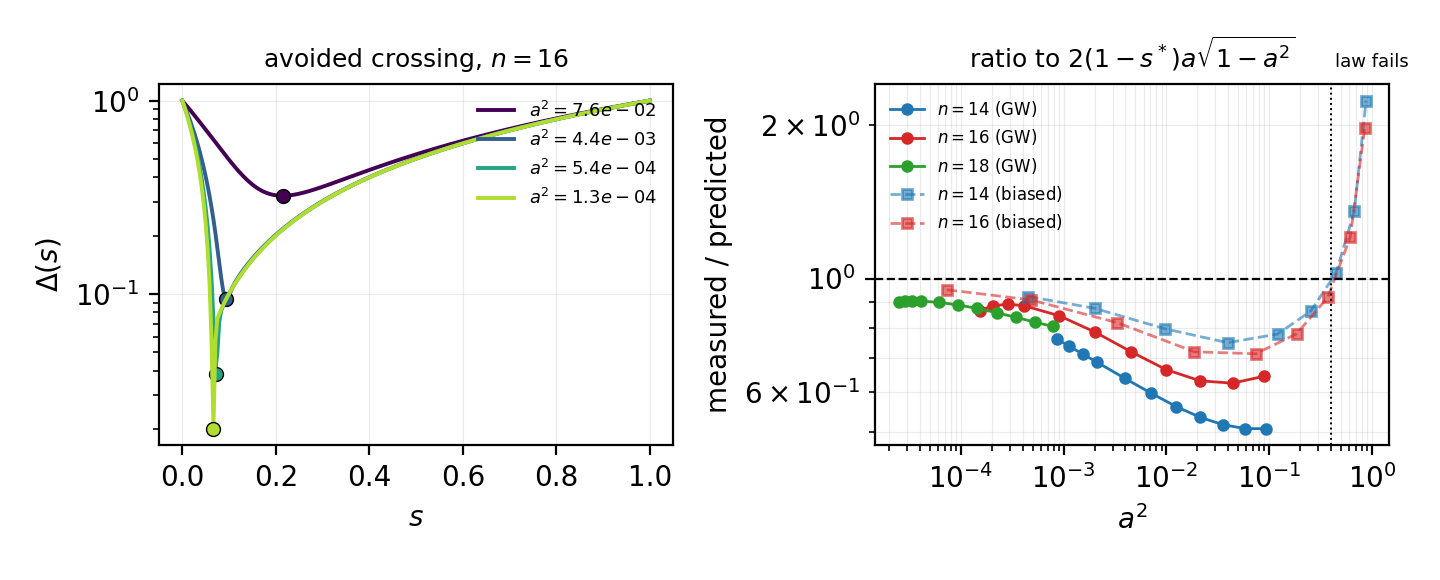}
\caption{\textbf{The gap law on Max-Cut.} \textbf{Left:} the gap $\Delta(s)$ along the
sweep for one $n=16$ instance at four warm-start overlaps, markers at $\Delta_{\min}$.
Lowering $a^{2}$ deepens the crossing and moves $s^\star$ toward $s=0$. \textbf{Right:}
measured $\Delta_{\min}$ divided by the prediction $2(1-s^\star)a\sqrt{1-a^{2}}$ evaluated
at the measured $s^\star$. Solid: warm starts from tempered Goemans--Williamson marginals.
Dashed: warm starts biased toward the known optimum, an independent construction that
reaches overlaps GW tempering cannot, and which agrees with it where the two overlap.
The prediction is an upper bound below $a^{2}\approx0.4$ (dotted) and fails above it,
where the factor $\sqrt{1-a^{2}}$ turns the prediction over while the true gap continues
to rise.}
\label{fig:mcgap}
\end{figure}

Figure~\ref{fig:mcgap} (left) shows the resulting $\Delta(s)$ on a single $n=16$
instance as $a^{2}$ is dialled down. The crossing sharpens and $s^\star$ migrates toward
the start of the sweep, the behaviour the two-level picture predicts. The right panel
tests the prediction quantitatively, comparing $\Delta_{\min}$ against
$2(1-s^\star)a\sqrt{1-a^{2}}$ with $s^\star$ taken from the same run, on random $G(n,1/2)$
instances at $n=14,16,18$. The prediction is an upper bound throughout this ensemble, whose overlaps all lie below $a^2\approx10^{-1}$, and it tightens
in exactly the regime where the two-level restriction should be justified: the ratio of
measured to predicted rises from $0.51$ at $a^{2}\!\approx\!10^{-1}$, where neglected
levels contribute, to $0.90$ at $a^{2}\!\approx\!10^{-4}$, where a single crossing
dominates. Pooling all $33$ measurements gives a log--log slope
$\d\log\Delta_{\min}/\d\log a^{2}=0.408$, rising to $0.462$ on the subset with
$a^{2}<10^{-3}$ and thus approaching the predicted $1/2$ as the two-level regime is
entered. We attribute no part of this to $n$. Because the accessible overlap falls
steeply with size, each $n$ populates a different window of $a^{2}$ (at $n=18$ the
tempering reaches only $a^{2}<10^{-3}$), so a per-$n$ slope would report the window and
not the size. The controlled statement is that the ratio is governed by $a^{2}$, the
three sizes agreeing wherever their windows overlap. The law therefore transfers intact
to a setting with no molecular structure, no spectral readout and a diagonal $H_P$,
which is evidence that the overlap-as-currency claim of this paper is a statement about
the anchored construction rather than about chemistry.

We draw no benchmarking conclusion from this. With the local-adiabatic schedule of
\S\ref{sec:schedule} the resulting time to solution is $T\sim1/a$ against the $1/a^{2}$
samples that rejection sampling from the same $|\chi\rangle$ would need, a Grover-like
square root over that particular baseline. But $a^{2}$ itself decays exponentially in
$n$, and the relevant classical baselines for Max-Cut are not rejection
samplers~\citep{myklebust2015sa,goto2021sbm}. The point here is the scaling law, not a
competitive claim.

\begin{figure}[t]
\centering
\includegraphics[width=\linewidth]{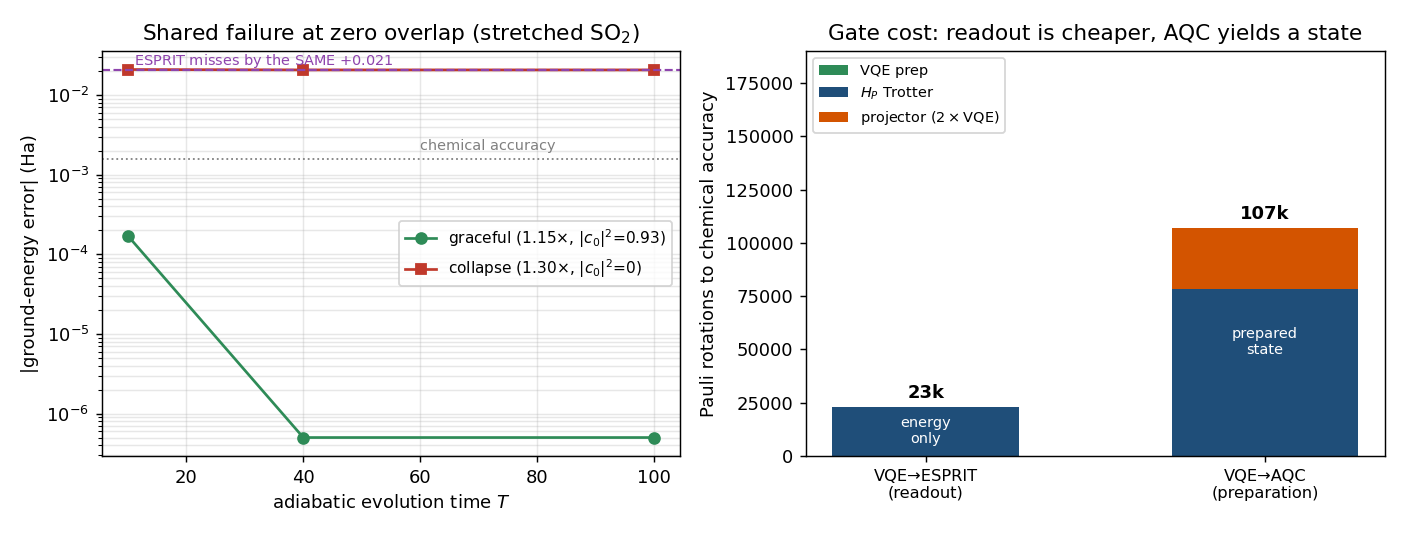}
\caption{\textbf{Shared failure and relative cost.} \textbf{Left:} at zero overlap
WA-ASP misses $E_0$ by the same $+0.021\,$Ha as ESPRIT, for every $T$; at
$1.15\times$ it recovers $E_0$. \textbf{Right:} gate budget to chemical accuracy on
SO$_2$. The readout is ${\sim}5\times$ cheaper, but WA-ASP outputs a state.}
\label{fig:aqccost}
\end{figure}

\paragraph{The same failure, and the same cause.} Because WA-ASP spends exactly the
overlap that the readout hybrid spends, it inherits exactly the readout failure. On the
zero-overlap stretched SO$_2$ point (\S\ref{sec:so2mr}) the true ground lies in a
symmetry sector disconnected from the closed-shell warm start. The path develops a
genuine level crossing ($\Delta_{\min}{=}0$), and WA-ASP converges, for \emph{every}
schedule length $T$, to the lowest state the warm start can reach, thus missing $E_0$ by
$+0.021\,$Ha. This is the \emph{identical} $+0.021\,$Ha by which ESPRIT missed the same
point (Fig.~\ref{fig:aqccost}, left): two different algorithms, one shared blind spot.
At the graceful $1.15\times$ stretch, where overlap survives, both succeed.

\paragraph{What it costs.} Compiling the explicit circuit on SO$_2$
(\S\ref{sec:trotter}), one WA-ASP step is $392$ off-diagonal $H_P$ rotations plus a
projector step of $2\times72=144$ rotations, twice the pair-UCCD circuit. The projector
adds only ${\sim}37\%$ over a bare $H_P$ layer, so the molecular Hamiltonian still
dominates. Reaching chemical accuracy ($T{\approx}15$--$20$, ${\sim}200$ Trotter steps)
costs ${\sim}10^5$ rotations, roughly $5\times$ the readout route's
$\sim\!2.3\times10^4$ (Fig.~\ref{fig:aqccost}, right). The trade is 
complementary rather than competitive. The readout is cheaper and returns only an
energy, plus an ancilla and controlled evolution. While WA-ASP is dearer but leaves a
\emph{prepared ground state}, reusable for any observable. Thus, spend overlap to read, or
spend more of it to prepare.

\subsection{Anchoring on a start that is not a circuit}
\label{sec:anchor}
The driver of \S\ref{sec:aqc} is $H_I=\mathbf 1-\ket\chi\bra\chi$, realised as
$U_\chi R_{\rm ref}U_\chi^\dagger$. It therefore needs a unitary $U_\chi$ with
$U_\chi\ket{\rm ref}=\ket\chi$. Two questions follow for a front-end that is not a
variational circuit: what if the start is not a circuit at all, and what if it is not
pure.

\paragraph{Compiling a classical description.} The driver needs a \emph{circuit}, not an
\emph{optimised} circuit, and that distinction disposes of most cases. A QAOA circuit at
any depth, an operator sequence emitted by a generative model, a Trotterised
imaginary-time or short adiabatic segment, a hardware-efficient ansatz at any parameter
setting. All are circuits already and serve as $U_\chi$ unchanged, optimality of the
parameters being irrelevant to the construction. Where the start arrives as a classical
description it is compiled. A matrix-product state of bond dimension $\beta$ on $n$ sites
is realised exactly by a sequential circuit in $\mathcal O(n\beta^2)$ two-qubit
gates~\citep{schon2005sequential}. A sparse vector of $K_{\rm sp}$ computational-basis
amplitudes, such as the leading configurations of a selected configuration-interaction (CI)
expansion or the best strings returned by a classical heuristic, is realised in
$\mathcal O(K_{\rm sp}n)$~\citep{gleinig2021sparse}. A single Slater determinant is realised by a
network of Givens rotations in $\mathcal O(n^2)$~\citep{kivlichan2018givens}, and a
multi-determinant reference by a linear combination of unitaries over such networks. What governs the schedule is then the
overlap of the \emph{compiled} state. Fitting a circuit at fidelity $F$ to a description
whose own overlap is $a$ leaves the back-end spending
$|\!\braket{E_0|\chi_{\rm fit}}\!|$, so compilation error is a budgeted term in the
resource accounting rather than a hidden loss.

\paragraph{A start that is not pure.}\label{sec:mixedstart} If the front-end returns a mixed
$\rho=\sum_i\nu_i\ket{\chi_i}\bra{\chi_i}$, ordered $\nu_1\ge\nu_2\ge\dots$ 
then $\ket\chi\bra\chi$ is undefined. Taking the driver to be $\mathbf 1-\rho$ is the
natural generalisation, and it is instructive because it fails in a specific and
quantifiable way. The eigenbasis of $H(0)=\mathbf 1-\rho$ is the eigenbasis of $\rho$, with
$\ket{\chi_1}$ the instantaneous ground state, $\ket{\chi_2}$ the first excited, and so on.
Perfect adiabatic following therefore maps $\ket{\chi_i}\mapsto\ket{E_{i-1}}$ level by
level, and the final ground-state population is
\begin{equation}
\lim_{T\to\infty}\operatorname{Tr}\!\big[\ket{E_0}\!\bra{E_0}\,\rho(T)\big]=\nu_1 .
\label{eq:purityceiling}
\end{equation}
This is a \emph{ceiling}, not a rate. It cannot be bought off with a longer schedule, and
it does not depend on $a_1=|\!\braket{E_0|\chi_1}\!|$, which continues to govern the
avoided crossing and hence the time to approach the ceiling. Mixedness and poor overlap
are thus different currencies. Overlap costs time, purity caps fidelity. A second and
independent penalty follows from the same construction, since $H(0)$ now carries its own
gap $\nu_1-\nu_2$, which for a high-entropy $\rho$ creates a bottleneck near
$s=0$ in addition to the usual crossing near $s^\star$.

Table~\ref{tab:mixed} confirms Eq.~\eqref{eq:purityceiling} on the $n=10$ TFIM. Three
starts share the same principal overlap $a_1^2=0.800$ and differ only in their spectra.
The pure case converges to $1$, and the two mixed cases converge to $0.7996$ and $0.5989$
against ceilings of $0.80$ and $0.60$, the approach in every case being from below and
governed by $a_1$.

\begin{table}[t]\centering\small
\setlength{\tabcolsep}{4pt}
\begin{tabular}{lccccc}
\toprule
$\{\nu_i\}$ & $T{=}5$ & $T{=}20$ & $T{=}80$ & $T{=}300$ & ceiling\\
\midrule
$(1,0,0)$ & $0.921$ & $0.976$ & $0.997$ & $0.9998$ & $1$\\
$(0.80,0.15,0.05)$ & $0.723$ & $0.770$ & $0.795$ & $0.7996$ & $0.80$\\
$(0.60,0.25,0.15)$ & $0.531$ & $0.565$ & $0.591$ & $0.5989$ & $0.60$\\
\bottomrule
\end{tabular}
\caption{\textbf{Purity caps fidelity; overlap sets the rate.} Ground-state population
after WARP from a mixed start on the $n=10$ TFIM, all three starts sharing
$a_1^2=0.800$. The limit is $\nu_1$ in each case, independent of $a_1$.}
\label{tab:mixed}
\end{table}

\paragraph{Four ways around the ceiling.} Eq.~\eqref{eq:purityceiling} is what motivates
the remedies. \emph{First}, and the default for a
noisy device, anchor the driver on the \emph{intended} unitary of the preparing circuit
rather than on the state the device produced. $U_\chi$ is well defined whatever the
fidelity of its execution, the decoherence is an error on the driver in common with every
other gate, and no ceiling arises because no mixed object enters the construction.
\emph{Second}, where $H_P$ is diagonal and the front-end is a classical or annealing
sampler, the members of $\rho$ are themselves basis states, $\ket{\chi_k}=\ket{z_k}$ for
sampled strings $z_k$ whose observed frequencies are the weights $\nu_k$. Re-prepare the
histogram coherently as $\ket\chi\propto\sum_{k\le K_{\rm sp}}\sqrt{\nu_k}\ket{z_k}$, with
$K_{\rm sp}$ the number of distinct strings retained, that is the sparsity of the
superposition. The eigenstates of a diagonal $H_P$ are themselves basis states, so this
carries the sampler's success probability across intact, $a^2=\nu_{\rm opt}$, and converts
an incoherent front-end into a pure one at the cost of
$\mathcal O(K_{\rm sp}n)$ gates. \emph{Third}, in general, prepare a purification
$\ket\Psi=\sum_i\sqrt{\nu_i}\ket{\chi_i}\ket{i}$ on system plus ancilla. Then
$\mathbf 1-\ket\Psi\bra\Psi$ is a genuine projector on the enlarged space, $H_P$ dilates
as $H_P\otimes\mathbf 1$, its ground eigenspace is $\ket{E_0}\otimes\mathbb C^r$ with $r=\operatorname{rank}\rho$ the dimension of the ancilla register, and the
ancilla is traced out at the end, the whole construction of \S\ref{sec:aqc} applying
unchanged. \emph{Fourth}, symmetry verification discards the component of $\rho$ violating
a conserved quantity of $H_P$ and so restores purity in part, at a sampling overhead,
before the driver is applied.

\paragraph{What the ensemble-averaged residue does not give.} An acquisition on a mixed start returns
$g(t)=\mathrm{Tr}[\rho\,\ee^{-iH_Pt}]=\sum_n(\sum_i\nu_i|\braket{E_n|\chi_i}|^2)\ee^{-iE_nt}$,
whose residue is the \emph{ensemble-averaged} weight. That residue is larger than the weight a
weak member carries alone, so one asks whether the mixture can serve the weak member as the
strong reference of \S\ref{sec:homodyne} serves the weak one. It cannot. The homodyne residue is a \emph{product} $c_S^*c_W$ and is
linear in the small amplitude, the strong reference multiplying the weak one. This residue is a
\emph{sum}, and a sum buries rather than amplifies. On three members carrying $0.90$, $0.08$ and
$10^{-5}$ on one line under mixture weights $0.70$, $0.25$, $0.05$, the averaged residue is
$0.65$ --- some $6.5\times10^{4}$ times the weak member's own weight --- and yet setting that
member's weight to \emph{zero} moves the averaged residue by eight parts in ten million. The
excess is other members' weight, carrying no information about the weak one. A mixture is a
background, not a gain.

Two things do follow. Where the mixture weights are adjustable and known, acquisitions at several
settings form a linear system in the individual weights and invert for them; the system is
well-conditioned, but every measured quantity is still a squared overlap, so the repetition count
still scales as $|c|^{-4}$ with a condition number on top. Separation, not sensitivity. And by
contrast the purification above \emph{does} restore linearity, because it restores coherence. On
$\ket\Psi$, an interference acquisition against a strong reference with the label register
projected onto $\ket{j}$ carries the residue $\sqrt{\nu_j}\,c_S^*c_j$, linear in the weak
amplitude and attenuated only by $\sqrt{\nu_j}$. On the ensemble above that is $8.9\times10^{-6}$
against a diagonal $10^{-8}$, a gain of nearly three orders of magnitude, and \S\ref{sec:homodyne}
then applies to it unchanged. The route from a mixed front-end to a weak overlap runs through the
purification, not through the averaged residue.

 Three families deserve particular emphasis. \emph{Multi-determinant references},
CASSCF/CASCI~\citep{roos1980casscf}, selected CI~\citep{huron1973cipsi},
symmetry-projected Hartree--Fock, or geminal/AGP states, compiled by sparse
multi-determinant preparation~\citep{tubman2018postponing,fomichev2024initial}, are the
direct remedy for the zero-overlap collapse of \S\ref{sec:so2mr}. This is an artifact of
the closed-shell \emph{reference}, not of the method: where a single determinant loses
the ground state at a crossing, a few determinants retain $\mathcal O(1)$ overlap.
\emph{Eigenstate filtering} (Chebyshev/Gaussian filters, LCU, or
QSP/QET~\citep{lin2020nearoptimal,ge2019faster}) is not a source but an overlap
\emph{amplifier}, buying $|c_0|^2$ with coherent depth, a third exchange rate beside the
QAOA-depth and adiabatic-time rates measured here. The  \emph{sample-based
diagonalisation}~\citep{robledo2024sqd}, \emph{embedding/downfolding}
(DMET~\citep{knizia2012dmet}), \emph{dissipative or Zeno
preparation}~\citep{verstraete2009dissipative}, and \emph{Hamiltonian continuation} from
a solvable limit of the same family each supply overlap by a different physical route.
This generality is exercised in Part~\ref{part:frontends}, where the protein-folding
example of \S\ref{sec:apps} seeds WA-ASP from an annealer-style state rather than a VQE. Each source plugs into
the same overlap accounting and the same gap law (App.~\ref{app:adiabatic}). The choice
of source changes the overlap one starts with, not the machinery that spends it.

\subsection{Shaping the path, accelerating the schedule}
\label{sec:schedule}
Two freedoms must be separated. The adiabatic \emph{path} is the curve $H(s)$ traced through
operator space as $s$ runs $0\!\to\!1$. The \emph{schedule} is the law $s(t)$ by which
$s$ is advanced. The linear interpolation used so far,
$H(s)=(1-s)H_I+s\tilde H_P$, is one path among many. More generally
$H(s)=f_I(s)H_I+f_P(s)\tilde H_P\,[+f_C(s)H_C]$ for any continuous $f_I,f_P$ with
$f_I(0)>0$, $f_I(1)=0$, $f_P(0)=0$ and $f_P(1)>0$, the subscripts naming the driver, the
problem and a catalyst, so that $g$ is left to the correlation function alone. The optional
catalyst term vanishes at both endpoints, $f_C(0)=f_C(1)=0$, and may be
non-stoquastic~\citep{albash2016nonstoquastic}.
Nonlinear $f_I,f_P$ reshape $\Delta(s)$ \emph{itself}. The schedule only sets the rate at
which a given $\Delta(s)$ is traversed. The distinction matters because the two are
often conflated.

\begin{figure*}[t]\centering
\includegraphics[width=\textwidth]{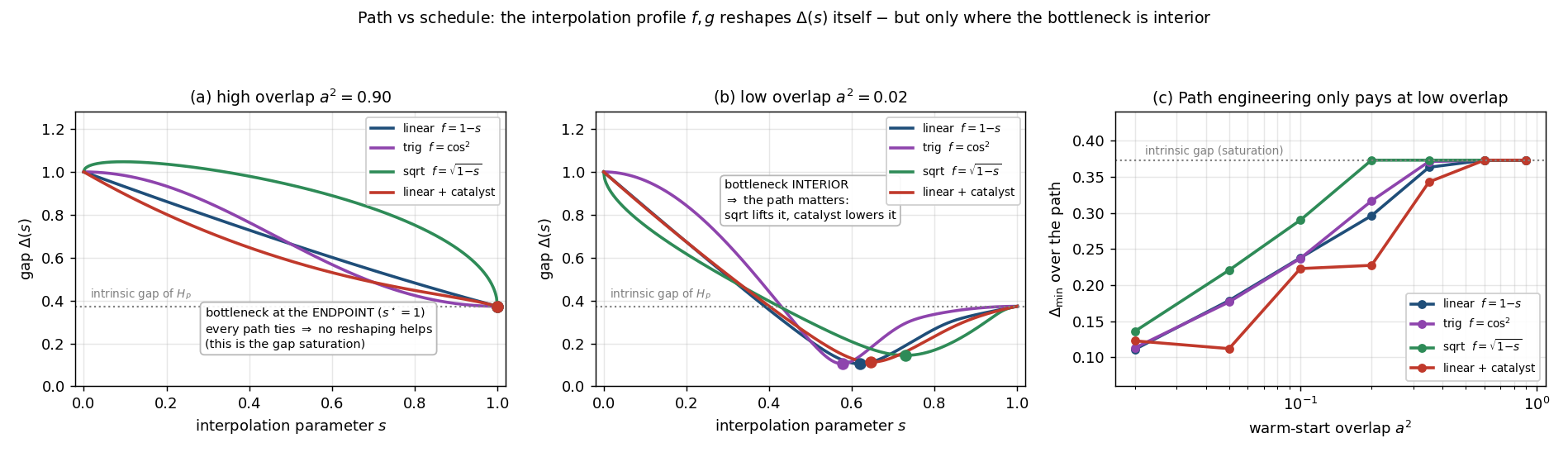}
\caption{\textbf{Path versus schedule.} (a) At high overlap the gap bottleneck sits at
the $s\!=\!1$ endpoint, where all paths coincide, the origin of gap saturation.
(b) At low overlap the bottleneck is interior and the interpolation profile matters; a
square-root path lifts it, a catalyst lowers it. (c) Path engineering pays only at low
overlap.}
\label{fig:path}
\end{figure*}

Whether path-shaping pays turns entirely on \emph{where the bottleneck sits}
(Fig.~\ref{fig:path}). At high overlap it sits at the endpoint, $\Delta_{\min}$ is
attained at $s^\star\!=\!1$, where every admissible path satisfies $H(1)\propto\tilde
H_P$ regardless of $f_I,f_P$. All paths so tie exactly ($\Delta_{\min}=0.373$ for all four
profiles at $a^2=0.90$), and no reshaping can help. This is the mechanism behind the
saturation of \S\ref{sec:gaplaw}. Once the warm start is good, the molecule's own gap,
not the interpolation, is the binding constraint. At low overlap the bottleneck moves
interior ($s^\star\!\approx\!0.6$) and the path becomes a real lever. A square-root
profile $f_I=\sqrt{1-s},\,f_P=\sqrt{s}$ raises $\Delta_{\min}$ by ${\sim}38\%$ ($0.149$ vs $0.108$) at $a^2=0.02$, and at $a^2=0.20$ removes the interior bottleneck altogether,
restoring the full intrinsic gap. A catalyst in the simple hopping form \emph{lowers the gap} ($0.089$ vs $0.108$ at $a^2=0.02$), a negative that echoes
the counterdiabatic result below. Terms bolted onto the projector driver tend to disturb
the structure that makes it work. The practical reading is that path engineering is
worth its complexity only in the low-overlap regime. This is also the regime where one
should question whether the warm start is good enough at all.

\begin{figure*}[t]\centering
\includegraphics[width=\textwidth]{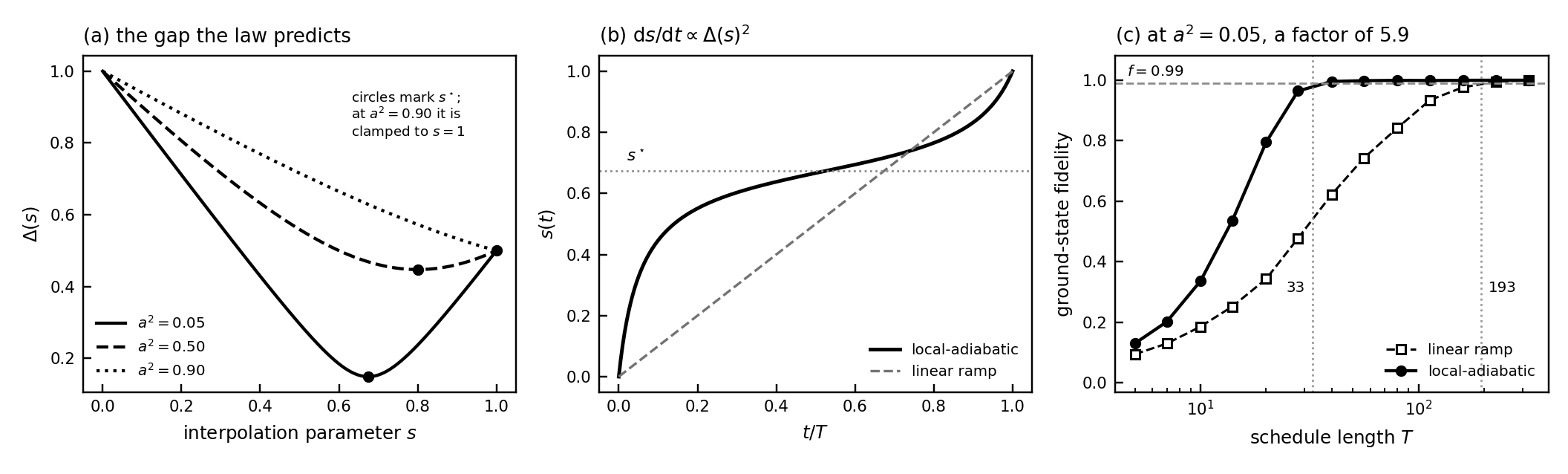}
\caption{\textbf{The schedule set from the measured gap.} (a) The closed-form gap profile
$\Delta(s)$ of Eq.~\eqref{eq:gaplaw} at three overlaps; circles mark $s^\star$, which is
interior at low overlap and clamped to the endpoint above the crossover. (b) The
local-adiabatic schedule generated from that profile dwells at the bottleneck, against the
linear ramp. (c) Exact integration of the two-level reduction at $\bar\mu=0.5$ and $a^2=0.05$:
the local-adiabatic schedule reaches $f=0.99$ at $T=33$, the linear ramp at $T=193$.}
\label{fig:sched}
\end{figure*}

The $5\times$ cost gap of \S\ref{sec:aqc} is a property of the schedule, not of the
construction, and Fig.~\ref{fig:warp}~(right), shown later with the decision rule it also
serves, tests two accelerants in a deliberately hard low-overlap bottleneck. A \emph{local-adiabatic} schedule that slows where the gap
is smallest ($\mathrm{d}s/\mathrm{d}t\propto\Delta(s)^2$) \citep{rolandcerf2002} gives a
robust improvement that grows with the severity of the bottleneck, and the size of it can be
stated exactly. Fig.~\ref{fig:sched} integrates the two-level reduction of
App.~\ref{app:adiabatic} exactly at $\bar\mu=0.5$. The gap profile of
Eq.~\eqref{eq:gaplaw} is computed from the measured pair $(a,\bar\mu)$ before any schedule is
run, so the rate is set from a gap that has been \emph{measured}. No
spectroscopy of the gap is required, and where the variance certificate of
\S\ref{sec:stopping} is available at each point the same rate is generated in closed loop from
the running variance. The resulting schedule dwells at the bottleneck and crosses the wide
regions quickly. At $a^2=0.05$ it reaches a ground-state fidelity of $0.99$ at $T=33$ where the
linear ramp needs $T=193$, a factor of $5.9$; at $T=40$ the two return $0.996$ and $0.623$. The
two agree only once $T$ is long enough that either succeeds, so the saving is realised precisely
in the small-overlap regime where the back-end is expensive --- the same regime in which
path-shaping above is worth its complexity.
\emph{Counterdiabatic} driving by the exact adiabatic gauge potential is transitionless,
error ${\sim}10^{-9}$ at \emph{any} $T$, even $T{=}1$, but that gauge potential is as
nonlocal as the problem is hard. The finding is that the cheap approximations do not reach that regime. A low-order nested-commutator (Krylov) gauge potential, or a
pool seeded from the VQE generators~\citep{sels2017}, injects error at exactly the fast
schedules where it would be needed, because the projector driver's gauge potential is
high-rank. The practical recommendation is therefore the local-adiabatic schedule. Transitionless driving stays a theoretical ideal rather than a near-term accelerant.

\subsection{Discrete amplification}
\label{sec:grover} 
\begin{figure}[t]\centering
\includegraphics[width=\linewidth]{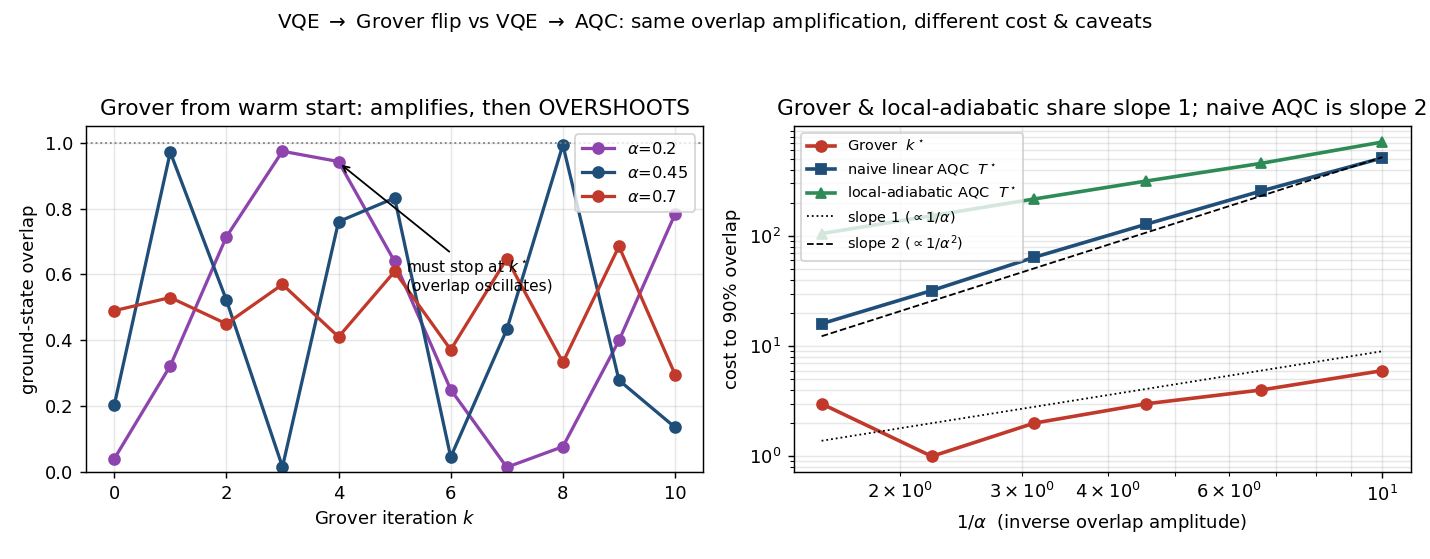}
\caption{\textbf{Discrete (Grover) versus continuous (adiabatic) amplification of a
warm start.} Left: the Grover overlap amplifies then overshoots and oscillates,
forcing a stop at $\ell^\star$. Right: cost to $90\%$ overlap on the model
spectrum, Grover and local-adiabatic AQC scale as $1/a$ (slope~$1$), naive
linear AQC as $1/a^2$ (slope~$2$).}
\label{fig:grover}
\end{figure}

Every preparation route amplifies the warm start's overlap. One may ask whether a discrete Grover
rotation~\citep{grover1996search,brassard2002amplitude} could replace the adiabatic
preparation. On a model spectrum, a ground state at zero energy and a band of $31$
excited states above a gap $\Delta=0.8$, with a warm start of tunable overlap amplitude
$a=|\!\braket{E_0|\chi}\!|$ as everywhere else, the Grover iteration $Q=(2\ket{\chi}\!\bra{\chi}-\mathbf
1)(\mathbf 1-2\ket{E_0}\!\bra{E_0})$ rotates the warm start toward the ground state in
$\sim\!1/a$ steps (Fig.~\ref{fig:grover}, right), quadratically fewer than the
$\sim\!1/a^2$ of a naive linear schedule. Three caveats stood against Grover when this
comparison was first made, and one of them has since fallen. First, the same $1/a$ scaling is recovered by a \emph{local-adiabatic}
schedule~\citep{rolandcerf2002} (Fig.~\ref{fig:grover}, right, slope~$1$), so the
quadratic speedup is not unique to Grover. Second, Grover \emph{overshoots}: the overlap
oscillates (Fig.~\ref{fig:grover}, left), so one must halt at exactly
$\ell^\star\!\approx\!\pi/(4a)$, which needs knowing $a$, and for large $a$
the discrete rotation is too coarse to land on unit overlap (a granularity issue cured
only by fixed-point variants~\citep{yoder2014fixedpoint}). This second caveat is the one that
turns out to be removable, and \S\ref{sec:homodyne} removes it. Needing $a$ is fatal only when
$a$ cannot be measured, which is exactly the sub-floor regime the diagonal channel cannot
reach; the homodyne channel measures it there, to a tolerance well inside what
Eq.~\eqref{eq:kstar} demands. The asymmetry is what makes
the measurement a prerequisite. An adiabatic schedule run too long
still succeeds, so a bound on $a$ suffices, whereas an amplification run too long returns
nothing, so only an estimate will do. Third and decisively, Grover needs an explicit
reflection about the ground state $\mathbf 1-2\ket{E_0}\!\bra{E_0}$. The ground state is
the unknown being computed. The adiabatic back-end never needs it. Its marking is
implicit in evolving toward the \emph{known} $H_P$. The one way to supply the oracle is
from the spectral back-end itself, ESPRIT returns $E_0$, and a threshold placed in the
gap turns ``phase-flip components with energy below $E_0+\Delta/2$'' into an exact
ground-state reflection, a \textsc{vqe}$\to$ESPRIT$\to$Grover composition in which the
readout enables the preparation's oracle, at a cost ${\sim}1/\Delta$ per flip to resolve
it. We therefore keep the adiabatic back-end as the default, with Grover amplification a
connected alternative wherever an energy oracle is cheap \textemdash{} and, since the second
caveat has fallen, a live one wherever the overlap can be metered as well.

\subsection[A recursive third route, and the test that admits it]{A recursive third route, and the measurement that admits it}
\label{sec:dbqite}

Amplification and adiabatic preparation both spend the warm start once. A third route
spends it \emph{recursively}. Double-bracket quantum imaginary-time evolution
(DB-QITE)~\citep{gluza2025dbqite} refines $\ket{\omega_m}$ by a group commutator of
evolution under the problem Hamiltonian with a reflection about the current state,
\begin{equation}
\ket{\omega_{m+1}}=\ee^{i\sqrt{\xi_m}\tilde H_P}\,
\ee^{i\sqrt{\xi_m}\ket{\omega_m}\bra{\omega_m}}\,
\ee^{-i\sqrt{\xi_m}\tilde H_P}\ket{\omega_m},
\label{eq:dbrec}
\end{equation}
started at $\ket{\omega_0}=\ket{\chi}$, the warm start itself. It belongs in this paper
for one reason: its cost is priced in our currency and in nothing else. The published
convergence guarantee gives $f_m\ge1-(1-a^2\Delta^2/12\|\tilde H_P\|^3)^m$, so the
iterations needed for a target infidelity $\epsilon$ are
\begin{equation}
m^\dagger\simeq\frac{12\|\tilde H_P\|^3\ln(1/\epsilon)}{a^2\Delta^2},
\label{eq:elldagger}
\end{equation}
fixed by the overlap and the gap together. Its depth grows as
$\mathcal O(3^m)$, each iteration nesting the last, so unlike the other two preparation
routes this one is not merely more or less expensive according to $a$: it is only operable when
$m$ is a small integer. That is a threshold, not a scaling,
and it has to be decided before the recursion is begun.

\begin{figure*}[t]\centering
\includegraphics[width=\linewidth]{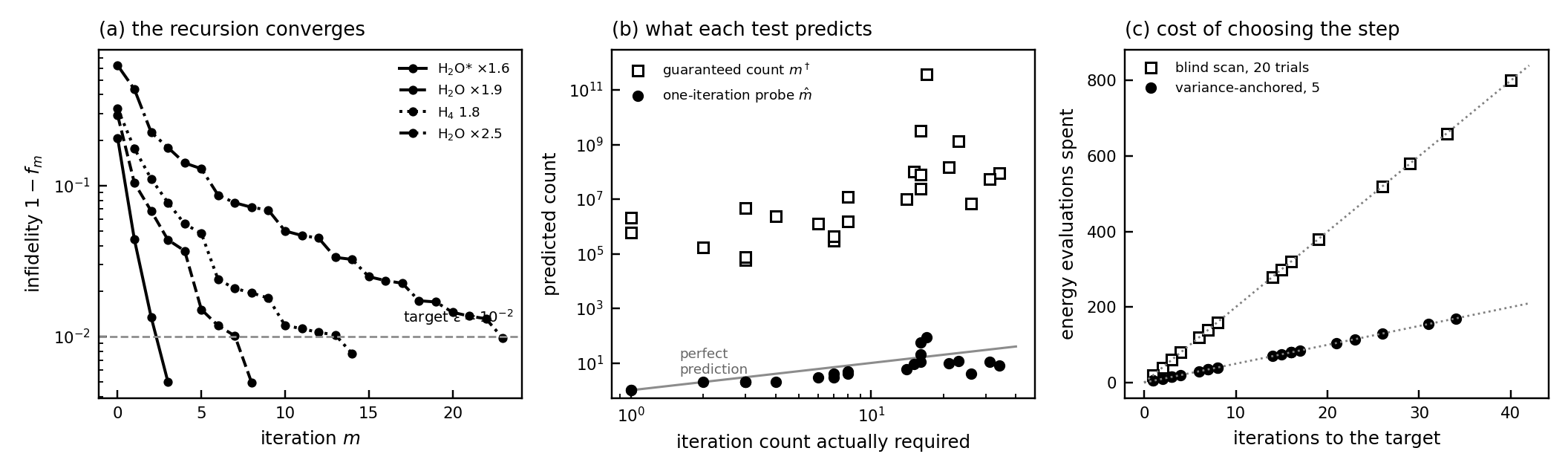}
\caption{\textbf{A recursive back-end, admitted by measurement rather than by
guarantee.} Twenty-three instances from H$_2$O, N$_2$, H$_4$ and LiH, $8$--$12$ qubits, CI
dimension $36$--$400$, exact diagonalisation and exact propagation of
Eq.~\eqref{eq:dbrec}. \textbf{(a)} Infidelity against iteration for four of them.
\textbf{(b)} What each admission test predicts against what was required. The convergence
guarantee, Eq.~\eqref{eq:elldagger}, overshoots by $1.9\times10^{4}$ to $2.2\times10^{10}$
(median $1.5\times10^{6}$); the one-iteration probe, Eq.~\eqref{eq:ellhat}, lands within a
factor of three on nineteen of twenty-three. \textbf{(c)} Anchoring the step size on the
measured moments, Eq.~\eqref{eq:sstar}, costs $1410$ energy evaluations against $5800$ for
a blind scan, at the same iteration count.}
\label{fig:dbqite}
\end{figure*}

We therefore asked what Eq.~\eqref{eq:elldagger} is worth as that decision. Twenty-three
instances were taken from H$_2$O, N$_2$, H$_4$ and LiH at active spaces of four to six
orbitals, that is at eight to twelve qubits, each geometry chosen to place the
Hartree--Fock overlap between $0.09$ and $0.99$; every Hamiltonian was diagonalised exactly
and every recursion propagated exactly, to $\epsilon=10^{-2}$. The matrices so diagonalised
have dimension $\binom{n_{\rm o}}{n_\alpha}\binom{n_{\rm o}}{n_\beta}$, the number of
determinants spanning an active space of $n_{\rm o}$ orbitals holding $n_\alpha$ and
$n_\beta$ electrons of each spin: $36$ to $400$ here. That is smaller than the $2^{2n_{\rm o}}$
of the register --- six active orbitals occupy twelve qubits but span only $400$ determinants
--- because particle number and spin projection are conserved and $H_P$ does not connect
their sectors. The
counts actually required ran from one iteration to thirty-four
(Fig.~\ref{fig:dbqite}a). The counts Eq.~\eqref{eq:elldagger} returns for the same
instances ran from $5.7\times10^4$ to $3.7\times10^{11}$, exceeding the truth by between
$1.9\times10^{4}$ and $2.2\times10^{10}$, median $1.5\times10^{6}$
(Fig.~\ref{fig:dbqite}b, squares)! Stretched N$_2$ converges in seventeen iterations where
the bound demands $3.7\times10^{11}$! A guarantee that loose can \emph{admit} a warm start and can never soundly refuse one, being sufficient but not necessary. Used as a rejection test it would nonetheless have refused all twenty-three, every one of which converged.

What does work is a measurement, and it is cheap because the metering of
\S\ref{sec:homodyne} is already in hand. Run one iteration. Meter the fidelity before and
after it against a strong reference, giving $f_0$ and $f_1$. Form the contraction of the
infidelity per iteration and extrapolate,
\begin{equation}
\lambda=\frac{1-f_1}{1-f_0},\qquad
\hat m=\left\lceil\frac{\ln\!\big(\epsilon/(1-f_0)\big)}{\ln\lambda}\right\rceil .
\label{eq:ellhat}
\end{equation}
The probe costs one iteration to decide a recursion of depth $3^{\hat m}$, and it lands
within a factor of three of the true count on nineteen of the twenty-three
(Fig.~\ref{fig:dbqite}b, circles) --- six orders of magnitude better than the guarantee on
the same instances!

It is an estimate, not a bound, and we report where it errs. Eq.~\eqref{eq:ellhat} takes
$\lambda$ to be constant, and it is not. The sequence $\lambda_m$ was non-monotone on
fifteen of the twenty instances reaching three iterations, and the first contraction was
the smallest of its sequence on only thirteen of twenty-one, so no direction of error is
guaranteed. The median ratio of true count to $\hat m$ is $1.67$, an
\emph{under}-prediction, which is the dangerous direction: it admits a recursion that then
overruns the depth budget. A stated safety factor of about two is thus introduced; taking the
largest contraction over a three-iteration probe instead moves the median to $0.87$ at the
cost of two more iterations.

The same two moments fix the step size. The energy decrease of an iteration is
$2\xi_m\mathcal V_m$ to first order in the step, with $\mathcal V_m$ the energy
variance --- a quantity the readout returns as the second moment of its own recovered
weights, at no extra cost. Since the rescaled ground energy is zero, the whole gain still available is the current
energy itself, $\mathcal E_m=\braket{\omega_m|\tilde H_P|\omega_m}$, so
\begin{equation}
\hat\xi_m=\frac{\mathcal E_m}{2\mathcal V_m}
\label{eq:sstar}
\end{equation}
is the step that would deliver all of it at the first-order rate. Searching five multiples
of $\hat\xi_m$ replaces the blind scan over five and a half decades that the method is
otherwise run with. Over the same twenty-three instances the scan reached the target in
$290$ iterations at $5800$ energy evaluations and the variance-anchored rule in $282$ at
$1410$ (Fig.~\ref{fig:dbqite}c); $4.1\times$ fewer evaluations at no cost in iterations,
or $3.4\times$ if the variance is re-acquired at every iteration rather than at an
interval.

Two other observations complete the picture. The reflection
$\ee^{i\sqrt{\xi_m}\ket{\omega_m}\bra{\omega_m}}$ is about the \emph{current}
state, so it is $W_m U_\chi R_{\rm ref} U_\chi^\dagger W_m^\dagger$ with $W_m$
the accumulated recursion --- and that conjugation is where the $3^m$
depth comes from. At $m=0$ it is exactly $U_\chi R_{\rm ref}U_\chi^\dagger$, the driver
of \S\ref{sec:aqc} and the anchor of \S\ref{sec:grover} at a different angle, so all three
preparation routes share one compiled primitive and the choice among them is a choice of
schedule. And the guarantee is a lower bound on a fidelity with no convergence test and a
presumption of a unique ground state. On the other hand, the two-sided bracket of \S\ref{sec:stopping}, formed
on $\ket{\omega_m}$ between iterations, supplies the excited-state detection it lacks.

None of this disturbs the two-back-end picture of Fig.~\ref{fig:selection}, which asks what
one wants --- an energy or a state --- and is answered by the readout or by preparation. What
it does is populate the preparation branch. There are now three ways down it, the adiabatic
schedule of \S\ref{sec:aqc}, the discrete amplification of \S\ref{sec:grover} and the
recursion here, and the choice among them is priced in the same currency: amplification costs
${\sim}1/a$, a linear adiabatic schedule ${\sim}1/a^2$, and a direct acquisition on the readout
branch ${\sim}1/a^4$. Those three are admissible at any overlap and differ only in price. The
recursion does not. It is the one route whose \emph{operability} turns on the measured pair
$(a,\Delta)$. The overlap
is not a figure of merit but a price, and a price can be too high to pay.

\warppart{Readout and preparation together}{Having described each family alone, we now
put them on one account: which to choose, how each certifies the other, and how the
currency itself is measured when it is scarce.}
\label{part:together}

\section{One account, two ways to spend it}
\label{sec:together}

Parts~\ref{part:readout} and~\ref{part:prep} described each back-end as though it were
alone, and each is usable alone. But they are not independent, because they draw on the
same account. The readout's shot cost runs as $1/a^4$ and the schedule length as $1/a^2$,
both in the one quantity the front-end supplied, so a statement about one of them is
always also a statement about the other. The two are not the same currency on hardware. The
readout spends its $1/a^4$ in repetitions of a short circuit, so it buys precision with
wall-clock time and is bounded by the coherence time only through the fixed acquisition
length $\tau$. The sweep spends its $1/a^2$ in coherent evolution, so its cost is charged
directly against the coherence time and cannot be paid in repetitions at all. Which of the
two is the scarce resource is therefore a property of the device and not of the problem. Three consequences follow, and they are the
subject of this part. There is a rule for deciding which to run
(\S\ref{sec:decision}). There is a certificate that needs no estimate of the minimum gap
(\S\ref{sec:stopping}). And
because $a$ prices the back-ends, it has to be measurable when it is small, which is
exactly when its own acquisition cannot see it---the subject of \S\ref{sec:homodyne}.

\paragraph{The back-ends diagnose each other.} The two branches of
Fig.~\ref{fig:selection} are drawn as alternatives, but each repairs the other's
characteristic failure, and the repairs run in opposite directions. When the readout
cannot resolve the ground line because $a^2$ is too small, a partial sweep buys overlap
and hands back. The sweep is deliberately short, so the state it returns is still
contaminated and its energy still biased, which is precisely why the readout rather than
a measurement must take delivery. When the sweep cannot certify what it has reached
because $\sigma$ is blind below the splitting it is asked to detect, the readout run on
the prepared state returns $|c_0|^2$ and names the eigenstate. Neither repair
manufactures overlap; the first spends time to concentrate it, the second spends samples
to observe it. What makes the pairing more than an accounting convenience is that the
regimes are complementary. Preparation is cheap and readout
expensive when the intrinsic gap is wide, readout is cheap and preparation expensive when
it is narrow, and the failure of either is the trigger to call the other.

\subsection{Read or prepare? A decision rule}
\label{sec:decision}
\begin{figure}[t]
\centering
\includegraphics[width=\linewidth]{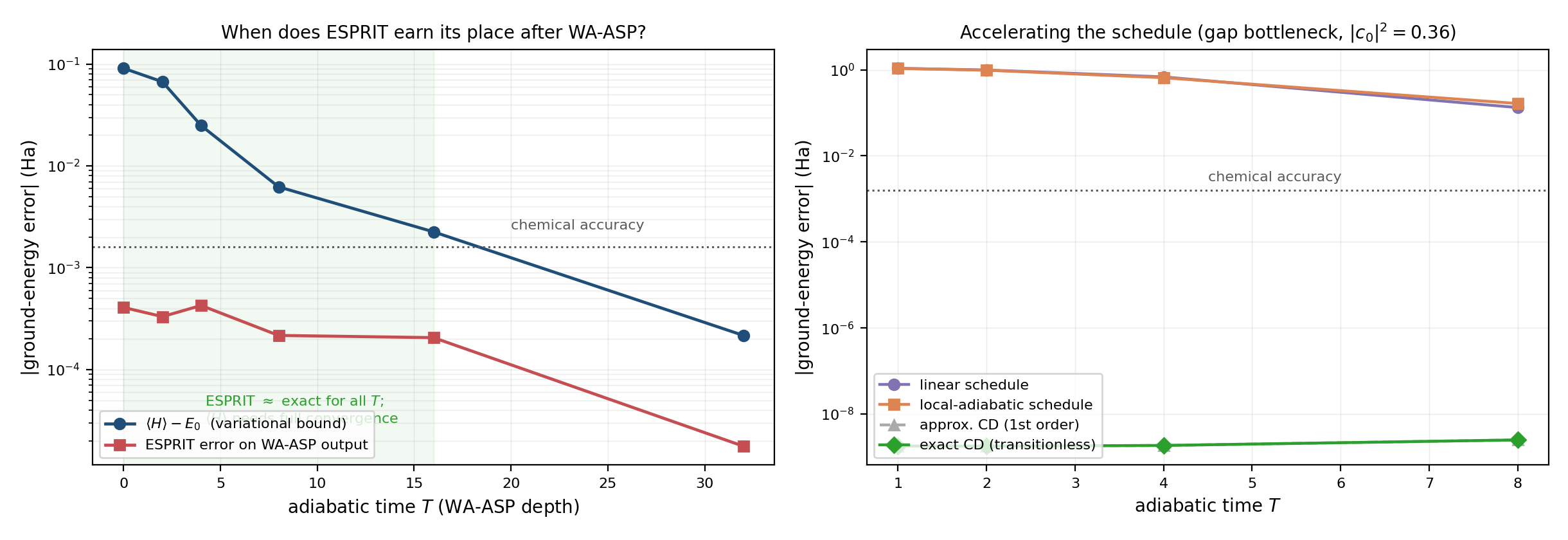}
\caption{\textbf{The \textsc{warp} decision rule and schedule.} \textbf{Left:}
ESPRIT reads $E_0$ to chemical accuracy at every adiabatic time, including
$T{=}0$, while $\langle H\rangle$ converges only as the state does, so spend
overlap on readout for an energy, on preparation for a state. \textbf{Right:} in a
hard gap bottleneck the local-adiabatic schedule helps and exact counterdiabatic
driving is transitionless, while low-order approximate CD does not help.}
\label{fig:warp}
\end{figure}

Once preparation has driven the overlap to unity the energy is free as
$\langle H\rangle$, so readout and preparation are not
rivals but answers to different questions, and Fig.~\ref{fig:warp}~(left) makes
the split precise on H$_2$. The variational estimate~(\ref{eq:deficit}) is an upper bound whose error
is \emph{linear in the leakage population} $1-|c_0|^2$, equivalently, \emph{quadratic}
in the residual state error $\big\||\psi\rangle-|E_0\rangle\big\|$. It reaches chemical accuracy only when the state
is nearly exact ($T\!\approx\!16$ here). ESPRIT instead reads the true eigenvalue
off the ground line and is \emph{already below chemical accuracy at}
$T{=}0$, on the bare VQE state. Hence, if we:
\begin{itemize}\itemsep2pt
\item want only the \emph{energy}, ground line resolvable (non-negligible, not
necessarily dominant) then use ESPRIT directly, no adiabatic stage;
\item want the \emph{state} (forces, dipoles, any observable) $\Rightarrow$ WA-ASP
to convergence, after which $\langle H\rangle$ is free;
\item \emph{marginal overlap or near-degeneracy} then employ a short WA-ASP to
lift the ground line into dominance, \emph{then} ESPRIT, the one genuinely
three-stage regime.
\end{itemize}

\paragraph{A fifth route, and the test that admits it.}
\label{sec:filter}
The four routes above are priced on one scale. Three of them carry an exponent in the
overlap, $a^{-1}$ to amplify, $a^{-2}$ to sweep and $a^{-4}$ to acquire, and the recursion of
\S\ref{sec:dbqite} enters as a threshold rather than a scaling. A fifth route is available, and
it is worth pricing on the same terms. A \emph{filter} applied to the warm start before the acquisition, of the kind used in
filtered phase estimation~\citep{lee2025fqpe} or in eigenvalue
transformation~\citep{qetu2025filter}, suppresses unwanted eigencomponents, multiplying the
amplitude on the wanted eigenstate by a factor $\kappa$ at the price of an acceptance probability
$p_f$. A dissipative preparation~\citep{dissipative2026rapid} would enter the same ladder
differently again, its cost depending on the initial overlap in a manner not of the power-law
form the other four share, and we do not price it here. Both are measurable on the device: $\kappa^2$ is the ratio of the filtered residue at that
eigenvalue to the unfiltered one, and $p_f$ is the observed acceptance fraction, so a short
calibration returns both before the budget is committed.

The acquisition needs repetitions $\propto a^{-4}$, so
\begin{equation}
\frac{M_{\rm filtered}}{M_{\rm bare}}=\frac{1}{p_f}\left(\frac{a}{\min\{\kappa a,1\}}\right)^{4},
\label{eq:filtertest}
\end{equation}
the minimum saying only that an amplitude cannot exceed one. While $\kappa a\le1$ the ratio is
$(\kappa^4p_f)^{-1}$, \emph{independent of the overlap}, so the filter pays whenever
$\kappa^4p_f>1$. Once $\kappa a>1$ the factor $\kappa$ cannot be
delivered, the ratio becomes $a^4/p_f$, and the postselection is pure loss. Combining the two,
the filter reduces the repetition count if and only if
\begin{equation}
p_0<\sqrt{p_f}\quad\text{and}\quad\kappa^4p_f>1 .
\label{eq:filtercond}
\end{equation}

\begin{figure*}[t]\centering
\includegraphics[width=0.86\textwidth]{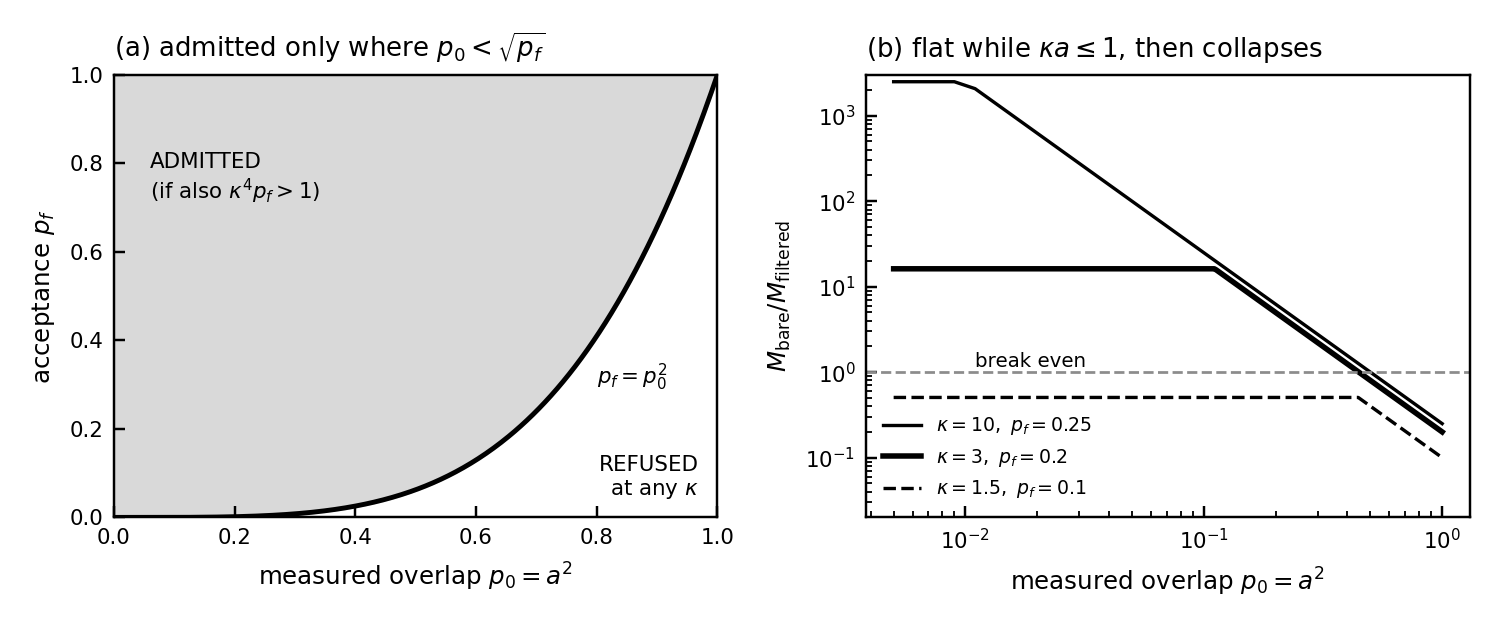}
\caption{\textbf{When a filtering stage pays.} (a) The admitted region of
Eq.~\eqref{eq:filtercond}. The boundary $p_f=p_0^2$ is a ceiling on the measured overlap that no
budget and no amplification factor overcome. (b) Inside the region the factor delivered is
$\kappa^4p_f$ and is flat in the overlap while the amplified amplitude has room to grow,
collapsing once it saturates at unity. Both $\kappa$ and $p_f$ are measured on the device.}
\label{fig:filter}
\end{figure*}

The first condition is a ceiling on the measured overlap (Fig.~\ref{fig:filter}), and it is of
the same character as the mixedness ceiling of \S\ref{sec:mixedstart}. No increase in the
measurement budget removes it, and no value of $\kappa$ overcomes it. \textbf{A warm start
already good enough is made worse by filtering it.} A stage with $\kappa=3$ and $p_f=0.2$ divides
the repetition count by $16.2$ at $p_0=0.05$ and by $5.0$ at $p_0=0.20$, and \emph{multiplies} it
by $1.8$ at $p_0=0.60$; a stage with $\kappa=1.5$ and $p_f=0.1$ has $\kappa^4p_f=0.51$ and is
refused at every overlap. This is the currency argument again, in its sharpest form. The value of
a stage that manufactures overlap depends on how much overlap you already have, and only a
measurement can tell.

A single warm-start front-end so feeds two overlap-spending back-ends. 

\subsection{Stopping without knowing the gap}
\label{sec:stopping}
When should the adiabatic back-end evolution halt? The obvious answer needs the minimum gap
$\Delta_{\min}$, and that is the expensive quantity we are trying to avoid.

\begin{figure}[t]\centering
\includegraphics[width=\linewidth]{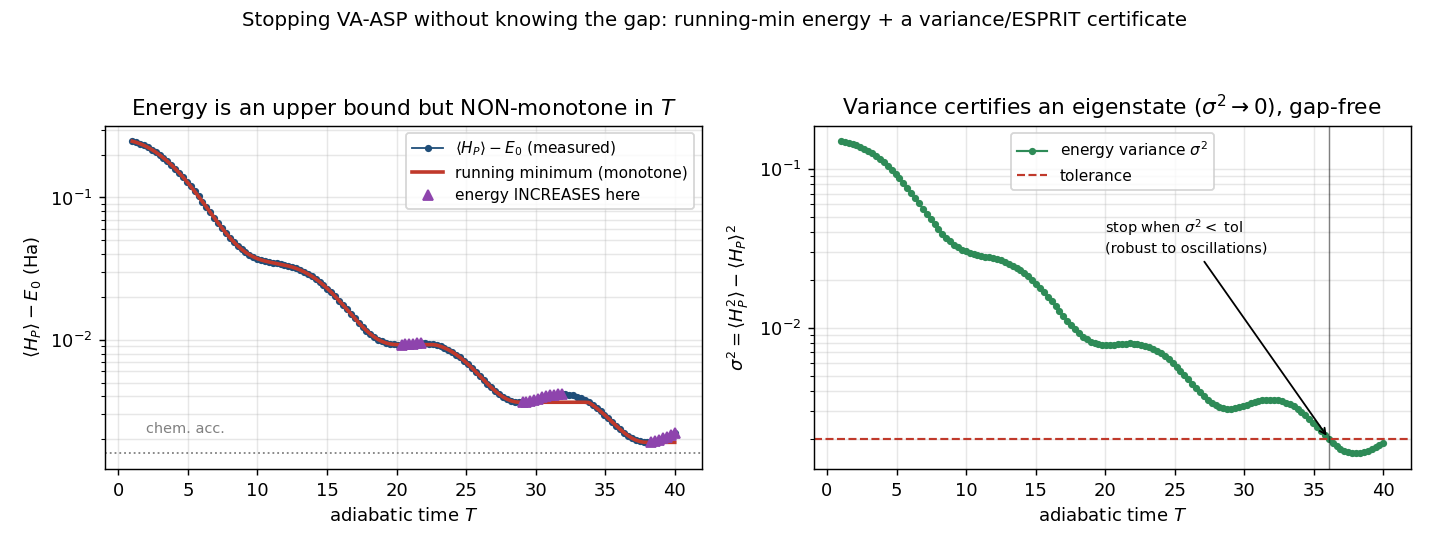}
\caption{\textbf{Gap-free stopping.} Left: $\braket{H_P}-E_0$ on H$_2$ is an upper
bound but non-monotone in $T$ (triangles mark steps where it \emph{rises}). The running
minimum is monotone. Right: the energy variance $\sigma^2\to0$ certifies an eigenstate
independent of the oscillations, stop when it falls below tolerance.}
\label{fig:stop}
\end{figure}

The energy expectation is measured anyway, and it is an unconditional upper bound.
$\braket{H_P}\ge E_0$ holds for any state. However, it is not monotone in $T$. The excited populations undergo
Landau--Zener--St\"uckelberg oscillations~\citep{shevchenko2010lzs}, so the energy and
the ground overlap both approach their limits through a decaying but oscillatory
envelope. On H$_2$ from a $0.7$-overlap warm start the energy rises on $17\%$ of
$T$-steps (Fig.~\ref{fig:stop}, left). A test that waits for the energy to flatten is unusable as a transient turning point can mimic convergence.

The running minimum of
$\braket{H_P}$ over a geometric ladder $T,2T,4T,\dots$ is
non-increasing, so it records the tightest bound thus far. However, a flat running minimum can also mean that a longer evolution has yet to beat the
record. It is a bound-tracker, not a stopping rule.

The following fix evaluates a certificate at the current $T$ instead of a
trend. The energy variance $\sigma^2=\braket{H_P^2}-\braket{H_P}^2$ vanishes if and only
if the state is an eigenstate. Together with $\braket{H_P}$ near the bottom of the
spectrum it certifies the ground state (Fig.~\ref{fig:stop}, right). The ESPRIT
ground-line weight $|c_0|^2$ is the stronger test, and is developed below. It measures the
fidelity rather than bounding it, and names the eigenstate reached rather than merely
detecting that one was. Both are threshold tests on the
state in hand, so both are robust to the oscillations. A stalled but contaminated state
keeps a large $\sigma^2$ and fat excited lines, so it fails the test. A transient energy
plateau cannot fake a small variance. The spectral back-end supplies the stopping
certificate for the adiabatic back-end, and the two halves of the method become
mutually diagnostic.

We use $\sigma^2$ here as an instrument, a threshold test on the state in hand. It is also a
property of that state, and \S\ref{sec:notpriced} returns to it in that second role, where it
turns out to price what a later acquisition will cost. The practical consequence is that a large
$\sigma^2$ says more than \emph{not converged yet}. It also says the state will be expensive to
sample, and it says so from two moments the readout returns in any case.

The gap returns only as a convenience. The local-adiabatic schedule of
\S\ref{sec:schedule} uses $\Delta(s)$ to allocate rate, but $\Delta(s)$ can be estimated
on the fly from $\sigma^2$ rather than known in advance.

A vanishing $\sigma$ certifies that the
state is an eigenstate; it does not certify \emph{which} one. A sweep that crosses a
small gap diabatically lands in $\ket{E_1}$, where $\sigma=0$ exactly, the pointwise test
passes, but the number reported is $E_1$.

Where controlled evolution is available the ambiguity does not arise, because the
spectral back-end answers the question directly. Running the readout on the prepared
state, $g(t)$ of Eq.~(\ref{eq:gt}), returns the pairs $(E_n,|c_n|^2)$:
the ground line and the first excited line appear separately, with their weights, and $|c_0|^2$ is the fidelity itself rather than a proxy for it.  Complete capture is when a state that is exactly $\ket{E_1}$ has no
component at $E_0$, so that line is absent from $g(t)$ altogether.
What identifies it is the comparison with $E_0$ from the front-end readout, which is
available precisely because $\ket{\chi}$ does carry ground weight; the sweep has returned
a state whose lowest line sits above the ground line already measured. Either way the
answer is a reading rather than a threshold test, and in the second case it is a reading
that only the pairing of readout and preparation supplies.

The remainder of this paragraph treats the case where that cost is not available and
$\sigma$ must carry the certificate alone. The clause ``$\braket{H_P}$ near the bottom of
the spectrum'' is then what excludes capture, and it has to be made operational, because
the bottom of the spectrum is the unknown.

\begin{figure}[t]\centering
\includegraphics[width=\linewidth]{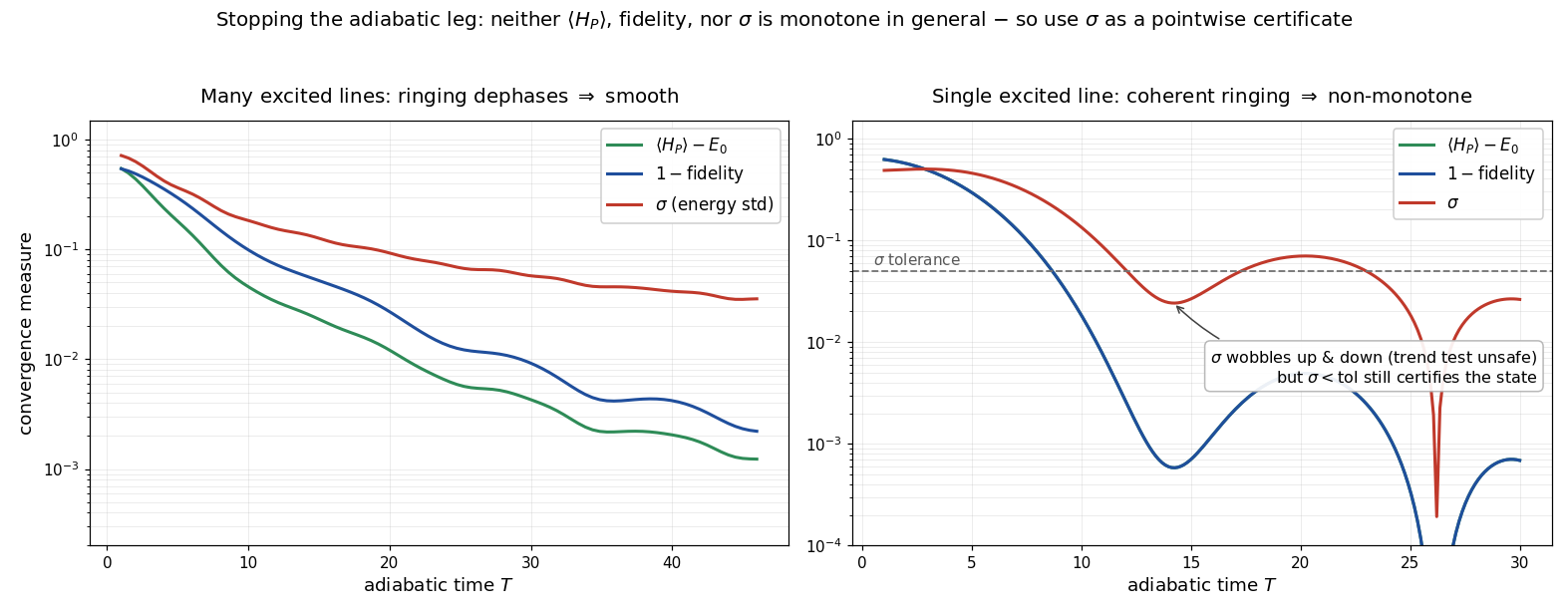}
\caption{\textbf{The variance is a certificate, not a trend.} Energy error,
infidelity, and $\sigma$ versus $T$ for a warm start with many populated excited lines
(left: ringing dephased, all smooth) and for a single dominant excited line (right:
coherent ringing, all three, including $\sigma$, non-monotone). A pointwise
$\sigma<\mathrm{tol}$ test certifies the current state in both regimes.}
\label{fig:stopsig}
\end{figure}

\paragraph{Is the variance monotone?}
It is not. When one excited line dominates the error, the
diabatic ringing is coherent and $\sigma$ oscillates with $T$. A two-level projector
sweep from a $0.6$-amplitude warm start sees $\sigma$ rise on $40\%$ of $T$-steps, more
often than the energy does (Fig.~\ref{fig:stopsig}, right). When many lines are
populated their phases are incommensurate, the ringing dephases, and $\sigma$ descends
smoothly (Fig.~\ref{fig:stopsig}, left). What matters is the kind of test, not the
monotonicity. A trend test asks whether $\sigma$ has stopped falling, and that is unsafe
on an oscillating signal. A pointwise test asks whether $\sigma<\mathrm{tol}$ right now,
and that certifies the state in hand whatever route it took to get there. The reason is
that $\sigma$ is a property of that state alone. 

The data already in hand contain the discriminator, provided each bound is used in the
direction in which it is valid. Variationally, $\braket{H_P}(T')\ge\Eo$ at every rung, so
the running minimum $U(T)=\min_{T'\le T}\braket{H_P}(T')$ is an upper bound on $\Eo$ that
improves monotonically and may be carried across rungs. Weinstein's
bound~\citep{weinstein1934lowerbound} is instead a statement about a single state and its
own variance, and comes in two forms. Unconditionally, some eigenvalue lies within
$\sigma(T)$ of $\braket{H_P}(T)$. And if in addition $\braket{H_P}(T)<E_1$, that
eigenvalue is $\Eo$, giving $\Eo\ge\braket{H_P}(T)-\sigma(T)$. Because the mean and the
spread must be measured on the same state, the pair $\braket{H_P}(T),\sigma(T)$ cannot be
broken apart: $\sigma(T)$ bounds the distance from $\braket{H_P}(T)$ to the nearest
eigenvalue, not the distance from $U(T)$, so the interval
$[U(T)-\sigma(T),\,U(T)+\sigma(T)]$ has no warrant even though it is narrower. Combining
the variational upper bound with the conditional lower bound gives a two-sided but
unequally sourced certified interval,
\begin{equation}
\Eo\in\big[\braket{H_P}(T)-\sigma(T),\;U(T)\big],
\label{eq:certint}
\end{equation}
its lower endpoint drawn from the state at rung $T$ and its upper endpoint from the whole
ladder. Its width $\sigma(T)+\big(\braket{H_P}(T)-U(T)\big)$ is never worse than the
$2\sigma$ of the Weinstein interval alone, and usually better.

The unconditional form supplies the missing test. If
$\braket{H_P}(T)-\sigma(T)>U(T)$, the interval is empty. The eigenvalue the sweep has
settled on lies strictly above a bound that $\Eo$ is known to respect, so it is not
$\Eo$, and the rung is rejected as excited-state capture. The test needs no quantity that
is not measured anyway, and it is applicable whenever the warm start's own energy lies below
$E_1$, which is what a converged front-end delivers.

Two limits bound what this can achieve. First, $\sigma$ cannot resolve structure finer
than the splitting it is asked to detect. For a two-level mixture across a gap
$\Delta_{01}$ carrying spectral weights $p_0$ and $p_1=1-p_0$,
$\sigma=\Delta_{01}\sqrt{p_0p_1}\le\Delta_{01}/2$, so a small $\sigma$ says nothing about
mixing when $\Delta_{01}$ is itself small, and $\mathrm{tol}$ must be read against
$\Delta_{01}$ rather than against chemical accuracy alone. 
Second, the failure is bounded by the same smallness that causes it. Diabatic
transfer requires
$\Delta_{\min}\lesssim\sqrt{\Omega/T}$ with $\Omega$ the spectral bandwidth, while the resulting error
matters only if $\Delta_{\min}>\epsilon_{\mathrm{chem}}$, below which the two states are
degenerate to within the accuracy claimed and either is acceptable. The band
$\epsilon_{\mathrm{chem}}<\Delta_{\min}<\sqrt{\Omega/T}$ is empty once $T>\Omega/\epsilon_{\mathrm{chem}}^{2}$, about
$4\times10^{5}\,\Ha^{-1}$ at $\Omega\!\sim\!1\Ha$.

\paragraph{Does more overlap help?} It depends which gap is binding, and the two regimes
of Fig.~\ref{fig:aqc} (right) separate cleanly. While $\Delta_{\min}$ is set by the warm
start, raising $a$ raises the gap directly, and the Landau--Zener weight
$\exp(-\pi\Delta_{\min}^{2}T/2v)$ suppresses capture exponentially in $a^{2}$: overlap is
the remedy. Once $\Delta_{\min}$ has saturated at the intrinsic gap of $H_P$, further
overlap buys nothing, because the bottleneck is no longer a property of the warm start.
That regime is not a failure of the method so much as a change of which back-end to use.
A small intrinsic gap is precisely the situation the spectral half of this paper handles well. ESPRIT resolves near-degenerate lines from short signals (\S\ref{sec:esprit}),
returning $E_0$ and $E_1$ separately together with their weights, where adiabatic
preparation must instead pay $T\sim\Delta_{\min}^{-2}$ to keep them apart. When the
intrinsic gap closes, read rather than prepare.

\paragraph{Measuring the certificate.}
The certificate is only useful if $\sigma$ is cheap to obtain. Three routes exist and
they are not equivalent.

For a diagonal $H_P$ the certificate is free. The eigenstates are basis strings and the
eigenvalues are the classical costs. Each shot
returns a string, and evaluating the cost on that string gives its exact eigenvalue.
Then $\braket{H_P}$ is the sample mean of the observed costs, $\sigma$ is their sample
standard deviation, and the ground fidelity $f$ is the observed frequency of the best
cost. One measurement setting gives all three, with no $\braket{H_P^2}$ decomposition
and no spectral readout. On the Petersen MaxCut instance of \S\ref{sec:qaoa},
$20\,000$ shots recover $\braket{H_P}$, $\sigma$ and $f$ to better than $2\times10^{-2}$,
and the certified bracket of App.~\ref{app:certified} holds at every QAOA depth.

\begin{figure}[t]\centering
\includegraphics[width=\linewidth]{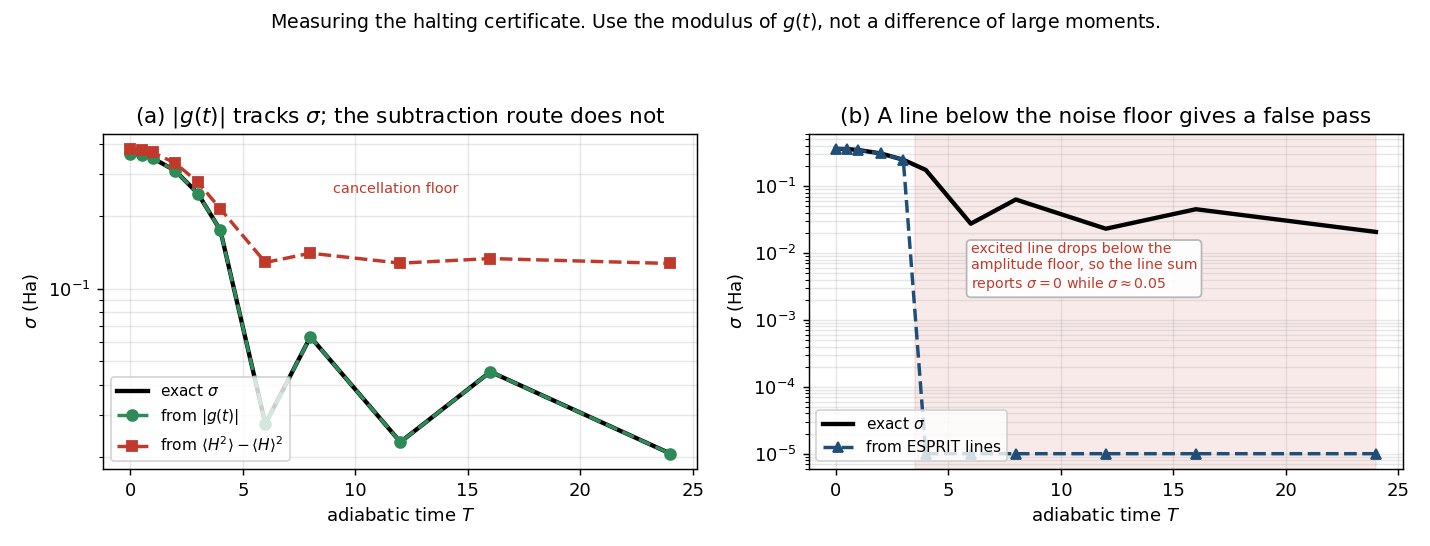}
\caption{\textbf{Measuring the halting certificate.} (a) On H$_2$ along a WA-ASP
evolution, $\sigma$ from $|g(t)|$ tracks the exact value to $2.2\times10^{-4}\Ha$,
while the $\braket{H_P^2}-\braket{H_P}^2$ route stalls at $1.1\times10^{-1}\Ha$ because
it subtracts two large nearly equal numbers. (b) The ESPRIT line sum agrees while both
lines are resolved, then reports $\sigma=0$ once the excited line falls below the
amplitude floor. That is a false pass.}
\label{fig:certest}
\end{figure}

For a non-commuting $H_P$, take $\sigma$ from the modulus of the correlation $g(t)$~(\ref{eq:gt}). Expanding
$|g(t)|^2=\sum_{n,m}p_np_m\cos\!\big((E_n-E_m)t\big)$ gives
\begin{equation}
\sigma^2=\frac{1-|g(t)|^2}{t^2}+\mathcal O(t^2).
\label{eq:sigmafromg}
\end{equation}
The modulus depends only on energy \emph{differences}, so it is invariant under a shift
of $H_P$ and carries no cancellation between $\braket{H_P^2}$ and $\braket{H_P}^2$. Near
convergence these are both large and nearly equal, and
their difference is the small number we want, so the relative error grows exactly where
the certificate must be sharp. Fig.~\ref{fig:certest}(a) shows the size of the effect on
H$_2$. Over $T=0$ to $24$ the modulus route tracks $\sigma$ to $2.2\times10^{-4}\Ha$.
The subtraction route stalls at an error of $1.1\times10^{-1}\Ha$, larger than $\sigma$
itself once $T>4$. Both estimates use the same $g(t)$ record that the spectral back-end already samples, so the certificate costs no extra circuits.

A third route sums the recovered ESPRIT lines,
$\sigma^2=\sum_n w_nE_n^2-\big(\sum_n w_nE_n\big)^2$. It is accurate while every
populated line stays above the amplitude floor, matching $\sigma$ to
$3\times10^{-3}\Ha$ for $T\le2$ on H$_2$. It then fails, and it fails in the unsafe
direction. Once the excited line drops below the floor the sum sees a single line and
reports $\sigma=0$, a false certificate of convergence, while the true $\sigma$ is
$0.175\Ha$ at $T=4$ (Fig.~\ref{fig:certest}b). Truncating the line sum always biases
$\sigma$ low, because a discarded line at distance $d$ from the mean contributes $w d^2$
to the variance. Eq.~\eqref{eq:sigmafromg} avoids this. It never needs the lines
resolved.

\section{Metering the currency}
\label{sec:homodyne}

A second reference state was introduced in \S\ref{sec:multistart} for coverage, to reach
lines that one warm start leaves dark. It buys something else as well, and the two must be
separated carefully. Coverage is about \emph{which} lines appear at all. What follows is about
\emph{how small} a weight can be measured on a line that does appear, and it is the
mechanism by which the currency of this paper is made observable at the point where it
runs out.

Write $c_W=\langle E_n|\chi_W\rangle$ and $c_S=\langle E_n|\chi_S\rangle$ for the amplitudes of
two references on one line. The autocorrelation of $\chi_W$ carries that line with residue
$|c_W|^2$, a square, while the correlation of $\chi_S$ against $\chi_W$,
\begin{equation}
g_{SW}(t)=\langle \chi_S|\ee^{-iH_Pt}|\chi_W\rangle=\sum_n c_S^{*}c_W\,\ee^{-iE_nt},
\label{eq:gsw}
\end{equation}
carries it with residue $c_S^{*}c_W$, a product. The two are tied by
$|c_S^{*}c_W|=\sqrt{|c_S|^2\,|c_W|^2}$, and that identity has two faces. Because a geometric mean
never exceeds the larger of its arguments, the cross channel can never make a line visible that
was invisible in both diagonals, which is why it is useless for detection and why nothing in
\S\ref{sec:multistart} rests on it. But because a geometric mean vastly exceeds the
\emph{smaller} of its arguments, the cross channel is an excellent instrument for measuring
$c_W$ itself once $c_S$ is known. The diagonal is quadratic in the small quantity and the cross
channel is linear in it, amplified by $|c_S|$. This is homodyne detection, with the strong
reference playing local oscillator, and it is the reason the two faces of the same identity
point in opposite directions.

The estimator follows the same split. Frequencies are the non-linear half of the problem and
residues the linear half, so we solve the hard half where the signal is strong. ESPRIT on
$g_{SS}$ returns the line positions and $|c_S|^2$; the weak signal is then projected onto those
fixed exponentials by ordinary least squares, and
$|c_W| = |\widehat{c_S^{*}c_W}|/\sqrt{\widehat{|c_S|^2}}$, with the relative phase coming free as
the argument of the same complex number. Fixing the eigenvector's phase so that $c_S$ is real
positive makes the result well defined.

That phase is where the single-sweep acquisition of \S\ref{sec:quadrature} earns its second
keep. Here, unlike in an autocorrelation, the residue is complex, and the phase we are reporting
is a ratio of the two quadratures. Acquiring them in separate passes exposes that ratio to any
contrast drift between the passes; acquiring them from the same repetitions cancels a common
factor exactly. The figures are in \S\ref{sec:quadrature}; the single-sweep value is flat in the drift where
the two-pass value grows in proportion to it. The modulus is not protected --- the common factor
survives there --- so the two acquisitions are chosen on which of the two is wanted.

\begin{figure*}[t]
\centering
\includegraphics[width=0.92\textwidth]{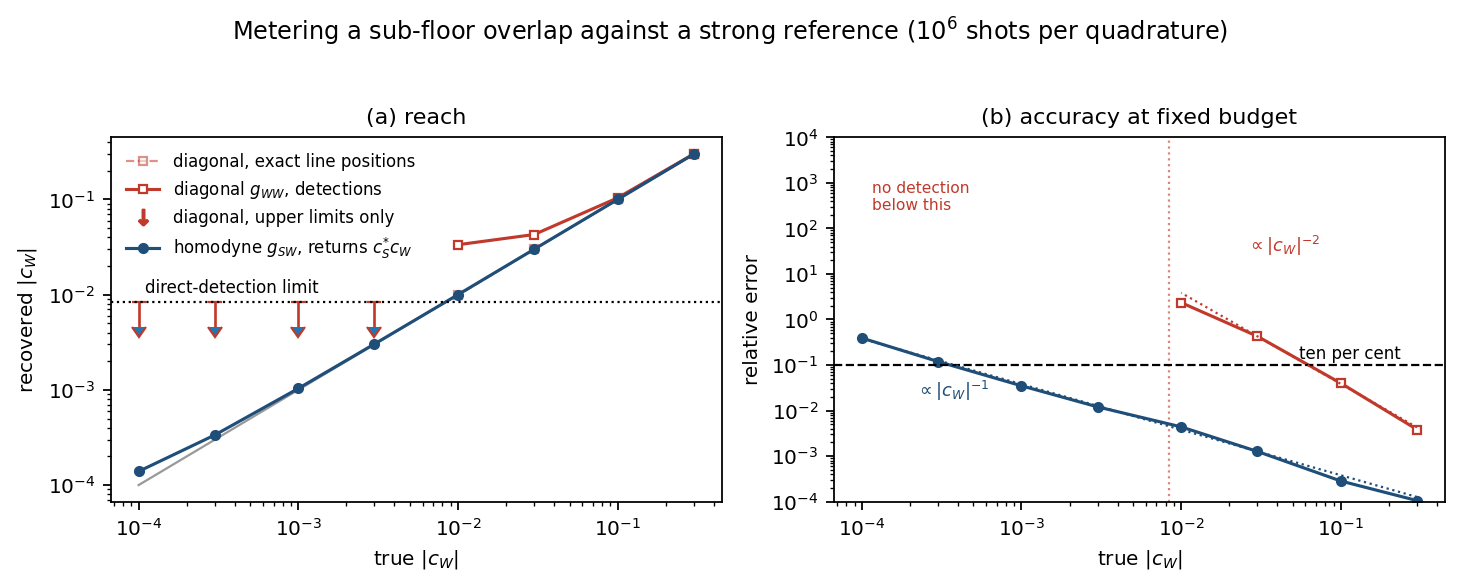}
\caption{\textbf{Metering a sub-floor overlap.} \textbf{Left:} recovered against true $|c_W|$ at
$10^6$ shots per quadrature. The diagonal channel returns a value only while its residue
$|c_W|^2$ stands above the noise floor, shown as open squares; below the direct-detection limit
its fitted residue is consistent with zero, so it yields no measurement of $|c_W|$ but only an
upper limit, drawn as downward arrows. The homodyne channel returns values throughout, following
the true overlap to $10^{-4}$. \textbf{Right:} relative error at fixed budget, defined only where
there is a detection, falling as $|c_W|^{-2}$ for the diagonal and as $|c_W|^{-1}$ for the
homodyne channel.}
\label{fig:homodyne}
\end{figure*}

Figure~\ref{fig:homodyne} measures this on the SO$_2$ active space, with a $24$-line spectral
support and $10^6$ shots per quadrature. The residue floor is $7.1\times10^{-5}$, so direct
detection should stall at $|c_W|\approx 8.4\times10^{-3}$, and the diagonal channel supplied with
exact line positions stalls at $8.3\times10^{-3}$. The homodyne channel tracks the true overlap down to $|c_W|=10^{-4}$, where
the residue $|c_W|^2=10^{-8}$ is some four orders of magnitude below the floor, and it holds
ten-per-cent accuracy to $|c_W|\approx3\times10^{-4}$ while returning the relative phase to a
hundredth of a radian at $3\times10^{-3}$. The shot scaling is the sharp statement. Direct
detection needs $N_s\propto|c_W|^{-4}$ shots per sample and the homodyne route needs
$N_s\propto|c_W|^{-2}$, the same exponent amplitude estimation buys, obtained here from
interference alone with no amplification circuit and no extra ancillae.

\paragraph{The first use: metering a scarce currency.} The overlap is what sets every cost in
this paper, and a candidate warm start with
$p_0\sim10^{-6}$ cannot be scored at all by its own acquisition at any tolerable budget, yet can
be scored here in the same run that reads the spectrum. Ranking front-ends, or watching $p_0$
climb as an ansatz is deepened, becomes possible in exactly the regime where it matters most. Of
the two, the tracking is the stronger: an absolute value of $10^{-8}$ tells one only that the
state is hopeless, whereas a \emph{derivative} tells one whether deepening the circuit is
working, and below the floor the diagonal returns zero at every depth and so no derivative at
all.
\paragraph{The second use: transition amplitudes.} Take the weak reference to be the strong one
with an operator applied, $\chi_W = O\chi_S/\|O\chi_S\|$. The residue of Eq.~\eqref{eq:gsw} is then
\begin{equation}
\begin{aligned}
R_n&=c_S^{*}c_W=\frac{\braket{\chi_S|E_n}\braket{E_n|O|\chi_S}}{\|O\chi_S\|},\\
\braket{E_n|O|\chi_S}&=\frac{R_n\,\|O\chi_S\|}{\braket{\chi_S|E_n}},
\end{aligned}
\label{eq:transamp}
\end{equation}
the denominator being the strong overlap the diagonal channel has already returned and the norm
coming from the normalisation of the preparation, or from a separate
$\braket{\chi_S|O^\dagger O|\chi_S}$, or from the heralding probability where $O$ is applied by a
linear combination of unitaries. What changes is the \emph{power} of the amplitude that is
measured. A conventional readout of a transition strength measures a probability, so an amplitude
$\alpha$ enters the record as $\alpha^2$ and is lost once $\alpha$ falls to about the square
root of the floor. Equation~\eqref{eq:transamp} is linear in it. Weak transitions are exactly
where that matters: spin-forbidden transitions under a spin--orbit
operator~\citep{marian2012soc}, symmetry-forbidden ones made weakly allowed by vibronic
borrowing~\citep{herzberg1933vibronic}, two-photon and higher multipole amplitudes,
charge-transfer amplitudes between weakly coupled diabats. In each the square is under the floor
and the amplitude is not. The phase comes with the modulus, which a squared readout cannot give at
all, so a set of lines from the one acquisition returns a complex transition spectrum rather than a
set of intensities \textemdash{} and relative phase is what distinguishes constructive from
destructive interference between pathways. Oscillator strengths
$f_n\propto(E_n-E_g)|\braket{E_n|O|\chi_S}|^2$, a polarisability
$\alpha(\omega)=\sum_n|\braket{E_n|O|\chi_S}|^2[(E_n-E_g-\omega)^{-1}+(E_n-E_g+\omega)^{-1}]$, and a
diabatic transfer rate all follow from the measured amplitudes, computed rather than measured as
probabilities.

\begin{figure*}[t]
\centering
\includegraphics[width=\textwidth]{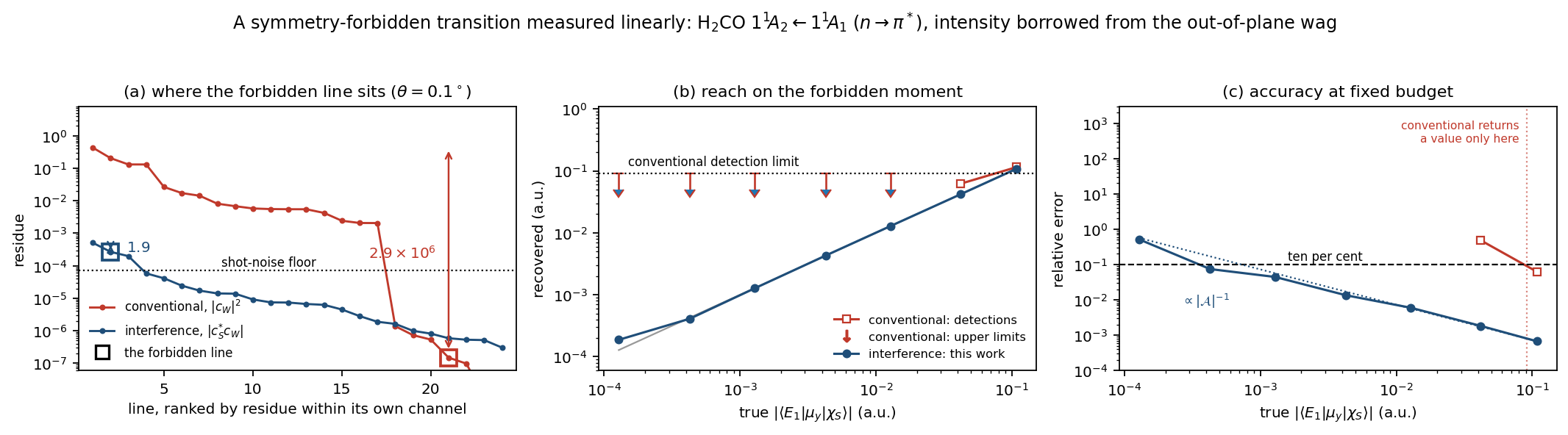}
\caption{\textbf{A symmetry-forbidden transition measured at linear sensitivity.} Formaldehyde
$1^1\!A_2\leftarrow1^1\!A_1$ ($n\rightarrow\pi^*$), CAS(6e,6o)/6-31G, the moment turned on from an
exact zero by the out-of-plane wag. \textbf{(a)} the lines of one acquisition at
$\theta=0.1^\circ$, each channel's residues sorted within that channel. The forbidden line ranks
$21$st of $24$ in the conventional record, $2.9\times10^{6}$ below the strongest line there, and
$2$nd of $24$ in the interference record, $1.9$ below it, because the cross channel weights each
line by the reference's own amplitude. \textbf{(b)} recovered against true moment; the
conventional readout returns a value at the top two geometries only and upper limits below.
\textbf{(c)} relative error, falling as $|A|^{-1}$ for the interference channel.}
\label{fig:forbidden}
\end{figure*}

\paragraph{A worked forbidden transition.} Formaldehyde's
$1^1\!A_2\leftarrow1^1\!A_1$ ($n\rightarrow\pi^*$) transition is electric-dipole forbidden in
$C_{2v}$ for every component, and the out-of-plane wagging mode of the CH$_2$ group,
$\nu_4$ of $b_1$ symmetry, is the vibronic
coordinate from which it borrows intensity in the Herzberg--Teller
sense~\citep{herzberg1933vibronic,clouthier1983formaldehyde}. In CAS(6e,6o)/6-31G the root sits at $5.14\,$eV and
at the planar geometry the moment is $1.5\times10^{-14}$ a.u., which is zero to machine
precision; pyramidalising the carbon by $\theta$ turns it on linearly. The wag preserves the
$y\!\rightarrow\!-y$ mirror, so $\braket{E_n|\mu_y|E_n}=0$ in every root at every geometry, and
$\mu_y$ therefore measures a pure transition moment with no permanent-dipole term to subtract.
The reference is the two-determinant state $\chi_S\propto\ket{\rm HF}+\ket{n\rightarrow\pi^*}$,
which a shallow circuit prepares and which carries $|\braket{E_1|\chi_S}|=0.69$ on the forbidden
root; the probe is $\chi_W=\mu_y\chi_S/\|\mu_y\chi_S\|$. Sweeping $\theta$ from $0.03^\circ$ to
$30^\circ$ moves the moment over three decades, from $1.28\times10^{-4}$ to
$1.08\times10^{-1}$ a.u., at $10^6$ shots per quadrature over $400$ samples
(Fig.~\ref{fig:forbidden}).

The conventional readout returns a value at only the top two geometries, saturating below
$9.1\times10^{-2}$ a.u. The interference channel tracks the true moment over the whole sweep,
to $1.28\times10^{-4}$ a.u.\ at $51\%$ and to $4.3\times10^{-4}$ at $7\%$, holding better than
$1.5\%$ from $4.3\times10^{-3}$ upward, with the error falling as $|A|^{-1}$ as predicted. In
oscillator strength that is $f$ recovered as $4.2\times10^{-7}$ against an exact
$4.3\times10^{-7}$, and $4.5\times10^{-4}$ against $4.5\times10^{-4}$ \textemdash{} strengths
five orders of magnitude below the bright transitions of the same molecule.

Panel (a) ranks the lines of one acquisition by residue within each
channel. The dipole operator scatters $\chi_S$ across the whole manifold, so $\chi_W$ carries its
weight at $10$ to $44\,$eV, and the forbidden line ranks $21$st of $24$ in the conventional
record, its residue $1.5\times10^{-7}$ sitting $2.9\times10^{6}$ below the strongest line there. No estimator separates that at any
tolerable budget, and the conventional limit above is set by that dynamic range rather than by
the shot floor alone. The interference weights every line by the reference's \emph{own}
amplitude there, so lines $\chi_S$ does not populate are suppressed in the same measurement that
amplifies the wanted one: the same line ranks $2$nd of $24$ and the ratio collapses from
$2.9\times10^{6}$ to $1.9$! This is a filter matched to the reference, and it is a distinct
benefit from the linear-versus-quadratic scaling, one that grows with how badly the operator
scatters the reference. The effect belongs to the interference itself rather than to the cross-correlation
geometry. The superposition realisation described below reaches the same line by a different
route, riding on a pedestal of the reference's own population $|c_S|^2/2$, and there the
ratio is $1.007$ --- no suppression of the unwanted lines at all.

\begin{figure*}[t]
\centering
\includegraphics[width=\textwidth]{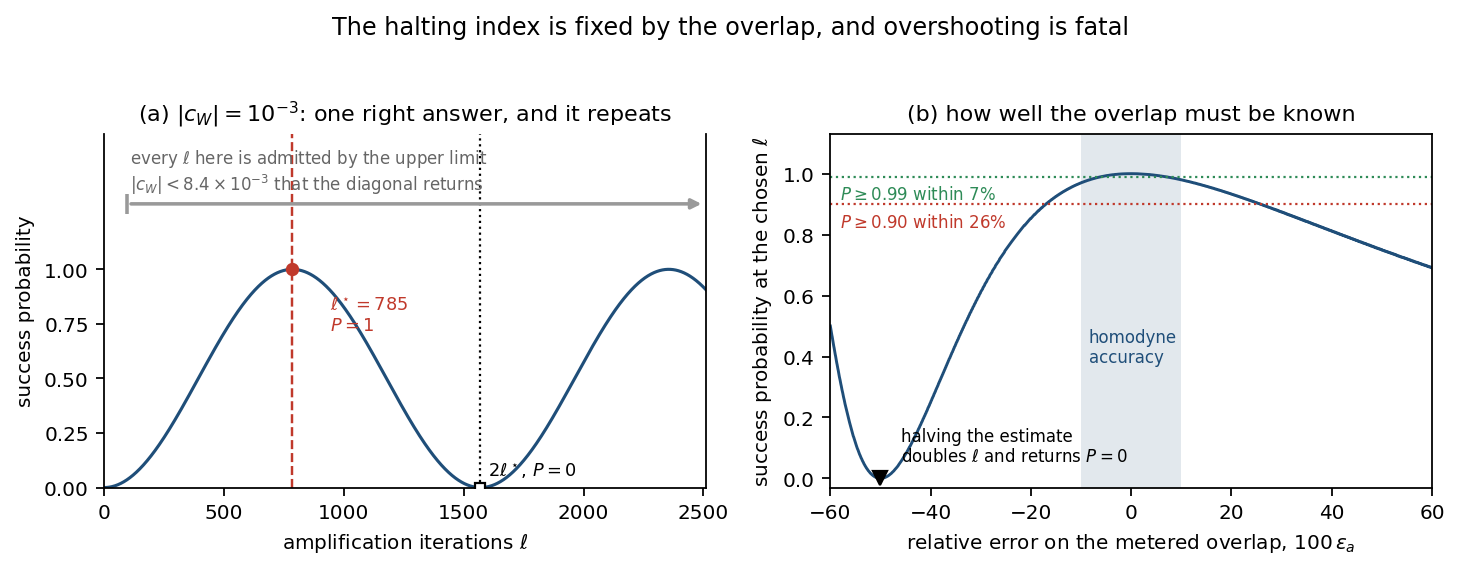}
\caption{\textbf{Why a metered overlap is a prerequisite for amplification, not a convenience.}
\textbf{Left:} success against iteration count at $|c_W|=10^{-3}$. The amplification is complete at
$\ell^\star=785$ and returns to zero at $2\ell^\star$, so overshooting is fatal and running longer is not
a safe course. The bar above marks every $\ell$ admitted by the upper limit
$|c_W|<8.4\times10^{-3}$ that the diagonal channel returns below its floor, which is the whole
range to the right of $\ell=93$. \textbf{Right:} success at the chosen $\ell$ against the
relative error on the metered overlap. The tolerance is $7\%$ for $P\ge0.99$ and $26\%$ for $P\ge0.90$; halving
the estimate doubles $\ell$ and returns nothing. The shaded band is the ten-per-cent accuracy the
homodyne channel holds to $|c_W|=3\times10^{-4}$.}
\label{fig:halting}
\end{figure*}

\paragraph{The third use: halting an amplification.} Here is the case where it is not a diagnostic but a
prerequisite. Amplitude amplification anchored on the warm-start reflection rotates by
$2\arcsin a$ per iteration, so after $m$ iterations the amplitude on the marked state is
$\sin\!\big((2m+1)\arcsin a\big)$ and it must be halted at
\begin{equation}
\ell^\star=\Big\lfloor\frac{\pi}{4\arcsin a}-\frac12\Big\rceil\approx\frac{\pi}{4a},
\label{eq:kstar}
\end{equation}
which depends on the overlap and on nothing else. Two features of that dependence make the overlap
a necessary input (Fig.~\ref{fig:halting}). First, \emph{there is no safe
direction of error}. An adiabatic schedule run longer than needed still succeeds, so a lower bound
on $a$ suffices to set it. An amplification run longer than needed \emph{fails}, the amplitude
returning to zero at $2\ell^\star$ and oscillating with that period thereafter. At $|c_W|=10^{-3}$,
$\ell^\star=785$ gives $P=1.0000$ and $\ell=1570$ gives $P=0.0000$ exactly. An upper limit on the
overlap, which is precisely what the diagonal returns below its floor, bounds $\ell^\star$ from below
and does not determine it. Secondly, the tolerance is finite, not tight, and not symmetric. Setting $\ell$ from
$\hat a=a(1+\epsilon_a)$ holds $P\ge0.99$ for $\epsilon_a\in[-6.0\%,+6.8\%]$ and $P\ge0.90$
for $\epsilon_a\in[-17\%,+26\%]$. The narrow side is under-estimation, which is the same
asymmetry again, since under-estimating $a$ lengthens $\ell$ and it is length that fails.
$P$ collapses to zero at $\epsilon_a=-\tfrac12$, where the halting index has been doubled. The homodyne channel holds ten per cent to $|c_W|=3\times10^{-4}$, inside the band
for $P>0.97$, while the diagonal of the same state returns no estimate at all below
$8.4\times10^{-3}$. So this is the sharpest answer to what metering buys: not a cheaper
computation, but a back-end that is otherwise \emph{inoperable}. Where
the accuracy does miss the tolerance, a fixed-point variant~\citep{yoder2014fixedpoint}, which does
not overshoot, or the local-adiabatic schedule of \S\ref{sec:grover}, which keeps the $1/a$
scaling and tolerates an overestimate, is chosen instead. And the same number arbitrates between
back-ends before any is run, the costs being $a^{-1}$ for amplification, $a^{-2}$ for a linear
adiabatic schedule and $a^{-4}$ for a spectral acquisition. That arbitration is a comparison of
\emph{overlap} costs only. Amplification carries a second precondition the other two do not,
the explicit ground-state reflection of \S\ref{sec:grover}, so a metered $a$ makes it
operable rather than sufficient.

Four limitations bound the claim, and they are not small. The local oscillator must have
amplitude on the line concerned, so a state dark to $\chi_S$ is beyond reach and none of the
coverage argument of \S\ref{sec:multistart} is displaced. The acquired unitary contains both
preparation circuits, so its depth is the sum of the two preparation depths and the Trotter
block, against one preparation depth and the Trotter block for the diagonal, and the quadratic
saving in shots is bought with depth. Only the phase relative to $c_S$ is observable,
the eigenvector's own phase being convention. And the interference is not unique to the cross
correlation, since the diagonal of $(\chi_S+\chi_W)/\sqrt2$ contains $2\,{\rm Re}(c_S^{*}c_W)$
linearly and gives the same scaling at the same preparation cost, with a subtraction of two
$\mathcal O(1)$ quantities where Eq.~\eqref{eq:gsw} has none. It escapes the dynamic-range trap too,
though by a different route. The wanted line sits on the reference's own $|c_S|^2/2$ pedestal
rather than being weighted by $c_S$, which on the formaldehyde record leaves it within a factor
$1.007$ of the strongest line. It is the interference that does the work, not the particular
way of arranging it.

The measurement itself is not new. Off-diagonal matrix elements
$\langle\chi_i|\ee^{-iH_Pt}|\chi_j\rangle$ are the standard currency of quantum Krylov and filter
diagonalisation, where they populate a generalised eigenproblem and are acquired by exactly this
interferometer~\citep{parrish2019filter,stair2020multireference,klymko2022real,cortes2022krylov}.
What is different here is the use. Those methods want the matrix in order to diagonalise it,
and read residues only as a by-product; the point above is that one residue of that same matrix
is a metrological instrument for a quantity the diagonal cannot reach, and that the quantity it
reaches is the one the architecture spends.

\newpage
\warppart{Front-ends}{Seven widely different sources of a warm start, and then a test of
whether the argument itself depends on the gate model. Nothing in Parts I to III changes as the
front-end is swapped.}
\label{part:frontends}

\section{The front-end is a swappable module}
\label{sec:frontends}

Swappability has been asserted since \S\ref{sec:intro} and used implicitly throughout.
This part is the experiment that earns it. Nothing in either back-end is tied to a
particular choice of front-end. The spectral readout needs only a state to time-evolve,
and WA-ASP needs only a state to anchor its projector driver. What both consume is the
spectral overlap distribution $p_n=|\langle E_n|\chi\rangle|^2$, whose leading component
is the target overlap $p_0$, together with the way the rest of the weight is spread. Since
that is all that crosses the algorithmic interface, the front-end can be swapped for
different preparation mechanisms. This does not imply equal end-to-end cost: the
preparation, compilation, sampling, and error-suppression costs of each front-end must
remain part of the overall resource accounting.

The sources we exercise below were chosen to be as unlike one another as possible, and the
back-ends are held fixed as they are swapped. This is a validation of interface independence,
not a ranking of the front-ends or a claim that their preparation costs are comparable. Counting the variational eigensolver that
served as the front-end throughout Parts~\ref{part:readout} to~\ref{part:together}, they
number seven: a tunable QAOA circuit (\S\ref{sec:qaoa}),
purely classical heuristics with no variational eigensolver anywhere in the pipeline
(\S\ref{sec:apps}), classical lattice reduction on a problem whose target is not the
ground state (\S\ref{sec:lattice}), a generative sequence model (\S\ref{sec:gen}), a
sample-based quantum diagonalisation (\S\ref{sec:sqd}), and a machine-learning trainer
whose loss is itself the Hamiltonian (\S\ref{sec:qml}). The catalogue is wider than
what we test. Also available are a quantum
annealer such as a D-Wave processor for combinatorial
$H_P$~\citep{kadowaki1998annealing,johnson2011dwave}; simulated annealing, a greedy or
local-search solution loaded into the device; a mean-field (Hartree--Fock) determinant,
coupled-cluster amplitudes, or a
tensor-network/DMRG state compiled to a circuit~\citep{white1992dmrg}; an imaginary-time
(QITE)~\citep{motta2020qite} or short-adiabatic prep; a physically motivated ansatz
(BCS, N\'eel, Slater). A machine-learning model that predicts good initial amplitudes, a
neural-network quantum state~\citep{carleo2017nqs}, say, or a \emph{generative sequence
model} that emits the preparation circuit outright
(\S\ref{sec:gen})~\citep{nakaji2024gqe,koziellpipe2026adaptgqe}; or a \emph{previous WARP run}
on a neighbouring instance (a nearby geometry, time step, or parameter), recycled by
continuation.

That last entry is one case of a general pattern. Villar-Rodriguez \emph{et
al.}~\citep{villarrodriguez2026transfer} set out the reuse of knowledge across related
problems under a single notation, drawing transfer learning and transfer optimisation
together and organising the available strategies into a taxonomy, and they instantiate it
on quantum routines through reverse annealing, a multitasking QAOA and a sequential
variational eigensolver. In the present language, a transfer of that kind is a
front-end whose preparation cost is amortised over a family of instances, and what it delivers across the interface is still $\{p_n\}$, priced on the
same terms as any other source.

Among these, sample-based quantum diagonalisation (SQD) is the case that makes the split sharp,
and \S\ref{sec:sqd} treats it separately. Every other front-end hands the
back-end one object, a state on the device. SQD hands it two, the sampled device state and
the classical vector obtained by diagonalising the projected Hamiltonian, and the two are
spent in different places. The state is read out, because it is already prepared and the
readout returns a true eigenvalue where the diagonalisation returns a Rayleigh--Ritz
bound. The vector is never prepared at all, because the certificates of
\S\ref{sec:stopping} need only $\braket{H_P}$ and $\braket{H_P^2}$ on the vector in hand,
and those are classical contractions over the sampled configurations. Reading out the free
state and certifying with the free vector is a feature that no single-object
front-end offers, and it is what turns the swappability argument from a convenience into a
concrete saving.

\subsection{QAOA, and the exchange rate made quantitative}
\label{sec:qaoa}
The warm start above was a VQE or, for the diagonal problems, a stand-in Boltzmann
state. Here we run the pipeline with a genuine, \emph{tunable} front-end (QAOA at
depth $D$~\citep{farhi2014qaoa,hadfield2019qaoa}) which turns the warm-start overlap
into a dial and lets us test this paper's central claim directly: overlap is the
conserved currency traded between front-end and back-end.

\begin{figure*}[t]\centering
\includegraphics[width=\textwidth]{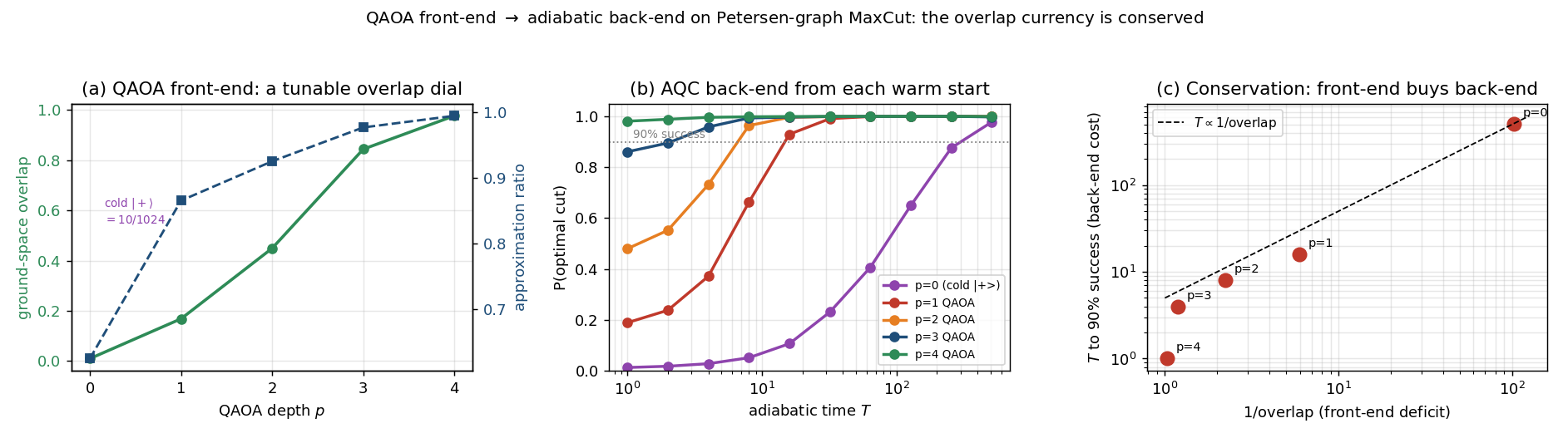}
\caption{\textbf{QAOA front-end $\to$ adiabatic back-end on Petersen-graph MaxCut.}
(a) QAOA depth is an overlap dial. (b) WA-ASP success versus adiabatic time from each
warm start. (c) The time to $90\%$ success scales as the inverse overlap. Front-end
effort buys back-end time at the gap law's exchange rate.}
\label{fig:qaoa}
\end{figure*}

\paragraph{Instance and front-end.} $H_P=-\sum_{(i,j)\in\mathcal E}\tfrac12(1-Z_iZ_j)$ encodes
MaxCut on the Petersen graph ($10$ vertices, $15$ edges, $3$-regular), so $10$ qubits;
QAOA performance on MaxCut is well characterised~\citep{crooks2018qaoa}. Brute force gives a maximum cut of $12$ and a $10$-fold ground space (including the
global $\mathbf{Z}_2$ complement symmetry). The QAOA state $\prod_{l=1}^{D}e^{-i\beta_l
B}e^{-i\gamma_l H_P}\ket{+}^{\otimes n}$ is optimised over
$(\boldsymbol\gamma,\boldsymbol\beta)$ by a classical outer loop (COBYLA, twelve
restarts, linear-ramp seed~\citep{zhou2020qaoa,sack2021annealinginit}). Depth buys quality. Writing the approximation ratio for the mean cut the state
returns divided by the maximum cut, so that $1$ is optimal, the ratios are
$0.866,0.925,0.977,0.995$ and the ground-space overlaps $0.168,0.449,0.844,0.977$ at
$D=1,2,3,4$ (Fig.~\ref{fig:qaoa}a), against $10/1024=0.0098$ for the uniform
$\ket{+}^{\otimes n}$.

\paragraph{WA-ASP Back-end, and the conservation law.} Feeding each state to WA-ASP, the
adiabatic time to $90\%$ success decreases as the front-end deepens: $T\!=\!512,16,8,4,1$
for $D=0,1,2,3,4$ (Fig.~\ref{fig:qaoa}b). Plotted against the inverse overlap these
collapse onto $T\propto1/a^2$ (Fig.~\ref{fig:qaoa}c), the linear-schedule prediction of
the gap law (App.~\ref{app:adiabatic}). This is the thesis in one picture. A  cold start
pays $512$ units of adiabatic time. One QAOA layer buys a $32\times$ reduction. Four
layers make the back-end nearly free. Front-end effort and back-end effort are
interchangeable at the exchange rate the gap law sets. The composition is what one
optimises, not either half alone.

\begin{figure}[t]\centering
\includegraphics[width=\linewidth]{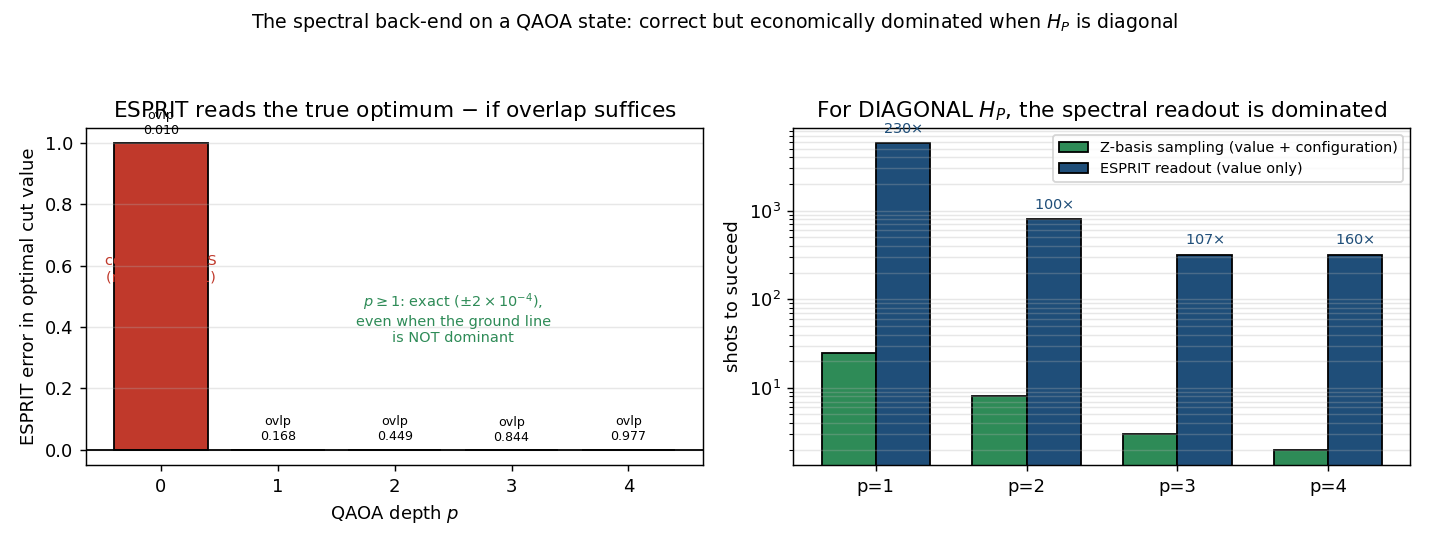}
\caption{\textbf{The spectral back-end on a QAOA state.} Left: ESPRIT recovers the
true optimal cut from every $D\ge1$ warm start (even at $D=1$, where the ground line
is subdominant) but fails from the cold start. Right: for diagonal $H_P$ direct
$Z$-basis sampling is $100$--$230\times$ cheaper and also returns the configuration.}
\label{fig:qaoa2}
\end{figure}

\paragraph{Spectral Back-end, revisited.} Two observations.
\emph{(i) It works, and non-trivially so.} A diagonal $H_P$ does not make the spectral
lines coincide, MaxCut on this graph having $11$ distinct integer cut values. ESPRIT
recovers the optimum exactly ($|{\rm err}|\!\le\!2\times10^{-4}$) from every QAOA state
with $D\ge1$ (Fig.~\ref{fig:qaoa2}, left). That includes $D=1$, where the ground line
carries weight $0.168$ while a \emph{different} line (cut $=11$) dominates at $0.397$.
Reading a subdominant line is exactly what a parametric estimator can do and a
bin-limited transform cannot. Integer-valued objectives also fix
the model order \emph{a priori} ($p=11$ here, read off the graph), removing ESPRIT's one
hyperparameter (\S\ref{sec:modelorder}). From the cold start the readout fails,
returning cut $=11$, consistent with the overlap precondition throughout. \emph{(ii) It
is dominated anyway.} When $H_P$ is diagonal its eigenstates are computational-basis
strings, so a single $Z$-basis measurement of the QAOA state returns a bitstring
\emph{and} its exact cost. Direct sampling so finds the optimum with $99\%$ confidence
in $\lceil\ln(0.01)/\ln(1-a^2)\rceil=25,8,3,2$ shots at $D=1,2,3,4$, while the readout
needs $\mathcal{O}(N/a^4)\approx5760,800,320,320$ shots (Fig.~\ref{fig:qaoa2}, right), a
$100$--$230\times$ penalty, and returns only the value, not the configuration. The spectral readout earns its keep on \emph{non-commuting} $H_P$ where no measurement returns
an eigenvalue directly. This is precisely the molecular setting of
\S\ref{sec:so2}--\ref{sec:h2o}.

\paragraph{Reading.} For combinatorial $H_P$ the operative composition is so
QAOA$\to$WA-ASP. The front-end supplies overlap, the adiabatic back-end converts it into
the optimal \emph{configuration}, and the spectral back-end, though correct, is outbid
by sampling. The pipeline's value here lies not in outrunning classical MaxCut, the instance being small enough to enumerate, but in the clean exhibition of the exchange rate between
front-end and back-end that the rest of this paper measures in less transparent
settings.

\subsection[Classical heuristics; readout after preparation]{Classical heuristics, and the readout after preparation}
\label{sec:apps}
Two end-to-end examples ground the method, one for each family, followed by a
question the two families raise together.

\paragraph{A drug-relevant molecule.} Carbon monoxide is pharmacologically active, it
binds heme iron, the basis of CO poisoning,  is an endogenous gasotransmitter, and is
delivered therapeutically by CO-releasing molecules~\citep{motterlini2002}, so its
ground-state energetics matter. On a 12-qubit CAS(6e,6o) Hamiltonian (STO-3G, frozen
core), the nine-parameter pair-UCCD bound is loose ($+0.065\,$Ha, $41\times$ chemical
accuracy, overlap $0.95$) yet ESPRIT reads the true $E_0=-111.331\,$Ha to $0.09\times$
chemical accuracy, recovering the bound's deficit. At a $1.33\times$ stretch modelling
bond homolysis (multireference onset, gap $0.120$, overlap $0.85$) the bound loosens to
$+0.115\,$Ha while ESPRIT still lands at $0.06\times$ chemical accuracy. The adiabatic
back-end independently prepares the state. At larger stretches the overlap collapses
($\sim\!0.1$) and the precondition of \S\ref{sec:so2mr} applies.

\begin{figure}[t]\centering
\includegraphics[width=\linewidth]{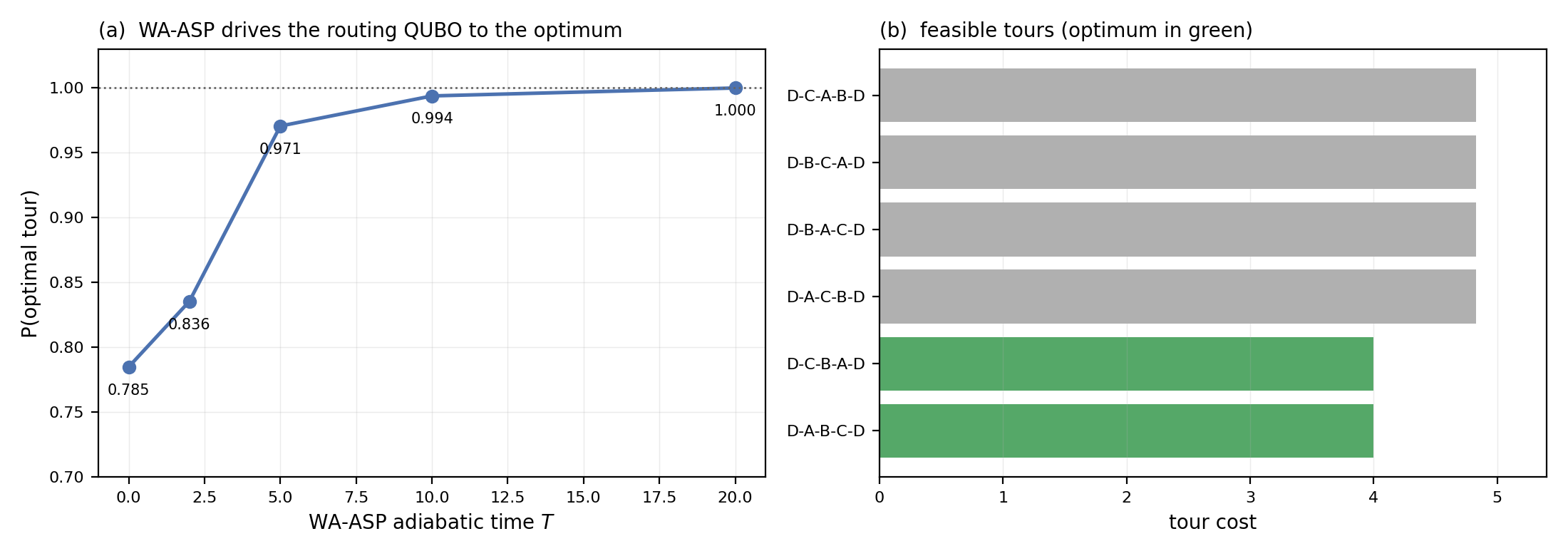}
\caption{\textbf{Routing application.} A delivery instance (left) and its optimal
tour. The adiabatic back-end drives the optimal-tour probability from the
variational warm start ($0.785$) to $0.9999$ (right).}
\label{fig:tsp}
\end{figure}

\paragraph{A routing problem.} A single-vehicle delivery tour over a depot and three
stops is encoded as a QUBO and mapped to a diagonal Ising $H_P$~\citep{lucas2014ising}
on $3\times3=9$ qubits (one-hot $x_{c,p}$, penalties for each stop and position used
once, objective the tour length). The optimal tour costs $4.0$. A variational warm start
(a Boltzmann-weighted feasible superposition standing in for a
QAOA~\citep{farhi2014qaoa} output) places probability $0.785$ on the optimum. Since
$H_P$ is diagonal the readout is dominated by direct computational-basis sampling
(\S\ref{sec:qaoa}), so the adiabatic back-end is operative. Projector-driven WA-ASP
amplifies the optimal-tour probability to $0.836,0.971,0.994,0.9999$ at $T=2,5,10,20$
(Fig.~\ref{fig:tsp}), the expected cost converging to $4.000$, showing the non-chemistry
branch end-to-end.

\begin{figure}[t]\centering
\includegraphics[width=\linewidth]{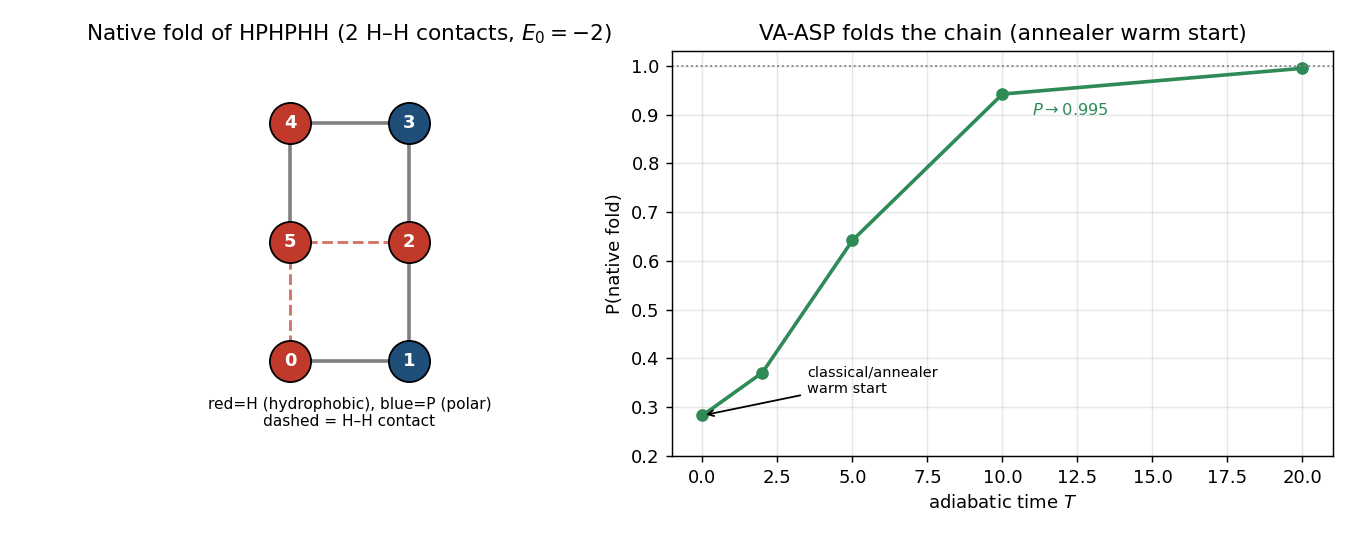}
\caption{\textbf{Protein-folding application.} The native fold of the HPHPHH chain
(left; red = hydrophobic, blue = polar, dashed = H--H contact) and the WA-ASP
convergence to it (right) from a non-variational (annealer/heuristic) warm start.}
\label{fig:protein}
\end{figure}

\paragraph{A protein-folding problem.} Lattice protein folding is a flagship hard
instance. The native structure minimises a contact energy over self-avoiding
conformations, and the problem maps to a diagonal Ising
$H_P$~\citep{perdomo2012folding,robert2021protein}. We fold the hydrophobic--polar chain
HPHPHH on a square lattice, encoding the bond directions in $8$ qubits, the first bond
fixed to remove translation/rotation,  with penalties for self-overlap and a reward $-1$
per non-adjacent hydrophobic (H--H) contact. The native fold is the compact loop of
Fig.~\ref{fig:protein} with two H--H contacts ($E_0=-2$). Here we deliberately seed
WA-ASP \emph{not} from a VQE state but from a Boltzmann-weighted feasible superposition
standing in for a classical heuristic or a quantum-annealer sample, placing probability
$0.283$ on the native fold, to underline that the warm start need not be variational.
Projector-driven WA-ASP raises the native-fold probability to $0.37,0.64,0.94,0.995$ at
$T=2,5,10,20$, with the expected energy converging to $-2$. As with routing, $H_P$ is
diagonal so the readout is dominated by direct sampling (\S\ref{sec:qaoa}) and the
adiabatic back-end is operative.

\begin{figure}[t]\centering
\includegraphics[width=\linewidth]{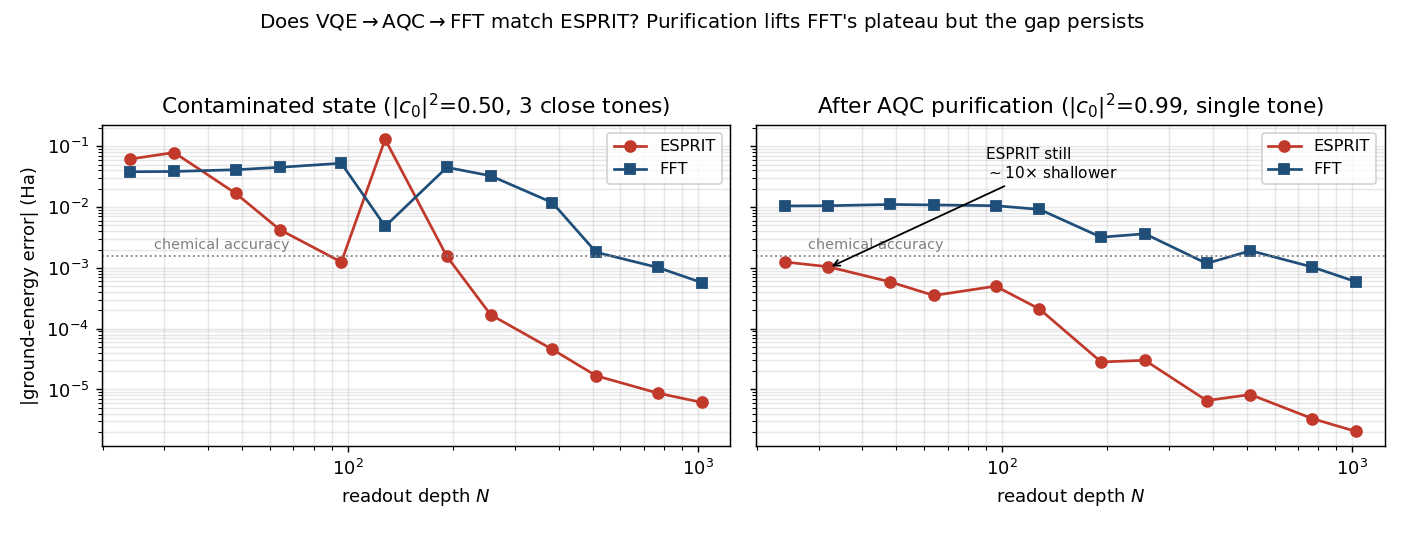}
\caption{\textbf{FFT vs ESPRIT after preparation.} Median ground-energy error vs
readout depth on a contaminated state (left) and after adiabatic purification to a
single tone (right). Purification lifts FFT's plateau but ESPRIT stays
${\sim}10\times$ shallower even on the pure tone.}
\label{fig:q3}
\end{figure}

\paragraph{Does preparation make the Fourier transform competitive?} Since adiabatic
preparation purifies the state toward the ground, one might expect a cheap FFT readout
to then match ESPRIT, removing ESPRIT's one liability, the \emph{a priori} model order.
We test this directly (Fig.~\ref{fig:q3}). Purification does lift the FFT's
contamination plateau. On a near-degenerate three-tone signal FFT needs depth
$N\!\sim\!500$ for chemical accuracy while a purified single tone needs $N\!\sim\!400$.
But the gap to ESPRIT does \emph{not} close, ESPRIT reaches chemical accuracy at
$N\!\sim\!100$ contaminated and at $N\!\sim\!25$ once purified, still ${\sim}10\times$
shallower than FFT \emph{on the same pure single tone}. The reason is that ESPRIT's edge
is not only multi-frequency resolution but parametric localisation of even one tone from
a short record, which the bin-limited FFT cannot match. The practical upshot is the
opposite of the premise: preparation helps the readout most not by enabling FFT but by
making ESPRIT's model order \emph{known}. A purified state has a single dominant line,
so $p\!\approx\!1$ is no longer a guess. These measurements identify the tested resource trade: the composition can be selected
according to whether coherent depth, model-order knowledge, or preparation cost is the
limiting resource. The hyperparameter-free FFT remains a viable fallback when its longer
acquisition is acceptable.

\subsection[A target that is not the ground state]{A lattice problem, and a target that is not the ground state}
\label{sec:lattice}
Every problem so far has asked for a ground state. The shortest-vector problem (SVP),
the hardness assumption underlying lattice-based post-quantum
cryptography~\citep{regev2009lwe}, does not. It asks for the \emph{first excited} state,
because its ground state is a known triviality. We use it to exercise one piece of
machinery the rest of the paper has not needed. Deciding which part
of the spectrum is visible is all.

\begin{figure*}[t]\centering
\includegraphics[width=\textwidth]{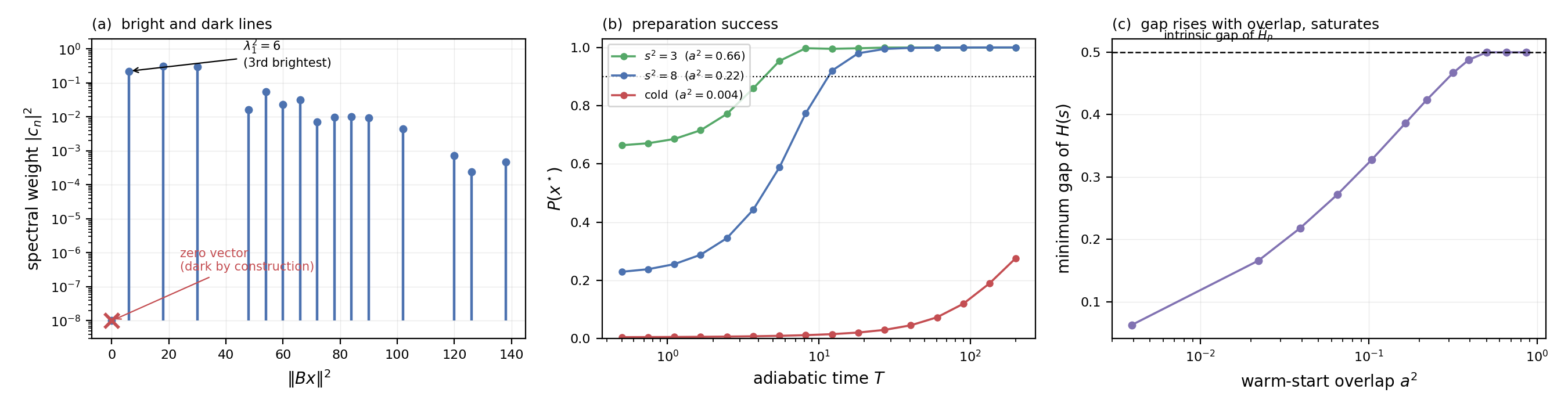}
\caption{\textbf{A rank-4 lattice.} (a) Spectral weights from the $\sigma_L^2=8$ warm start. The zero-vector line is dark by construction and the sought line at $\lambda_1^2=6$ is
only the third brightest. (b) WA-ASP success against schedule time from each warm start.
(c) The minimum gap rises with warm-start overlap and saturates at the intrinsic gap, as
in Fig.~\ref{fig:aqc}.}
\label{fig:svp}
\end{figure*}

\paragraph{Instance and encoding.} For a lattice basis $\mathbf B$ with Gram matrix
$\mathbf G=\mathbf B\mathbf B^{\!\top}$, a lattice vector is $\mathbf Bx$ with $x$ an integer coefficient vector, and
$\|\mathbf Bx\|^2=x^{\!\top}\mathbf Gx$ is a quadratic form. Bounding each $|x_i|$ and encoding it in
qubits gives a diagonal $H_P$ whose eigenvalues are the squared norms and whose
eigenstates are computational-basis
strings~\citep{joseph2021twoising,albrecht2023svpvqe}. We take the rank-4 lattice
\begin{equation}
\mathbf B=\begin{pmatrix}3&0&3&0\\3&-3&-6&0\\5&-4&5&6\\-2&-3&-1&4\end{pmatrix},\qquad
|\det\mathbf B|=108,
\end{equation}
with coefficients bounded to $x_i\in\{-1,0,1,2\}$ and encoded in two qubits each,
$x_i=2q_{i1}+q_{i0}-1$, so $8$ qubits and $256$ configurations collapsing onto $99$
distinct integer lines. See Fig.~\ref{fig:svp}. The shortest non-zero vector is $x^\star=(2,0,-1,1)$, so that the first successive minimum $\lambda_1$ of the lattice, the length of a shortest non-zero lattice vector, satisfies $\lambda_1^2=6$, unique in the box, a genuine combination of three basis vectors and
three times shorter in squared norm than the shortest basis vector of $\mathbf B$ itself
(which is $18$). But $x=0$ gives squared norm $0$. The true ground state is the zero vector, at
the known basis string $\ket{z_0}=\ket{01010101}$, and the state we want is the one
above it. This obstruction is the standing difficulty of the SVP-as-Hamiltonian
literature, met previously by non-adiabatic or excited-state annealing, by
folded-spectrum constructions, or by penalty terms that reintroduce auxiliary binary
variables~\citep{joseph2020notso,albrecht2023svpvqe}.

\paragraph{The front-end as a selection rule.} It is inexpensive here, and the reason is
the bright/dark partition of \S\ref{sec:excited}. The zero vector is \emph{one known
basis string}, so a warm start built over non-zero coefficient vectors carries zero
amplitude on it by construction. The trivial line is dark, absent from $g(t)$ at any
depth, and the lowest line the estimator recovers is the one we want. For the warm start
we take a discrete Gaussian over the lattice,
$c_x\propto\ee^{-\|\mathbf Bx\|^2/4\sigma_L^2}$ for $x\neq0$ and $c_0=0$, standing in for a
sampler seeded by a reduced basis~\citep{gpv2008}. The width $\sigma_L^2$, written with a
subscript to keep it distinct from the schedule parameter $s$ of \S\ref{sec:aqc}, from the
energy spread $\sigma$ of \S\ref{sec:stopping} and from the shot-noise $\sigma$ of
App.~\ref{app:shot}, is set in practice from the
Gram--Schmidt norms of an LLL- or BKZ-reduced basis~\citep{lll1982}, so that the
classical preprocessing lattice algorithms already perform is precisely what manufactures
the overlap. Here $\sigma_L^2$ is the dial, exactly as QAOA depth is in
\S\ref{sec:qaoa}.

Table~\ref{tab:svpread} reports a single acquisition at each width. At $\sigma_L^2=8$ the
target is only the \emph{third} brightest of twelve bright lines, because a Gaussian
over a lattice places more total mass on the more numerous longer vectors, and the
estimator returns it anyway, to $3\times10^{-4}$ in the median and never worse than
$10^{-3}$. This is the ``lowest line, not dominant line'' selection
rule of \S\ref{sec:qaoa}. And
because $\mathbf G$ is integral the spectrum is integer-valued, which fixes the model order
\emph{a priori} (\S\ref{sec:modelorder}) and lets a readout error below one half be
rounded. The estimate is not approximate but exact.

\begin{table}[t]\centering\small
\setlength{\tabcolsep}{3.5pt}
\begin{tabular}{lccccc}
\toprule
warm start & $a^2$ & bright & rank of & median & worst\\
& & lines & target & $|{\rm err}|$ & $|{\rm err}|$\\
\midrule
$\sigma_L^2=3$ & $0.658$ & $3$ & 1st & $3.4\times10^{-5}$ & $1.2\times10^{-4}$\\
$\sigma_L^2=8$ & $0.222$ & $12$ & 3rd & $3.0\times10^{-4}$ & $1.0\times10^{-3}$\\
$\sigma_L^2=8$$^\dagger$ & $0.222$ & $11$ & 3rd & $2.7\times10^{-4}$ & $9.0\times10^{-4}$\\
zero vector & $0$ & --- & --- & \multicolumn{2}{c}{dark}\\
\bottomrule
\end{tabular}
\caption{\textbf{Reading the shortest-vector length.} Error in the lowest recovered line
against $\lambda_1^2=6$, over $21$ independent shot-noise realisations of an acquisition
with $N=200$ samples at $2^{17}$ shots per point.
Here $^\dagger$ marks a second acquisition at the same width with the samples
traded against shots, $N=300$ at $2^{15}$ shots per point, showing the readout is not tuned to one budget.
Every realisation rounds to the exact value $6$. The trivial ground line is dark by
construction, so the lowest recovered line is the sought first excited state, and it
need not be the brightest for the estimator to return it.}
\label{tab:svpread}
\end{table}

\paragraph{A rank-one penalty for the preparation back-end.} WA-ASP drives to the ground
state of whatever it is given. We therefore give it
\begin{equation}
H_P'=H_P+\Lambda\,\ket{z_0}\!\bra{z_0},\qquad \Lambda>\lambda_1^2,
\label{eq:rank1}
\end{equation}
with $\Lambda=12$, which lifts the zero vector above the target and makes $x^\star$ the
ground state. A rank-one
projector onto a single basis string is the same object as the driver of
\S\ref{sec:aqc}. A phase on one computational-basis state, the very
$\ee^{i\theta\ket{\rm ref}\bra{\rm ref}}$ that already sits at the centre of the
projector sandwich, conjugated by single-qubit $X$ gates on the qubits where $z_0$ is
zero. No ancilla, no auxiliary binary variables, and no re-expression of $H_P'$ as a sum
of Pauli strings, which is what makes the penalty route unattractive when the constraint
must be pushed back through a QUBO~\citep{albrecht2023svpvqe}. Where $\Lambda$ is not known in advance, the spectral back-end supplies it, having resolved both lines in the same
acquisition.

Table~\ref{tab:svpprep} is then the gap law of \S\ref{sec:gaplaw} on a lattice. The
minimum gap rises with overlap and saturates at $0.500$, the intrinsic gap of
$\tilde H_P'$ in the normalisation where the first gap is unity, and the schedule times
of the warm starts collapse onto $T\propto1/a^2$ with $T\!\cdot\!a^2\approx2.5$--$2.9$.
A Gaussian at $\sigma_L^2=8$ reaches $90\%$ in $T=11.2$ against $1378$ for a uniform start over
the non-zero configurations, a $123\times$ reduction bought by classical lattice
reduction, the same trade the QAOA layer buys in \S\ref{sec:qaoa}. Following the
schedule at $\sigma_L^2=8$, the optimal-vector probability rises to
$0.311,0.550,0.862,0.987,0.9988$ at $T=2,5,10,20,50$ and the expected squared norm
converges $26.71\to21.14\to14.08\to8.44\to6.39\to6.06$ on the optimum $6$.

\begin{table}[t]\centering\small
\begin{tabular}{lccc}
\toprule
warm start & overlap $a^2$ & min gap & $T$ to $90\%$\\
\midrule
exact $\ket{x^\star}$ & $1.000$ & $0.500$ & ---\\
$\sigma_L^2=3$ & $0.658$ & $0.500$ & $4.3$\\
$\sigma_L^2=5$ & $0.393$ & $0.488$ & $7.0$\\
$\sigma_L^2=8$ & $0.222$ & $0.424$ & $11.2$\\
$\sigma_L^2=20$ & $0.065$ & $0.272$ & $38.0$\\
cold (uniform, $x\neq0$) & $0.0039$ & $0.063$ & $1378$\\
\bottomrule
\end{tabular}
\caption{\textbf{Sampler width as an overlap dial.} The minimum gap rises with overlap
and saturates at the intrinsic gap. The warm starts collapse onto $T\propto1/a^2$
($T\!\cdot\!a^2=2.86,2.76,2.49,2.47$), while the cold start sits in the Grover-like
regime at $5.40$.}
\label{tab:svpprep}
\end{table}

\paragraph{Reading.} As with routing and folding, $H_P$ is diagonal, so for
\emph{finding} the vector direct sampling dominates the readout (\S\ref{sec:qaoa}) and
the adiabatic half is the operative one. The example is deliberately compact, rank $4$ being small enough to enumerate exactly for verification, while cryptographically relevant ranks are of order $400$, requiring
$\sim\!10^3$ qubits on the tightest published bounds~\citep{albrecht2023svpvqe}. It is
that the machinery extends without modification to a target that is not the ground
state, that the choice of front-end is what makes the trivial solution invisible and the
wanted one reachable, and that a standard classical preprocessing step is a warm-start
source in exactly the sense of \S\ref{sec:frontends}.

\subsection{A generative front-end, and what it does not supply}
\label{sec:gen}
The QAOA front-end of \S\ref{sec:qaoa} made the currency tunable through circuit depth.
A recent and quite different front-end makes it tunable through a learned model, and it
is worth examining here because its designers optimise, and score, the  quantity which is not quite essential.

\paragraph{The construction.} Generative sequence models can be trained to emit a
state-preparation circuit directly. A transformer is trained by next-token prediction on
reference circuits produced by an adaptive variational
construction~\citep{grimsley2019adapt}, and conditioned on the problem Hamiltonian as
multimodal input. For a fixed active space the instance is specified by its vector of
canonically ordered Pauli coefficients, which a learned encoder maps into the model's
embedding and injects at designated token positions. The model then emits an ordered
list of operator-pool indices and rotation coefficients in a single autoregressive pass,
with no gradient evaluation, no iterative operator selection and no per-instance
re-optimisation. Optional reinforcement-learning post-training against an energy reward,
with self-distillation on the model's own best outputs, pushes it past the accuracy of
its training data. Such a front-end has been demonstrated on $12$--$16$ qubit active
spaces of a pharmaceutical molecule, generating circuits three to four orders of
magnitude faster than the procedure that produced its training data, with the generated
circuits compiled and executed on trapped-ion
hardware~\citep{koziellpipe2026adaptgqe}; related constructions generate optimisation
circuits from a Hamiltonian conditioning~\citep{nakaji2024gqe,tyagin2025qaoagpt}.

Three properties matter for what follows, and none of them is accuracy. The output is an
operator sequence with coefficients, that is, a circuit acting on a reference, so the
projector driver of \S\ref{sec:aqc} is available without further construction, unlike a
warm start delivered as a fixed state. The model returns a variational bound, and its
residual error is reported to grow with the length of the emitted sequence and with
active-space size, because next-token errors accumulate autoregressively. And the
training cost is paid once for a family of Hamiltonians and amortised over its members,
which is the ``previous run on a neighbouring instance'' of \S\ref{sec:frontends} with the transfer learned and stored~\citep{villarrodriguez2026transfer}.

\begin{figure*}[t]\centering
\includegraphics[width=\textwidth]{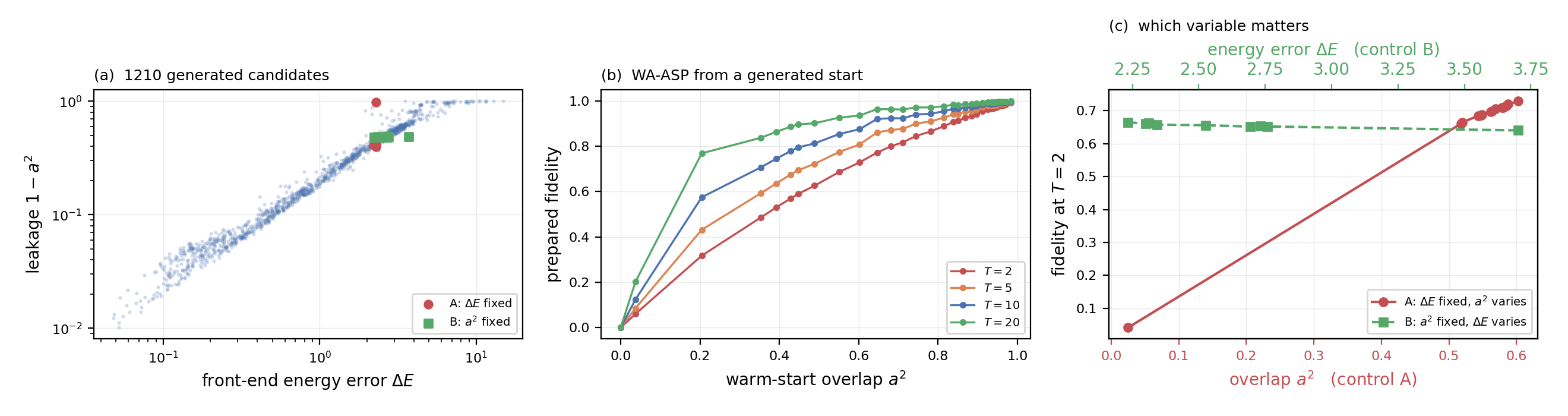}
\caption{\textbf{A generative front-end on the TFIM benchmark.} (a) $1210$ surrogate
candidates in the energy-error/leakage plane, with the two rule-selected control bands
marked. (b)
Fidelity WARP reaches at fixed schedule time, against warm-start overlap. (c) The
controls: at fixed energy error, fidelity tracks overlap (solid); at fixed overlap, a
$66\%$ change in energy error moves it barely (dashed).}
\label{fig:gen}
\end{figure*}

\paragraph{A surrogate we can measure.} What the present purpose requires is not a
trained model but the statistics of the circuits such a model emits, the published models
being directed at molecular active spaces. We build a surrogate
candidate pool on the TFIM benchmark of \S\ref{sec:scaling} at $n=10$, $h/J=1.1$
($E_0=-13.1417$, gap $0.451$; the smallness of that gap matters below), reproducing the
two error modes these models report: truncated or extended operator sequences, and
perturbed coefficients from per-digit generation. A converged ADAPT reference circuit on
this Hamiltonian reaches depth $26$ at $\Delta E=2.2\times10^{-3}$ and $a^2=0.9992$, the
analogue of the training data. Twenty-two structurally distinct circuits are then produced
by a randomised ADAPT that chooses uniformly among the three largest gradients at each
step, as a temperature-sampled model would, and each is perturbed $55$ times by
truncation, Gaussian coefficient noise and occasional operator substitution, giving
$1210$ candidates spanning $\Delta E\in[0.048,14.9]$ and $a^2\in[0,0.990]$
(Fig.~\ref{fig:gen}a).

\paragraph{Energy is not the currency.} The back-ends are paid in $a^2$, but a
generative front-end that samples a batch of candidates per instance has no direct access to it
and ranks them by variational energy instead. The two rankings are not the same. Writing the bound's deficit as in Eq.~(\ref{eq:deficit}), the energy is not a
function of the overlap alone. It is the overlap contaminated by the mean excitation energy of the leaked population $\bar\mu$, a nuisance
parameter that varies by $6.4\times$ across our pool. Best-of-$16$ by energy and by
overlap accordingly disagree on $23\%$ of draws ($27\%$ at best-of-$32$), and the energy
rule delivers mean leakage $0.0337$ against $0.0320$ for the overlap rule, some $5\%$
worse. Energy is a serviceable proxy; and Eq.~\eqref{eq:deficit} states the condition under
which it stops being one.

The controlled statement is sharper (Fig.~\ref{fig:gen}c). We select two control bands by
a stated rule: among all narrow bands of one variable, take the one spanning the widest
range of the other. Holding the energy error fixed to within $1\%$ (at $\Delta
E\approx2.28$), the pool still contains candidates whose overlap ranges from $a^2=0.025$
to $0.602$, and the fidelity WA-ASP reaches at $T=2$ ranges with it, from $0.042$ to
$0.729$. Holding the overlap fixed to within $0.7\%$ (at $a^2\approx0.52$) and letting
the energy error vary by $66\%$, from $2.23$ to $3.70$, the fidelity moves by $0.025$,
from $0.665$ to $0.640$. Two circuits with the \emph{same} variational energy can
therefore be worth a lot to the back-end; while a $66\%$
difference in that energy at matched overlap is worth only two points of fidelity.

The same nuisance parameter accounts for both bands, read in opposite directions. In the
first, $\Delta E$ is held fixed while $\bar\mu$ differs by $2.45\times$ ($2.3$ against
$5.8$). The product can only stay fixed if the leaked populations differ by the same
factor, so a $24$-fold spread in $a^2$ hides behind a single energy. In the second, $a^2$
is held fixed while $\bar\mu$ differs by $1.66\times$ ($4.6$ against $7.7$), and that
alone opens the $66\%$ spread in $\Delta E$ at constant value to the back-end. Fidelity
tracks $a^2$ and nothing else, while energy tracks $a^2$ multiplied by where the leakage
happens to sit.

There is no need to settle for a proxy, because the readout back-end measures the
currency directly. ESPRIT returns pairs $(E_n,|c_n|^2)$: the residue of the lowest line
\emph{is} $a^2$. A candidate can therefore be ranked by the quantity the preparation
back-end will actually spend, and ranking is a far weaker demand than spectroscopy, so a
short acquisition at low model order suffices to order a pool, with the full sampling
budget reserved for the winner. The front-end's figure of merit is thus not visible to
the front-end, and is an output of the back-end.

Where a spectrum is already in hand, energy does bound the overlap. Eckart's
inequality~\citep{eckart1930theory} gives
\begin{equation}
a^2\;\ge\;1-\frac{\Delta E}{E_1-E_0},
\label{eq:eckart}
\end{equation}
requiring only $E_0$ and $E_1$, which are properties of the instance rather than of the
candidate and so are obtained once and reused across the pool. The bound is monotone in
$\Delta E$, so ranking by energy does optimise something real: a guaranteed floor on
$a^2$. On this pool, however, it is almost everywhere empty. The floor is positive only
for $\Delta E<E_1-E_0=0.451$, while the pool spans $\Delta E\in[0.048,14.9]$ and both
control bands lie at $\Delta E\in[2.23,3.70]$, five to eight times the gap, where
Eq.~\eqref{eq:eckart} returns floors of $-3.9$ and $-7.2$ and so cannot certify even
$a^2>0$. A strongly correlated instance is precisely one with a small gap, and a small
gap is what makes the energy floor vacuous. The regime in which overlap most needs to be
known is the regime in which energy is least able to report it. Direct measurement is
then the only recourse.

Two remarks on the front-end itself follow. Minimising $\braket{H_P}$ minimises
$(1-a^2)\bar\mu$ and not $(1-a^2)$, so a variational optimiser is not pursuing overlap
even incidentally except when $\bar\mu$ happens to be constant across the candidates it
compares. Deeper ans\"atze and better optimisers buy energy, and buy overlap only through
that contaminated channel. Optimising the variance instead of the energy would target an
eigenstate directly, but it targets the \emph{nearest} eigenstate rather than the lowest,
reintroducing the excited-state capture of \S\ref{sec:stopping}, and it requires
$\braket{H_P^2}$. The two control bands are set out in Table~\ref{tab:bands}.

\begin{table}[t]\centering
\caption{\textbf{The two control bands.} Every row satisfies
$\Delta E=(1-a^2)\bar\mu$. The first pair fix the product and let its two factors vary
together, so a $24$-fold spread in $a^2$ sits behind one energy; the second pair fix
$a^2$ and let $\bar\mu$ alone open a $66\%$ spread in $\Delta E$. Fidelity is what
WA-ASP reaches at $T=2$.}
\label{tab:bands}
\setlength{\tabcolsep}{4pt}
\begin{tabular}{llrrrr}
\toprule
band & candidate & $\Delta E$ & $a^2$ & $\bar\mu$ & $F(T{=}2)$\\
\midrule
fixed $\Delta E$ & low overlap  & 2.28 & 0.025 & 2.34 & 0.042\\
                 & high overlap & 2.28 & 0.602 & 5.73 & 0.729\\
\addlinespace
fixed $a^2$      & low $\Delta E$  & 2.23 & 0.52 & 4.65 & 0.665\\
                 & high $\Delta E$ & 3.70 & 0.52 & 7.71 & 0.640\\
\bottomrule
\end{tabular}
\end{table}

\paragraph{Composing the two laws.} Equation~\eqref{eq:deficit} and the gap law
compose, and the composition is instructive about what energy can and cannot buy.
Writing $x=1-a^2$ for the leaked population, Eq.~\eqref{eq:deficit} reads $x=\Delta
E/\bar\mu$, and substituting into $\Delta_{\min}\approx2(1-s^\star)a\sqrt{1-a^{2}}$ gives
$\Delta_{\min}\approx2(1-s^\star)\sqrt{x(1-x)}$, the same two-level form as the variance
$\sigma=\Delta_{01}\sqrt{p_0p_1}$ of \S\ref{sec:stopping}. With the local-adiabatic schedule,
\begin{equation}
T\;\gtrsim\;\frac{1}{2(1-s^\star)}\Big(1-\frac{\Delta E}{\bar\mu}\Big)^{-1/2},
\label{eq:Tofdeltae}
\end{equation}
so the adiabatic time diverges as the inverse square root of the energy headroom
$\bar\mu-\Delta E$; the linear schedule replaces the exponent by $-1$. Two restrictions
apply. The gap law is an upper bound only for $a^{2}\lesssim0.4$. Beyond that the
factor $\sqrt{1-a^{2}}$ turns the prediction over at $a^{2}=1/2$ while the true
$\Delta_{\min}$ continues to rise toward the intrinsic scale, and the measured-to-predicted
ratio reaches $3.1$ at $a^{2}=0.94$. And $\bar\mu$ is not measurable from the energy,
being the same nuisance parameter that separates the two rankings above, so
Eq.~\eqref{eq:Tofdeltae} predicts a time only once $\bar\mu$ is fixed. What it does
without $\bar\mu$ is bracket one. Since $E_1-E_0\le\bar\mu\le\Omega$ with $\Omega$ the spectral
bandwidth, the lower end reproduces Eq.~\eqref{eq:eckart}, and the upper end gives a
guaranteed minimum cost $T\gtrsim[2(1-s^\star)]^{-1}(1-\Delta E/\Omega)^{-1/2}$ from the
measured energy alone. The bracket is as wide as $\Omega/(E_1-E_0)$, which is largest for the
strongly correlated instances the method targets \textemdash{} the same reason the energy
floor is vacuous there.

\paragraph{The residual is the back-end's job.} A candidate at $a^2=0.930$ carries a
variational bound $0.322$ above $E_0$, at $-12.819842$. ESPRIT on that state returns the
exact $E_0=-13.141677$ to a median error of $5.6\times10^{-6}$ over $21$ shot-noise
realisations from $N=200$ samples, and $1.3\times10^{-5}$ from $N=120$. The front-end's residual error, which is the quantity
these models are trained to reduce and the quantity that degrades as their sequences
lengthen, is removed entirely by the readout. This is the hybrid of \S\ref{sec:hybrid}
with a learned front-end in place of VQE, and it inherits that hybrid's properties
unchanged, including the precondition.

\paragraph{Reading.} Two practical consequences follow, whose value at scale is a matter
for further work. First, candidates should be scored by
the ground-line weight the readout already returns, or by the variance certificate of
\S\ref{sec:stopping}, rather than by the energy, since the energy is quadratically
insensitive to precisely the residual the back-end must remove. Second, the reward such
models optimise during reinforcement post-training is ordinarily built from a classical
reference energy, CASCI or CCSD, which is expensive or unavailable at the sizes where a
generative front-end is wanted.  Spectral back-end supplies the true eigenvalue, and the
ground-line weight supplies the overlap itself, so the model can be trained on the
resource its consumer spends. The composition, here as in \S\ref{sec:qaoa}, is what one
optimises, not either half alone.

The same reading applies to a front-end that amortises the classical optimisation itself.
Karim \emph{et al.}~\citep{karim2026} train a network on the intermediate steps of earlier VQE
runs to predict near-optimal angles across a family of Hamiltonians at no additional quantum
cost, and report that on four-qubit HeH$^+$ the result stalls $0.019\Ha$ short of the exact
energy and outside chemical accuracy, on states whose ground-line weight is some $0.99$, so that
what limits that number is the instrument used to read them and not the overlap they supply, a
case we take up quantitatively in separate work.

\subsection{Sample-based diagonalisation, and its two outputs}
\label{sec:sqd}

The front-ends exercised so far all hand the back-end a single object, a state on the
device. Sample-based quantum diagonalisation (SQD) is different, and it is the front-end most likely to be reached for in chemistry
at scale~\citep{robledo2025chemistry,kanno2026qsci}.

SQD prepares a shallow correlated ansatz, samples it in the computational basis, collects
the sampled configurations into a subspace $\mathcal{S}$, and diagonalises $H_P$ projected
onto $\mathcal{S}$ classically, iterating with a configuration-recovery step that repairs
samples of the wrong particle number. What it returns is a Rayleigh--Ritz value, so an
\emph{upper bound} on $E_0$, together with a classical vector of configuration-interaction
coefficients supported on $\mathcal{S}$. A pipeline that has run SQD therefore holds two different objects, and the
architecture of this paper uses them in two different ways.

The first is the \emph{device state} that was sampled. It is already present on the processor
after the SQD sampling stage, so the WARP readout requires no additional state-preparation
circuit beyond that already incurred by SQD. This is a reuse of an existing quantum state,
not a claim that preparing the state was free. The second is the \emph{classical vector}. It never has to be prepared at all in
order to be useful, because every certificate of \S\ref{sec:stopping} needs only
$\langle H_P\rangle$ and $\langle H_P^2\rangle$ on the vector in hand, and those are
classical contractions over $\mathcal{S}$. Preparing that vector on the device, by
contrast, costs a multi-determinant state-preparation circuit, so the sensible division is
to read out from the state that is free and certify with the vector that is free.

We test this on N$_2$ in a CAS(10e,8o) active space in the 6-31G basis, $3136$
determinants, at equilibrium and at a stretched geometry where the single-reference
picture has collapsed. The SQD front-end is emulated. The
ansatz is the first-order perturbative vector of the CAS Hamiltonian, which is the object
an LUCJ circuit's coupled-cluster amplitudes encode, and it is sampled under multinomial
shot statistics. No configuration recovery is modelled, since that repairs hardware noise
we are not simulating. The signal $g(t)$ is then sampled under the binomial statistics of
the interferometer as elsewhere in this paper.

\begin{table}[t]
\centering
\footnotesize
\setlength{\tabcolsep}{4pt}
\begin{tabular}{lrr}
\hline
 & $R=1.10$\,\AA & $R=2.00$\,\AA \\
\hline
reference determinant $a^2$        & 0.9341    & 0.0433 \\
device-state $a^2$                 & 0.9790    & 0.0903 \\
gap $E_1-E_0$ (Ha)                 & 0.29316   & 0.01069 \\
$|\mathcal{S}|$ at $10^6$ shots    & 51        & 34 \\
SQD vector $a^2$                   & 0.9974    & 0.3840 \\
$r_{\rm eff}$ of the device state  & 1.0       & 3.7 \\
\hline
SQD bound error (Ha)               & $5.1\times10^{-3}$ & $7.9\times10^{-2}$ \\
WARP readout error (Ha)            & $7.1\times10^{-7}$ & $1.4\times10^{-6}$ \\
\hline
\end{tabular}
\caption{An SQD front-end on N$_2$ CAS(10e,8o). The readout is taken on the same device
state the sampling came from, at $10^6$ shots, and returns a true eigenvalue where SQD
returns a bound.}
\label{tab:sqd}
\end{table}

\begin{figure*}[t]
\centering
\includegraphics[width=0.92\textwidth]{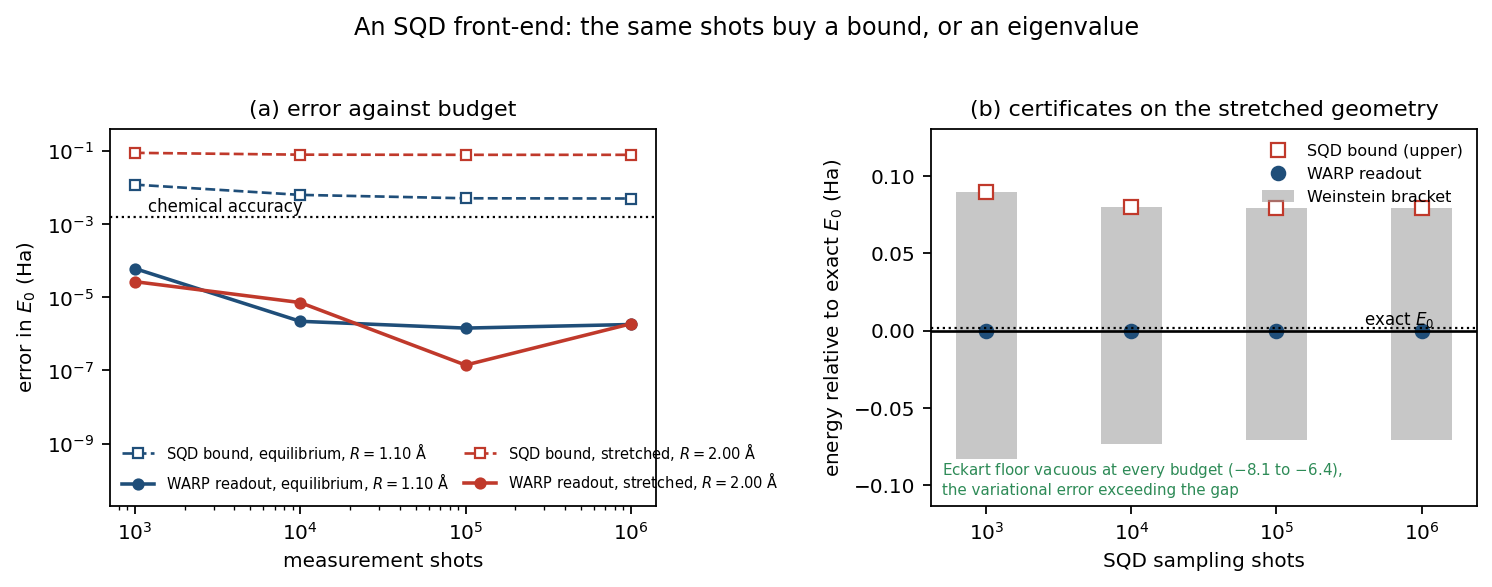}
\caption{\textbf{An SQD front-end.} \textbf{Left:} error in $E_0$ against measurement
budget. The SQD bound plateaus once the subspace has absorbed everything the ansatz
samples with appreciable probability, while the readout of the same device state
continues to improve. \textbf{Right:} on the stretched geometry, the Weinstein bracket
built from the SQD vector at no quantum cost, the SQD bound as its upper endpoint, and
the readout, all relative to the exact $E_0$.}
\label{fig:sqd}
\end{figure*}

Table~\ref{tab:sqd} and Fig.~\ref{fig:sqd} report it. At equilibrium SQD converges
quickly and then plateaus at $5.1\,$mHa, three times chemical accuracy, because the
subspace has absorbed everything the ansatz samples with appreciable probability and a
Rayleigh--Ritz value cannot do better than the span of what it sampled. The readout of the
same device state returns $E_0$ to $7\times10^{-7}\,$Ha, four orders of magnitude closer,
at model order six. Stretched, the reference determinant carries $a^2=0.043$ and the
ansatz $0.090$, the single-reference picture having collapsed; SQD is then $79\,$mHa high,
fifty times outside chemical accuracy, while the readout of that same $9\%$ overlap
returns $E_0$ to $1.4\times10^{-6}\,$Ha. In both cases the readout lands below the bound,
as it must, being non-variational.

The mechanism has since been measured on hardware, which we had not expected to see so soon.
Seitz \emph{et al.}~\citep{seitz2026workflows} optimise a sixteen-site transverse-field Ising
state by VQE on a superconducting processor, re-prepare it on a trapped-ion processor, and read
it there by phase estimation. The variational energy sits $2.47\pm0.10$ above the exact one in
units of the coupling, and reading that same state returns $0.61\pm0.36$. Their noiseless runs
separate the two front-ends at a fixed back-end and a fixed shot count, a classically optimised
product state giving $0.8\pm1.1$ against $0.25\pm0.21$ for the variational one. That is an
exchange rate on the readout side, priced in the width of an error bar rather than in record
length, and it is the pairing of \S\ref{sec:resource} arrived at by measurement.

The certificates come free with the classical vector. The Weinstein bracket
$[\langle H_P\rangle-\sigma,\,E_{\rm SQD}]$ contained the true $E_0$ at every geometry and
budget tested, with a width of $0.11$--$0.16\,$Ha. The Eckart floor at equilibrium returns
$a^2\ge0.983$ against a true $0.997$, valid and tight. Stretched it returns $-6.4$, that
is, vacuous, because the variational error $79\,$mHa exceeds the gap $10.7\,$mHa. That is
the regime limitation of \S\ref{sec:stopping} behaving exactly as stated, and it is
informative rather than merely negative, since it reports that the SQD vector is not close
enough for the guarantee to bite while the bracket still localises the answer.

Two findings are also worth recording. First, the example turned out to require the
adaptive machinery of \S\ref{sec:adaptive}, in two separate
ways. A fixed model
order failed at both ends, being too small on the stretched geometry and overfitted at
high shot counts, and selecting the order by reconstruction residual against the
independently known noise floor fixed it. And the hard unit-circle admissibility
test rejected every root at $10^3$ shots, where the noise displaces roots from the circle
by more than five per cent. The noise-calibrated region recovered them. Second, and
of practical importance to benchmarking SQD, the active space is defined only up to a
unitary rotation among its orbitals. Any such rotation leaves the CAS spectrum exactly
invariant but changes the expansion over determinants, hence both the reference overlap and
the sampled subspace, hence the bound. With plain Hartree--Fock orbitals the degenerate $\pi$
orbitals of N$_2$ mix arbitrarily and the device-state overlap moved by about half a per
cent between runs \textemdash{} more than the difference the method is used to resolve.

The fix is what we call a \emph{deterministic canonicalisation}. Given the
molecule, the basis and the active space, returns one and only one ordered set of active
orbitals, so two runs on the same input agree to within a sign. Ours has three parts. Solve the
self-consistent field to a tolerance tighter than the accuracy claimed of the answer, here
$10^{-12}$, so the solver has converged well inside the region where the orbitals are still
moving. Adapt to the point group and label by irreducible representation, which removes the
freedom to mix across symmetries and confines what remains to rotations within each block.
Order within each block by a stated key, here ascending orbital energy, so a near-degenerate
pair is presented in a fixed order rather than the order the solver happened to produce. A
localisation with a geometric key~\citep{foster1960canonical,pipek1989localization}, or natural
orbitals ordered by occupation~\citep{lowdin1955natural}, serve equally well What matters is not which rule but that a rule is stated, since it is the arbitrariness
and not any particular choice that destroys reproducibility. The step costs nothing on the
device, and makes the subspace, the bound, the reference overlap and $r_{\rm eff}$ functions of
the molecule and active space alone, so that a change between runs is attributable to the
sampling.

\subsection[A better basis is not a cheaper readout]{The encoding is a choice too: a better basis is not a cheaper readout}
\label{sec:basis}

Everything so far has held $H_P$ fixed and varied what is handed to it. But $H_P$ is the Hamiltonian projected onto a chosen single-particle basis, and that
basis is optimised. A recognised line of work re-optimises the exponents and contraction
coefficients of a minimal basis so as to return a lower variational energy \emph{at a fixed
spin-orbital count}, and hence at a fixed qubit count~\citep{deb2026basis}. Better energy for
the same qubits is exactly what a resource-limited device wants.

We report a negative observation about that programme, in the currency of this paper. A basis
chosen for its energy is not thereby the basis that is cheap to read out, because energy and
acquisition cost are independent figures of merit, and improving one carries no information
about the other.

\paragraph{The mechanism, stated carefully.} Two cases must be kept apart, and the paper needs
both. A rotation among the orbitals of a \emph{fixed} one-particle span leaves the many-body
operator, its spectrum and its eigenstates untouched, and changes only the expansion of the
reference determinant over them. That is the gauge freedom \S\ref{sec:sqd} pins by
canonicalisation. A change of the basis \emph{set} --- different exponents, different
contraction coefficients --- is a change of the one-electron space, so $H_P$ becomes a different
operator with a different spectrum and different eigenstates, over which a reference expands
differently again. In both cases the distribution $\{p_n\}$ moves while the spin-orbital count
does not, so two bases of equal qubit count can present the same $p_0$ but  acquisition costs
differing by orders of magnitude, for the reason \S\ref{sec:resource} gives. Neither case is a
rotation of the eigenbasis. It is what makes the second case a
different Hamiltonian rather than the same one seen differently.

\paragraph{Pricing.} We price an acquisition as $r_{\rm eff}/p_0^{2}$, the inverse square in the
overlap amplitude that App.~\ref{app:shot} derives together with the model order that
$r_{\rm eff}$ sets. It is a \emph{stated pricing} and not a measured runtime, and every
statement below is a statement about it.

\begin{figure*}[t]\centering
\includegraphics[width=\linewidth]{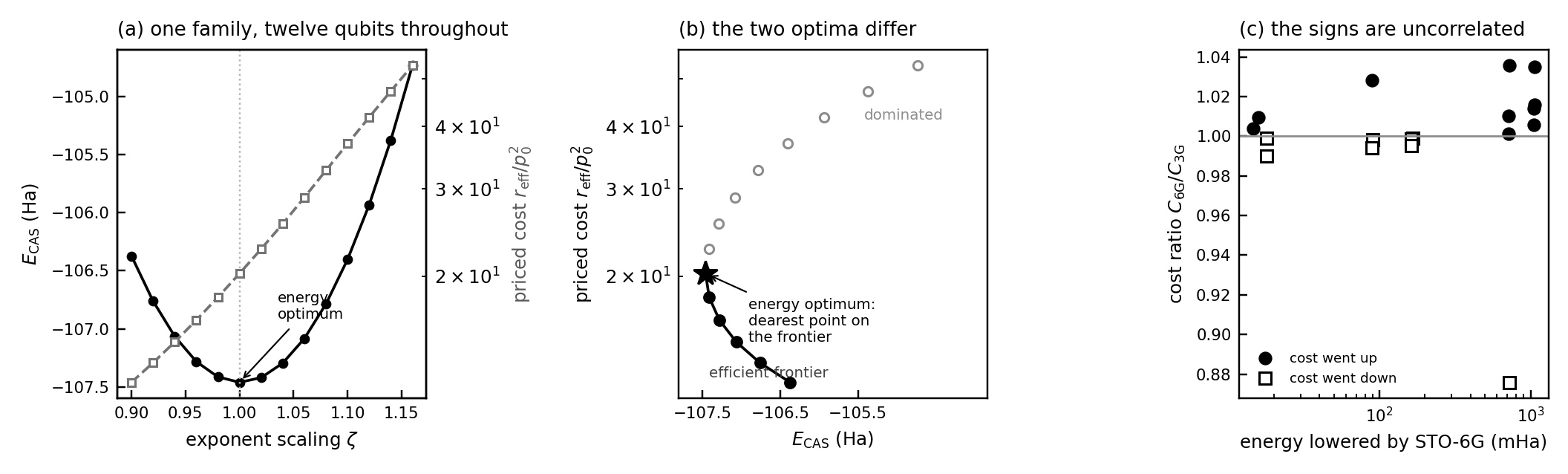}
\caption{\textbf{A better basis is not a cheaper readout.} Acquisition is priced as
$r_{\rm eff}/p_0^{2}$, a stated pricing rather than a measured runtime.
\textbf{(a)} Scaling every Gaussian exponent of a minimal basis by $\zeta$ on N$_2$ at
$1.8$\,\AA, at twelve qubits throughout. The energy has a minimum, the cost does not, and the
cost spans a factor of $4.3$ across the family. \textbf{(b)} The same family as cost against
energy. Open circles are dominated --- some other member is both lower in energy and cheaper.
The energy optimum (star) is the dearest point on the efficient frontier. \textbf{(c)} Eighteen
systems, STO-3G against STO-6G at identical qubit count. The larger basis lowers the energy in
all eighteen, by $14.6$ to $1060$ mHa, and \emph{raises} the acquisition cost in ten of them, so
the two figures of merit are independent in sign.}
\label{fig:basis}
\end{figure*}

\paragraph{The signs are uncorrelated.} Eighteen systems were taken from N$_2$, H$_2$O, LiH,
H$_4$ and BeH$_2$, each at several geometries, and the two minimal bases STO-3G and STO-6G
compared at identical spin-orbital count, that is at an identical eight to twelve qubits. The
Hartree--Fock determinant was expanded over the exact eigenstates of the active-space
Hamiltonian in each basis and the pair $(p_0, r_{\rm eff})$ read off. The larger basis returned
the lower energy in \emph{all eighteen}, by between $14.6$ and $1060$ mHa --- the improvement the
optimisation is performed to obtain. It returned the \emph{higher} acquisition cost in ten of
the eighteen (Fig.~\ref{fig:basis}c). The sign of the change in cost carries no information
about the sign or the size of the change in energy.

\paragraph{The two optima differ.} On N$_2$ at $1.8$\,\AA{} in six active orbitals, so at twelve
qubits throughout, we scaled every Gaussian exponent by a common factor $\zeta$ --- the
one-parameter case of the exponent optimisation of Ref.~\citep{deb2026basis}. Across
$\zeta\in[0.90,1.16]$ the overlap $p_0$ ran from $0.51$ to $0.31$ and $r_{\rm eff}$ from $3.19$
to $5.20$, so the priced cost spanned $12.2$ to $53.1$, a factor of $4.3$, with the qubit count
fixed by construction (Fig.~\ref{fig:basis}a). Eight of the fourteen members are strictly dominated. For each of them some
other member is both lower in energy and cheaper to read. On the efficient frontier the
energy-optimal basis is the \emph{dearest} point (Fig.~\ref{fig:basis}b), and moving along the
frontier away from it buys cost with energy at a measurable rate --- $46$ mHa for ten per cent,
$178$ mHa for nineteen, $395$ mHa for twenty-seven.

\paragraph{What we do and do not claim.} The effect in the like-for-like comparison is small
and asymmetric. Where the larger basis raises the cost it raises it by at most $3.6\%$, and where
it lowers it the largest fall is $12\%$, on stretched H$_2$O. It is the \emph{sign}, not the size,
that is the finding. The $4.3\times$ range of panel (a) is real and is
at fixed qubit count, but reaching its cheap end costs of order a hartree, which no one would
pay; the useful part of that curve is the first few tens of millihartree. And the selection this
suggests --- choose among candidate bases on $(p_0,r_{\rm eff})$ and the energy jointly, rather
than on the energy alone --- costs one scouting acquisition per candidate basis, so it is
worthwhile when the candidates are few and the campaign in the chosen basis is long, and not
otherwise. We have not demonstrated an end-to-end saving, but we have shown that the quantity being
optimised is not the quantity that sets the readout cost.

The question is live elsewhere and the answer bears on it. Viswanathan \emph{et
al.}~\citep{viswanathan2026pools}, building operator pools for adaptive variational algorithms,
leave the selection of the basis and of the starting operators as an open problem and propose
cheap energy sorting as a way forward. Figure~\ref{fig:basis}(c) is a direct negative for that
proposal at fixed qubit count. A larger minimal basis lowered the variational energy in all
eighteen systems and raised the acquisition cost in ten of them, so an ordering by energy is not
an ordering by cost, and a heuristic that sorts on the first will misorder on the second.

\paragraph{A stochastic basis must be pinned, and that is not a canonicalisation.} One
consequence is a matter of hygiene, and it compounds the one \S\ref{sec:sqd} already reports.
The bases of Ref.~\citep{deb2026basis} are produced by a memetic search --- a genetic algorithm
with local refinement --- so the exponents and coefficients returned depend on the seed, the
population and the termination. Two runs on the same molecule give different numbers, and since
a different basis set is a different one-electron space, $H_P$ itself is then a different
operator between runs, and with it the spectrum, the overlap, the sampled subspace and
$r_{\rm eff}$. A difference between runs is no longer attributable to the sampling. The remedy
is not the canonicalisation of \S\ref{sec:sqd}, which removes a rotation, because here there is
no rotation to remove. The basis must instead be \emph{pinned}: record the optimiser with its
seed and termination criterion and the coefficients it returned, and treat those recorded
coefficients, not the procedure that produced them, as the definition of the basis --- so the
basis is reproduced by reading a table rather than by re-running a search. The canonicalisation
is then performed within the basis so pinned. The two operations act on different objects and in
that order: pinning fixes which one-electron space is used, canonicalisation fixes which ordered
set of orbitals spans the active space within it.

\subsection{A machine-learning loss Hamiltonian}
\label{sec:qml}
Many variational quantum-machine-learning losses are expectation values
$\langle\psi(\theta)|H_{\rm loss}|\psi(\theta)\rangle$ of a Hermitian operator whose
ground state is the optimal model, formally a VQE with $H_{\rm loss}$ in place of a
molecular Hamiltonian. Training is then ground-state preparation, and the projector
back-end of \S\ref{sec:aqc} applies directly. Train variationally to a warm start,
then drive that state to the true minimum of $H_{\rm loss}$ with WA-ASP. Because the
adiabatic evolution is \emph{gradient-free}, it does not see the vanishing gradients
of a barren plateau~\citep{mcclean2018barren,cerezo2021cost}. Its success is governed
instead by the warm start's overlap.

Which losses qualify is settled by their algebra. Embed a labelled dataset
$\{(x_i,y_i)\}$, $y_i\in\{\pm1\}$, by a feature map $x\mapsto\ket{\phi(x)}=U_\phi(x)\ket0$
with induced quantum kernel $k(x,x')=|\!\braket{\phi(x)|\phi(x')}\!|^2$
\citep{havlicek2019,schuld2019}. The energy-based objective
$\mathcal L(\zeta)=-N_{\rm data}^{-1}\sum_i y_i|\!\braket{\zeta|\phi(x_i)}\!|^2$ is then
$\bra{\zeta}H_{\rm loss}\ket{\zeta}$ exactly, with $H_{\rm loss}$ the class-weighted
second-moment operator of the feature ensemble. It is a Rayleigh quotient. Minimised over
all unit vectors it returns the lowest eigenvector of $H_{\rm loss}$, the optimal
class-separating direction in feature space and a quantum analogue of Fisher's linear
discriminant. Restricting $\ket\zeta$ to a variational manifold makes the trained minimum
an upper bound on $\lambda_{\min}(H_{\rm loss})$, which is the VQE situation of
\S\ref{sec:hybrid} with a data-assembled operator in place of a molecular one. Losses
quadratic in $\ket\zeta$ therefore transfer without modification. Cross-entropy losses over
measurement outcomes do not, and would need an effective-Hamiltonian recasting first.

\begin{figure}[t]\centering
\includegraphics[width=\linewidth]{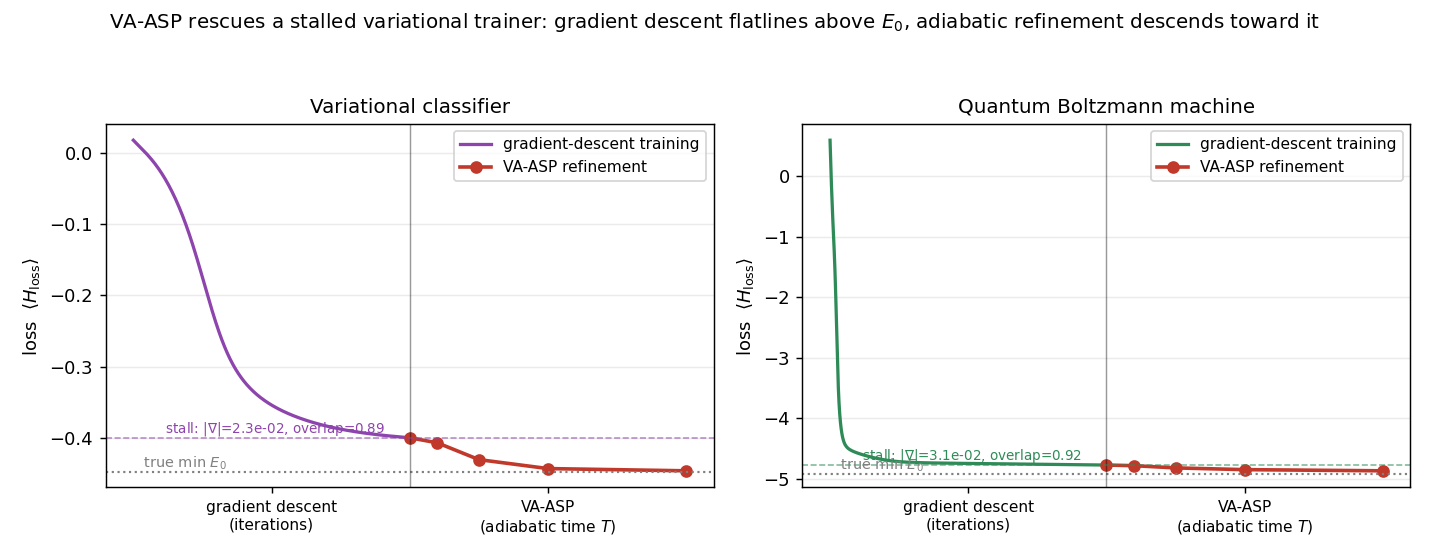}
\caption{\textbf{Gradient-free rescue of a stalled trainer.} Gradient descent
flatlines above the true minimum $E_0$ (dotted) at a small-gradient stall (dashed);
WA-ASP, being gradient-free, descends from the stall toward $E_0$.}
\label{fig:qmlrescue}
\end{figure}

We test this on two models at $n=5$. A \emph{variational classifier} uses that $H_{\rm
loss}=-\tfrac1N\sum_i y_i\,|\phi(x_i)\rangle\langle\phi(x_i)|$ on a feature-map encoding
of two labelled clusters. A
\emph{quantum Boltzmann machine}~\citep{amin2018quantum} uses $H_{\rm loss}=-\sum_i
b_iZ_i-\sum_{i<j}W_{ij}Z_iZ_j-\Gamma\sum_iX_i$, with random $b$ and $W$ and $\Gamma=0.6$,
the transverse field supplying the non-commuting character. That model trains on its Gibbs
state $\ee^{-\beta H_\theta}/Z_\theta$, whose expectations are the hard term in the
gradient. As $\beta\to\infty$ the Gibbs state collapses onto the ground-state projector, so
ground-state preparation is the mode-finding primitive of the generative model and the seed
for low-temperature sampling, and the adiabatic schedule doubles as a controllable
annealing path. For each, plain gradient descent on a
hardware-efficient ansatz~\citep{kandala2017hardware} stalls (Fig.~\ref{fig:qmlrescue}): the loss flatlines above
$E_0$ with a small gradient ($\sim\!2$--$3\times10^{-2}$), yet the stalled
state retains overlap $0.89$ (classifier) and $0.92$ (QBM). WA-ASP from that state
descends toward $E_0$, to within $+0.0017$ (classifier) and $+0.045$ (QBM) by $T=20$,
finishing the descent the optimiser could not.

\begin{figure}[t]\centering
\includegraphics[width=\linewidth]{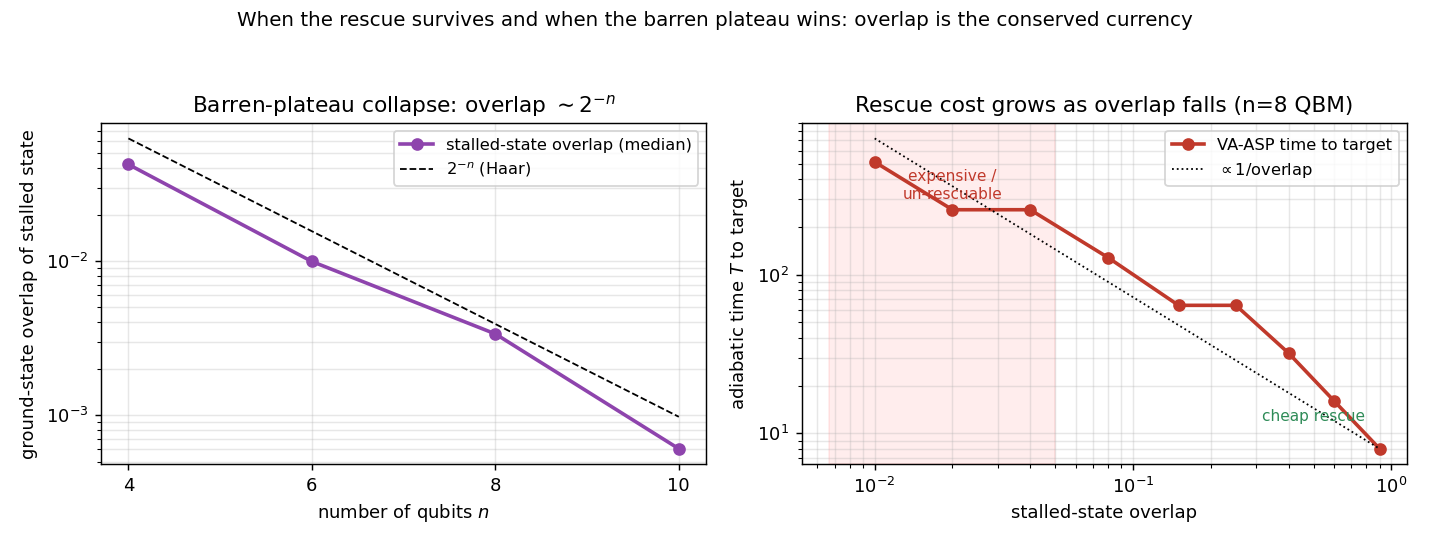}
\caption{\textbf{When the rescue survives.} Left: the stalled-state overlap collapses
as $2^{-n}$ under a barren plateau. Right: the WA-ASP time to a fixed target grows as
the inverse overlap. Together they bound the rescue to the moderate-overlap regime.}
\label{fig:qmlscaling}
\end{figure}

The rescue is real but bounded. The bound is again overlap. A barren plateau drives the
stalled overlap toward $2^{-n}$ (Fig.~\ref{fig:qmlscaling}, left), while the WA-ASP time
to a fixed target grows roughly as the inverse overlap (Fig.~\ref{fig:qmlscaling},
right). Composing the two, the cost to rescue a \emph{true} exponential plateau is
exponential.
What it does rescue is the common practical failure. Local-minimum and flat-region
stalls that retain $\mathcal{O}(1)$ overlap, where gradients have vanished but the state
has not yet wandered far from the target.

\section{The argument off the gate model}
\label{sec:revanneal}

Everything above spends the warm start on a circuit. If the currency argument is about the
resource rather than about the circuit model, it should survive a change of hardware
altogether. A clean test is available and reported below.

On an annealer the natural analogue of WA-ASP is a \emph{reverse} anneal: begin at a
computational-basis state, run backwards to a point where a transverse field is present, hold,
then anneal forward again~\citep{pelofske2023reverse}. That route consumes a warm start in
exactly the sense used throughout this paper, the state at which the reversal begins being
$\ket{\chi}$, and a classical annealing schedule stopped at a temperature $T_{\rm SA}$ is a
natural front-end for it. Its Boltzmann ensemble, re-prepared coherently as in
\S\ref{sec:mixedstart}, has overlap equal to the probability that the schedule returns the
optimum, so the temperature is the knob that sets the overlap supplied.

What the annealing literature does not do is measure that overlap and configure the route from
it, and there is a reason to want to. A reverse anneal too short to be adiabatic does not
preserve what it was given. It relaxes toward a distribution characteristic of the schedule, and
it therefore \emph{destroys} a warm start already better than that distribution.

We tested that on a ten-spin random Ising instance, fourteen couplings and small random fields,
propagated exactly. Six warm starts were taken from classical schedules stopped at
$T_{\rm SA}=4.0$ down to $0.3$, supplying overlaps from $0.008$ to $0.84$. Each was run over a
grid of eight reversal points $s_r$ from $0.9$ to $0.2$ and six durations $t_{\rm rev}$ from
$5$ to $160$, the hold set to $t_{\rm rev}/2$, so forty-eight exact propagations per warm start
and $288$ in all. The best start, of overlap $0.84$, was recovered at no point of that grid. Its
best run returned $0.44$ and its median $0.04$. The start of overlap $0.034$ was improved
instead, its best run returning $0.10$, a gain of $3.0\times$.

We do not quote a crossover overlap, although one lies between $0.12$, where the gain is still
$1.2\times$, and $0.28$, where it is $0.49\times$. The reason is that the surface is not
converged. The gain is non-monotone in both variables and sharply so. For the $0.84$ start the
best point of the whole grid sits at the \emph{shortest} duration, at $s_r=0.6$, returning
$0.44$, while the two adjacent reversal points at that same duration return $0.002$ and
$0.028$. That is coherent ringing rather than an adiabatic regime, and a crossover read off it
would not survive a convergence study.

The qualitative statement is robust. The route needs an
\emph{admission} test and not merely a cost estimate, and the test cannot be run on the anneal's
output alone, an output of overlap $0.3$ being unremarkable unless one knows the input was
better. This is the same structure as the recursion of \S\ref{sec:dbqite}, reached on hardware
that shares none of the recursion's circuit structure, and it is the clearest evidence we have
that the architecture is about the currency rather than about gate-model circuits.

\warppart{Summary, discussion and outlook}{}
\label{part:summary}

\section{Summary and discussion}
\label{sec:summdisc}
\subsection{The architecture in one picture}
\label{sec:arch}
Figure~\ref{fig:arch} puts all the separate pieces in one place. The front-end supplies exactly one object. Everything to its right differs only in how that one
object is spent: as the state whose weights are read, as the projector that drives a sweep, as the
reflection an amplification iterates, or as the start of a recursion that re-spends its
preparation unitary at every step. That is the currency thesis stated structurally.

The second feature is that the arrows run both ways, which is the content of Part~\ref{part:frontends}. What the
readout measures configures every other route and decides between all four before any is run;
but a partial sweep or an amplification hands the readout an overlap its own acquisition could
not see and a state carrying one dominant line, so the model order stops being a hyperparameter,
and the readout's certificates are what supply the recursion with the convergence and wrong-root
tests it does not carry. Neither side is upstream of the other.

The third is the handover among the preparation routes themselves. A raised overlap lowers the
amplification's halting index, and it removes iterations of a recursion whose depth grows as
$3^m$, so a polynomial expenditure on one route buys an exponential reduction in another. That
relation is the reason the four are worth having together rather than as a menu of alternatives.

\subsection{Summary table of every case studied}
\label{sec:summary}

Table~\ref{tab:cases} lists every case studied in this work: the component of the
architecture it exercises, the conventional method it is measured against, and the number
that came out. The comparisons are made on equal terms by construction. Where a variational
bound is the baseline, the bound and the readout are computed from the \emph{same} state,
so the difference is attributable to the readout and not to a change of ansatz; where the
Fourier transform is the baseline, both estimators are handed the identical sampled signal.
Every system is diagonalised exactly and every acquisition is sampled under the binomial
statistics of the interferometer, so each recovered quantity has a known answer behind it.

The final block of the table records the conditions under which the architectural
claims cease to be useful or require qualification. A commuting cost operator is read more
cheaply by direct sampling. A dense many-body spectrum erodes the spectral margin as the
system grows, while a congested band imposes a ceiling on the number of lines that can be
resolved from a finite record. A selection rule over bases changes a priced cost by a few
per cent in the examples studied here, not an end-to-end runtime. The probe that admits the
recursive back-end is an estimate rather than a bound and therefore requires a safety factor.
Finally, no cross channel can reveal a spectral line that is absent from both reference
states. These rows are not exceptions to the architecture; they delimit its domain of use.

\clearpage
\onecolumn
\begin{landscape}
\begingroup
\footnotesize
\setlength{\tabcolsep}{5pt}
\renewcommand{\arraystretch}{1.15}
\newcolumntype{L}[1]{>{\raggedright\arraybackslash}p{#1}}

\begin{longtable}{@{}L{40mm}L{46mm}L{44mm}L{86mm}@{}}
\caption{\textbf{Summary of the numerical and architectural tests of WARP.}
Each row tests a specific component or claim of the architecture and reports the baseline
used for comparison and the measured outcome. Comparisons against a variational bound use
the \emph{same} state for both quantities, while comparisons against the Fourier transform
use the identical sampled signal. Every system is diagonalised exactly and every acquisition
is sampled under the binomial statistics of the interferometer. Chemical accuracy is
$1.6\times10^{-3}\Ha$ throughout. The final block records conditions and limitations rather
than additional performance claims.}
\label{tab:cases}\\
\toprule
\textbf{Case} & \textbf{WARP component} & \textbf{Baseline} & \textbf{Measured result}\\
\midrule
\endfirsthead

\multicolumn{4}{@{}l}{\emph{Table~\ref{tab:cases}, continued.}}\\[2pt]
\toprule
\textbf{Case} & \textbf{WARP component} & \textbf{Baseline} & \textbf{Measured result}\\
\midrule
\endhead

\midrule
\multicolumn{4}{r@{}}{\emph{continued overleaf}}\\
\endfoot
\bottomrule
\endlastfoot

\multicolumn{4}{@{}l}{\textbf{A\quad Spectral readout} \ \ \emph{-- a true eigenvalue
out of a state someone else prepared}}\\[3pt]

H$_2$, STO-3G, 2 qubits; gate-level Trotter, Hadamard test, finite shots
& Parametric estimation (ESPRIT) of $g(t)$
& the discrete Fourier transform of the same signal
& $\sim\!50\times$ shorter signal and shallower circuit: $N=40$ against $2048$, depth
  $39$ against $2047$ steps, $2.6\times10^{6}$ against $3.3\times10^{7}$ shots. All $200$
  noise realisations inside chemical accuracy.\\[2pt]

H$_2$, STO-3G; $2\times10^{7}$ shots, $N=96$, a stated differential contrast between the
two quadrature passes
& The two quadrature settings interleaved within one sweep, no extra ancilla
& the ordinary two-pass acquisition of the same signal
& The two-pass error is $5.5\times10^{-4}\Ha$ to $2\%$ imbalance, $3.4\times10^{-2}$ at $3\%$
  and $0.89\Ha$ beyond $5\%$, the estimator locking onto a mirror line. Doubling the order to
    $p=6$ restores $6.5\times10^{-4}$; interleaving holds $p=3$ and $5\times10^{-4}$ at every
  imbalance, at no cost in depth, ancillas or shots. Equal statistical efficiency throughout, so
  what is bought is model order and not variance. Sharing a shot between the quadratures is
  neither necessary nor sufficient --- a diagonal-basis pair run as two passes fails
  identically at $0.89\Ha$.\\[2pt]

H$_2$ from a deliberately loose VQE state, $|c_0|^2=0.948$
& Hybrid warm start $\to$ ESPRIT
& the variational bound on the \emph{identical} state
& $91\,$mHa recovered below the bound. Bound $+0.091\Ha$ ($57\times$ chemical accuracy);
  readout error $-1.0\times10^{-6}\Ha$.\\[2pt]

Transverse-field Ising model, 16 qubits, $h/J=1.1$
& Hybrid warm start $\to$ ESPRIT at scale
& a five-layer VQE bound, and the FFT
& Bound error $+0.109\Ha$ ($\sim\!70\times$ chemical accuracy) $\to$ readout $+0.005$.
  Depth saving over the FFT $\sim\!25\times$ at 12 qubits.\\[2pt]

SO$_2$, CAS(6e,6o), 12 qubits
& Hybrid warm start $\to$ ESPRIT
& Hartree--Fock and pair-UCCD bounds; the FFT
& $0.033\Ha$ below the VQE bound and inside chemical accuracy (HF bound is
  $+0.120\Ha$). Depth $\sim\!59$ steps against $\sim\!255$.\\[2pt]

SO$_2$ stretched $1.15\times$, near-degenerate, gap $0.023\Ha$
& ESPRIT super-resolution below the Fourier limit
& the FFT, which cannot resolve the pair at all
& $10^{-5}$--$10^{-6}\Ha$. The regime where super-resolution is decisive rather than
  merely cheaper.\\[2pt]

H$_2$O, CAS(8e,6o)
& Hybrid warm start $\to$ ESPRIT
& HF and pair-UCCD bounds; the FFT
& $\sim\!10^{-4}\Ha$ by depth $40$, below both bounds; the FFT plateaus near
  $3\times10^{-3}\Ha$.\\[2pt]

H$_2$ bright spectrum, one record at $N=200$, $2^{17}$ shots
& Reading every line of one acquisition
& a separate calculation per excited state
& Excited states at no extra circuit cost. Excitation energy $1.748553\Ha$ returned as
  $1.748796$, an error of $2.4\times10^{-4}\Ha$.\\[2pt]

Two warm starts of identical overlap, both $p_0=0.20$
& Spectral complexity $r_{\rm eff}=(\sum_n p_n^2)^{-1}$ as a second resource
& selecting a front-end on ground-state overlap alone
& A $2\times$ cost difference the overlap cannot see: $r_{\rm eff}=1.47$ against $16.30$,
  needing $N=40$ against $N=80$ to reach $10^{-4}\Ha$.\\[2pt]

N$_2$ SQD study at $10^{3}$ shots
& Adaptive model order and noise-calibrated admissibility
& a fixed model order and the hard unit-circle test
& Load-bearing, not a refinement. A fixed order failed at both ends; the unit-circle test
  rejected \emph{every} root, the noise displacing them by over five per cent.\\[4pt]

\midrule
\multicolumn{4}{@{}l}{\textbf{B\quad Adiabatic preparation} \ \ \emph{-- the same overlap
spent on a state rather than an eigenvalue}}\\[3pt]

H$_2$ and H$_2$O
& WA-ASP, driver anchored on the warm start
& textbook adiabatic evolution: transverse-field driver from the uniform superposition
& $T\!\approx\!16$ against $T\!\approx\!128$, an $8\times$ reduction. The cold start
  overlaps the ground state by only $0.15$, opening a gap bottleneck the anchored driver
  never has.\\[2pt]

Stretched H$_2$O
& WA-ASP from a VQE-anchored start
& the same sweep anchored on Hartree--Fock
& $T\!\approx\!20$ against $T\!\approx\!40$, and the advantage grows with correlation as
  the reference overlap drops.\\[2pt]

Projector interpolation at $a^2=0.02$
& Conditional path shaping, decided before the schedule is run
& the linear interpolation
& Minimum gap raised $\sim\!38\%$, $0.149$ against $0.108$. At $a^2=0.20$ the interior
  bottleneck vanishes entirely, which is why the test is conditional.\\[2pt]

Filtering stage, $\kappa$ and $p_f$ measured
& Admission test of Eq.~\eqref{eq:filtercond}, against the unfiltered acquisition
& running the stage regardless
& $\kappa=3$, $p_f=0.2$: repetitions $\div16.2$ at $p_0=0.05$, $\div5.0$ at $0.20$,
  $\times1.8$ at $0.60$. The ceiling $p_0<\sqrt{p_f}$ is not removable by budget.\\[2pt]

Ten-spin random Ising, $8\times6$ grid of reversal points and durations, exact
& Reverse anneal from a coherently re-prepared SA ensemble
& the overlap the anneal was given
& Overlap $0.84$ recovered at no grid point, best of $48$ runs $0.44$, median $0.04$; overlap
  $0.034$ improved to $0.10$. Needs an admission test, not a cost estimate.\\[2pt]

Two-level reduction at $a^2=0.05$, $\bar\mu=0.5$
& Local-adiabatic schedule generated from the \emph{measured} gap law
& the linear ramp over the same path
& $f=0.99$ at $T=33$ against $T=193$, a factor of $5.9$; at $T=40$, $0.996$ against $0.623$.
  The rate is set from a computed gap, so no spectroscopy of the gap is
  required.\\[2pt]

H$_2$ sweep from $0.7$ overlap, minimum gap unknown
& Pointwise variance certificate with a running minimum
& a convergence-trend test on the energy
& A trend test is unusable: the energy \emph{rises} on $17\%$ of schedule steps. The
  certificate needs no gap estimate and identifies \emph{which} eigenstate was
  reached.\\[4pt]

Admitting a recursive (DB-QITE) back-end (\S\ref{sec:dbqite})
& The published convergence guarantee, Eq.~\eqref{eq:elldagger}, against a one-iteration
  metered probe, Eq.~\eqref{eq:ellhat}. 23 instances, H$_2$O/N$_2$/H$_4$/LiH, 8--12 qubits,
  determinant-space dimension 36--400, exact propagation
& the guarantee, used as an admission test
& True counts $1$--$34$. The guarantee overshoots by $1.9\times10^{4}$ to $2.2\times10^{10}$
  (median $1.5\times10^{6}$), so it can admit and never refuse; the probe lands within
  $3\times$ on $19$ of $23$. Depth $\mathcal O(3^m)$ makes this a threshold, not a
  scaling. The probe is an estimate and not a bound, its median $1.67$ in the direction of
  under-prediction, so it carries a safety factor.\\[2pt]

Step size of the recursive back-end (\S\ref{sec:dbqite})
& Anchoring on the measured moments, $\hat\xi=\mathcal E/2\mathcal V$, then five multiples
& a blind scan over $5.5$ decades, 20 trials
& $282$ iterations at $1410$ energy evaluations against $290$ at $5800$: $4.1\times$ fewer
  evaluations at no cost in iterations ($3.4\times$ if the variance is re-acquired every
  iteration).\\[2pt]

\midrule
\multicolumn{4}{@{}l}{\textbf{C\quad Front-ends} \ \ \emph{-- the swappability claim
tested}}\\[3pt]

MaxCut, Petersen graph, 10 qubits, QAOA at depth $D$
& QAOA $\to$ WA-ASP, overlap as the exchange rate
& a cold start, $D=0$
& One QAOA layer buys $32\times$: $T=512,16,8,4,1$ at $D=0\ldots4$, collapsing onto
  $T\propto1/a^2$.\\[2pt]

Same instance, read rather than prepared
& ESPRIT on the QAOA state
& a bin-limited transform
& The optimum to $|{\rm err}|\le2\times10^{-4}$ from every $D\ge1$, including $D=1$ where
  the ground line carries $0.168$ against a dominant line at $0.397$.\\[2pt]

Lattice shortest-vector problem, rank-4; target is the \emph{first excited} state
& Rank-one penalty from the driver's own primitive; lattice reduction as the front-end
& folded-spectrum constructions, excited-state annealing, or penalties with auxiliary
  variables
& $123\times$ in schedule time, with no ancillas. On readout the target is only the
  \emph{third} brightest of twelve lines at $a^2=0.222$ and still returns to a median
  $3.0\times10^{-4}$.\\[2pt]

Generative front-end, $1210$-candidate pool
& Ranking candidates on overlap, the quantity the back-ends spend
& ranking by variational energy, which such models are trained on
& The rankings differ: best-of-16 disagree on $23\%$ of draws. At fixed overlap a $66\%$
  energy change moves fidelity two points, while circuits of identical energy differ by
  $0.69$.\\[2pt]

N$_2$, CAS(10e,8o), SQD front-end at $10^{6}$ shots
& Read out the sampled device state; certify with the classical vector
& the Rayleigh--Ritz bound the same calculation returns
& Four orders of magnitude, no new circuit: $5.1\times10^{-3}\to7.1\times10^{-7}\Ha$ at
  equilibrium, $7.9\times10^{-2}\to1.4\times10^{-6}$ stretched, where the reference
  overlap has fallen to $0.043$.\\[2pt]

Stalled machine-learning trainer
& WA-ASP driven from the stalled state
& continued gradient descent
& A stall is not a lost state. Gradient $2$--$3\times10^{-2}$ and the loss flat, yet
  overlap $0.89$ and $0.92$ survive for a gradient-free back-end.\\[4pt]

Single-particle basis at fixed qubit count (\S\ref{sec:basis})
& STO-3G against STO-6G at identical spin-orbital count on 18 systems from N$_2$, H$_2$O, LiH,
  H$_4$, BeH$_2$; and a scaled-exponent family on N$_2$ at $1.8$\,\AA\ at 12 qubits throughout
& choosing a basis on the variational energy alone
& The larger basis lowers the energy in $18$ of $18$, by $14.6$ to $1060$ mHa, and \emph{raises}
  the priced cost $r_{\rm eff}/p_0^2$ in $10$ of them: the signs are uncorrelated. Within the
  family the cost spans $4.3\times$ at fixed qubits, eight of the fourteen members are dominated,
  and the energy optimum is the dearest point on the efficient frontier.\\[2pt]

\midrule
\multicolumn{4}{@{}l}{\textbf{D\quad What follows from taking overlap seriously} \ \
\emph{-- if overlap prices everything, it must be measurable when scarce}}\\[3pt]

SO$_2$, CAS(6e,6o), $10^{6}$ shots per quadrature, $400$ samples
& Homodyne interference of a weak reference, amplitude $c_W$ on the line, against a
  strong one of amplitude $c_S$
& the diagonal autocorrelation, whose residue is $|c_W|^2$
& An overlap four orders below the floor. Floor $7.1\times10^{-5}$, diagonal stalls at
  $8.3\times10^{-3}$, interference tracks to $|c_W|=10^{-4}$. Repetitions scale as
  $a^{-2}$ in place of $a^{-4}$, phase obtained for free.\\[2pt]

H$_2$CO $1^1\!A_2\!\leftarrow\!1^1\!A_1$ ($n\!\to\!\pi^*$), CAS(6e,6o)/6-31G,
symmetry-forbidden
& Transition amplitude from the interference residue, linear in the moment
& a conventional readout of the transition \emph{probability}, quadratic in it
& Dynamic range $2.9\times10^{6}\to1.9$. The conventional channel saturates at
  $9.1\times10^{-2}$ a.u.; interference tracks to $1.28\times10^{-4}$. Oscillator strength
  $4.2\times10^{-7}$ against an exact $4.3\times10^{-7}$.\\[2pt]

Purified mixed front-end, three members, $\nu_j=0.05$
& Homodyne against the purification, label register projected on $\ket{j}$
& the diagonal residue of that member alone
& Residue $\sqrt{\nu_j}c_S^*c_j = 8.9\times10^{-6}$ against a diagonal $10^{-8}$, a gain of
  $886\times$, linear in the weak amplitude where the diagonal is quadratic.\\[2pt]

Amplitude amplification from a sub-floor warm start
& Halting index $\ell^\star$ set from the metered overlap
& an upper limit on the overlap, all the diagonal returns
& Otherwise \emph{inoperable}, not merely expensive. With $P$ the success probability at
  index $\ell$, at $|c_W|=10^{-3}$ the halting index $\ell^\star=785$ gives $P=1.0000$ and
  $\ell=1570$ gives $P=0.0000$ exactly, so a bound will not serve. Setting $\ell$ from
  $\hat a=a(1+\epsilon_a)$ holds $P\ge0.99$ only for $\epsilon_a\in[-6\%,+7\%]$.\\[2pt]

Stretched N$_2$, variational error exceeding the gap
& The Eckart floor read together with the two-sided bracket
& a bound quoted without its validity condition
& The floor returns $-6.4$, from a $79\,$mHa error against a $10.7\,$mHa gap. It declares
  itself vacuous rather than returning a number that is merely wrong, so the method reports
  the condition it has violated and identifies the vector as not close enough to be the
  ground state. The bracket still localises $E_0$ to $0.11$--$0.16\Ha$, and the readout of
  that same $9\%$ overlap returns it to $1.4\times10^{-6}\Ha$.\\[2pt]

Linear vibronic coupling models, dipole autocorrelation
& Parametric estimation in place of the transform
& the Fourier transform of the same signal, gate cost $\propto t^{3/2}$
& $8\times$ the gates on a sparse band: $200\,$fs of propagation falls to $50$. On
  congested bands $1600\,$fs falls to $800$, a factor of two in time and under three in
  gates.\\[4pt]

\midrule
\multicolumn{4}{@{}l}{\textbf{E\quad Scope and limitations} \ \ \emph{-- reported alongside the results,
and defining where the architecture does and does not apply}}\\[3pt]

Diagonal cost operators: MaxCut, lattice SVP
& Spectral readout of $g(t)$
& direct $Z$-basis sampling
& $100$--$230\times$ cheaper the other way, and it returns the configuration as well as
the value. The readout earns its keep only on \emph{non-commuting} $H_P$.\\[2pt]

Dense many-body spectra, TFIM at 12 and 16 qubits
& ESPRIT against the FFT as size grows
& nothing --- but the margin shrinks
& $\sim\!25\times$ at 12 qubits reducing to $\sim\!4$--$6\times$ at 16. A many-body start
populates a dense forest of small amplitudes that act as correlated noise. This is the
central scaling caution.\\[2pt]

Congested vibronic bands, $r_{\rm eff}$ large
& Super-resolution of a crowded spectrum
& a hard ceiling, not a competing method
& The gain falls to $2\times$ in time. Super-resolution buys the accuracy of a line and
never the number of lines: at most $N/2$ modes, whatever the budget.\\[2pt]

Choosing a basis on the priced cost
& $r_{\rm eff}/p_0^2$ as a selection rule over bases
& the variational energy, which is not displaced
& The like-for-like effect is small and asymmetric: the STO-3G/STO-6G cost ratio runs from
  $0.88$ to $1.04$, the rises never exceeding $3.6\%$. The $4.3\times$ range needs of order a hartree of energy to traverse. And the
  rule costs one scouting acquisition per candidate basis. We show the optimised quantity is not
  the quantity that sets the readout cost; we do not show an end-to-end saving.\\[2pt]

A line dark in both diagonals
& Cross-correlation between two references
& nothing can --- it is forbidden
& $|c_S^{*}c_W|=\sqrt{|c_S|^2|c_W|^2}$, so no cross channel reveals a line invisible in both
autocorrelations. The multiple-warm-start argument rests on coverage, not
interference.\\

\end{longtable}
\endgroup
\end{landscape}
\twocolumn

\input{figarch}

\noindent The architectural claims have now been made and measured. What remains is to place
them --- against the estimator they replace, against the variational method they are usually
compared with, against the existing literature on time-series eigenvalue estimation, and against
the classical algorithms that would have to be beaten for any of it to matter.
\addtocontents{toc}{\protect\setcounter{tocdepth}{1}}
\subsection{What the combined tests establish}
The individual demonstrations in Table~\ref{tab:cases} support a more specific conclusion
than the performance of any one back-end. First, a warm start carries information that is
not exhausted by its variational energy or by its ground-state overlap. The spectral weight
$\{p_n\}$ determines which eigenvalues are present in the measured signal, while $p_0$ controls
the leading overlap dependence of the preparation procedures. Second, the same supplied state
can therefore be interrogated or refined without changing the front-end. Third, the back-ends
are coupled by measurable quantities: readout can expose spectral complexity and overlap,
partial preparation can increase the useful ground-state weight, and the resulting state can be
returned to the readout. The experiments consequently support WARP as a modular interface and as a measurable
resource that the back-ends demonstrably spend. They do not establish that any one back-end is universally
optimal, nor that the architecture removes the cost of producing a useful warm start.

\subsection{Resource trade-offs and what is actually being priced}
The compact description $(p_0,\mathcal C_{\rm spec})$ is useful only as a summary of the
front-end resource; it is not a complete hardware cost model. The total cost of a particular
workflow also contains the preparation and compilation cost of $|\chi\rangle$, controlled or
Trotterised evolution, measurement repetitions, classical post-processing, and, for WA-ASP,
the implementation of the projector onto the warm start. The experiments show why these costs
should be kept separate. ESPRIT trades coherent acquisition time for model-order selection and
classical estimation; WA-ASP trades an overlap-dependent gap for an anchoring construction;
amplitude amplification trades depth against a precisely metered stopping index; and recursive
refinement trades repeated evolution for a measured estimate of its convergence scale. Thus WARP supplies a common denomination in which those trades can be stated and measured,
not a universal scalar cost or a claim of equal cost across front-ends.

\subsection{Scope, limitations and negative results}
\label{sec:scope}
Several results in Table~\ref{tab:cases} are deliberately negative. Direct sampling is
preferable for diagonal cost operators, dense spectra reduce the readout margin, and
super-resolution cannot increase the number of spectral lines that a finite record can support.
The basis-selection study changes a priced cost without demonstrating an end-to-end runtime
saving, and the recursive admission probe is an estimate rather than a certificate. Likewise,
the absence of a line from both autocorrelations cannot be repaired by cross-correlation. These
results are important because they prevent the currency picture from being read as a promise
that every problem benefits from a spectral back-end.

We have directly tested the first entry in that list.
On a diagonal cost operator with a product warm start, every coordinate of the resource
description is available in closed form. For a state with single-site magnetisations $m_j$
the effective spectral rank is $r_{\rm eff}=\prod_j 2/(1+m_j^2)$, the whole R\'enyi family
factorises in the same way, and the first two energy moments follow from
$\langle\prod_{i\in S}\sigma^z_i\rangle=\prod_{i\in S}m_i$ at a cost linear in the number
of terms. We verified these against the enumeration at sixteen qubits to machine precision.
What is \emph{not} available in closed form is $p_0$, together with any inserted tail weight
$P(E\le E_t)$, as both require the ground configuration. The
priced cost $r_{\rm eff}/p_0^{2}$ therefore splits, with the denominator on the wrong side of
the line.

Consequently, taking the bias-field construction
of Romero \emph{et al.}~\citep{romero2025bfdcqo} on sixty instances of their nearest-neighbour
three-body spin glass, at four register sizes and fifteen seeds each, the bias strength that
maximises $p_0$ has median $3.98$ and varies from below unity to the top of the range tested.
Minimising $\langle H_P\rangle$ instead drives the bias to the edge of the grid on more than
half the instances and discards a median of $2.3$ orders of magnitude of ground-state weight.
No closed-form coordinate recovers the right value. A selection rule built on the tail weight
locates it within a factor of two on ninety per cent of instances when the target energy is
within two per cent of $E_0$, and fails to beat a fixed hand-tuned bias when the target is the
best energy a ten-thousand-shot record actually returns, which sits a median of $7.6$ per cent
above $E_0$. Separately, $r_{\rm eff}$ is estimable from a shot record alone by counting coincident
bitstrings, unbiased to $0.03$ bits in our tests, but the estimator needs
$N_s\gtrsim\sqrt{r_{\rm eff}}$ and so reports only once the distribution has already concentrated.

The pattern across those three tests is the boundary we record here. On a diagonal problem,
each coordinate of the currency becomes cheaply available in the regime where it has stopped
being informative. The readout is redundant because a basis measurement returns the
configuration and the objective together, the selection rule needs a target already close to
the answer, and the rank estimator needs a rank already small. We record this as a boundary of
the architecture rather than of the particular algorithms tested, and it is the reason that the
front-end studies of Part~\ref{part:frontends} are drawn from problems whose eigenvalue
information is not available from a single basis measurement.

Another limitation is the implementation. The numerical studies include the stated Trotter and
shot-noise effects, but they do not constitute a hardware demonstration with simultaneous
coherent evolution, gate errors, decoherence, and readout errors. The observed depth reductions
therefore identify a potentially useful hardware regime rather than establishing a hardware
advantage. In the same way, the front-end experiments demonstrate interface independence, not
equal preparation cost or equal scalability. A front-end may fail to provide sufficient overlap,
may incur a large compilation cost, or may produce a spectrally congested state. Those costs
remain part of the problem and are not hidden by the WARP abstraction.

\subsection{ESPRIT versus Fourier transform.} The measured comparison is
Table~\ref{tab:fftvsesprit}; what it shows is a resource trade rather than a universal dominance.
 In identical simulated signals, ESPRIT reaches the stated accuracy with fewer samples
because it fits the exponential model instead of resolving a grid, and it compensates for that with
a model-order hyperparameter and greater sensitivity to noise per sample (\S\ref{sec:modelorder}). The Fourier transform is nonparametric and needs no order,
and pays in evolution time. That asymmetry has a second edge which \S\ref{sec:quadrature}
exposes. Because the parametric estimator commits to a line count, it is the one that breaks
when the measurement adds lines of its own, and a differential quadrature contrast does exactly
that. The transform would merely show the mirror lines; the estimator, at an order chosen for
the true ones, returns a mirror line in place of a true one. Choosing a parametric method is
therefore also choosing to control what the apparatus adds to the spectrum. That the mechanism
is old elsewhere~\citep{valkama2001iq} and the circuit is known
already~\citep{imamura2026parallel} is the point. Which side of that trade one wants is set by the hardware. Where depth is the binding constraint, as it is in the early fault-tolerant era, the
parametric estimator is the favourable one, and where samples are cheap and coherence is
long it need not be.

\subsection{Readout versus VQE}
VQE and spectral readout provide different information about the same supplied
state. The variational expectation value satisfies
$\braket{H_P}\ge E_0$, whereas spectral readout can identify an eigenvalue carrying
nonzero spectral weight in that state. When a warm start has useful ground-state
overlap but a loose variational energy, as in Fig.~\ref{fig:hybrid}, the two
procedures therefore report different quantities and spectral readout can locate
$E_0$ below the variational value. This does not make spectral readout cheaper than
VQE or replace the optimisation that produced the warm start. It adds controlled
evolution, an ancilla-based measurement, coherent acquisition time, and classical
spectral estimation. The hybrid should consequently be viewed as a trade between
variational bias and additional readout resources: a warm start supplies the overlap,
and the back-end spends that overlap to identify the eigenvalue.

\subsection{Relation to existing methods.} Time-series eigenvalue estimation has a
substantial lineage~\citep{somma2019time,obrien2019qpe} and underpins the early
fault-tolerant QPE algorithms that achieve Heisenberg scaling with a single
ancilla~\citep{lin2022heisenberg,ding2023even,wan2022randomized,zhang2022computing,dong2022ground}.
Our ESPRIT pipeline is classical signal processing in the time domain, and in that
respect it shares its setting with those works rather than departing from it. The
difference is the estimator (rotational-invariance subspace fitting rather than
density-matrix or Bayesian inference of the spectral measure) and the \emph{a priori}
model order of \S\ref{sec:modelorder}. It is
also closely related to quantum Krylov / filter diagonalisation methods, which build a
small generalised eigenproblem from real-time-evolved states and likewise return true
eigenvalues~\citep{parrish2019filter,stair2020multireference,klymko2022real};
imaginary-time variants exist as well~\citep{motta2020qite}. The connection is closer
than a shared setting. The Krylov subspace
$\mathcal K_k(A,v)=\operatorname{span}\{v,Av,\dots,A^{k-1}v\}$ underlies all three
estimators in this work's orbit. Lanczos~\citep{lanczos1950} tridiagonalises a Hermitian
$A$ on an orthonormal Krylov basis and approximates its extreme eigenvalues by matching the
moments $\bra vA^m\ket v$; conjugate gradient~\citep{hestenes1952} is that same recursion
run on a linear system, its residuals the Lanczos vectors up to scaling. Quantum Krylov
methods build the subspace from real-time-evolved states and solve a generalised
eigenproblem in their overlaps. The ESPRIT readout is the moment-matching dual of that
construction. Its samples $g(t_m)=\sum_n p_n\ee^{-iE_nt_m}$ are the real-time power moments
of the spectral measure, the Hankel matrix of $\{g(t_m)\}$ is their moment matrix, and the
estimator recovers the support $\{E_n\}$ by fitting exponentials to it. The three are
faces of one problem, extracting spectral information from a Krylov sequence. WARP uses the
last of them for the readout, and an adiabatic path rather than a Krylov subspace for the
preparation. The contrast with canonical textbook
QPE~\citep{nielsen2010quantum,abrams1999quantum} is sharper. There, the transform is performed \emph{coherently on the device}, an inverse
QFT acting on an ancilla register that holds the phase. The achievable resolution is
therefore fixed in advance by the register width. Some $m$ ancillae buy $m$ bits of phase
and a grid of spacing $2\pi/2^m$, and the eigenvalue must be read off that grid. Estimating in the time
domain moves the transform off the device entirely. Resolution is then limited by
signal-to-noise rather than by register width. And because a subspace estimator solves an
algebraic problem instead of binning, it is not tied to any grid at all, which is what
allows resolution below the $2\pi/\tau$ Fourier limit from a signal that short. The price is
repetitions, and the benefit is that ancilla count and coherent transform depth both
leave the circuit.

\subsection{Bounds on and measuring the overlap.}
The certificate of App.~\ref{app:certified} has a close relative in recent work that bounds
eigenstate overlap from Hamiltonian moments by linear and semidefinite
programming~\citep{lin2025moments}, generalising the Eckart inequality to arbitrary moment order
and using the result to decide whether a state is worth running QPE on at all. Two differences are worth noting.

Their moments are \emph{computed}, from a classical representation of the state, by exact
diagonalisation or a matrix-product contraction. Ours are \emph{measured}. The $k$th moment of
the warm start is $\braket{\chi|H_P^k|\chi}=\sum_n p_nE_n^k$, a sum over the line positions
$E_n$ and the weights $p_n$ that a single acquisition has already returned, so every moment
their programme asks for follows by summation and costs no further measurement. A front-end whose output has no compact classical description, which
covers the sampled device states of \S\ref{sec:sqd} and any start that is not pure, admits no
moment computation of that kind at all, and the measured route is still available.

The second difference is the one \S\ref{sec:grover} turns on. A bound is not a value. The
amplification's halting index is fixed by the overlap value alone and its success probability oscillatorily returns
to zero if the index is overrun, so a lower bound on the overlap gives a lower bound on the
index and does not give the index. The two are complementary. A certified
bound screens a candidate before any acquisition is run, and the measurement sets the parameter
once one has been.

\subsection{A guarantee is not an admission test.}
\label{sec:guarantee}
The recursive back-end of \S\ref{sec:dbqite} produced the sharpest methodological lesson in
this paper, and it generalises past the particular algorithm. Its convergence guarantee is
correct, proved, and tight in the regime it assumes --- a step size small enough for the
first-order bound to hold. It is also, as an operational instrument, not practically useful. On
twenty-three exactly propagated instances it overshoots the iteration count actually
required by a median factor of $10^6$, because the step size it assumes is orders of
magnitude below practicality. Used as a rejection test it refuses every warm start that in fact converges, and every such refusal is uninformative. The asymmetry it leaves is worth stating in general terms: a worst-case bound can \emph{admit} a computation and can almost never \emph{soundly refuse} one, being sufficient but not necessary, and
the gap between those two uses widens with the looseness of the bound. Where the decision
is a threshold rather than a scaling --- and the $\mathcal O(3^m)$ depth here makes it a
threshold --- that gap is the whole of the decision.

What replaces it is not a better bound but a measurement, and the measurement is available
only because the architecture already meters overlaps below the floor of their own
acquisition. One iteration and two metered fidelities predict the count within a factor of
three where the guarantee errs by $10^6$. We would put this forward as a general pattern for
warm-start algorithms whose cost is set by quantities the algorithm does not itself return:
probe, meter, extrapolate. The cost of the pattern is that the result is an estimate,
and we report its failure mode rather than its median --- $\lambda$ is not constant across
iterations, the error is two-sided, and the median direction is under-prediction, which is
the direction that admits a run that then overruns.

The pattern is not ours alone, and it has since been reached from an entirely different
direction. Viswanathan \emph{et al.}~\citep{viswanathan2026pools} verify the Lie-algebraic
completeness of an operator pool by a rank condition on a binary matrix, replacing an exponential
search by a polynomial one, and then report that completeness so verified does not deliver what
it appears to promise. A strictly minimal complete pool gives their worst convergence, and the
pool must be augmented with physically motivated operators before the algorithm performs. Their
words are that ``completeness alone does not guarantee efficient convergence'' and that ``the
theoretical completeness criterion alone is insufficient to guarantee an efficient algorithm''.
That is a structural guarantee admitting a route without pricing it, established algebraically
rather than by propagation, on a front-end rather than a back-end. Two independent instances of
one asymmetry is what makes it a pattern rather than an anecdote.

\subsection{What the currency does not price.}
\label{sec:notpriced}
The
architecture prices everything downstream of the front-end in the overlap distribution, and that
is the paper's thesis. But quantities \emph{upstream} of it are set by other criteria entirely,
and the two need not agree. The single-particle basis is optimised for variational energy
(\S\ref{sec:basis}); a recursive back-end is characterised by a worst-case convergence bound
(\S\ref{sec:dbqite}). Both are the right figure of merit for the question they were designed
against, and both are silent about, or actively misleading concerning, the cost the acquisition
will actually incur. In the basis case, the two figures move independently in sign. In the
back-end case the bound errs by six orders of magnitude in the direction that would refuse
everything. Neither is a defect of the other work. It is a consequence of taking a currency
seriously enough to notice that most of the quantities the front-ends optimise are denominated in
something else.

A third quantity is missing, and recent work locates it. The CASH-QSE construction of
Izmaylov~\citep{izmaylov2026cash} partitions a wave function into a classically tractable
active-space component and quantum states built exactly orthogonal to it, and reports that the
sampling cost of the result is not determined by the quality of the classical reference. That
diagnostic, the squared reference/FCI overlap, is our $p_0$, and the spread it leaves is wide.
Two of the reference states in that work have overlaps agreeing to eight per cent ($0.917$ and
$0.995$) and reported sampling costs differing by a factor of $87$. We reproduced those overlaps
independently and then asked what accounts for the difference.

The first candidate is the one that work names without calling it a resource. Its \emph{killer}
operators leave the matrix element exactly invariant and change only the variance of the
estimator measuring it. That route we can now exclude. A killer $O(\hat N-N)$ with
$[O,\hat N]=0$ annihilates a reference of definite particle number, so it leaves
$\mathrm{Var}_\phi(H_P)$ exactly invariant, and under a fermionic number-conserving
fragmentation it leaves every fragment variance invariant as well. In the qubit picture, where
the measurable fragments do break particle number, it is not inert, but it is small. Optimising a
one-body killer family over a qubit-wise commuting grouping of N$_2$/STO-3G moves
$\sum_g\sqrt{\mathrm{Var}_\phi(H_g)}$ by between $0.35$ and $3.3$ per cent across nine
references, hence the shot count by at most a factor of $0.94$. A decomposition effect of that
size cannot supply a factor of $87$.

What does track the spread is the energy variance of the reference itself,
\begin{equation}
\sigma^2_\phi \;=\; \mathrm{Var}_\phi(H_P)\;=\;\sum_n p_n\,(E_n-\bar E)^2 ,
\label{eq:sigma2}
\end{equation}
and this is the point of the section. The quantity is not a new one. It is the same
$\sigma^2=\braket{H_P^2}-\braket{H_P}^2$ that \S\ref{sec:stopping} already uses as a stopping
certificate, and that \S\ref{sec:sqd} evaluates on the classical vector to form a two-sided
bracket. What is new is the role. There it is an instrument, read to decide whether to stop.
Here it is a resource. The two moments that certify an eigenstate are the same two that price
the acquisition. Equation~\eqref{eq:sigma2} is a functional of the currency,
because it is built from the weights $\{p_n\}$ \emph{and} the eigenvalues $\{E_n\}$. It is not a
functional of the weights alone. Both $p_0$ and $r_{\rm eff}$ discard the eigenvalues, and so
does every other summary of $\{p_n\}$ on its own. Two references carrying identical weights over
differently spaced spectra are indistinguishable to $r_{\rm eff}$ and distinguishable to
$\sigma^2_\phi$. On the pair above, $r_{\rm eff}$ would need an exponent of $27.6$ to account for
the factor of $87$, where $\sigma^2_\phi$ needs between $1.24$ and $1.88$. To leading order
$\sigma^2_\phi\simeq\Delta E\,\bar\delta$, with $\Delta E=\bar E-E_0$ the variational error and
$\bar\delta$ a weighted mean excitation energy that varies by a factor of $1.5$ where
$\sigma^2_\phi$ varies by a factor of $10^2$. The quantity doing the work is therefore the
variational error, which is exactly the $79\,$mHa that \S\ref{sec:sqd} already reports.

We state the limits, because on eight reported cost points they matter more than the result.
Under leave-one-out cross-validation $\sigma^2_\phi$ predicts about half the out-of-sample
variance in the logarithm of the reported cost, with a median held-out miss of a factor of six,
and every model built from the weights alone falls to or below zero on the same test. The fitted
exponent is near $1.7$ rather than the value $1$ that estimating $\langle H_P\rangle$ to fixed
precision would give, and it differs between same-geometry and cross-geometry comparisons. All
eight points are one molecule. So the correct statement is not that the currency was incomplete.
It is that we had taken the wrong functional of it, that the energy weighting of the residual
spectrum is the part $r_{\rm eff}$ throws away, and that recovering it accounts for much of what
the two state coordinates could not.

\subsection{Classical versus quantum.}
\label{sec:classvq}
We stress that every pedagogical result here was produced on a laptop by exact diagonalisation and matrix
exponentiation. ESPRIT applied to a classically computed $g(t)$ is only reading
eigenvalues off a matrix one already has. The methods become genuinely quantum
algorithms only when $g(t)$ is generated by a device propagating a Hamiltonian too large
to store ($2^n$ amplitudes), and they become genuinely \emph{useful} only in the regime
where the best classical heuristics (coupled cluster, DMRG, selected configuration
interaction, quantum Monte Carlo) fail, i.e.\ strong correlation. That regime is also
where the two preconditions of spectral readout, a resolvable model order and a
non-vanishing ground overlap, are hardest to satisfy. So the approach relocates rather
than dissolves the field's central difficulty.

\addtocontents{toc}{\protect\setcounter{tocdepth}{2}}
\section{Conclusion}
\label{sec:conclusion}

This work has identified the resource a warm start carries, shown how to measure it when it
is too small to see directly, and spent it on four back-ends and a set of results that a
single-number description of state quality does not reach. The central object is not the identity of the front-end, nor a
single scalar measure of state quality, but the spectral-weight distribution
$
    p_n=|\langle E_n|\chi\rangle|^2.
$
The ground-state weight $p_0$ determines the leading overlap dependence of
the preparation procedures considered here, while the remaining spectral
weight determines how difficult it is to resolve the spectrum. For this
reason the useful resource description is the distribution itself, with
$(p_0,\mathcal C_{\rm spec})$ providing a compact cost-oriented summary when
appropriate. Equal ground-state overlap does not imply equal spectral-readout
cost, and a successful spectral readout does not by itself manufacture the
overlap needed for state preparation.

Where that pair is not enough, \S\ref{sec:notpriced} locates what it leaves out. Against the
sampling costs reported for a subspace method, neither $p_0$ nor $r_{\rm eff}$ survives
cross-validation, while the energy variance
$\sigma^2_\phi=\sum_np_n(E_n-\bar E)^2$ does. That quantity is a functional of the weights
\emph{and} the eigenvalues. The
distribution remains the currency. What changes is which functional of it one should read, and
the readout returns the two moments it needs in any case.

The readout results provide the most direct demonstration of this
architecture. Sampling $g(t)=\langle\chi|e^{-iH_Pt}|\chi\rangle$ and estimating
its frequencies parametrically with ESPRIT gives, in the tested simulations,
chemical accuracy from a signal roughly $50\times$ shorter than the
corresponding Fourier acquisition. The reduction is accompanied by a
shorter coherent circuit, at the price of model-order selection and the
conditions required for parametric spectral estimation. The study of the
two Hadamard-test quadratures also exposed a practical failure mode. When
the quadratures are acquired in separate passes, differential imbalance can
introduce conjugate spectral components and corrupt the recovered energies.
Interleaving the quadratures removes this mirror-line mechanism in the tested
protocol without requiring an additional ancilla. ESPRIT itself is an existing classical tool. What is new is the acquisition it is given, the
defect that acquisition was found to carry, and the fact that the same measurement returns the
weights that every other back-end in the paper needs.

The same supplied state can then be used on the preparation side. The
warm-start-anchored adiabatic construction converts the overlap with the
target ground state into a corresponding gap law, while discrete amplitude
amplification uses the overlap to determine its stopping index. Recursive
double-bracket imaginary-time evolution provides a third route whose
operability is itself overlap- and metering-dependent. These back-ends have
different coherent-depth and measurement requirements, so WARP should not
be read as declaring one route universally preferable. They provide
different ways of spending the same supplied resource. The numerical
examples show how partial preparation can increase usable ground-state
weight and hand the state back to spectral readout, and how readout can in
turn meter the weight produced by preparation.

Two constructions carry that economy. On the preparation side one object is spent three ways: $\ket\chi$ is the
state the readout reads, the anchor of the driver $\mathbf 1-\ket\chi\bra\chi$, and the
reflection $\mathbf 1-2\ket\chi\bra\chi$ the amplification iterates. It is the same state
each time, so a front-end that improves it improves all three at once, and the closed-form gap
law is what prices that improvement. On the acquisition side one interferometer returns
everything: the two Hadamard-test settings give the overlap, the gap, the mean excitation, the
energy and its variance, and the line positions and weights, from a single record. The
architecture is not a list of four back-ends that happen to share a notation. It is one resource
and one measurement, and the economy follows from there being only one of each.

That the circuits are standard is what makes the analysis of them worth reporting. Because the
two quadrature settings are separated in time, their interferometric contrasts need not agree,
and the reconstruction is then $g(t)$ plus a conjugate copy of itself rather than a distorted $g(t)$.
A transform absorbs that harmlessly. An estimator holding a fixed line count cannot, and it
fails abruptly and without a signature. We know of no statement of this in the quantum
eigenvalue-estimation literature, and the remedy costs a shot order.

This exchange is particularly clear when the variational energy is not the
quantity of interest. In the stretched-N$_2$ sample-based calculation,
the variational estimate remains about $79$~mHa above the ground-state
energy  (that is, 50 times above the chemical accuracy); whereas spectral readout of the already prepared state recovers
$E_0$ to approximately $1.4\times10^{-6}$~Ha (three orders of magnitude below the chemical accuracy). The molecular examples also
show the same mechanism beyond the smallest two-qubit demonstration,
including the $0.033$~Ha improvement over the variational bound reported for
the SO$_2$ case. These comparisons do not establish a universal quantum
advantage over classical methods. They establish that a prepared quantum
state can contain eigenvalue information that is not represented by its
variational expectation value.

The front-end studies test whether this interface is tied to a particular
state-preparation method. It is not. Variational states, QAOA states,
classical heuristics, lattice-reduction constructions, generative models,
sample-based diagonalisation, and machine-learned states can all supply the
same type of warm-start object. The sample-based diagonalisation example is
especially illustrative because the quantum state and a classical vector
are already available from the same preparation procedure; spectral readout
can interrogate the state without requiring a new preparation front-end.
The lattice and other examples also show that the target need not always be
the ground state. At the same time, interchangeability must not be confused
with equal cost. Each front-end has its own overlap, preparation,
compilation, sampling, and classical-processing costs, and WARP relocates
rather than removes those costs.

The metering results strengthen this resource interpretation. When the
wanted component is too weak to be reliably measured in the warm start's
own acquisition, interference with a stronger reference can recover the
small overlap linearly in the weak amplitude, extending the measurable range
in the tested examples. The same information can determine whether an
amplification route is operable and can provide a stopping or admission
criterion for a preparation procedure. The architecture therefore has a
closed-loop character: preparation changes the spectral weights, readout
measures them, and the measured information can determine the next
back-end operation.

Several limitations delimit the claim. Super-resolution improves the
accuracy with which a spectral line is located but does not increase the
number of physically resolvable lines. Parametric readout requires an
appropriate model order and sufficient spectral coverage. The recursive
and amplification procedures remain conditional on useful overlap, and
WA-ASP requires an implementable preparation of the projector state  $|\chi\rangle\langle\chi|$. The basis study
is a statement about the specified resource pricing, not a measured
end-to-end runtime advantage. More generally, the simulations include the
stated Trotter and shot-noise effects but do not constitute a full hardware
demonstration with decoherence, gate errors, and readout errors. These
limitations define the regime in which the present resource comparisons
should be interpreted.

The resulting picture is deliberately architectural rather than a claim to
replace the individual methods used within it. WARP provides a common
interface for combining VQE, QAOA, adiabatic preparation, amplitude
amplification, recursive refinement, ESPRIT, and sample-based diagonalisation:
a front-end prepares a state with a measurable spectral-weight distribution,
and a back-end uses that distribution according to the task it is asked to
perform. The practical question consequently becomes how to divide the
available budget among state preparation, coherent evolution, measurement,
and classical post-processing for a particular Hamiltonian and hardware
platform. The present results show that this question can be made
quantitative, and that the same warm-start resource can support both
eigenvalue readout and state preparation.

One use of the architecture lies outside the tasks demonstrated here. Because the weights can
be metered before they are spent, candidate warm starts can be screened on the quantity the
chosen back-end will actually consume rather than on a variational energy. The loop is to
generate candidates, meter the relevant spectral weight, select on the downstream cost, and only
then commit to the expensive preparation or refinement. Interference metering makes that
practical for states that would otherwise appear dark, since it recovers overlaps below the
direct-detection floor. We state this as an architectural proposal and not as a demonstrated
screening advantage, because its value depends on the cost of generating candidates and on what
the meter itself consumes. The same reasoning marks the boundary of the spectral side. For a
diagonal cost operator the readout is unnecessary, as direct sampling in the computational basis
already returns both the objective and the configuration, and the spectral component earns its
place only where eigenvalue information is not available from a single basis measurement.

The natural next step is experimental validation. The present simulations
identify relevant resource scalings and failure modes. Hardware studies must
determine how those scalings change under decoherence, imperfect controlled
evolution, measurement error, and the compilation overhead of implementing
the warm-start projector or controlled time evolution. They must also
determine where a reduction in coherent acquisition length translates into
a meaningful end-to-end benefit after preparation and measurement costs are
included. 

\section*{Acknowledgements}
We thank Peter Drummond and 
Sebastian Romero for constructive feedback.

\section*{Declaration}
\subsection*{Use of AI tools}
Generative AI was used the preparation of this work.
As a numerical collaborator, it wrote and executed some simulation code. As an
adversarial reader, it checked derivations and stated claims against numerical
counterexamples. As an editor, it drafted and revised prose. All scientific claims, the choice of
methods, and the interpretation of every result are the author's own, and all numerical
results reported have been verified by the author. The author accepts full
responsibility for the content.

\subsection*{Patents}
Provisional patent applications covering the warm-start-anchored construction  described in this work have been filed (numbers 2026906721 and 2026907713, Australia, 2026). The author declares no other competing interests.

\warppart{Appendices}{}
\label{part:appendices}

\appendix

\section[Shot noise, SNR and the Cram\'er--Rao bound]{Shot noise, signal-to-noise ratio, and the Cram\'er--Rao bound}
\label{app:shot}

\paragraph{Hadamard-test estimator.}
A single shot of the real-part circuit (Fig.~\ref{fig:hadamard}) returns an
ancilla bit $b\in\{0,1\}$ with $\Pr(b{=}0)=q=\tfrac12\big(1+\mathrm{Re}\,g\big)$, where $q$ is
written for the outcome probability so as not to collide with the ground-state weight $p_0$.
Over $N_s$ independent shots, $n_0\sim\mathrm{Binomial}(N_s,q)$ and the estimator
$\widehat{\mathrm{Re}}=2n_0/N_s-1$ is unbiased, $\mathbf E[\widehat{\mathrm{Re}}]=\mathrm{Re}\,g$, with
\begin{equation}
\operatorname{Var}\!\big(\widehat{\mathrm{Re}}\big)
=\frac{4q(1-q)}{N_s}=\frac{1-(\mathrm{Re}\,g)^2}{N_s}.
\end{equation}
The imaginary part (with the $S^{\dagger}$ gate) obeys the analogous relation, so
the complex estimate $\hat g=\widehat{\mathrm{Re}}+i\,\widehat{\mathrm{Im}}$ is
unbiased with per-sample variance
\begin{equation}
\sigma_g^2\;:=\;\operatorname{Var}(\hat g)=\frac{2-|g|^2}{N_s}\in\Big[\tfrac1{N_s},\tfrac2{N_s}\Big].
\end{equation}
Distinct sample times use separate circuits, so the noise is independent across
$k$ (white) and Gaussian for large $N_s$.

\paragraph{Signal-to-noise ratio.}
The ground mode contributes amplitude $A=w_0=|c_0|^2$ (the squared overlap) to
$g_k$. Its per-sample power SNR is
\begin{equation}
\varrho=\frac{A^2}{\sigma_g^2}=\frac{|c_0|^4\,N_s}{2-|g|^2}\;\approx\;\tfrac12\,|c_0|^4\,N_s,
\end{equation}
so holding $\varrho$ fixed costs $N_s\propto|c_0|^{-4}$ shots, the steep inverse-overlap
penalty that a VQE warm start mitigates by maximizing $|c_0|^2$.

\paragraph{Cram\'er--Rao bound on the energy.}
For a dominant, well-separated tone of amplitude $A$ in white noise $\sigma_g^2$ over
$N$ samples, the bound on the mode's angular frequency $\omega=E\delta$
is~\citep{stoica1989music}
\begin{equation}
\operatorname{Var}(\hat\omega)\;\ge\;\frac{6\,\sigma_g^2}{A^2\,N(N^2-1)} .
\end{equation}
With $E=-\omega/\delta$ and $A=|c_0|^2$, for large $N$,
\begin{equation}
\operatorname{Std}(\hat E)\;\gtrsim\;\frac{\sqrt6\,\sigma_g}{|c_0|^2\,N^{3/2}\,\delta}
\;\xrightarrow{\;\sigma_g^2=2/N_s\;}\;\frac{2\sqrt3}{|c_0|^2\,\sqrt{N_s}\,N^{3/2}\,\delta}.
\label{eq:crbstd}
\end{equation}
Equation~\eqref{eq:crbstd} makes the central trade-offs quantitative.
\emph{Depth beats shots:} at fixed $N_s$, $\operatorname{Std}(\hat E)\propto N^{-3/2}$,
but at fixed $N$ it falls only as $N_s^{-1/2}$; doubling $N$ should shrink the spread
by $2^{3/2}\!\approx\!2.8$ and quadrupling $N_s$ by $2$, matching the $\sim\!3\times$
and $\sim\!2\times$ observed in \S\ref{sec:modelorder}. \emph{Overlap:}
$\operatorname{Std}(\hat E)\propto|c_0|^{-2}$ at fixed budget, equivalently the
shot cost to hold a target precision scales as $|c_0|^{-4}$.

Writing the total budget $\mathcal M=2NN_s$ (real and imaginary parts at all times)
and $\tau=N\delta$ for the acquisition length, Eq.~\eqref{eq:crbstd} becomes
\begin{equation}
\operatorname{Std}(\hat E)\;\gtrsim\;\frac{2\sqrt6}{|c_0|^2\,\tau\,\sqrt{\mathcal M}} .
\label{eq:crbbudget}
\end{equation}
Precision improves \emph{linearly} in the coherent evolution time $\tau$ but only as
$\sqrt{\mathcal M}$ in the total measurement budget. Thus, within this single-tone
coherent-acquisition model, increasing the record length can be a more efficient way
of improving frequency precision than increasing the number of shots. This scaling
should not be interpreted as a claim of quantum metrological advantage over the
standard quantum limit; it is the resource trade specific to the time-series
estimation protocol considered here. When $\tau$ is capped by hardware coherence,
additional precision must instead be purchased through the measurement budget.

\paragraph{Systematic floor and the optimal step.} The CRB is a statistical lower bound
and adds in quadrature to the systematic Trotter bias $b(\delta)\sim\kappa\,n\,\delta^2$
(\S\ref{sec:esprit}), $\mathrm{MSE}(\hat E)\approx b(\delta)^2+\operatorname{Var}(\hat
E)$. At fixed $N,N_s$ the two terms pull oppositely in $\delta$ (bias grows as $\delta^2$,
variance falls as $\delta^{-2}$), minimised at $\delta^\star\propto N^{-1/2}$. Finally,
Eqs.~\eqref{eq:crbstd}--\eqref{eq:crbbudget} assume a single dominant, well-separated
tone; in the dense-spectrum regime of \S\ref{sec:scaling} the ``frequency forest''
lowers the effective single-tone SNR and the bound becomes optimistic. This is the
quantitative origin of ESPRIT's shrinking advantage at larger $n$.

\section{ESPRIT: signal model and estimator}
\label{app:esprit}

\paragraph{Signal model.}
Sampling the autocorrelation function of Eq.~\eqref{eq:gt} at $t_k=k\delta$ gives
a finite sum of complex exponentials,
\begin{equation}
g_k=\sum_{m=1}^{p} w_m\,z_m^{\,k},\qquad k=0,\dots,N-1,
\label{eq:expmodel}
\end{equation}
with \emph{modes} $z_m=\ee^{-iE_m\delta}$ lying on the unit circle ($|z_m|=1$),
positive weights $w_m=|c_m|^2$, and $p$ the number of \emph{distinct} populated
eigenvalues. Degenerate levels share a mode and merge into a single exponential
(\S\ref{sec:modelorder}). ESPRIT recovers $\{z_m\}$, and so
$E_m=-\arg(z_m)/\delta$, from the samples $g_k$.

\paragraph{Hankel structure.}
Choose window lengths $L,M$ with $L+M=N+1$ and $L,M>p$, and form the
$L\times M$ Hankel matrix $\mathbf H$ with $[\mathbf H]_{rc}=g_{r+c}$
($0\le r<L,\ 0\le c<M$). Substituting Eq.~\eqref{eq:expmodel},
\begin{equation}
\begin{gathered}
\mathbf H=\mathbf V_L\,\mathbf W\,\mathbf V_M^{\!\top},\\
[\mathbf V_m]_{rm}=z_m^{\,r},\qquad
\mathbf W=\mathrm{diag}(w_1,\dots,w_p),
\end{gathered}
\end{equation}
a product of two Vandermonde matrices and a diagonal weight matrix, since
$\sum_m z_m^{r}\,w_m\,z_m^{c}=\sum_m w_m z_m^{r+c}=g_{r+c}$. As the $z_m$ are
distinct and $L,M>p$, both Vandermonde factors have full column rank $p$, so
$\operatorname{rank}\mathbf H=p$ exactly in the noiseless case.

\paragraph{Rotational invariance.}
Write $\mathbf V_L^{\uparrow}$ ($\mathbf V_L^{\downarrow}$) for $\mathbf V_L$ with
its last (first) row deleted. Advancing the row index by one multiplies column
$m$ by $z_m$, i.e.\ the shift property
$\mathbf V_L^{\downarrow}=\mathbf V_L^{\uparrow}\boldsymbol\Phi$ with
$\boldsymbol\Phi=\mathrm{diag}(z_1,\dots,z_p)$. Let
$\mathbf U_s\in\mathbf C^{L\times p}$ hold the $p$ leading left singular vectors
of $\mathbf H=\mathbf U\boldsymbol\Sigma\mathbf V^{\dagger}$. Then
$\operatorname{col}\mathbf U_s=\operatorname{col}\mathbf V_L$, so
$\mathbf U_s=\mathbf V_L\mathbf T$ for some invertible
$\mathbf T\in\mathbf C^{p\times p}$. Deleting rows,
$\mathbf U_s^{\uparrow}=\mathbf V_L^{\uparrow}\mathbf T$ and
$\mathbf U_s^{\downarrow}=\mathbf V_L^{\downarrow}\mathbf T
=\mathbf V_L^{\uparrow}\boldsymbol\Phi\,\mathbf T$, whence
\begin{equation}
\mathbf U_s^{\downarrow}=\mathbf U_s^{\uparrow}\,\boldsymbol\Psi,\qquad
\boldsymbol\Psi=\mathbf T^{-1}\boldsymbol\Phi\,\mathbf T .
\end{equation}
$\boldsymbol\Psi$ is similar to $\boldsymbol\Phi$. Its eigenvalues are thus exactly the modes, $\operatorname{eig}(\boldsymbol\Psi)=\{z_m\}$. In the noiseless
case $\boldsymbol\Psi=(\mathbf U_s^{\uparrow})^{+}\mathbf U_s^{\downarrow}$
(least-squares ESPRIT). Total-least-squares ESPRIT balances perturbations in both
$\mathbf U_s^{\uparrow}$ and $\mathbf U_s^{\downarrow}$ and is preferable under
noise.

\paragraph{Weights, order, identifiability.} Given the modes, $E_m=-\arg(z_m)/\delta$
and the weights solve the Vandermonde least-squares problem $\min_{\mathbf
w}\lVert\mathbf g-\mathbf V\mathbf w\rVert_2$ with $[\mathbf V]_{km}=z_m^{k}$.
Noiselessly $\operatorname{rank}\mathbf H=p$; with shot noise the singular values split
into $p$ ``signal'' values above a noise floor and the remainder near it, so the order
$p$ is read from the spectral gap, or an information criterion such as MDL/AIC~\citep{wax1985detection}.
Identifiability needs $N\ge 2p+1$ and distinct modes. Genuine modes satisfy
$|z_m|\approx1$ and carry weight above the floor, while spurious roots of an
over-specified model drift off the unit circle with vanishing weight, the pruning rule
of \S\ref{sec:modelorder}. Prony's method is the exactly-determined limit ($N=2p$,
root-finding on a linear-prediction polynomial). The matrix-pencil method uses the
pencil $\mathbf H^{\downarrow}-\lambda\mathbf H^{\uparrow}$~\citep{hua1990matrix}; MUSIC
is the noise-subspace dual. ESPRIT's reliance on the SVD signal subspace makes it the
most noise-robust of these at the moderate orders used here.

\smallskip
\noindent\rule{\columnwidth}{0.4pt}\\[2pt]
{\footnotesize\textsc{Algorithm}~LS-ESPRIT$\big(\{g_k\}_{k=0}^{N-1},\,p,\,\delta\big)$}
\begin{enumerate}[leftmargin=1.4em,itemsep=0pt,topsep=2pt]\footnotesize
\item Form the Hankel matrix $[\mathbf H]_{rc}=g_{r+c}$, $L=\lfloor N/2\rfloor$, $M=N-L+1$.
\item Compute the SVD $\mathbf H=\mathbf U\boldsymbol\Sigma\mathbf V^{\dagger}$; set $\mathbf U_s=\mathbf U[:,1{:}p]$.
\item Form $\boldsymbol\Psi=(\mathbf U_s^{\uparrow})^{+}\mathbf U_s^{\downarrow}$ and diagonalise: $\{z_m\}=\operatorname{eig}\boldsymbol\Psi$.
\item Energies $E_m=-\arg(z_m)/\delta$; weights from $\min_{\mathbf w}\lVert\mathbf g-\mathbf V\mathbf w\rVert_2$.
\item Keep modes with $|z_m|\approx1$ and weight above the noise floor. Return the dominant- (warm start) or lowest- (sparse spectrum) energy survivor.
\end{enumerate}
\noindent\rule{\columnwidth}{0.4pt}

\section{The adiabatic theorem and the overlap--gap law}
\label{app:adiabatic}
For $H(s)$ with $s=t/T$, evolving from the instantaneous ground state at $s=0$, the
leading diabatic error obeys
\begin{equation}
\varepsilon\;\lesssim\;\frac1T\,\max_s\frac{|\bra{1(s)}\partial_s H(s)\ket{0(s)}|}{\Delta(s)^2},
\end{equation}
so a fidelity target needs $T\gtrsim1/\Delta_{\min}^2$ with
$\Delta_{\min}=\min_s\Delta(s)$; a local-adiabatic schedule $\mathrm
ds/\mathrm dt\propto\Delta(s)^2$ concentrates the evolution at the bottleneck and
improves the constant. Restricting the projector-driven interpolation
$H(s)=(1-s)(\mathbf 1-\ket{\chi}\!\bra{\chi})+s\widetilde H_P$ to
$\operatorname{span}\{\ket{\chi},\ket{E_0}\}$ with overlap $a=|\!\braket{E_0|\chi}\!|$ and
diagonalising the resulting $2\times2$ can be done in closed form. Writing
$b=\sqrt{1-a^2}$ and $\bar\mu=\bra{\chi_\perp}\widetilde H_P\ket{\chi_\perp}$ for the mean
excitation energy of the leaked component, and $\eta=1-2a^2+\bar\mu$, the minimum gap and its
location are
\begin{equation}
\Delta_{\min}=\frac{2ab\,\bar\mu}{\sqrt{\eta^2+4a^2b^2}},
\qquad
s^\star=\min\Big\{1,\;1-\frac{\eta\bar\mu}{\eta^2+4a^2b^2}\Big\}.
\label{eq:gaplaw}
\end{equation}
The clamp is not cosmetic. The stationary point of the two-level gap is interior only while
$\eta\ge0$, that is while $2a^2\le1+\bar\mu$; above that overlap the unclamped expression
returns $s^\star>1$, which lies off the path. At $\bar\mu=0.5$ the crossover sits at $a^2=0.75$,
and beyond it the bottleneck is the endpoint $s=1$, where $\Delta=\bar\mu$. Fig.~\ref{fig:sched}(a)
shows the three regimes.
For small $a$ this reduces to $\Delta_{\min}\simeq2a\bar\mu/(1+\bar\mu)$, linear in the overlap
amplitude, attained at $s^\star\to1/(1+\bar\mu)$. The bottleneck sits at a fixed point of the
schedule set by the ratio of problem to driver scale, and does \emph{not} migrate toward
$s=1$ as $a$ falls. The convenient approximation $\Delta_{\min}\approx2(1-s^\star)a\,b$
used with a measured $s^\star$ in \S\ref{sec:gaplaw} follows from
Eq.~\eqref{eq:gaplaw} on dropping a factor $\sqrt{1+4a^2b^2/\eta^2}$; it is accurate to
$2\%$ at $a^2\le0.04$ and degrades as $a$ grows, which is the origin of the
measured-to-predicted ratio crossing unity in Fig.~\ref{fig:mcgap}. The two-level
reduction itself ceases to describe the full problem once $a=\mathcal O(1)$, where the
gap saturates at the intrinsic gap of $\widetilde H_P$. Hence $T\sim1/\Delta_{\min}^2\sim1/a^2$ at small overlap and
$T\sim1/\Delta_{\rm intrinsic}^2$ (overlap-independent) once well warm-started, the
saturation of \S\ref{sec:aqc}. The QML scaling of \S\ref{sec:qml} measures a
time-to-target growing about as $1/a$, looser than the worst-case $1/a^2$ because the
tolerance is met before full convergence. Either power, composed with the
barren-plateau decay $a\sim2^{-n}$, gives an exponential rescue cost for a true
plateau.

\section{A certified lower bound on the ground energy}
\label{app:certified}
The variance certificate of \S\ref{sec:stopping} becomes a two-sided, gap-free bracket
on $E_0$ once the ground fidelity is known. For a prepared state $\ket\psi$ write
\begin{equation}
\bar E=\braket{\psi|H_P|\psi},\quad
\sigma^2=\braket{\psi|H_P^2|\psi}-\bar E^2,\quad
f=|\!\braket{E_0|\psi}\!|^2,
\end{equation}
with $\bar E\ge E_0$ (variational principle) and $f$ the ground-state overlap
\emph{probability}. Decompose $\ket\psi=\sqrt f\,\ket{E_0}+\sqrt{1-f}\,\ket{\psi_\perp}$
with $\braket{E_0|\psi_\perp}=0$, and set
$\bar E_\perp=\braket{\psi_\perp|H_P|\psi_\perp}$. Because $(H_P-E_0)\ket{E_0}=0$ the cross
terms vanish, giving
\begin{align}
\bar E-E_0&=(1-f)\,(\bar E_\perp-E_0),\label{eq:B-energy}\\
\sigma^2&=f\,(\bar E-E_0)^2+(1-f)\,\sigma_\perp^2,\label{eq:B-var}\\
\sigma_\perp^2&=\braket{\psi_\perp|(H_P-\bar E_\perp)^2|\psi_\perp}.\nonumber
\end{align}
The parallel-axis identity gives
$\sigma_\perp^2=\operatorname{Var}_\perp(H_P)+(\bar E_\perp-\bar E)^2\ge(\bar E_\perp-\bar E)^2$,
while $\bar E=fE_0+(1-f)\bar E_\perp$ with \eqref{eq:B-energy} gives
$\bar E_\perp-\bar E=\tfrac{f}{1-f}(\bar E-E_0)$. Substituting into \eqref{eq:B-var},
\begin{equation}
\sigma^2\;\ge\;f(\bar E-E_0)^2+\frac{f^2}{1-f}(\bar E-E_0)^2
=\frac{f}{1-f}\,(\bar E-E_0)^2,
\end{equation}
and, since $\bar E\ge E_0$,
\begin{equation}
\bar E-\sigma\sqrt{\tfrac{1-f}{f}}\;\le\;E_0\;\le\;\bar E.
\label{eq:B-bracket}
\end{equation}
Equality in the lower bound holds iff $\operatorname{Var}_\perp(H_P)=0$, i.e.\ the
excited part is a single eigenstate. The bound is thus exact for any two-level
(ground plus one excited) mixture, and is strictly tighter than the cruder
$E_0\ge\bar E-\sigma/\sqrt f$ because $\sqrt{(1-f)/f}<1/\sqrt f$.

The bracket \eqref{eq:B-bracket} needs three inputs, all supplied by the method: $\bar E$
and $\sigma$ from the $\braket{H_P}$ and $\braket{H_P^2}$ measurements of this appendix,
and the fidelity $f$ from the ESPRIT ground-line weight $|c_0|^2$ (\S\ref{sec:hybrid}).
It collapses as $\sigma\to0$ (an eigenstate) or as $f\to1$ (where $\sqrt{(1-f)/f}\to0$
and $E_0\to\bar E$ from below). Because $\sqrt{(1-f)/f}$ is decreasing in $f$, any rigorous
lower bound $f\ge f_{\min}$, for instance a conservatively error-barred ESPRIT weight,
gives a valid certificate $E_0\ge\bar E-\sigma\sqrt{(1-f_{\min})/f_{\min}}$. Absent any handle on $f$ the bound is uninformative ($f\to0\Rightarrow-\infty$), which states precisely that certification requires the overlap the front-end is there to supply.

\small
\bibliographystyle{unsrt}
\bibliography{refs}

\end{document}

%% file: figcircle.tex
\begin{figure}[t]\centering
\begin{tikzpicture}[scale=0.62,font=\scriptsize\sffamily,
  lead/.style={gray,thin,-},]
\draw[thick,-Latex] (-4.2,0) -- (4.2,0) node[right,font=\tiny\sffamily]{$\mathrm{Re}\,z$};
\draw[thick,-Latex] (0,-4.2) -- (0,4.2) node[above,font=\tiny\sffamily]{$\mathrm{Im}\,z$};
\begin{scope}
  \clip (0,0) circle (3.42);
  \fill[black,opacity=0.08] (0,0) circle (3.42);
  \fill[white] (0,0) circle (2.72);
\end{scope}
\draw[dashed] (0,0) circle (2.72);
\draw[dashed] (0,0) circle (3.42);
\draw[thick] (0,0) circle (3.2);
\foreach \ang/\r in {52/2.90, 96/3.02, 140/2.86, 214/3.08, 318/2.94}
  {\fill ({\r*cos(\ang)},{\r*sin(\ang)}) circle (2.6pt);}
\draw[thick] (1.28,-1.28) circle (3.2pt);
\draw[thick] (1.16,-1.40) -- (1.40,-1.16);
\draw[thick] (1.16,-1.16) -- (1.40,-1.40);
\draw[-Latex,thick] (0,0) -- ({2.90*cos(52)},{2.90*sin(52)});
\node[anchor=west,font=\tiny\sffamily,align=left] (l1) at (3.7,3.0) {unit circle $|z|=1$};
\draw[lead] (l1.west) -- ({3.2*cos(38)},{3.2*sin(38)});
\node[anchor=west,font=\tiny\sffamily,align=left] (l2) at (3.7,1.4) {genuine\\ damped roots};
\draw[lead] (l2.west) -- ({2.94*cos(-42)},{2.94*sin(-42)});
\node[anchor=west,font=\tiny\sffamily,align=left] (l3) at (3.7,-1.5) {spurious root,\\ rejected};
\draw[lead] (l3.west) -- (1.55,-1.28);
\node[anchor=east,font=\tiny\sffamily,align=right] (l4) at (-3.7,-2.6) {confidence region from\\ calibration or bootstrap};
\draw[lead] (l4.east) -- ({3.07*cos(232)},{3.07*sin(232)});
\node[anchor=east,font=\tiny\sffamily,align=right] (l5) at (-3.7,2.7) {$|z_n|=\ee^{-\gamma_n\delta}$};
\draw[lead] (l5.east) -- ({1.55*cos(52)},{1.55*sin(52)});
\end{tikzpicture}
\caption{\textbf{Noise-aware admissibility.} Under the damped model of
Eq.~\eqref{eq:dampedmodel} the poles lie strictly inside the unit circle by construction, so
the hard test $|z_n|\approx1$ would reject every genuine root. It is replaced by a confidence
region calibrated from the measured noise or from bootstrap resampling, shown shaded. A root
inside the circle but within that region is retained, and the radius of a retained root is returned as the decay rate $\gamma_n$. A root outside the region
and carrying vanishing weight is rejected as an artefact of an over-specified model order.}
\label{fig:admiss}
\end{figure}

%% file: figarch.tex
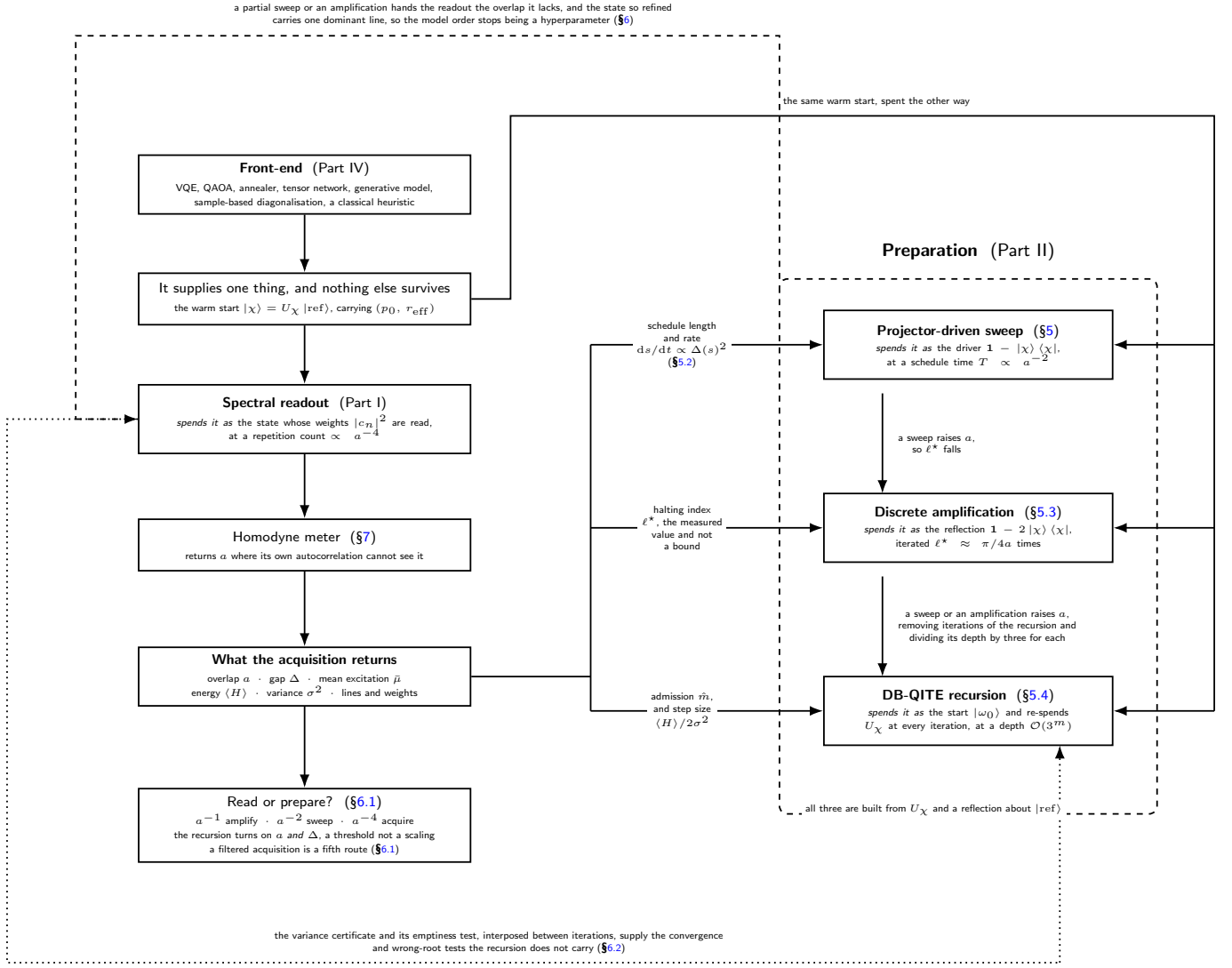
\begin{figure*}[!t]\centering
\resizebox{\textwidth}{!}{%
\begin{tikzpicture}[font=\scriptsize\sffamily,
  box/.style={draw,thick,align=center,inner sep=3.5pt,minimum height=9mm},
  be/.style={draw,thick,align=center,inner sep=3.5pt,minimum height=12mm,text width=48mm},
  lab/.style={font=\tiny\sffamily,align=center,fill=white,inner sep=2pt},
  ar/.style={-Latex,thick}]

\node[box,minimum width=58mm] (fe) at (0,5.0)
  {\textbf{Front-end} \ (Part~IV)\\[1pt]
   \tiny VQE, QAOA, annealer, tensor network, generative model,\\[-1pt]
   \tiny sample-based diagonalisation, a classical heuristic};
\node[box,minimum width=58mm] (ws) at (0,3.0)
  {It supplies one thing, and nothing else survives\\[1pt]
   \tiny the warm start $\ket{\chi}=U_\chi\ket{\mathrm{ref}}$, carrying $(p_0,\,r_{\rm eff})$};
\draw[ar] (fe.south) -- (ws.north);

\node[be,minimum width=58mm] (rd) at (0,0.9)
  {\textbf{Spectral readout} \ (Part~I)\\[1pt]
   \tiny \emph{spends it as} the state whose weights $|c_n|^2$ are read,\\[-1pt]
   \tiny at a repetition count $\propto a^{-4}$};
\draw[ar] (ws.south) -- (rd.north);

\node[box,minimum width=58mm] (mt) at (0,-1.3)
  {Homodyne meter \ (\S\ref{sec:homodyne})\\[1pt]
   \tiny returns $a$ where its own autocorrelation cannot see it};
\draw[ar] (rd.south) -- (mt.north);

\node[box,minimum width=58mm] (hub) at (0,-3.6)
  {\textbf{What the acquisition returns}\\[1pt]
   \tiny overlap $a$ \ $\cdot$ \ gap $\Delta$ \ $\cdot$ \ mean excitation $\bar\mu$\\[-1pt]
   \tiny energy $\langle H\rangle$ \ $\cdot$ \ variance $\sigma^2$ \ $\cdot$ \ lines and weights};
\draw[ar] (mt.south) -- (hub.north);

\node[box,minimum width=58mm] (dec) at (0,-6.2)
  {Read or prepare? \ (\S\ref{sec:decision})\\[1pt]
   \tiny $a^{-1}$ amplify \ $\cdot$ \ $a^{-2}$ sweep \ $\cdot$ \ $a^{-4}$ acquire\\[-1pt]
   \tiny the recursion turns on $a$ \emph{and} $\Delta$, a threshold not a scaling\\[-1pt]
   \tiny a filtered acquisition is a fifth route (\S\ref{sec:filter})};
\draw[ar] (hub.south) -- (dec.north);

\node[be] (sw) at (11.6,2.2)
  {\textbf{Projector-driven sweep} \ (\S\ref{sec:aqc})\\[2pt]
   \tiny \emph{spends it as} the driver $\mathbf 1-\ket{\chi}\bra{\chi}$,\\[-1pt]
   \tiny at a schedule time $T\propto a^{-2}$};
\node[be] (am) at (11.6,-1.0)
  {\textbf{Discrete amplification} \ (\S\ref{sec:grover})\\[2pt]
   \tiny \emph{spends it as} the reflection $\mathbf 1-2\ket{\chi}\bra{\chi}$,\\[-1pt]
   \tiny iterated $\ell^\star\!\approx\!\pi/4a$ times};
\node[be] (db) at (11.6,-4.2)
  {\textbf{DB-QITE recursion} \ (\S\ref{sec:dbqite})\\[2pt]
   \tiny \emph{spends it as} the start $\ket{\omega_0}$ and re-spends\\[-1pt]
   \tiny $U_\chi$ at every iteration, at a depth $\mathcal O(3^{m})$};
\node[anchor=south,font=\small\sffamily] at (11.6,3.55) {\textbf{Preparation} \ (Part~II)};

\draw[thick] (ws.east) -- (3.6,3.0) -- (3.6,6.2) -- (15.9,6.2) -- (15.9,2.2);
\draw[ar] (15.9,2.2) -- (sw.east);
\draw[thick] (15.9,2.2) -- (15.9,-4.2);
\draw[ar] (15.9,-1.0) -- (am.east);
\draw[ar] (15.9,-4.2) -- (db.east);
\node[lab,anchor=south] at (10.0,6.30) {the same warm start, spent the other way};

\draw[thick] (hub.east) -- (5.0,-3.6);
\draw[thick] (5.0,2.2) -- (5.0,-4.2);
\draw[ar] (5.0,2.2) -- (sw.west);
\draw[ar] (5.0,-1.0) -- (am.west);
\draw[ar] (5.0,-4.2) -- (db.west);
\node[lab] at (6.6,2.2) {schedule length\\ and rate\\ $\d s/\d t\propto\Delta(s)^2$\\ (\S\ref{sec:schedule})};
\node[lab] at (6.6,-1.0) {halting index\\ $\ell^\star$, the measured\\ value and not\\ a bound};
\node[lab] at (6.6,-4.2) {admission $\hat m$,\\ and step size\\ $\langle H\rangle/2\sigma^2$};

\draw[thick,dashed] (8.3,3.35) -- (8.3,7.6) -- (-4.0,7.6) -- (-4.0,0.9);
\draw[ar,dashed] (-4.0,0.9) -- (rd.west);
\node[lab,anchor=south] at (2.6,7.70)
  {a partial sweep or an amplification hands the readout the overlap it lacks, and the state so
   refined\\ carries one dominant line, so the model order stops being a hyperparameter
   (\S\ref{sec:together})};

\draw[thick,dotted] (rd.west) -- (-5.2,0.9) -- (-5.2,-8.6) -- (13.2,-8.6);
\draw[ar,dotted] (13.2,-8.6) -- (13.2,-4.8);
\node[lab,anchor=south] at (3.4,-8.52)
  {the variance certificate and its emptiness test, interposed between iterations, supply the
   convergence\\ and wrong-root tests the recursion does not carry (\S\ref{sec:stopping})};

\draw[ar] (10.1,1.35) -- (10.1,-0.40);
\node[lab,anchor=west] at (10.30,0.48) {a sweep raises $a$,\\ so $\ell^\star$ falls};
\draw[ar] (10.1,-1.85) -- (10.1,-3.60);
\node[lab,anchor=west] at (10.30,-2.72)
  {a sweep or an amplification raises $a$,\\ removing iterations of the recursion and\\
   dividing its depth by three for each};

\draw[thick,dashed,rounded corners] (8.3,3.35) rectangle (14.9,-6.0);
\node[lab,anchor=north] at (11.0,-5.75)
  {all three are built from $U_\chi$ and a reflection about $\ket{\mathrm{ref}}$};
\end{tikzpicture}}
\caption{\textbf{One account, and everything that spends it.} The front-end supplies a single
object, the warm start, and the four routes differ only in how they spend it. What the readout
measures then configures the other three and decides between all four before any is run. The
traffic is not one-way: a partial sweep or an amplification hands the readout an overlap it could
not otherwise reach and a state of lower spectral complexity, the readout's certificates are
interposed between the iterations of the recursion, and the preparation routes hand over to one
another, a raised overlap removing recursion iterations whose depth grows as $3^m$.}
\label{fig:arch}
\end{figure*}